\documentclass[journal=jacsat,manuscript=article,a4]{achemso}

\usepackage{amsmath}
\usepackage[table]{xcolor}
\usepackage[version=4]{mhchem}
\usepackage{hyperref}
\hypersetup{
	colorlinks   = true, 
	urlcolor     = blue,
	linkcolor    = blue, 
	citecolor   = blue 
}
\newcommand{\etal}{\textit{et al.}}

\usepackage{xr}
\author{Benjamin X. Shi}
\affiliation{Yusuf Hamied Department of Chemistry, University of Cambridge, Lensfield Road, Cambridge CB2 1EW, United Kingdom}%
\alsoaffiliation{Initiative for Computational Catalysis, Flatiron Institute, 160 5th Avenue, New York, NY 10010}

\author{Andrew S. Rosen}
\affiliation{Department of Chemical and Biological Engineering, Princeton University, Princeton, NJ 08540 USA}

\author{Tobias Sch\"{a}fer}
\affiliation{Institute for Theoretical Physics, TU Wien, Wiedner Hauptstra{\ss}e 8-10/136, 1040 Vienna, Austria}

\author{Andreas Gr\"{u}neis}
\affiliation{Institute for Theoretical Physics, TU Wien, Wiedner Hauptstra{\ss}e 8-10/136, 1040 Vienna, Austria}

\author{Venkat Kapil}
\affiliation{Yusuf Hamied Department of Chemistry, University of Cambridge, Lensfield Road, Cambridge CB2 1EW, United Kingdom}
\alsoaffiliation{Department of Physics and Astronomy, University College London, 7-19 Gordon St, London WC1H 0AH, UK}
\alsoaffiliation{Thomas Young Centre and London Centre for Nanotechnology, 9 Gordon St, London WC1H 0AH}

\author{Andrea Zen}
\affiliation{Dipartimento di Fisica Ettore Pancini, Universit\`{a} di Napoli Federico II, Monte S. Angelo, I-80126 Napoli, Italy}
\alsoaffiliation{Department of Earth Sciences, University College London, Gower Street, London WC1E 6BT, United Kingdom}

\author{Angelos Michaelides}
\affiliation{Yusuf Hamied Department of Chemistry, University of Cambridge, Lensfield Road, Cambridge CB2 1EW, United Kingdom}%
\email{am452@cam.ac.uk}

\title[Accurate enthalpies]{An accurate and efficient framework for modelling the surface chemistry of ionic materials}

\abbreviations{IR,NMR,UV}
\keywords{American Chemical Society, \LaTeX}

\begin{document}
\clearpage

\begin{abstract}
Quantum-mechanical simulations can offer atomic-level insights into chemical processes on surfaces.
This understanding is crucial for the rational design of new solid catalysts as well as materials to store energy and mitigate greenhouse gases.
However, achieving the accuracy needed for reliable predictions has proven challenging.
Density functional theory (DFT), the workhorse quantum-mechanical method, can often lead to inconsistent predictions, necessitating accurate methods from correlated wave-function theory (cWFT).
However, the high computational demands and significant user intervention associated with cWFT have traditionally made it impractical to carry out for surfaces.
In this work, we address this challenge, presenting an automated framework which leverages multilevel embedding approaches, to apply accurate cWFT methods to the surfaces of ionic materials with computational costs approaching DFT.
With this framework, we have reproduced experimental adsorption enthalpies for a diverse set of 19 adsorbate--surface systems.
Moreover, we resolve debates on the adsorption configuration of several systems, while offering  benchmarks to assess DFT.
This framework is  open-source, making it possible to more routinely apply cWFT to complex problems involving the surfaces of ionic materials.
\end{abstract}

\section{Introduction}

Understanding the chemical processes occurring on surfaces is critical to applications ranging from the production of fuels via heterogeneous catalysis~\cite{greeleyComputationalHighthroughputScreening2006,norskovComputationalDesignSolid2009b} to the storage of energy~\cite{jainComputationalPredictionsEnergy2016} and sequestration of greenhouse gases.
The adsorption and desorption of molecules from surfaces is a crucial process within all of these applications, and the strength of this binding is dictated by the adsorption enthalpy $H_\textrm{ads}$, making it a fundamental quantity to accurately predict~\cite{pablo-garciaFastEvaluationAdsorption2023,gaoDeterminingAdsorptionEnergies2020a,andersenAdsorptionEnthalpiesCatalysis2021,lanAdsorbMLLeapEfficiency2023}.
For example, candidate materials for \ce{CO2} or \ce{H2} gas storage are screened based upon their $H_\textrm{ads}$ value, often to within tight energetic windows~\cite{patelCarbonDioxideCapture2017} (${\sim}150\,$meV).
High accuracy on $H_\textrm{ads}$ is also needed to compare the competitive adsorption between two molecular species for the separation of flue gases~\cite{rosenTuningRedoxActivity2020}.
Finally, $H_\textrm{ads}$ is a necessary quantity within any (microkinetic) model of a surface chemical reaction, with an empirical dependence between the reaction rate and $H_\textrm{ads}$ according to the well-established volcano plots~\cite{bligaardBronstedEvansPolanyi2004a,falsigTrendsCatalyticCO2008,chengUtilizationThreeDimensionalVolcano2008,montoyaMaterialsSolarFuels2017}.

The rational design of new materials for the above applications relies on an atomic-level understanding of surface processes, together with an accurate $H_\textrm{ads}$.
Determining the adsorption configuration --- the geometry a molecule adopts on a surface --- is particularly important, as it underpins all subsequent processes~\cite{ertlElementaryStepsHeterogeneous1990}.
Quantum-mechanical simulation techniques can provide the atomic-level detail needed to study the adsorption configuration.
They have become widely used to complement experimental techniques~\cite{hammes-schifferIntegrationTheoryExperiment2021a,sehCombiningTheoryExperiment2017a}, where such detail is typically hard to obtain.
However, achieving reliable agreement between theory and experiments in determining $H_\textrm{ads}$ is challenging~\cite{plataCaseStudyMechanism2015} due to limitations/inaccuracies in the theoretical methods that are commonly employed and the frequent neglect of thermal contributions.
Moreover, these inaccuracies can affect the predicted adsorption configuration, leading to incorrect identification of the most stable configuration, or a fortuitous match to experimental $H_\textrm{ads}$ for a metastable configuration.

\begin{figure}[p]
    \centering
    \includegraphics[width=4.92in]{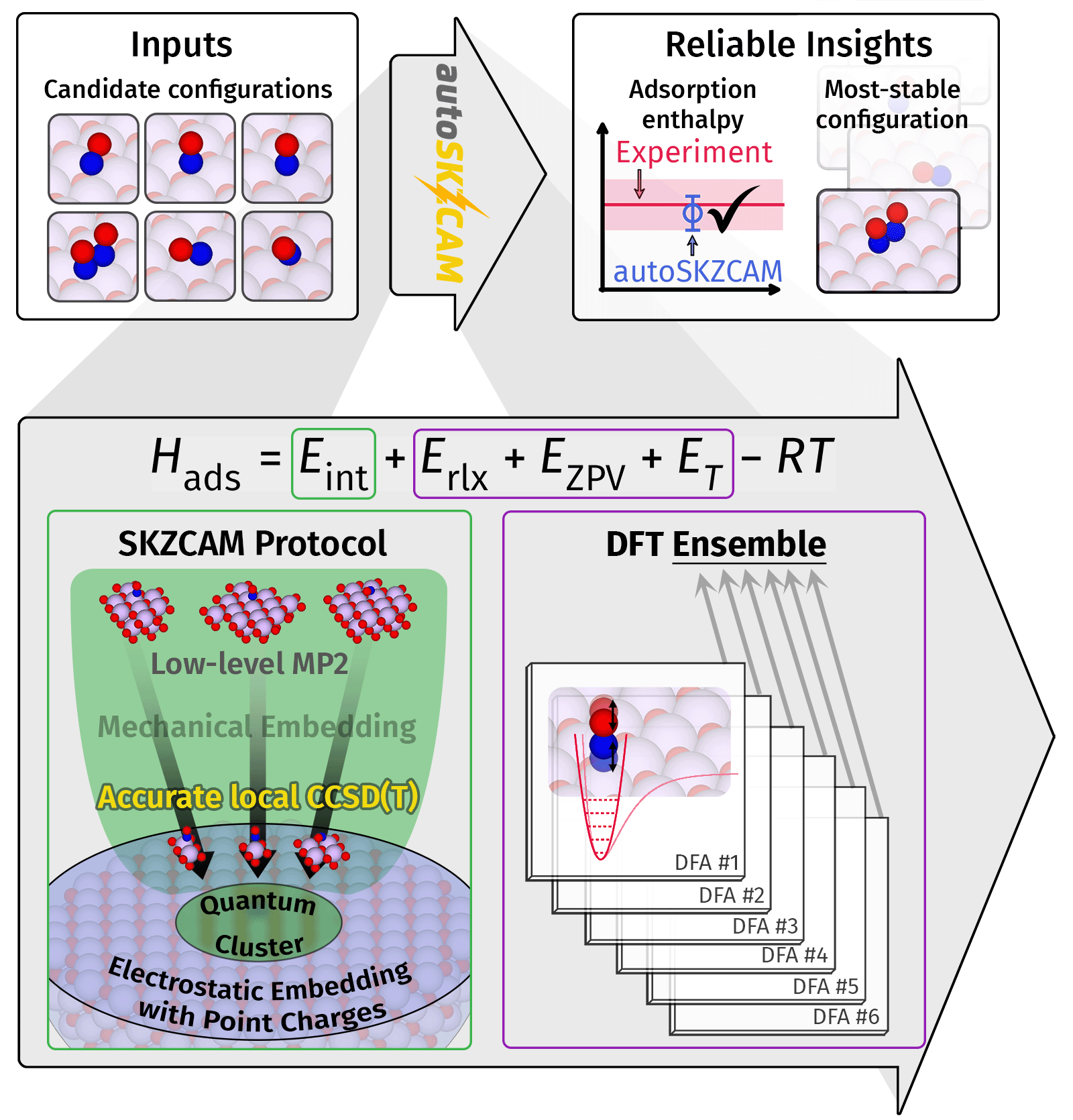}
    \caption{\label{fig:autoskzcam_framework} \textbf{Reliable insights into the surface chemistry of ionic materials with the autoSKZCAM framework.} Schematic description of the open-source autoSKZCAM framework. From a set of adsorbate--surface configurations, this framework can identify the most stable configuration and calculate an adsorption enthalpy $H_\textrm{ads}$ that reproduces experiment. It partitions $H_\textrm{ads}$ via a divide-and-conquer scheme. The dominant contribution --- the interaction energy $E_\textrm{int}$ --- is treated up to the ``gold-standard'' coupled cluster theory with single, double and perturbative triple excitations [CCSD(T)] level through the SKZCAM protocol. This protocol ensures low cost on $E_\textrm{int}$ by employing a multilevel approach, where CCSD(T) with a local approximation is embedded within more affordable levels of theory such as second-order M{\o}ller-Plesset perturbation theory (MP2). The remaining relaxation energy $E_\textrm{rlx}$, zero-point vibrational energy $E_\textrm{ZPV}$ and thermal contributions $E_\textrm{T}$ are treated through an ensemble of 6 widely-used density functional approximations in DFT, enabling an (averaged) estimate with a corresponding error prediction.}
\end{figure}

To extend simulations beyond just a tool to complement experiments, new techniques are needed that surpass the traditional cost--accuracy trade-off; they must achieve high accuracy on $H_\textrm{ads}$ (rivalling experiments) while being fast enough to sample multiple adsorption sites and configurations to correctly identify the most stable configuration.
Density functional theory (DFT) is the current workhorse technique, playing an important role in identifying the reactivity trends (e.g., Br{\o}nsted–Evans–Polanyi relationships~\cite{norskovUniversalityHeterogeneousCatalysis2002,michaelidesIdentificationGeneralLinear2003a,abild-pedersenScalingPropertiesAdsorption2007,calle-vallejoPhysicalChemicalNature2012,lansfordScalingRelationshipsTheory2017} and volcano plots~\cite{bligaardBronstedEvansPolanyi2004a,falsigTrendsCatalyticCO2008,chengUtilizationThreeDimensionalVolcano2008,montoyaMaterialsSolarFuels2017}) that now form pivotal tools for the \textit{in silico} design of new solid catalysts~\cite{norskovComputationalDesignSolid2009b}.
Despite these successes, the density functional approximations (DFAs) to the exchange-correlation functional and dispersion interactions within DFT are not systematically improvable, presenting ongoing challenges~\cite{ruzsinszkySpuriousFractionalCharge2006,mori-sanchezManyelectronSelfinteractionError2006} in making reliable predictions~\cite{schimkaAccurateSurfaceAdsorption2010a,plessowHowAccuratelyApproximate2020}.
For example, 6 different adsorption configurations have been proposed by different DFT studies for NO adsorbed on the MgO(001) surface (see Fig.~\ref{fig:autoskzcam_framework}).

Correlated wave-function theory (cWFT) provides a systematically improvable hierarchy of methods, where coupled cluster (CC) theory with single, double and perturbative triple excitations~\cite{raghavachariFifthorderPerturbationComparison1989} [CCSD(T)] is widely considered the method of choice.
Its high cost and steep computational scaling, however, limits its direct application to adsorbate--surface systems.
To bypass this cost, CCSD(T) is often applied within an embedding approach, where it is treated as either local fragments in periodic supercells~\cite{schaferLocalEmbeddingCoupled2021b,lauRegionalEmbeddingEnables2021c} or by approximating the surface as a finite embedded cluster~\cite{tosoniAccurateQuantumChemical2010,boeseAccurateAdsorptionEnergies2013a,alessioChemicallyAccurateAdsorption2019a,boeseAccurateAdsorptionEnergies2016,kubasSurfaceAdsorptionEnergetics2016c,araujoAdsorptionEnergiesTransition2022c}.
The latter embedded cluster approaches have demonstrated great success in reproducing experimental $H_\textrm{ads}$ estimates, while new methods~\cite{masiosAvertingInfraredCatastrophe2023c,schimkaAccurateSurfaceAdsorption2010a,boothExactDescriptionElectronic2013b} and approximations~\cite{mihmShortcutThermodynamicLimit2021b,yeAdsorptionVibrationalSpectroscopy2024a,yeInitioSurfaceChemistry2024} for the former are further enhancing its applicability.

The most common embedding approach for adsorbate--surface systems involving ionic materials is electrostatic embedding, where the system is modelled as a central `quantum' cluster surrounded by a field of point charges representing the long-range interactions from the rest of the surface.
This approach has been applied to systems ranging from simple ionic materials~\cite{sauerMolecularModelsInitio1989a} to challenging quantum materials~\cite{bogdanovEnhancementSuperexchangeDue2022}, not only on their surfaces, but in the bulk~\cite{dittmerAccurateBandGap2019b} as well as on steps, edges and kinks~\cite{sushkoRelativeEnergiesSurface2000e,chizalletAssignmentPhotoluminescenceSpectra2008}.
However, applying electrostatic embedding is challenging as designing efficient quantum clusters~\cite{kickTransferableDesignSolidstate2019} amenable to methods such as CCSD(T) while being converged to the bulk limit is not trivial, requiring significant manual effort and chemical intuition.
Consequently, studies until now have been mostly limited to one or two systems.
There have been advances~\cite{dittmerAccurateBandGap2019b, dittmerComputationNMRShielding2020a,shiGeneralEmbeddedCluster2022b, shiManyBodyMethodsSurface2023a} towards addressing these challenges --- for example, the SKZCAM protocol~\cite{shiGeneralEmbeddedCluster2022b, shiManyBodyMethodsSurface2023a} by some of the authors provides systematic rubrics for reaching the bulk limit at cluster sizes amenable to CCSD(T) with local approximations.
More recently, it has been combined with a mechanical embedding (i.e., ONIOM~\cite{chungONIOMMethodIts2015}) procedure, where the effort of reaching the bulk limit is performed with the more affordable second-order M{\o}ller-Plesset perturbation theory (MP2) on larger clusters, while CCSD(T) is performed on smaller clusters to correct MP2.
These improvements, however, come at the cost of significant additional complexity and overhead, such that it remains a challenge to tackle broad sets of adsorbate–surface systems or adsorption configurations --- currently routine with DFT.
To make these tools usable by the broader community, they need to be streamlined, automatised and simplified into ``black-box'' tools, providing reliable insights from simple inputs (as illustrated in the top panel of Fig.~\ref{fig:autoskzcam_framework}).

In this work, we introduce the autoSKZCAM framework, which delivers CCSD(T)-quality predictions to surface chemistry problems involving ionic materials at a cost and ease approaching DFT.
With this framework, we have been able to reproduce the experimental $H_\textrm{ads}$ for a set of 19 adsorbate--surface systems.
These systems (visualised in Fig.~\ref{fig:many_molecules} and Sec.~\ref{si-sec:adsorbate-surface_systems} of the Supplementary Information) include a diverse array of molecules adsorbed onto MgO(001) as well as TiO\textsubscript{2} anatase(101) and rutile(110).
Its low cost has been leveraged to study multiple adsorption configurations for some of the adsorbate--surface systems, aiding in resolving debates between experiments and simulations on the most stable adsorption configuration.
Furthermore, it has provided new benchmarks for assessing the performance of DFAs in DFT, facilitating future advances in DFA development.

As summarised in the bottom panel of Fig.~\ref{fig:autoskzcam_framework}, the autoSKZCAM framework automates the computation of an accurate yet efficient $H_\textrm{ads}$.
It partitions this quantity into separate contributions (discussed in the Methods and Sec.~\ref{si-sec:adsorption_enthalpy_def} of the Supplementary Information) that are addressed with appropriate techniques within a divide-and-conquer scheme.
The principal contribution, the adsorbate--surface interaction energy $E_\text{int}$, is calculated up to the CCSD(T) level using the SKZCAM protocol~\cite{shiGeneralEmbeddedCluster2022b,shiGoingGoldStandardAttaining2024,shiManyBodyMethodsSurface2023a} together with new local correlation approximations [i.e., LNO-CCSD(T)~\cite{nagyOptimizationLinearScalingLocal2018,nagyApproachingBasisSet2019} and DLPNO-CCSD(T)~\cite{riplingerEfficientLinearScaling2013,riplingerNaturalTripleExcitations2013,riplingerEfficientLinearScaling2013a,riplingerSparseMapsSystematic2016}].
The SKZCAM protocol has been automated within the present work, eliminating any manual intervention and enabling a wide range of systems to be tackled.
Moreover, this automation has allowed CCSD(T) to be mechanically embedded within additional ONIOM layers corresponding to more affordable, levels of theory.
This has reduced its cost by one order of magnitude compared to previous works (see Sec.~\ref{si-sec:cost_benchmark} of the Supplementary Information), making it now competitive with periodic hybrid DFT.
We show in Sec.~\ref{si-sec:dft_geom_error} of the Supplementary Information that the remaining relaxation, zero-point vibrational and thermal contributions to $H_\textrm{ads}$ can be estimated effectively with DFT by employing an ensemble of 6 widely-used density functional approximations.
This open-source computational framework is available on GitHub~\cite{shiBenshi97AutoSKZCAM2025a} and we anticipate it will facilitate the exploration of more complex surface phenomena for future energy-relevant technologies~\cite{norskovComputationalDesignSolid2009b}.


\section{Results}
\subsection{Agreement across diverse systems}

\begin{figure}[p]
    \includegraphics[width=\textwidth]{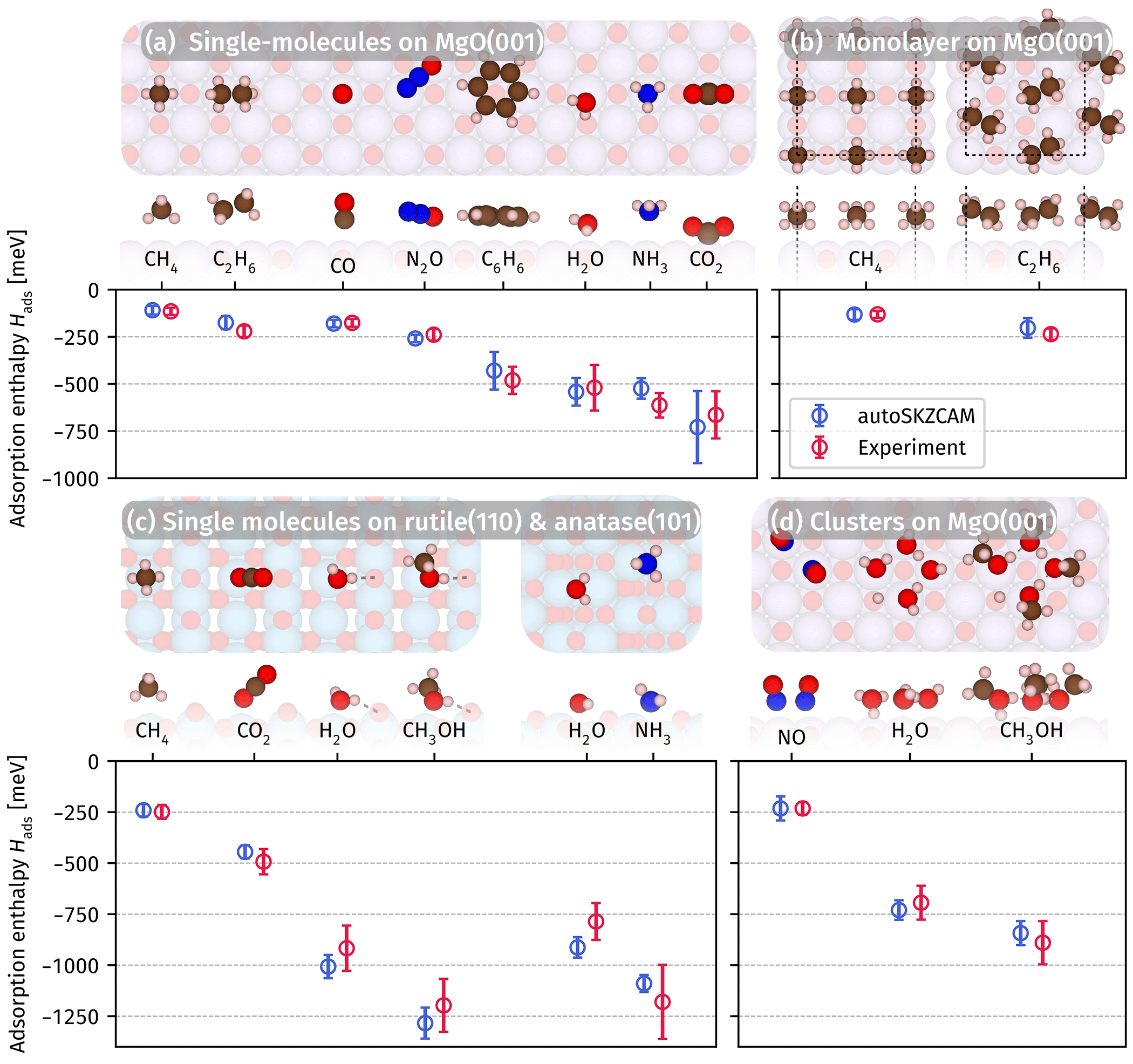}
    \caption{\label{fig:many_molecules}\textbf{Consensus with experiments for a range of adsorbates on ionic surfaces.} Comparison of adsorption enthalpies computed with the autoSKZCAM framework against high-quality temperature programmed desorption experiments for a set of 19 adsorbate--surface combinations. These include (a) single molecules adsorbed on the MgO(001) surface, (b) monolayers adsorbed on MgO(001), (c) single molecules adsorbed on \ce{TiO2} rutile(110) and anatase(101) as well as (d) clusters adsorbed on MgO(001). We discuss how we calculate the error bars (corresponding to 95\% confidence intervals or more) on $H_\textrm{ads}$ for the simulations and experiments (including references to the experimental data) in Secs.~\ref{si-sec:final_hads} and~\ref{si-sec:exp_redhead_analysis}, respectively. A top and side view of the most stable geometry for each system is shown above each label, with C, H, N, O, Mg and Ti atoms corresponding to the brown, white, dark blue, red, purple and light blue spheres respectively.}
\end{figure}

The 19 adsorbate--surface systems studied within this work are visualised in Fig.~\ref{fig:many_molecules}, where the $H_\text{ads}$ computed by the autoSKZCAM framework is evaluated against experiment.
In all of the systems, the autoSKZCAM framework has been able to reproduce experimental $H_\text{ads}$ measurements (tabulated and visualised in greater detail in Sec.~\ref{si-sec:final_hads} of the Supplementary Information), lying within their respective errors bars.
This range of systems cover an $H_\text{ads}$ of almost $1.5\,$eV, spanning weak physisorption to strong chemisorption and including a diverse set of molecules (CO, NO, \ce{N2O}, \ce{NH3}, \ce{H2O}, \ce{CO2}, \ce{CH3OH}, \ce{CH4}, \ce{C2H6} and \ce{C6H6}) on common surfaces of ionic materials [MgO(001) as well as TiO\textsubscript{2} anatase(101) and rutile(110)].
Besides the adsorption of small single molecules, some of which have been tackled before, this work also studies monolayers [Fig.~\ref{fig:many_molecules}b] and larger molecules such as \ce{C6H6} or molecular clusters of \ce{CH3OH} and \ce{H2O} [Fig.~\ref{fig:many_molecules}d].

The experimental estimates were largely taken from single-crystal TPD measurements compiled by Campbell and Sellers~\cite{campbellEnthalpiesEntropiesAdsorption2013}, where the effects of surface disorder or defects are expected to be minimal.
The error bars on these measurements correspond to a 95\% confidence interval on the experimental pre-exponential ($\nu$) factor, coming from predictions for the standard entropy of the adsorbate by Campbell and Sellers~\cite{campbellEntropiesAdsorbedMolecules2012}.
A similar confidence interval has been calculated for the the individual terms in the autoSKZCAM framework and the overall $H_\text{ads}$ estimate, as discussed in Secs.~\ref{si-sec:exp_redhead_analysis} and~\ref{si-sec:final_hads} of the Supplementary Information, respectively.
We connect static adsorption energies to $H_\text{ads}$ using the quasi-rigid-rotor harmonic oscillator (quasi-RRHO) method~\cite{liImprovedForceFieldParameters2015}, which improves over the standard harmonic approximation for treating low-lying vibrational modes of the adsorbate.
We expect the error bars (from the DFT ensemble) we have computed on these thermal contributions to be greater than or comparable to remaining anharmonic contributions~\cite{boeseAccurateAdsorptionEnergies2013a}.

The ability to study large systems, including molecular clusters on the surface, with the autoSKZCAM framework has been crucial towards reproducing experiments.
For example, we have studied several adsorption configurations of \ce{CH3OH} on MgO(001), including hydrogen (H-)bonded and partially dissociated clusters of \ce{CH3OH}.
We find that agreement to experiment in Fig.~\ref{fig:many_molecules} can only be obtained when considering partially dissociated clusters.
As discussed in Sec.~\ref{si-sec:ch3oh_discuss} of the Supplementary Information and in~\ref{fig:ch3oh_xc_hads}, other studied structures predict less stable adsorbates (i.e., a weaker absolute $H_\textrm{ads}$).
We show in~\ref{fig:h2o_xc_hads} that these insights are transferable to \ce{H2O}, which also forms partially dissociated clusters on MgO(001).

\subsection{Reliable insights at the atomic-level}

\begin{figure}[h!]
    \includegraphics[width=0.55\textwidth]{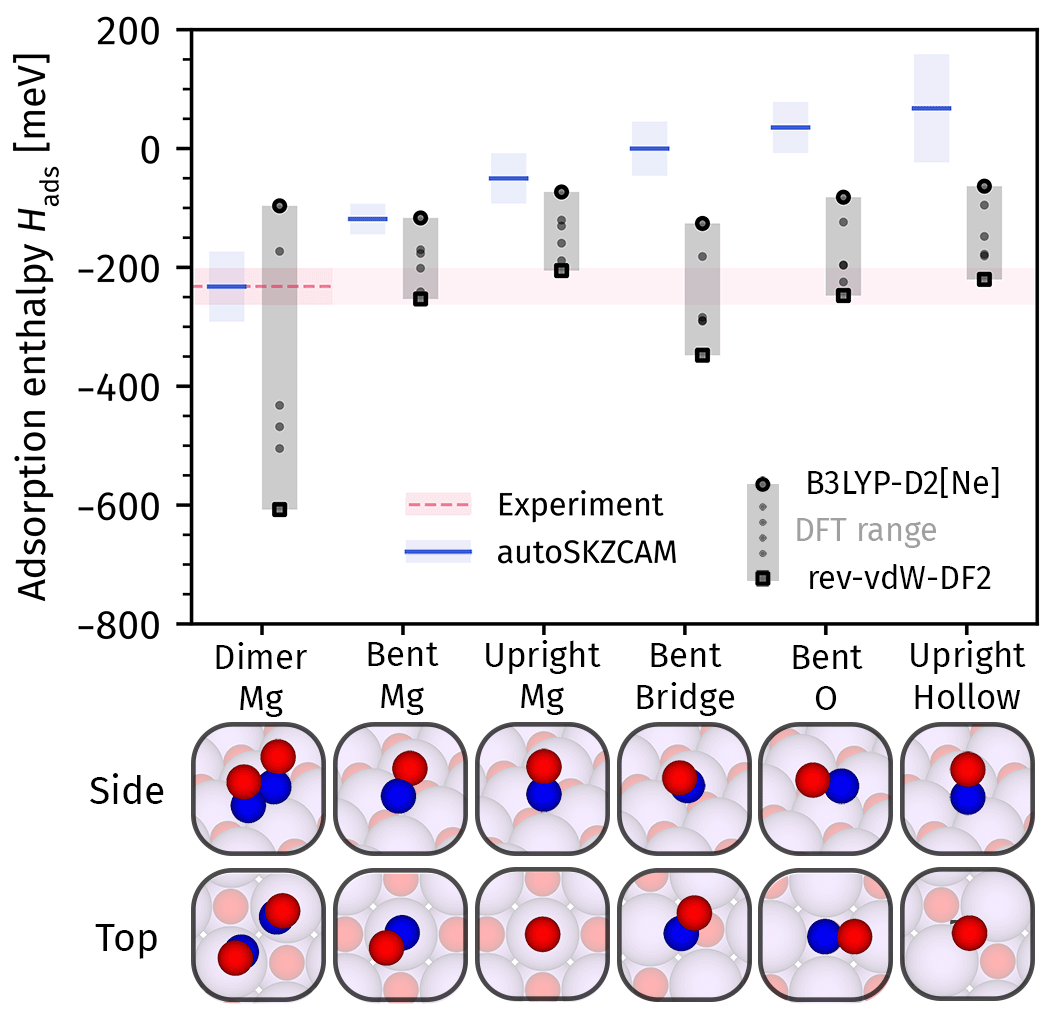}    \caption{\label{fig:no_struct_compare}\textbf{Correct identification of the NO on MgO(001) adsorption configuration.} For NO on MgO(001), six adsorption configurations have been proposed: ``Dimer Mg'', ``Bent Mg'', ``Upright Mg'', ``Bent Bridge'', ``Bent O'' and ``Upright Hollow''. These names reflect their orientation and binding sites on the surface. The adsorption enthalpy $H_\textrm{ads}$ is calculated for each configuration with the autoSKZCAM framework and a set of 6 density functional approximations (DFAs) in DFT. The estimates from the 6 DFAs are plotted as a DFT ``range'' between the smallest and largest values. Experiments~\cite{plateroDipoleCouplingChemical1985,divalentinNOMonomersMgO2002a} indicate that the Dimer Mg configuration is the most stable. The autoSKZCAM framework is the only method that correctly identifies this configuration while reproducing the experimental $H_\textrm{ads}$ measurement by Wichtendahl \etal{}~\cite{wichtendahlThermodesorptionCONO1999}. Conservative error estimates (corresponding to a 95\% confidence interval or more) are given for both experiment and autoSKZCAM, as discussed in Secs.~\ref{si-sec:final_hads} and~\ref{si-sec:exp_redhead_analysis}, respectively. The DFAs used are: PBE-D2[Ne], revPBE-D4, vdW-DF2, rev-vdW-DF2, PBE0-D4 and B3LYP-D2[Ne], with B3LYP-D2[Ne] and rev-vdW-DF2 explicitly indicated, as these sit at either end of the DFT range. The autoSKZCAM and DFT estimates are tabulated in Sec.~\ref{si-sec:no_si_config} of the Supplementary Information.}
\end{figure}

The automated nature and affordable cost of the autoSKZCAM framework allows us to compare the $H_\text{ads}$ across several configurations which the adsorbate can adopt in each adsorbate--surface system. 
Beyond \ce{H2O} and \ce{CH3OH}, we have used this framework to identify the most stable adsorption configuration of N$_2$O, CO$_2$ and NO on MgO(001) as well as CO$_2$ on \ce{TiO2} rutile(110) --- systems where there have been debate within the literature.
Here, inaccuracies in the DFAs within DFT can lead to ambiguities when determining the stable adsorption configuration through two possibilities: (1) the DFA predicts the wrong stable adsorption configuration; or (2) a metastable adsorption configuration fortuitously matches the experimental $H_\textrm{ads}$.
As verified in Fig.~\ref{fig:many_molecules}, the autoSKZCAM framework ensures that the experimental $H_\textrm{ads}$ is only reproduced when the correct stable adsorption configuration (corresponding to the most negative $H_\textrm{ads}$) has been identified.

The ambiguities from utilising DFT are particularly evident in the adsorption of NO on MgO(001), where different DFAs (and procedures) have led to the identification of multiple ``stable'' geometries, comprising 6 broad classes.
In Fig.~\ref{fig:no_struct_compare}, we present $H_\textrm{ads}$ estimates by several widely-used DFAs for these six adsorption configurations.
For all six configurations, there are DFAs that yield $H_\textrm{ads}$ values that agree with experiment.
Notably, the rev-vdW-DF2 DFA~\cite{hamadaVanWaalsDensity2014} predicts $H_\textrm{ads}$ values that agree with experiments for the ``Bent Mg'', ``Upright Mg'', ``Bent O'' and ``Upright Hollow'' adsorption configurations.
On the basis of such fortuitous agreement, prior studies (see Sec.~\ref{si-sec:comp_literature} of the Supplementary Information) have mis-identified several of these configurations to be the most stable.
The autoSKZCAM framework identifies the (covalently-bonded) dimer \textit{cis}-(NO)$_2$ (dubbed ``Dimer Mg'') configuration to be most stable, with an $H_\textrm{ads}$ consistent with experiment while all other (monomer) configurations are predicted to be less stable by more than $80\,$meV.
This prediction is commensurate with findings from Fourier-Transform Infrared Spectroscopy (FTIR)~\cite{plateroDipoleCouplingChemical1985} and electron paramagnetic resonance~\cite{divalentinNOMonomersMgO2002a} (EPR) experiments, both of which suggest that NO exists as a dimer on MgO(001), aside from a small number of monomers adsorbed on defect sites.

In many cases, debates on the most stable adsorption configuration cannot be resolved from experiments alone.
For example, techniques like FTIR spectroscopy, low-energy electron diffraction (LEED), or X-ray and ultraviolet photoelectron spectroscopy (XPS and UPS) only provide indirect evidence.
Moreover, while scanning tunneling microscopy (STM) offers real-space images, its resolution is often insufficient for definitive interpretation~\cite{e.hamlynImagingOrderingWeakly2018}.
The autoSKZCAM framework can be valuable within such contexts.
Notably, both experiments~\cite{meixnerKineticsDesorptionAdsorption1992b,chakradharCarbonDioxideAdsorption2013a} and simulations~\cite{pacchioniInitioClusterModel1994b,jensenCO2SorptionMgO2005b,downingReactivityCO2MgO2013,mazheikaNiSubstitutionalDefects2016a} have debated between a chemisorbed or physisorbed configuration [\ref{fig:co2_configurations} and Sec.~\ref{si-sec:co2_configurations} of the Supplementary Information] of \ce{CO2} on MgO(001).
With the autoSKZCAM framework, we show that it takes on a chemisorbed carbonate configuration in agreement with previous TPD measurements~\cite{chakradharCarbonDioxideAdsorption2013a,yanagisawaInteractionCO2Magnesium1995b}.
Similarly, the adsorption of \ce{CO2}~\cite{sorescuCO2AdsorptionTiO22011,kubasSurfaceAdsorptionEnergetics2016c} on \ce{TiO2} rutile(110) [\ref{fig:rutile_co2_configurations} and Sec.~\ref{si-sec:co2_rutile_config} of the Supplementary Information] as well as \ce{N2O}~\cite{huesgesDispersionCorrectedDFT2014} on MgO(001) [\ref{fig:n2o_configurations} and Sec.~\ref{si-sec:n2o_configurations_si} of the Supplementary Information] have been debated to take on either a tilted or parallel geometry; the autoSKZCAM framework predicts the tilted geometry to be most stable for the former and the parallel geometry for the latter.
Ultimately, it is the free energy of adsorption which dictates the relative stability of the geometries, but we expect missing entropic contributions from $H_\text{ads}$ to be small and within the error estimates of $H_\text{ads}$ for the systems studied here, only becoming prominent for large molecules or under confinement~\cite{collingeEffectCollectiveDynamics2020}.

\subsection{``Gold-standard'' benchmarks}

The predictions from the autoSKZCAM framework for the systems studied in this work can be valuable as a benchmark dataset for non-covalent interactions, which are crucial for modelling the binding of adsorbates to surfaces.
These interactions are physically reflected within the interaction energy $E_\textrm{int}$ contribution to $H_\textrm{ads}$ (see Methods), where it quantifies the strength of this binding.
Previous studies~\cite{wellendorffBenchmarkDatabaseAdsorption2015,r.rehakIncludingDispersionDensity2020,caldeweyherExtensionEvaluationD42020,shiGoingGoldStandardAttaining2024} have shown that DFAs struggle to consistently describe these interactions for adsorbate--surface systems and different DFAs can vary over a range exceeding $500\,$meV on $E_\textrm{int}$, even for a simple system like CO on MgO(001).
Here, CCSD(T) is considered a widely-trusted approach for treating non-covalent interactions.
However, while it has become common to generate reference datasets at the CCSD(T) level for small molecule interaction energies~\cite{rezacS66WellbalancedDatabase2011,donchevQuantumChemicalBenchmark2021b}, it has not been possible for adsorbate--surface systems so far.
These datasets are commonly used to e.g., parametrise the exchange--correlation functional or dispersion corrections in many modern DFAs.
Their poor performance for adsorbate--surface systems arises in part from the lack of available references, particularly those involving metal oxides~\cite{campbellEnthalpiesEntropiesAdsorption2013}; this gap can be addressed with CCSD(T)-level references provided by the autoSKZCAM framework.

\begin{figure}[ht!]
    \includegraphics[width=\textwidth]{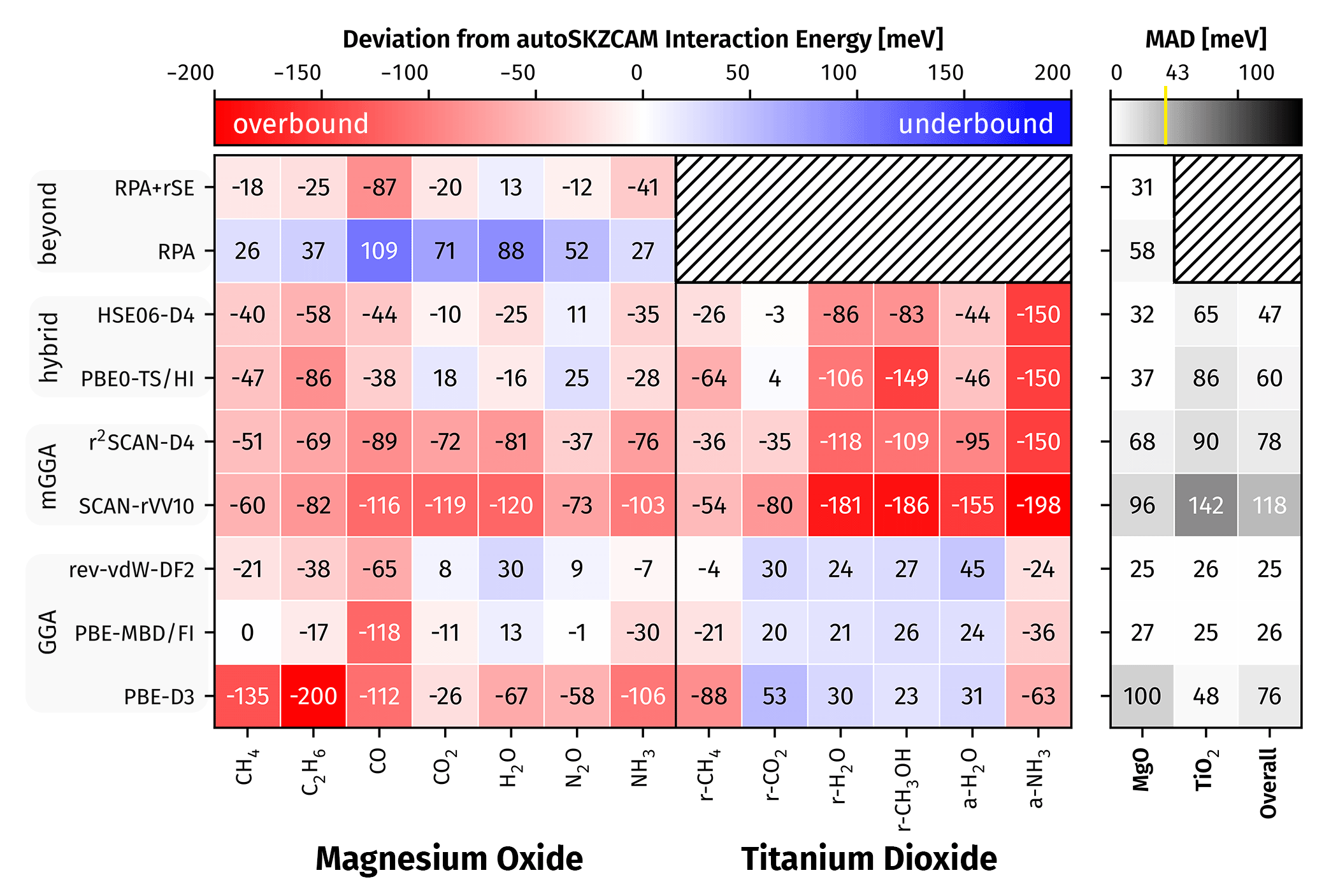}
    \caption{\label{fig:dft_compare}\textbf{A benchmark for lower-level theories.} The autoSKZCAM framework interaction energy benchmarks are used to assess a selection of density functional approximations along Jacob's ladder as well as the random phase approximation. The deviation from the autoSKZCAM estimate is given as a colour map, with red and blue indicating overbinding and underbinding, respectively. We consider a range of molecules physisorbed on the MgO(001) and TiO\textsubscript{2} surfaces, with rutile(110) indicated by `r-' and anatase(101) indicated by `a-'. The mean absolute deviations (MADs) across all of the systems (labelled `Overall'), as well as the subsets involving the MgO(001) and \ce{TiO2} surfaces, are given in grey on the right panel. We indicate the typical `chemical accuracy' of $43\,$meV in yellow on the colour bar. The autoSKZCAM and DFT interaction energies are tabulated in Sec.~\ref{si-sec:final_eint_estimates} of the Supplementary Information.}
\end{figure}

In Fig.~\ref{fig:dft_compare}, a set of DFAs selected from recent benchmark studies~\cite{r.rehakIncludingDispersionDensity2020,shiGoingGoldStandardAttaining2024} has been compared against the autoSKZCAM $E_\text{int}$ benchmarks; the values are tabulated in Sec.~\ref{si-sec:final_eint_estimates} of the Supplementary Information.
We do not aim here to provide a comprehensive overview of current DFAs nor make definitive statements about the performance of different rungs of Jacob's ladder of exchange-correlation (XC) functionals.
However, a broad set of XC functionals has been considered, starting from the generalised gradient approximation (GGA) all the way up to the state-of-the-art random phase approximation (RPA).
Each of these XC functionals (besides RPA) has been further paired with a wide range of dispersion corrections to improve their description of adsorbate--surface systems.
The resulting selection of DFAs includes the workhorse PBE-D3~\cite{perdewGeneralizedGradientApproximation1996d,grimmeConsistentAccurateInitio2010a}, newly developed DFAs such as r$^2$SCAN-D4~\cite{furnessAccurateNumericallyEfficient2020a,ehlertSCAND4DispersionCorrected2021} or SCAN-rVV10~\cite{sunStronglyConstrainedAppropriately2015,ningWorkhorseMinimallyEmpirical2022} as well as sophisticated hybrids and RPA.
Out of the investigated DFAs, we observe that two GGA-based DFAs (PBE-MBD/FI~\cite{tkatchenkoAccurateEfficientMethod2012,buckoExtendingApplicabilityTkatchenkoScheffler2014} and rev-vdW-DF2~\cite{hamadaVanWaalsDensity2014}) perform the best, with a mean absolute deviation (MAD) of $26\,$ and $25\,$meV across all of the systems (labelled `Overall' in Fig.~\ref{fig:dft_compare}).
On the other hand, RPA --- considered the current state-of-the-art method for surface chemistry --- has a MAD of $58\,$meV for the subset of MgO(001) adsorbate--surface systems that were studied.
These errors arise from a well-known systematic underbinding~\cite{renRandomPhaseApproximationElectron2011} of RPA, which is improved by incorporating the renormalised singles energy~\cite{klimesSinglesCorrelationEnergy2015} (rSE), instead overbinding with an MAD of $31\,$meV for the MgO(001) adsorbate--surface systems.
Unfortunately, the higher cost of RPA prevented its application to the TiO\textsubscript{2} surfaces and thus, these insights for RPA are limited to only the specific set of molecules adsorbed on MgO(001); broader comments require a more complete dataset involving more surfaces.

These benchmarks provide important insights towards designing improved DFAs.
For example, the PBE-D3~\cite{perdewGeneralizedGradientApproximation1996d,grimmeConsistentAccurateInitio2010a} with zero-damping, SCAN-rVV10~\cite{ningWorkhorseMinimallyEmpirical2022}, and r$^2$SCAN-D4~\cite{ehlertSCAND4DispersionCorrected2021} DFAs are all found to significantly overbind $E_\textrm{int}$ for systems involving MgO(001), with the latter two also overbinding for the \ce{TiO2} surfaces.
These observations are commensurate with previous findings, where the overbinding in PBE-D3 has been attributed to an overestimated Mg $C_6$ dispersion coefficient in the D3 dispersion correction~\cite{ehrlichSystemDependentDispersionCoefficients2011}.
SCAN and r$^2$SCAN have recently been shown to overbind solids~\cite{kothakondaTestingR2SCANDensity2023} and this work confirms that this overbinding persists for adsorbate--surface systems, with the rVV10 dispersion correction further exacerbating this overbinding.

\section{Discussion}

The agreement achieved in $H_\textrm{ads}$ between experiment and the autoSKZCAM framework is not trivial.
For example, we show in Fig.~\ref{fig:cost_comparison}a and Sec.~\ref{si-sec:comp_literature} of the Supplementary Information that DFT estimates collated from the literature can span a range of nearly $1000\,$meV for \ce{CO2} adsorbed on MgO(001) and \ce{H2O} on \ce{TiO2} rutile(110).
Beyond the errors in $E_\textrm{int}$ arising from the choice of DFA (as highlighted in Fig.~\ref{fig:dft_compare}), these errors also stem from the use of unconverged structural models in the embedded cluster or periodic slab approaches.
Additionally, electronic structure parameters such as the basis set size, treatment of frozen cores, and pseudopotentials must be carefully controlled, with the majority of studies neglecting thermal and vibrational contributions to $H_\textrm{ads}$.
These challenges become more pronounced for methods from cWFT, where ensuring converged electronic structure parameters or structural models are limited by their high computational cost.
We highlight this challenge in Fig.~\ref{fig:cost_comparison}a, where a range of $528\,$meV has been observed across cWFT based studies of \ce{H2O} on rutile(110), with similar discrepancies noted for CO on MgO(001)~\cite{shiManyBodyMethodsSurface2023a}.

Despite the challenges in applying cWFT methods like CCSD(T) to adsorbate--surface systems, there have been other successful applications within the literature besides this work.
Notably, the collection of molecules adsorbed on MgO(001) as well as zeolites and metal--organic frameworks studied by Sauer and co-workers~\cite{tosoniAccurateQuantumChemical2010,boeseAccurateAdsorptionEnergies2013a,alessioChemicallyAccurateAdsorption2019a,boeseAccurateAdsorptionEnergies2016} alongside the landmark study by Kubas \etal{}~\cite{kubasSurfaceAdsorptionEnergetics2016c} tackling 5 molecules on the rutile(110) surface, both with an embedded cluster approach.
More recently, this success has been extended towards periodic slab approaches with CCSD(T)~\cite{yeAdsorptionVibrationalSpectroscopy2024a,yeInitioSurfaceChemistry2024,al-hamdaniPropertiesWaterBoron2017a,tsatsoulisReactionEnergeticsHydrogen2018,brandenburgPhysisorptionWaterGraphene2019,schaferSurfaceScienceUsing2021a,schaferLocalEmbeddingCoupled2021b,lauRegionalEmbeddingEnables2021c,shiManyBodyMethodsSurface2023a} thanks to new algorithmic and methodological developments.    
Besides being in agreement with these previous estimates, we also find that our autoSKZCAM estimates are in agreement with previous DMC estimates for CO on MgO(001)~\cite{shiManyBodyMethodsSurface2023a} as well as H$_2$O on MgO(001)~\cite{karaltiAdsorptionWaterMolecule2012,alessioChemicallyAccurateAdsorption2019a}, upon inclusion of missing geometrical relaxation contributions in the latter.
The key advance in the present study is the breadth, size and number of adsorbate--surface systems that can now be tackled, driven by the (1) low-cost, (2) general applicability, and (3) automated nature of the autoSKZCAM framework.
These qualities are enabled by combining the mechanical embedding approach of Sauer and co-workers~\cite{sauerInitioCalculationsMolecule2019b} with the electrostatic embedding procedure pioneered by the co-workers of Catlow~\cite{luMultiscaleQMMM2023,buckeridgePolymorphEngineeringTiO22015a,scanlonBandAlignmentRutile2013a} and Reuter~\cite{bergerFirstprinciplesEmbeddedclusterCalculations2015a,kubasSurfaceAdsorptionEnergetics2016c}.
It is made more economical by making use of CCSD(T) with local approximations~\cite{nagyApproachingBasisSet2019,riplingerEfficientLinearScaling2013,maExplicitlyCorrelatedLocal2018}.

\begin{figure}[ht!]
    \centering
    \includegraphics[width=0.5\textwidth]{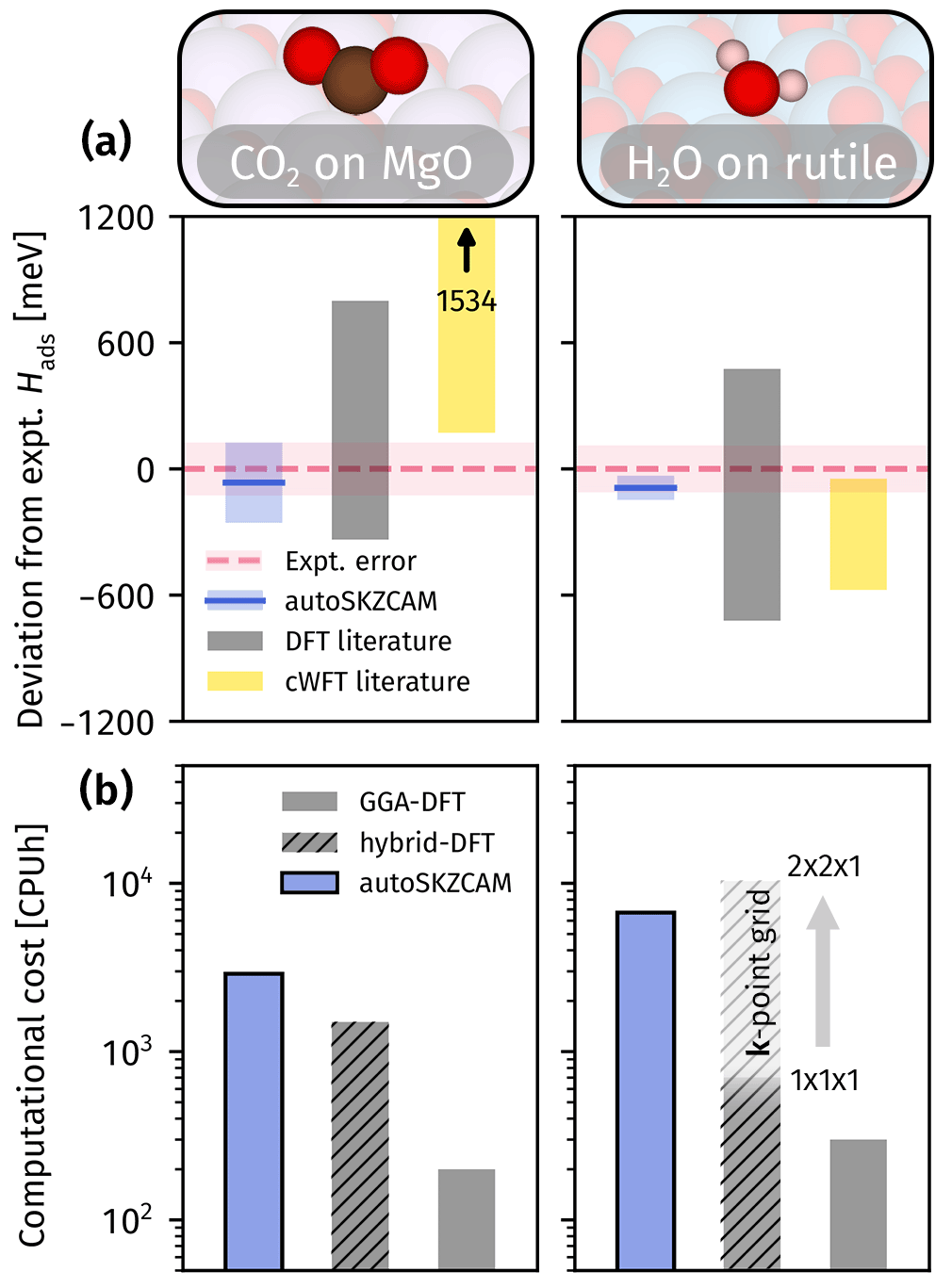}
    \caption{\label{fig:cost_comparison}\textbf{High accuracy at comparable cost to periodic hybrid DFT.} For the chemisorbed \ce{CO2} on MgO(001) and \ce{H2O} on \ce{TiO2} rutile(110), we demonstrate the (a) improved agreement to experimental TPD measurements~\cite{chakradharCarbonDioxideAdsorption2013a,dohnalekPhysisorptionN2O22006} for the autoSKZCAM framework relative to previous DFT (in grey) and cWFT (in yellow) literature, both of which are plotted as a range of values collated in Sec.~\ref{si-sec:comp_literature} of the Supplementary Information. We show in (b) that its computational cost to calculate the interaction energy is competitive with periodic hybrid DFT. For H$_2$O on TiO$_2$ rutile(110), the cost of periodic hybrid DFT can vary significantly depending on the choice of $k$-point grid (as discussed in Sec.~\ref{si-sec:cost_benchmark} of the Supplementary Information), highlighted by the faded region, with changes of the order of $20\,$meV. Conservative error estimates (corresponding to a 95\% confidence interval or more) are given for both experiment and autoSKZCAM, as discussed in Secs.~\ref{si-sec:final_hads} and~\ref{si-sec:exp_redhead_analysis} of the Supplementary Information, respectively.}
\end{figure}

The key limitation of the autoSKZCAM framework is that it can only treat the surfaces of ionic materials.
Going beyond electrostatic embedding towards (quantum) embedding approaches~\cite{sunQuantumEmbeddingTheories2016} that couple the environment to the quantum cluster through variables such as the density, the single-particle Green’s function or the single-particle density matrix, would allow covalently-bonded and metallic systems~\cite{libischEmbeddedCorrelatedWavefunction2014b} to be treated.
This improved treatment of the boundaries can also enable smaller and more efficient (high-level) quantum clusters to be used.
However, the coupling of the quantum cluster to the environment is currently non-trivial to get right, often involving several parameters to converge~\cite{nusspickelSystematicImprovabilityQuantum2022}.
Furthermore, these calculations~\cite{lauRegionalEmbeddingEnables2021c,schaferLocalEmbeddingCoupled2021b} build on top of a prior DFT or HF calculation that has to be performed on the full adsorbate--slab system --- typically a periodic model which can be computationally expensive.
Recent work~\cite{huangAdvancingSurfaceChemistry2025} has highlighted that it is possible to overcome this need for periodic calculations through a multilevel embedding procedure by combining the quantum embedding approaches with efficient (cluster) surface models that can be generated by e.g., the SKZCAM protocol.

There are also limitations with using CCSD(T) as the target level of theory for the autoSKZCAM framework.
While it is trusted for studying small and weakly correlated systems, there are open questions on the applicability of CCSD(T) towards more challenging systems.
For example, it cannot treat systems without a bandgap~\cite{kawasakiSimpleElectronCounting2018,neufeldHighlyAccurateElectronic2023c} (notably metals) and it performs poorly for open-shell molecules of radical character.
More recently, it is shown to disagree with quantum diffusion Monte Carlo -- another widely trusted method -- when studying large dispersion-bound molecules of $\pi$-$\pi$ character~\cite{al-hamdaniInteractionsLargeMolecules2021,schaferUnderstandingDiscrepanciesWavefunction2024} and medium-sized H-bonded molecules~\cite{shiSystematicDiscrepanciesReference2024}.
Moreover, beyond the adsorbate, there can be challenges in describing the surface of some transition metal oxides as they may exhibit antiferromagnetic~\cite{rothMagneticStructuresMnO1958} or more exotic (strongly-correlated) properties~\cite{tokuraOrbitalPhysicsTransitionMetal2000}.
To study such systems accurately, CCSD(T) must be replaced with a more appropriate level of theory, such as multi-reference approaches~\cite{heZerothorderActivespaceFrozenorbital2020,heCOInversionNaCl1002023a} or quantum embedding approaches (e.g., density matrix embedding theory~\cite{cuiSystematicElectronicStructure2022} and dynamical mean-field theory~\cite{georgesDynamicalMeanfieldTheory1996}).

We highlight the low computational cost of the autoSKZCAM framework in Fig.~\ref{fig:cost_comparison}b, where it is compared to periodic DFT -- performed using sensible electronic structure settings~\cite{jainCommentaryMaterialsProject2013c} -- with a hybrid and GGA DFA.
The cost to perform the autoSKZCAM framework is competitive with periodic hybrid DFT for both \ce{CO2} on MgO(001) and \ce{H2O} on \ce{TiO2} rutile(110).
Importantly, this cost does not change significantly with the increase in complexity from a MgO(001) to a \ce{TiO2} rutile(110) surface as the embedding procedure ensures that the largest system (i.e., embedded cluster) tackled remains consistent in size across these surfaces.
In Sec.~\ref{si-sec:cost_benchmark} of the Supplementary Information, we also compare our autoSKZCAM framework costs for CO on MgO(001) to previous work.
Its automation in the present work (as discussed in Sec.~\ref{si-sec:improvements_skzcam} of the Supplementary Information) has enabled further levels of mechanical embedding, which has led to an overall cost of ${\sim}600\,$CPUh to compute $E_\textrm{int}$ for CO on MgO(001) with the autoSKZCAM framework.
This is almost one order of magnitude lower than RPA, and nearly two orders of magnitude lower than a previous (un-automated) application of the SKZCAM protocol~\cite{shiManyBodyMethodsSurface2023a} and an efficient periodic CCSD(T) calculation~\cite{shiManyBodyMethodsSurface2023a,yeAdsorptionVibrationalSpectroscopy2024a}, while being more than three orders of magnitude cheaper than periodic quantum diffusion Monte Carlo~\cite{shiManyBodyMethodsSurface2023a}.

Beyond the applications demonstrated so far, we discuss here the potential of the autoSKZCAM framework to validate experimental results.
As shown in Fig.~\ref{fig:many_molecules}, it provides conservative error estimates for $H_\textrm{ads}$ that are lower than experimental uncertainties for the majority of systems.
These experimental values were re-analysed from previous temperature programmed desorption (TPD) measurements, following Campbell and Sellers~\cite{campbellEnthalpiesEntropiesAdsorption2013}, using more accurate system-specific~\cite{campbellEntropiesAdsorbedMolecules2012} pre-exponential factors $\nu$.
The majority of the experimental error arises from uncertainties in $\nu$ with additional minor contributions discussed in Sec.~\ref{si-sec:exp_redhead_analysis} of the Supplementary Information.
We show that using the original analysis procedure (setting $\nu$ to a default value of $10^{13}$) leads to $H_\textrm{ads}$ values which are in worse agreement with autoSKZCAM, with a root mean squared deviation of $102\,$meV compared to $58\,$meV for the system-specific procedure.
Going further, the autoSKZCAM framework can shed light on discrepancies between different temperature-programmed desorption experiments, as we have found for \ce{CO2} on MgO(001).
Two measurements exist: one reports a weak $H_\textrm{ads}$ characteristic of physisorption~\cite{meixnerKineticsDesorptionAdsorption1992b}, and the other reports a strong $H_\textrm{ads}$ indicative of chemisorption~\cite{chakradharCarbonDioxideAdsorption2013a}.
For this particular case, our autoSKZCAM estimates agree with the chemisorption data but cannot replicate the physisorption results.
In Sec.~\ref{si-sec:co2_configurations} of the Supplementary Information, we highlight inconsistencies in the physisorption experiment that cast doubt on these measurements.
When these inconsistencies are accounted for, the experimental data aligns with our chemisorption estimates.

\section{Conclusion}

We have developed and implemented the autoSKZCAM framework to calculate accurate yet low cost adsorption enthalpies $H_\textrm{ads}$ of molecules on the surfaces of ionic materials.
This has enabled agreement to experimental measurements for a diverse set of 19 adsorbate--surface systems, beyond the accuracy of any density functional approximation considered, while being at a cost comparable to hybrid periodic DFT.
We have revealed new insights into several of these systems, notably: \ce{CO2} must bind on MgO(001) in the long-debated chemisorbed state and that it can take on a tilted configuration on \ce{TiO2} rutile(110), albeit close in stability to a horizontal parallel configuration; \ce{N2O} binds in a horizontal parallel fashion; \ce{NO} exists as bound dimers on MgO(001); and that \ce{CH3OH} and \ce{H2O} form partially dissociated H-bonded clusters on top of MgO(001).
In addition, we show that this dataset can be used to benchmark DFT exchange-correlation functionals and dispersion corrections, providing direct insights into their performance for adsorbate--surface systems.

This framework has been coded into an open-source package on Github~\cite{shiBenshi97AutoSKZCAM2025a}, making it  a readily available tool to compute accurate reference data for adsorbate--surface systems to facilitate reliable surface chemistry studies (elaborated in Sec.~\ref{si-sec:quacc} of the Supplementary Information).
This reliability will be paramount in complementing experiments towards understanding important catalytic reaction processes, serving to unlock new directions for improving such processes.
Furthermore, its automated nature means that it can serve as a standalone tool within a computational catalyst discovery pipeline to screen for new solid catalysts. 
Similarly, it can be used to provide large databases containing $E_\textrm{int}$ benchmarks that can be used to directly parametrise improved (machine-learned) DFAs~\cite{kirkpatrickPushingFrontiersDensity2021,bystromCIDERExpressiveNonlocal2022} and electronic structure methods.

Due to their technological relevance and the ready availability of experimental data, this work focuses on metal-oxide surfaces, but we expect the autoSKZCAM framework to be applicable to the surfaces of most ionic materials (possessing a band gap).
Evidence in support of this statement is provided in Sec.~\ref{si-sec:lih_nacl_test} of the Supplementary Information, where we have calculated the $E_\text{int}$ for both H$_2$O on LiH(001) and acetylene on NaCl(001), reaching agreement to available theoretical [DMC and CCSD(T)] estimates~\cite{tsatsoulisComparisonQuantumChemistry2017} in the former and experimental measurements~\cite{dunnInfraredSpectraStructure1992,cabello-cartagenaStructureInfraredAbsorption2010} in the latter.
Although this work provides new tools for answering questions about the surface chemistry of ionic materials, there are many important classes of systems that it cannot tackle. 
It will be important to develop new embedded cluster approaches that can treat transition metal surfaces~\cite{araujoAdsorptionEnergiesTransition2022c,sheldonHybridRPADFT2024,carboneCOAdsorptionPt1112024} and covalent materials like metal--organic frameworks (MOFs)~\cite{sillarInitioStudyHydrogen2009a,sillarInitioPredictionAdsorption2012} and zeolites~\cite{sauerInitioCalculationsMolecule2019b,bergerAdsorptionCrackingPropane2021a}.
%
%
Furthermore, it is desirable to go beyond a simple (local) harmonic description of $H_\textrm{ads}$ to treat anharmonicities and non-localised phenomena (i.e., adsorption at high temperatures).
Towards this end, there will be exciting prospects in integrating embedded cluster models with machine-learned interatomic potentials~\cite{daruCoupledClusterMolecular2022,chenDataEfficientMachineLearning2023a} to extend the system sizes tackled and enable finite temperature effects to be incorporated.

\section{\label{sec:methods_main}Methods}

The autoSKZCAM framework partitions the adsorption enthalpy $H_\textrm{ads}$ into several key contributions~\cite{sauerInitioCalculationsMolecule2019b}:
\begin{equation}
    H_\textrm{ads} = E_\textrm{int} + E_\textrm{rlx} + E_\textrm{ZPV} + E_\textrm{T} -  RT.
\end{equation}
The interaction energy $E_\textrm{int}$ is defined as the energetic difference between the adsorbate--surface complex and the separate adsorbate and surface, both of which are fixed at their geometries in the complex.
This term is treated up to the CCSD(T) level through the SKZCAM protocol~\cite{shiGeneralEmbeddedCluster2022b,shiGoingGoldStandardAttaining2024,shiManyBodyMethodsSurface2023a} developed by Shi, Kapil, Zen, Chen, Alavi and Michaelides~\cite{shiGeneralEmbeddedCluster2022b}.
The relaxation energy $E_\textrm{rlx}$ is the energy for the fixed adsorbate and surface to relax into their equilibrium geometries, while the zero-point vibrational and thermal contributions are given by $E_\textrm{ZPV}$ and $E_\textrm{T}$ respectively. 
These remaining terms can be treated adequately with DFT, where we utilise an ensemble of 6 widely-used DFAs.
When studying clusters or monolayers, there are additional terms for the lateral interaction energy between the molecules which are treated at the CCSD(T) level as discussed in Sec.~\ref{si-sec:wft_ecoh_econf} of the Supplementary Information.
For chemisorbed \ce{CO2} on MgO(001), there is also an additional term (coming out of $E_\textrm{rlx}$) for the large conformational energy change in the \ce{CO2} molecule.
For the dissociated \ce{H2O} and \ce{CH3OH} clusters, we include a dissociation contribution $E_\textrm{diss}$ which accounts for the energy change arising from dissociation of the parent molecularly adsorbed cluster.

The autoSKZCAM framework has been coded within an open-source package on Github~\cite{shiBenshi97AutoSKZCAM2025a}, with examples and documentation provided within.
It makes extensive use of the QuAcc workflow library~\cite{rosenQuaccQuantumAccelerator2024}, which can be used to generate the relaxed adsorbate--surface geometries starting from just a molecule and crystal unit cell, as discussed in Sec.~\ref{si-sec:quacc} of Supplementary Information.
If the adsorption configuration is not known, scripts are also provided to perform a random structure search to obtain candidate adsorption configurations.

\subsection{Accurate interaction energies with the SKZCAM protocol}

The dominant contribution to the final $H_\textrm{ads}$ from the autoSKZCAM framework is the CCSD(T) level $E_\textrm{int}$ computed with the SKZCAM protocol.
It addresses previous limitations in generalising embedded cluster approaches to different systems, defining rubrics for a converging series of clusters that can be generalised to the adsorption of molecules on diverse sets of ionic crystals and their surface terminations.
There is typically a smoother convergence of $E_\text{int}$ along the series of clusters, allowing the bulk infinite-size limit $E_\textrm{int}$ to be reached by extrapolating from a set of small clusters, as described in Sec.~\ref{si-sec:skzcam_si_sec} of the Supplementary Information.
We use the lower-level second-order M{\o}ller-Plesset perturbation theory (MP2) to perform this (bulk) extrapolation with moderately sized clusters (${<}75$ atoms).
This MP2 prediction can then be elevated to the CCSD(T) level from smaller (${<}35$ atom) clusters through an ONIOM-like~\cite{chungONIOMMethodIts2015} mechanical embedding approach.
Specifically, we utilise new local approximations to CCSD(T), resulting in $E_\textrm{int}$ estimates that are competitive with DFT in computational cost (see Sec.~\ref{si-sec:cost_benchmark} of the Supplementary Information).
While previously requiring significant user intervention, this protocol has been automated within the present work, now facilitating routine application of CCSD(T) to adsorbate--surface systems involving ionic materials.
Moreover, it enables a further lowering in computational cost by allowing for more levels/layers (i.e., smaller basis sets or a bigger frozen core) in the mechanical embedding, as discussed in Sec.~\ref{si-sec:improvements_skzcam} of the Supplementary Information.

The electrostatic embedding environment was constructed using py-Chemshell 20.0~\cite{luMultiscaleQMMM2023}, setting formal point charges [i.e., Ti(4+), Mg(2+), O(2-)] in a $50\,$\AA{} ($60\,$\AA{}) field around the quantum cluster center for the MgO(001) (\ce{TiO2} rutile(110) or anatase(101)) surface.
A further region of effective core potentials was placed on the positive point charges within $4\,$\AA{} ($6\,$\AA{}) of the quantum cluster to prevent spurious charge leakage.
MP2 was performed within ORCA 5.0.3~\cite{neeseORCAQuantumChemistry2020} with the resolution-of-identity approximation while CCSD(T) was performed within MRCC~\cite{kallayMRCCProgramSystem2020} using the local natural orbital (LNO) approximation~\cite{nagyOptimizationLinearScalingLocal2018,nagyApproachingBasisSet2019}.
A two-point (double-zeta/triple-zeta) complete basis set extrapolation~\cite{neeseRevisitingAtomicNatural2011}, together with counterpoise corrections, was used to calculate the MP2 and LNO-CCSD(T) $E_\textrm{int}$.
Subsequent basis set and core-valence correlation contributions are added at the MP2 level, as discussed in  Sec.~\ref{si-sec:cwft_details} of the Supplementary Information.

\subsection{Robust geometrical and vibrational contributions with an ensemble of DFAs}

The remaining terms (i.e., $E_\textrm{rlx}$, $E_\textrm{ZPV}$ and $E_\textrm{T}$) form a small overall contribution to $H_\textrm{ads}$ that can be adequately treated with DFT.
These terms are estimated by employing an ensemble~\cite{wellendorffDensityFunctionalsSurface2012,ruiBestDFTFunctional2024} of 6 widely-used DFAs of differing exchange-correlation functionals (up to hybrids) and dispersion corrections.
The 6 DFAs used for MgO(001) were PBE-D2[Ne], revPBE-D4, vdW-DF, rev-vdW-DF2, PBE0-D4, B3LYP-D2[Ne], where the [Ne] indicates the use of neon $C_6$ parameters for the Mg atoms.
The TiO$_2$ rutile(110) and anatase(101) surfaces used PBE-TS/HI, revPBE-D4, vdW-DF, rev-vdW-DF2, r$^2$SCAN-rVV10, HSE06-D4.
Through an averaging, this choice can provide estimates with corresponding $2\sigma$ standard deviations as an error measurement -- typically much better than chemical accuracy (see Sec.~\ref{si-sec:dft_geom_error} of the Supplementary Information).
We discuss in Sec.~\ref{si-sec:egeom_errors} how this ensemble can be further used to assess inaccuracies from using a DFT geometry for the CCSD(T) treatment.
As a result, this DFA ensemble allows for conservative errors bars on the final $H_\textrm{ads}$ estimate when comparing against experiments.

The DFT calculations were performed in the Vienna \textit{ab-initio} simulations package (VASP) 6.3.0~\cite{kresseEfficiencyAbinitioTotal1996a,kresseEfficientIterativeSchemes1996a}.
Of the 13 systems involving MgO(001), we used a ($4{\times}4$) supercell for all systems except for \ce{C6H6}, \ce{CH3OH} cluster and \ce{H2O} cluster, where a ($8{\times}8$) supercell was used.
The MgO(001) surface slab consisted of 4 layers, with the bottom two layers fixed.
The TiO\textsubscript{2} rutile(110) surface slab consisted of a ($4 {\times} 2$) supercell with 5 tri-layers (and the bottom three fixed), while the anatase(101) surface slab consisted of a ($3 {\times} 1$) supercell with 4 tri-layers and the bottom layer fixed.
All surfaces incorporated $15\,$\AA{} of vacuum with geometrical relaxation performed with a force convergence cutoff of $0.01\,$eV/\AA{}.
A plane-wave kinetic energy cutoff of $600\,$eV was used, which was reduced to $520\,$eV for the hybrid HSE06-D4 calculations on the TiO\textsubscript{2} surface systems.
A $2{\times}2{\times}1$ $\Gamma$-centered Monkhorst-Pack $k$-point mesh was used for the MgO(001) surface (reduced to only the $\Gamma$-point for the larger surface), $2{\times}2{\times}1$ mesh for the rutile(110) surface and $3{\times}3{\times}1$ mesh for anatase(101) surface.
To calculate $E_\textrm{ZPV}$ and $E_\textrm{T}$ contributions, the contributions from individual vibrational modes were computed with the quasi rigid-rotor harmonic oscillator (quasi-RRHO) approximation~\cite{grimmeSupramolecularBindingThermodynamics2012,liImprovedForceFieldParameters2015}.
Further details, particularly the parameters for the benchmarks in Fig.~\ref{fig:dft_compare}, are given in Sec.~\ref{si-sec:dft_details} and Sec.~\ref{si-sec:cost_benchmark} of the Supplementary Information.

\clearpage

\section{Extended Data}
\setcounter{figure}{0} 
\renewcommand{\thefigure}{Extended Data Fig. \arabic{figure}}

\begin{figure}
    \includegraphics[width=\textwidth]{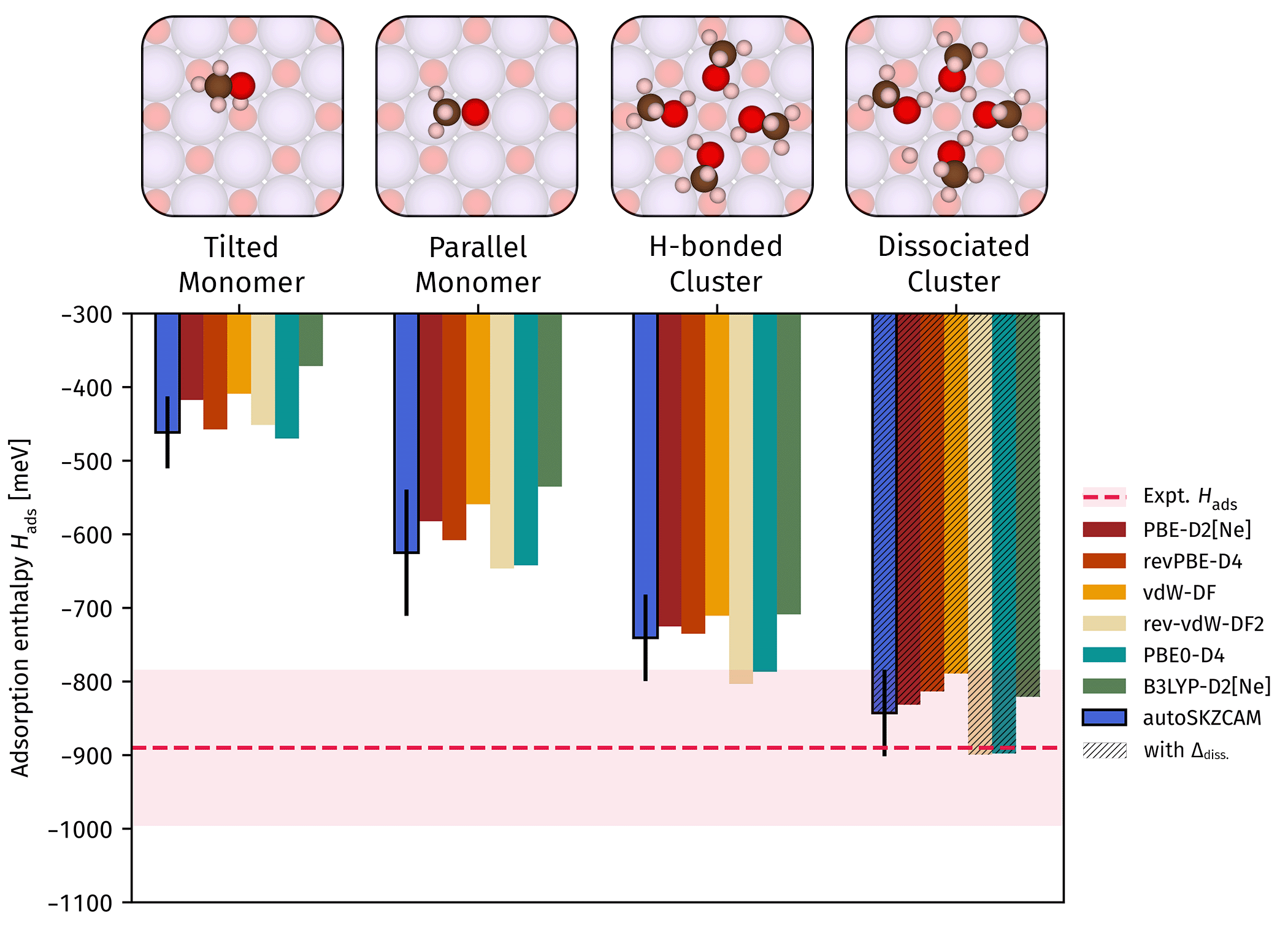}
    \caption{\label{fig:ch3oh_xc_hads}\textbf{Stabilisation of \ce{CH3OH} clusters on MgO(001) through H-bonding and dissociation.} Adsorption enthalpy $H_\textrm{ads}$ (per molecule) calculated with the autoSKZCAM framework and 6 different density functional approximations for \ce{CH3OH} on MgO(001). These are compared against TPD measurements by G{\"u}nster \etal{}~\cite{gunsterInteractionMethanolWater1999}. We consider the tilted and parallel adsorption configuration of a single \ce{CH3OH} molecule on MgO(001) as well as a molecularly adsorbed and dissociated tetramer. The $H_\textrm{ads}$ of the latter was computed by adding an additional term, $E_\textrm{diss}$, onto the molecularly adsorbed tetramer, as discussed in Sec.~\ref{si-sec:ediss_h2o_ch3oh} of the Supplementary Information. Conservative error estimates (corresponding to a 95\% confidence interval or more) are given for both experiment and autoSKZCAM, as discussed in Secs.~\ref{si-sec:final_hads} and~\ref{si-sec:exp_redhead_analysis} of the Supplementary Information, respectively.}
\end{figure}

\begin{figure}
    \includegraphics[width=0.8\textwidth]{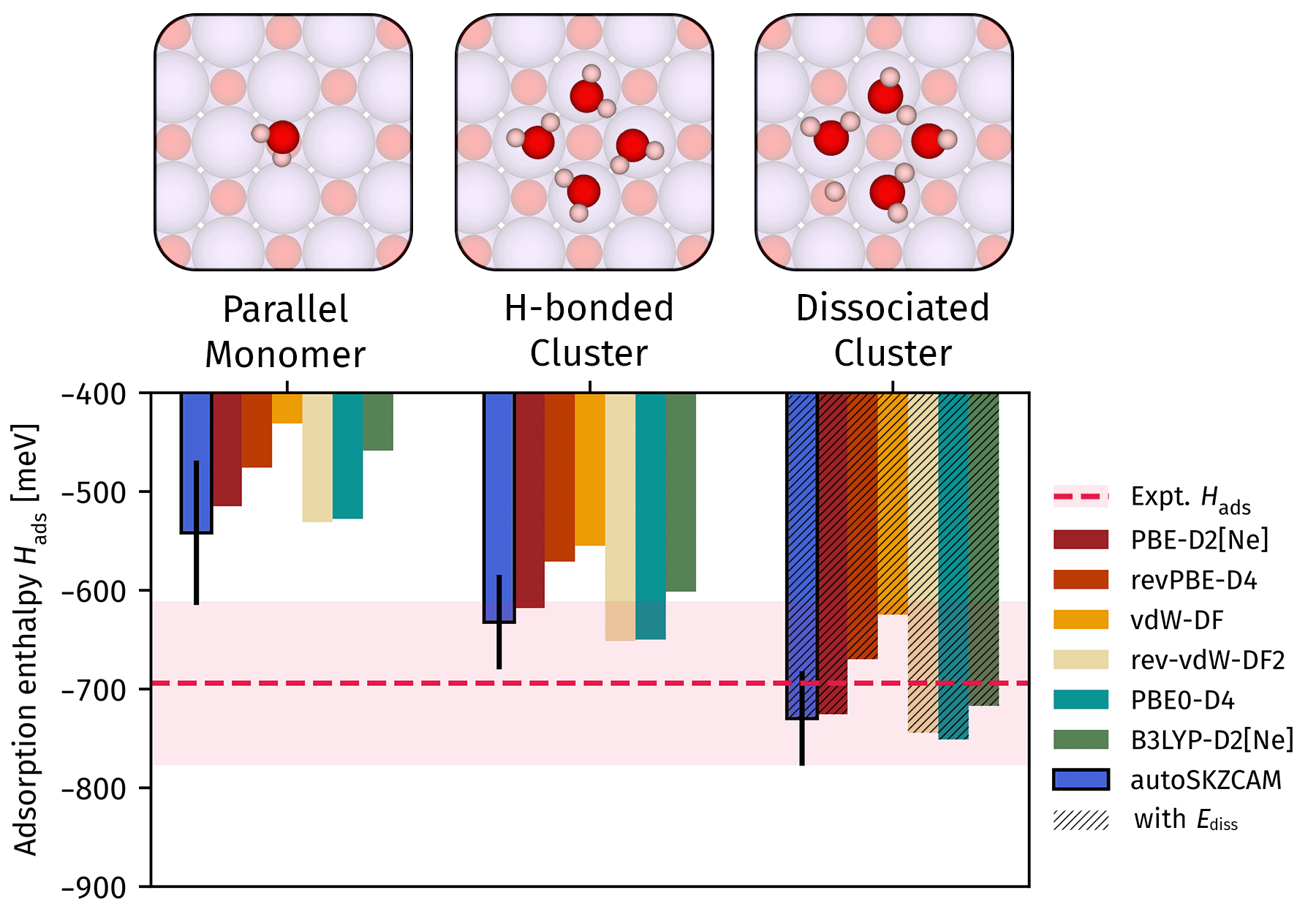}
    \caption{\label{fig:h2o_xc_hads}\textbf{Stabilisation of \ce{H2O} clusters on MgO(001) through H-bonding and dissociation.} Adsorption enthalpy $H_\textrm{ads}$ (per molecule) calculated with the autoSKZCAM framework and 6 different density functional approximations for \ce{H2O} on MgO(001). These are compared against TPD measurements by Stirniman \etal{}~\cite{stirnimanAdsorptionDesorptionWater1996}. We consider the adsorption of a single molecule, as well as a molecularly adsorbed and dissociated tetramer.  The $H_\textrm{ads}$ of the latter was computed by adding an additional term, $E_\textrm{diss}$, onto the molecularly adsorbed tetramer, as discussed in Sec.~\ref{si-sec:ediss_h2o_ch3oh} of the Supplementary Information. Conservative error estimates (corresponding to a 95\% confidence interval or more) are given for both experiment and autoSKZCAM, as discussed in Secs.~\ref{si-sec:final_hads} and~\ref{si-sec:exp_redhead_analysis} of the Supplementary Information, respectively.}
\end{figure}

\begin{figure}
    \includegraphics[width=\textwidth]{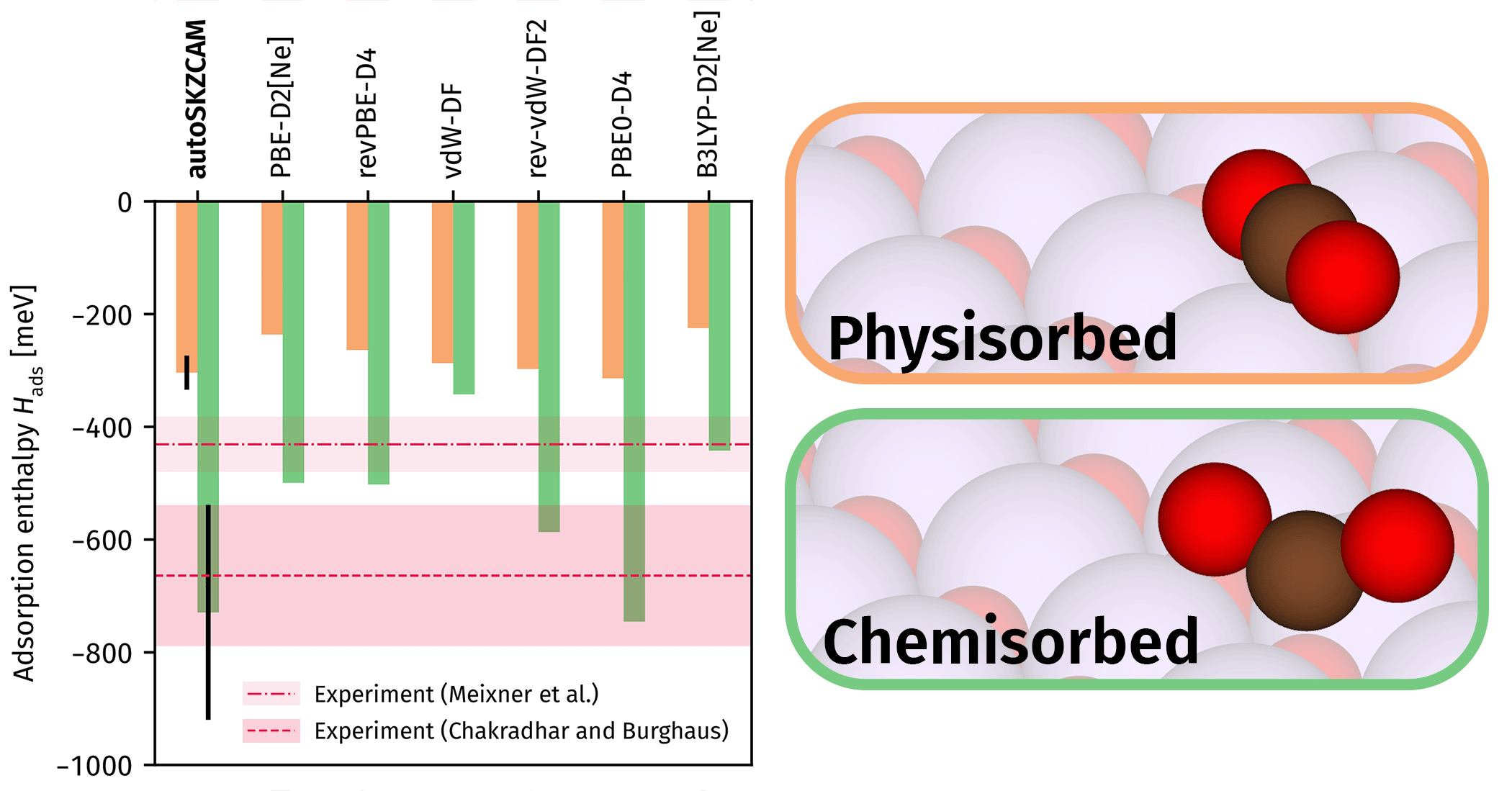}
    \caption{\label{fig:co2_configurations}\textbf{Resolving the chemisorbed state of \ce{CO2} on MgO(001).} Adsorption enthalpy $H_\textrm{ads}$ calculated with the autoSKZCAM framework and 6 different density functional approximations for the physisorbed (orange) and chemisorbed (green) state of \ce{CO2} on MgO(001). These are compared against TPD measurements by Meixner \etal{}~\cite{meixnerKineticsDesorptionAdsorption1992b} and Chakradhar and Burghaus~\cite{chakradharCarbonDioxideAdsorption2013a}, which propose a physisorbed and chemisorbed state, respectively. Conservative error estimates (corresponding to a 95\% confidence interval or more) are given for both experiment and autoSKZCAM, as discussed in Secs.~\ref{si-sec:final_hads} and~\ref{si-sec:exp_redhead_analysis} of the Supplementary Information, respectively.}
\end{figure}

\begin{figure}
    \includegraphics[width=\textwidth]{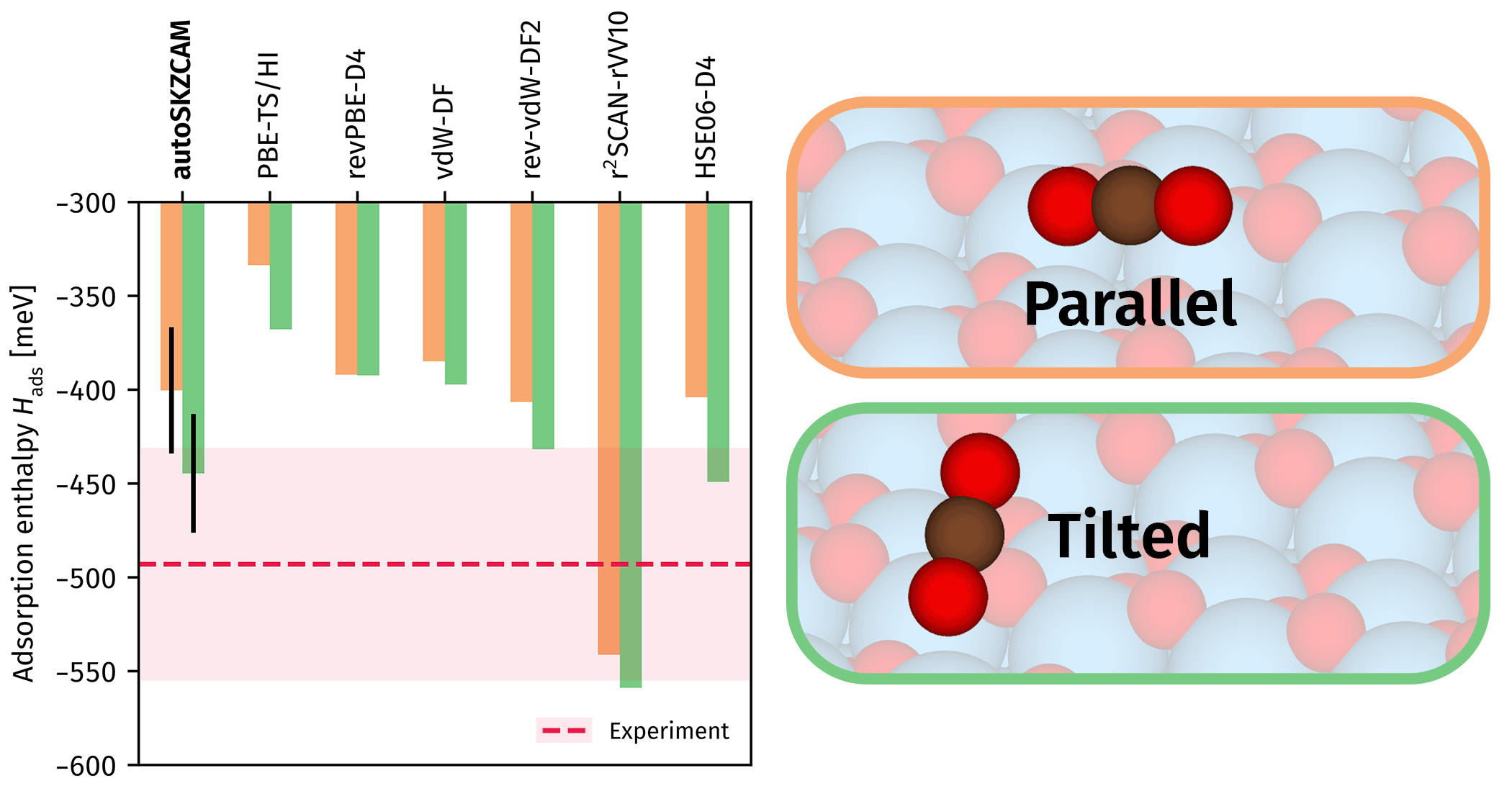}
    \caption{\label{fig:rutile_co2_configurations}\textbf{Preference towards the tilted configuration of \ce{CO2} on the \ce{TiO2} rutile(110).} Adsorption enthalpy $H_\textrm{ads}$ calculated with the autoSKZCAM framework and 6 different density functional approximations for the parallel (orange) and tilted (green) adsorption configuration of \ce{CO2} on \ce{TiO2} rutile(110). These are compared against TPD measurements by Thompson \etal{}~\cite{thompsonCO2ProbeMonitoring2003}.  Conservative error estimates (corresponding to a 95\% confidence interval or more) are given for both experiment and autoSKZCAM, as discussed in Secs.~\ref{si-sec:final_hads} and~\ref{si-sec:exp_redhead_analysis} of the Supplementary Information, respectively.}
\end{figure}

\begin{figure}
    \includegraphics[width=\textwidth]{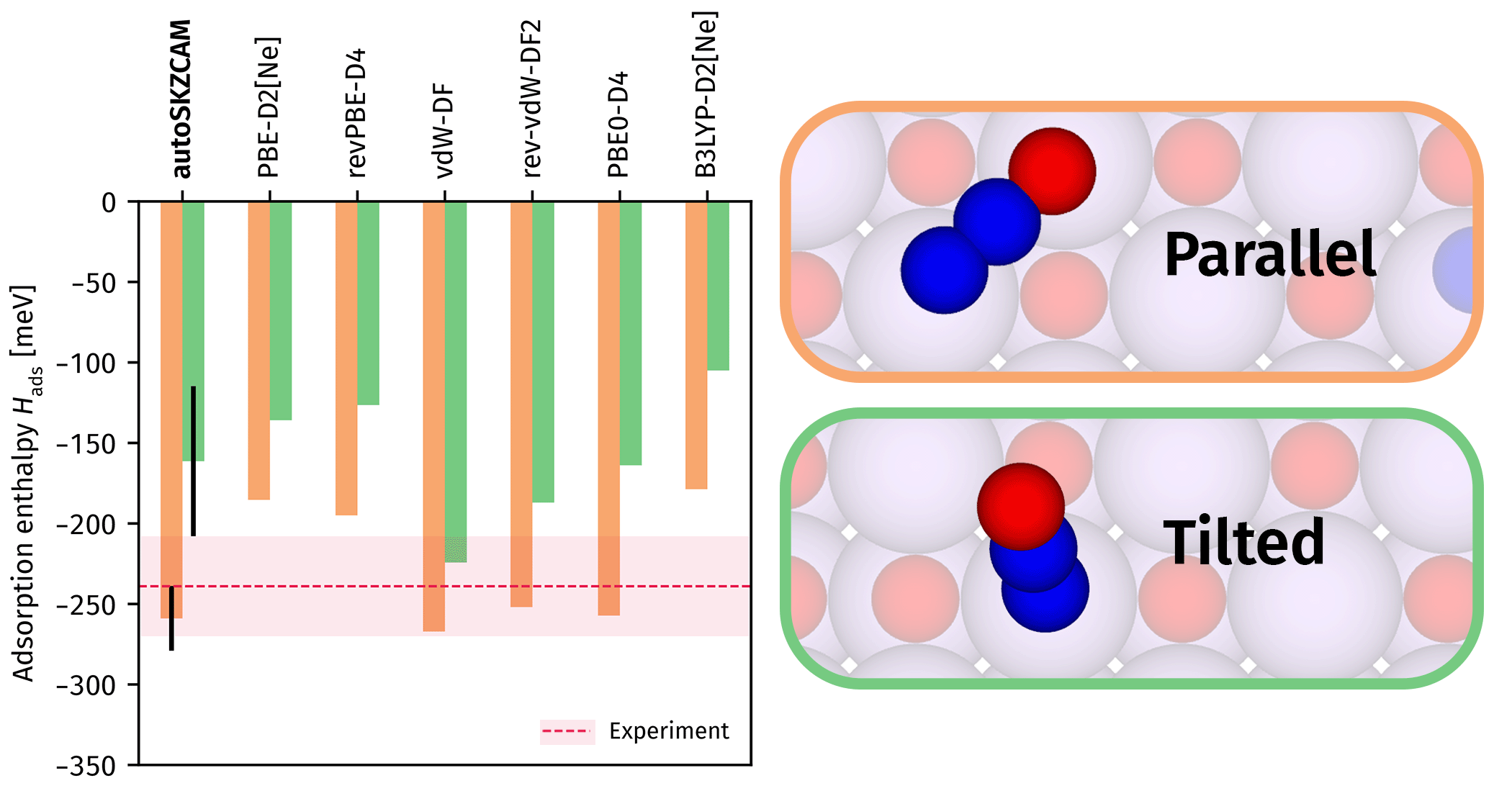}
    \caption{\label{fig:n2o_configurations}\textbf{\ce{N2O} on MgO(001) adopts a parallel configuration.} Adsorption enthalpy $H_\textrm{ads}$ calculated with the autoSKZCAM framework and 6 different density functional approximations for the parallel (orange) and tilted (green) adsorption configuration of \ce{N2O} on MgO(001). These are compared against TPD measurements by Lian \etal{}~\cite{lianN2OAdsorptionSurface2010b}. Conservative error estimates (corresponding to a 95\% confidence interval or more) are given for both experiment and autoSKZCAM, as discussed in Secs.~\ref{si-sec:final_hads} and~\ref{si-sec:exp_redhead_analysis} of the Supplementary Information, respectively.}
\end{figure}

\clearpage

\section{Code Availability}
The autoSKZCAM framework is freely available at \href{https://github.com/benshi97/autoSKZCAM}{https://github.com/benshi97/autoSKZCAM}, with documentation found at \href{https://www.benjaminshi.com/autoSKZCAM/}{https://www.benjaminshi.com/autoSKZCAM/} containing instructions and examples on how to run the code.

\begin{suppinfo}

See the supplementary information for a detailed compilation of the obtained results as well as further data and analysis to support the points made throughout the text. The input and output files associated with this work and all analysis can be found on GitHub at \href{https://github.com/benshi97/Data_autoSKZCAM}{benshi97/Data\_autoSKZCAM} or viewed (and analysed) online on \href{https://colab.research.google.com/github/benshi97/Data_autoSKZCAM/blob/master/analyse.ipynb}{Colab}.

\end{suppinfo}

\begin{acknowledgement}

The Flatiron Institute is a division of the Simons Foundation.
We thank Timothy Berkelbach for fruitful discussions on quantum embedding.
AZ acknowledges support from the European Union under the Next generation EU (projects 20222FXZ33 and P2022MC742).
TS acknowledges support from the Austrian Science Fund (FWF) [DOI:10.55776/ESP335].
BXS acknowledges support from the EPSRC Doctoral Training Partnership (EP/T517847/1).
AM and BXS acknowledges support from the European Union under the ``n-AQUA'' European Research Council project (Grant No.\ 101071937).
The authors are grateful for: resources provided by the Cambridge Service for Data Driven Discovery (CSD3) operated by the University of Cambridge Research Computing Service (\href{www.csd3.cam.ac.uk}{www.csd3.cam.ac.uk}), provided by Dell EMC and Intel using Tier-2 funding from the Engineering and Physical Sciences Research Council (capital grant EP/P020259/1), and DiRAC funding from the Science and Technology Facilities Council (\href{www.dirac.ac.uk}{www.dirac.ac.uk}); the Cirrus UK National Tier-2 HPC Service at EPCC (\href{http://www.cirrus.ac.uk}{http://www.cirrus.ac.uk}) funded by the University of Edinburgh and EPSRC (EP/P020267/1); the Vienna Scientific Cluster (VSC); computational resources granted by the UCL Myriad and Kathleen High Performance Computing Facility (Myriad@UCL and Kathleen@UCL), and associated support service; computational support from the UK Materials and Molecular Modelling Hub, which is partially funded by EPSRC (EP/P020194 and EP/T022213); and computational support from the UK national high performance computing service, ARCHER 2.
Both the UK Materials and Molecular Modelling Hub and ARCHER 2 access was obtained via the UKCP consortium and funded by EPSRC grant ref EP/P022561/1.

\end{acknowledgement}

\providecommand{\latin}[1]{#1}
\makeatletter
\providecommand{\doi}
  {\begingroup\let\do\@makeother\dospecials
  \catcode`\{=1 \catcode`\}=2 \doi@aux}
\providecommand{\doi@aux}[1]{\endgroup\texttt{#1}}
\makeatother
\providecommand*\mcitethebibliography{\thebibliography}
\csname @ifundefined\endcsname{endmcitethebibliography}
  {\let\endmcitethebibliography\endthebibliography}{}

\end{document}



\title{Supplementary Information for: \\ An accurate and efficient framework for modelling the surface chemistry of ionic materials}

\author{Benjamin X. Shi}
\affiliation{Yusuf Hamied Department of Chemistry, University of Cambridge, Lensfield Road, Cambridge CB2 1EW, United Kingdom}%
\affiliation{Initiative for Computational Catalysis, Flatiron Institute, 160 5th Avenue, New York, NY 10010}

\author{Andrew S. Rosen}
\affiliation{Department of Chemical and Biological Engineering, Princeton University, Princeton, NJ 08540 USA}

\author{Tobias Sch\"{a}fer}
\affiliation{Institute for Theoretical Physics, TU Wien, Wiedner Hauptstra{\ss}e 8-10/136, 1040 Vienna, Austria}

\author{Andreas Gr\"{u}neis}
\affiliation{Institute for Theoretical Physics, TU Wien, Wiedner Hauptstra{\ss}e 8-10/136, 1040 Vienna, Austria}

\author{Venkat Kapil}
\affiliation{Yusuf Hamied Department of Chemistry, University of Cambridge, Lensfield Road, Cambridge CB2 1EW, United Kingdom}
\affiliation{Department of Physics and Astronomy, University College London, 7-19 Gordon St, London WC1H 0AH, UK}
\affiliation{Thomas Young Centre and London Centre for Nanotechnology, 9 Gordon St, London WC1H 0AH}

\author{Andrea Zen}
\affiliation{Dipartimento di Fisica Ettore Pancini, Universit\`{a} di Napoli Federico II, Monte S. Angelo, I-80126 Napoli, Italy}
\affiliation{Department of Earth Sciences, University College London, Gower Street, London WC1E 6BT, United Kingdom}

\author{Angelos Michaelides}
\affiliation{Yusuf Hamied Department of Chemistry, University of Cambridge, Lensfield Road, Cambridge CB2 1EW, United Kingdom}%
\email{am452@cam.ac.uk}

\date{\today}

\maketitle
\newpage

\tableofcontents
\newpage

We provide additional supporting data as well as contextual information to the main text here. 
All output files are provided on \href{https://github.com/benshi97/Data_autoSKZCAM}{Github}~\cite{shiResearchDataSupporting2024}, with a corresponding Jupyter Notebook file that analyses all of the data.
This data can also be viewed and analysed on the browser with \href{https://colab.research.google.com/github/benshi97/Data_autoSKZCAM/blob/master/analyse.ipynb}{Google Colab}. 

Within this supplementary data, we start by providing more detail on two key developments of this work:
\begin{itemize}
    \item Firstly, we highlight the atomic-level insights provided by the autoSKZCAM framework into the adsorption configuration for several of the studied systems in Sec.~\ref{sec:system_discuss}, together with a discussion of the prior literature.
    \item Then, we tabulate the CCSD(T)-level dataset of references on the interaction energy for the set of adsorbate--surface system studied within this work, provided in Sec.~\ref{sec:final_eint_estimates}. These values can serve as a useful benchmark tool for assessing the performance of newly developed density functional approximations as well as approximate correlated wave-function methods.
\end{itemize}

In the second part to the supplementary information, we provide more elaborate details and concrete numbers to support the claims made within the main text:
\begin{itemize}
    \item We discuss the approach we take towards computing the adsorption enthalpy in Sec.~\ref{sec:adsorption_enthalpy_def} which forms the basis of the autoSKZCAM framework developed within the present work.
    \item We show the adsorbate--surface complexes studied within this work in Sec.~\ref{sec:adsorbate-surface_systems}.
    \item  We give the details of the correlated wave-function methods [MP2 and CCSD(T)] used in this work in Sec.~\ref{sec:cwft_details}.
    \item We describe the SKZCAM protocol -- used to calculate the interaction energy contribution to the adsorption enthalpy -- in Sec.~\ref{sec:skzcam_si_sec}. The set of clusters generated by this protocol and the MP2 and CCSD(T) estimates for each cluster are also provided.
    \item  We discuss additional contributions calculated at the CCSD(T)-level, such as conformational energy and cohesive energy (for the clusters and monolayers) contributions to the final adsorption enthalpy in Sec.~\ref{sec:wft_ecoh_econf}.
    \item We describe how the remaining contributions to the adsorption enthalpy -- geometrical relaxation, zero-point vibrational and thermal -- are calculated using an ensemble of density functional approximations in Sec.~\ref{sec:dft_geom_error}.
    \item We provide the final adsorption enthalpy estimates made with the autoSKZCAM framework in Sec.~\ref{sec:final_hads} and make some additional validation tests on its reliability in Sec.~\ref{sec:error_validation}.
    \item We describe how the autoSKZCAM framework has been automated using the QuAcc computational materials science workflow library in Sec.~\ref{sec:quacc}.
    \item We describe how we analyse accurate experimental adsorption enthalpies with reliable error bars in Sec.~\ref{sec:exp_redhead_analysis}.
    \item The final autoSKZCAM adsorption enthalpy estimates are compared to experiments in Sec.~\ref{sec:autoskzcam_exp_compare}.
    \item These estimates are further compared to the previous literature in Sec.~\ref{sec:comp_literature}.
    \item Finally, we benchmark and highlight the low-cost of the autoSKZCAM framework relative to DFT in Sec.~\ref{sec:cost_benchmark}.
    
\end{itemize}

\cleardoublepage 
\thispagestyle{empty} 
\phantomsection 
\addcontentsline{toc}{section}{Part I: Additional results and discussion} 

\begin{center}
    \vspace*{\fill} 
    {\LARGE Part I: Further insights and discussion}
    \vspace*{\fill}
\end{center}

\clearpage 

\section{\label{sec:system_discuss} Insights into adsorption configuration}

We now discuss specific subset of systems within the 19 studied adsorbate--surface systems (see Sec.~\ref{sec:adsorbate-surface_systems}) where there have been debates on the adsorption configuration.
We apply the autoSKZCAM framework to compute $H_\textrm{ads}$ for each of these systems and show that the lowest energy configuration has an $H_\textrm{ads}$ that agrees with experiment; with all other configurations predicted to be less stable with an $H_\textrm{ads}$ that cannot reproduce experiment.

\subsection{\label{sec:ch3oh_discuss}Cluster \ce{CH3OH} and \ce{H2O} on MgO(001)}

\begin{figure}[h]
    \includegraphics[width=\textwidth]{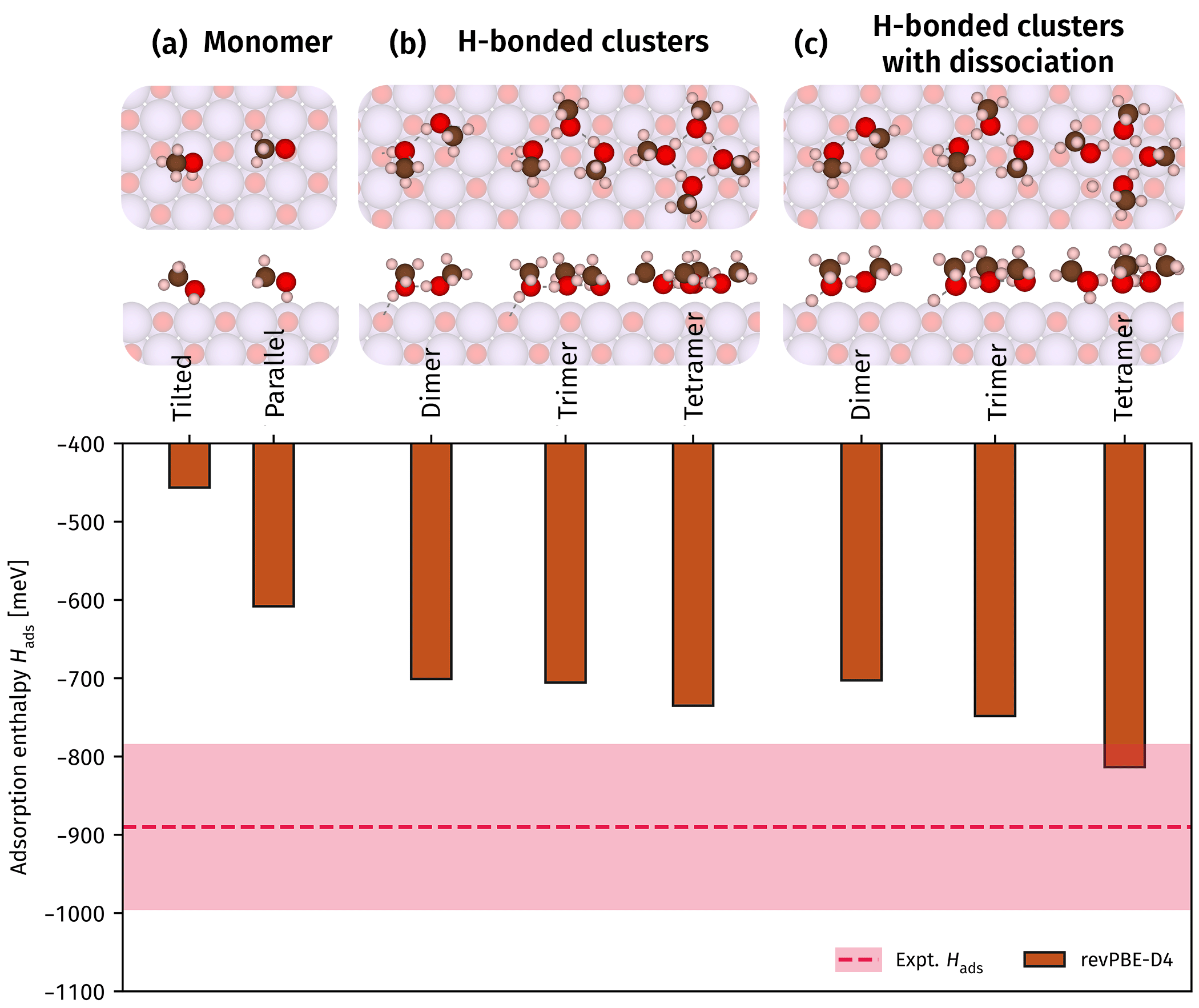}
    \caption{\label{fig:ch3oh_dissociation} Comparison of the $H_\textrm{ads}$ (per molecule) calculated with the revPBE-D4 functional for the adsorption of (a) monomer \ce{CH3OH} on MgO(001), (b) H-bonded clusters and (c) H-bonded \ce{CH3OH} cluster with partial dissociation. We consider the lowest energy dimer, trimer and tetramer adsorption configurations. These are compared against experimental TPD measurements by G{\"u}nster \etal{}~\cite{gunsterInteractionMethanolWater1999}. Conservative error estimates (corresponding to a 95\% confidence interval or more) are given for both experiment and autoSKZCAM, as discussed in Secs.~\ref{sec:final_hads} and~\ref{sec:exp_redhead_analysis}, respectively.}
\end{figure}

For both \ce{CH3OH} and \ce{H2O} on MgO(001), we have observed that it is necessary to account for H-bonded clustering of the molecules on the surface together with partial dissociation of the cluster to achieve agreement on $H_\textrm{ads}$ with experiments.
In Fig.~\ref{fig:ch3oh_dissociation}, we compare the $H_\textrm{ads}$ with the revPBE-D4~\cite{zhangCommentGeneralizedGradient1998,caldeweyherGenerallyApplicableAtomiccharge2019} density functional approximation (DFA) for several adsorption configurations of \ce{CH3OH} on MgO(001), involving a single molecule up to clusters adsorbed on the surface.
This DFA was chosen because it can successfully reproduce autoSKZCAM estimates in Extended Data Fig.~1 and~2 of the main text.
For single molecules, we considered both a tilted~\cite{petitjeanQuantitativeInvestigationMgO2010a,a.sainnaCombinedPeriodicDFT2021a} and parallel~\cite{brandaTheoreticalStudyCharge2002b,trabelsiThermodynamicStructuralStudy2004b} adsorption configuration.
These have been proposed within previous literature and we find that the tilted configuration is more stable, albeit unable to reproduce experiment.
This hints at missing contributions found in e.g., H-bonded and dissociated clusters.
For the H-bonded clusters, we have considered the lowest-energy, as discovered through a random structure search, geometries for the dimer, trimer and tetramer, while dissociation is induced by moving an H-atom onto a nearby O atom.
There is significant stabilisation for the H-bonded tetramer of \ce{CH3OH} relative to the dimer or trimer, as it can form a complete H-bonded network commensurate with the underlying geometry of the MgO(001) surface.
However, this is still insufficient for reproduce the experimental $H_\textrm{ads}$.
There is further stabilisation when dissociation is induced.
While dissociation is isoenergetic (to within $2\,$meV) to the molecular form for the dimer, there is stabilisation of $43\,$meV for the trimer, which goes up to $78\,$meV for the tetramer.
This stabilisation brings the dissociated tetramer $H_\textrm{ads}$ into agreement with experiments.
Such stabilisation upon dissociation has also been observed for clusters of \ce{H2O}, where we find that dissociation stabilises the \ce{H2O} tetramer by $81\,$meV (Extended Data Fig.~2 of the main text), which would bring the autoSKZCAM estimate into excellent agreement with experiment.


In Extended Data Figs.~1 and~2 of the main text, we plot the $H_\textrm{ads}$ for \ce{H2O} and \ce{CH3OH} on MgO(001) for the monomer and tetramer configurations using the autoSKZCAM framework and a set of density functional approximations.
As discussed in Sec.~\ref{sec:ediss_h2o_ch3oh}, we predict the $H_\textrm{ads}$ for the dissociated form of the tetramer by using the DFT ensemble to calculate $E_\textrm{diss}$ -- the stabilisation energy to form the dissociated cluster from the (molecular) H-bonded cluster.
This enables excellent agreement to the experimental $H_\textrm{ads}$ for both \ce{H2O} and \ce{CH3OH}.


Our insights are in agreement with experimental evidence, which suggests that even at low coverage limits, \ce{CH3OH} molecules will form 2D islands~\cite{rudbergAdsorptionMethanolMgO1002004}.
In particular, we suggest that these 2D islands contain H-bonded networks together with partial dissociation.
It should be noted that while an $H_\textrm{ads}$ value was obtained for a single-monomer of \ce{H2O} in Refs.~\citenum{ferryPropertiesTwodimensionalWater1997a,ferryWaterMonolayersMgO1001998}, this was done by subtracting the lateral interactions (including both H-bonding and partial dissocation) of a 2D monolayer of water from the monolayer $H_\textrm{ads}$ of \ce{H2O}.
Physically, we expect a similar behaviour for \ce{H2O} molecules on MgO(001), forming clusters even at the low coverage limits.

\subsection{\label{sec:co2_configurations} \ce{CO2} on MgO(001)}

The adsorption of \ce{CO2} on \ce{MgO} has been the subject of many theoretical and experimental studies over the years.
These have largely revolved around a physisorbed or chemisorbed (`monodentate') state.
Within experiments, early work by Meixner \etal{}~\cite{meixnerKineticsDesorptionAdsorption1992b} used laser-induced thermal desorption experiments to predict an $H_\textrm{ads}$ of $-431\,$meV, which was attributed to a physisorbed state due to its low (absolute) value.
However, a TPD experiment by Chakradhar and Burghaus~\cite{chakradharCarbonDioxideAdsorption2013a} came to an estimate of $-664\,$meV on $H_\textrm{ads}$ and attributed this to the formation of surface carbonates (i.e., chemisorption) as confirmed via XPS, with no evidence of a physisorbed state.
On the simulations front, there are several studies which have found the chemisorbed state to be completely unstable~\cite{pacchioniInitioClusterModel1994b,jensenCO2SorptionMgO2005b}, predicting a physisorbed state~\cite{pacchioniPhysisorbedChemisorbedCO21993a,hammamiCO2Adsorption0012008a,cornuLewisAcidoBasicInteractions2012b,manaeActivationCO2CH42022}, while many have also predicted a chemisorbed state to show significant stability~\cite{baltrusaitisPeriodicDFTStudy2012a,downingReactivityCO2MgO2013,mazheikaNiSubstitutionalDefects2016a}.


We use the autoSKZCAM framework to calculate the $H_\textrm{ads}$ for both the physisorbed and chemisorbed structure in Extended Data Fig.~3 of the main text.
We find that the autoSKZCAM framework comes into good agreement with the experiment of Chakradhar and Burghaus~\cite{chakradharCarbonDioxideAdsorption2013a} for the chemisorbed state while it does not come into agreement for the physisorbed state.
The DFAs all predict the chemisorbed state to be more stable than the physisorbed state by more than $200\,$meV, except for vdW-DF, where the differences are less than $30\,$meV.
It can be seen that while the chemisorbed state is predicted to be the most stable by all DFAs, most of the DFAs are unable to reach agreement with experiment.
Similarly, most DFAs are unable to match with the experiment by Meixner \etal{} for the physisorbed state.
This suggests potential errors within the original measurements by Meixner \etal{}.
For example, Chakradhar and Burghaus have surmised that the temperature reading (${\sim}120\,$K) of Meixner \etal{} was `un-calibrated' as the low desorption temperatures do not agree with previous TPD measurements~\cite{yanagisawaInteractionCO2Magnesium1995b,itoRolesLowcoordinatedSurface1997} including their own (${\sim}230\,$K).
In fact, data by Meixner \etal{} indicates a `lack of surface mobility' of the \ce{CO2}, which points towards a chemisorbed state.
If the measurements by Meixner \etal{} were re-analysed with a desorption temperature of $230\,$K, their experiment would predict $H_\textrm{ads}$ of $-826{\pm}94\,$meV, in good agreement with our autoSKZCAM predictions.
Beyond previous TPD measurements, there has been significant recent interest in MgO for \ce{CO2} storage~\cite{mcqueenAmbientWeatheringMagnesium2020b,donatProspectsMgObasedSorbents2022}, with the chemisorbed state now supported with new evidence from NMR experiments~\cite{fuSimultaneousCharacterizationSolid2018,duIdentificationCO2Adsorption2022}.

\subsection{\label{sec:no_si_config}\ce{NO} on MgO(001)}

\begin{figure}[h]
    \includegraphics[width=\textwidth]{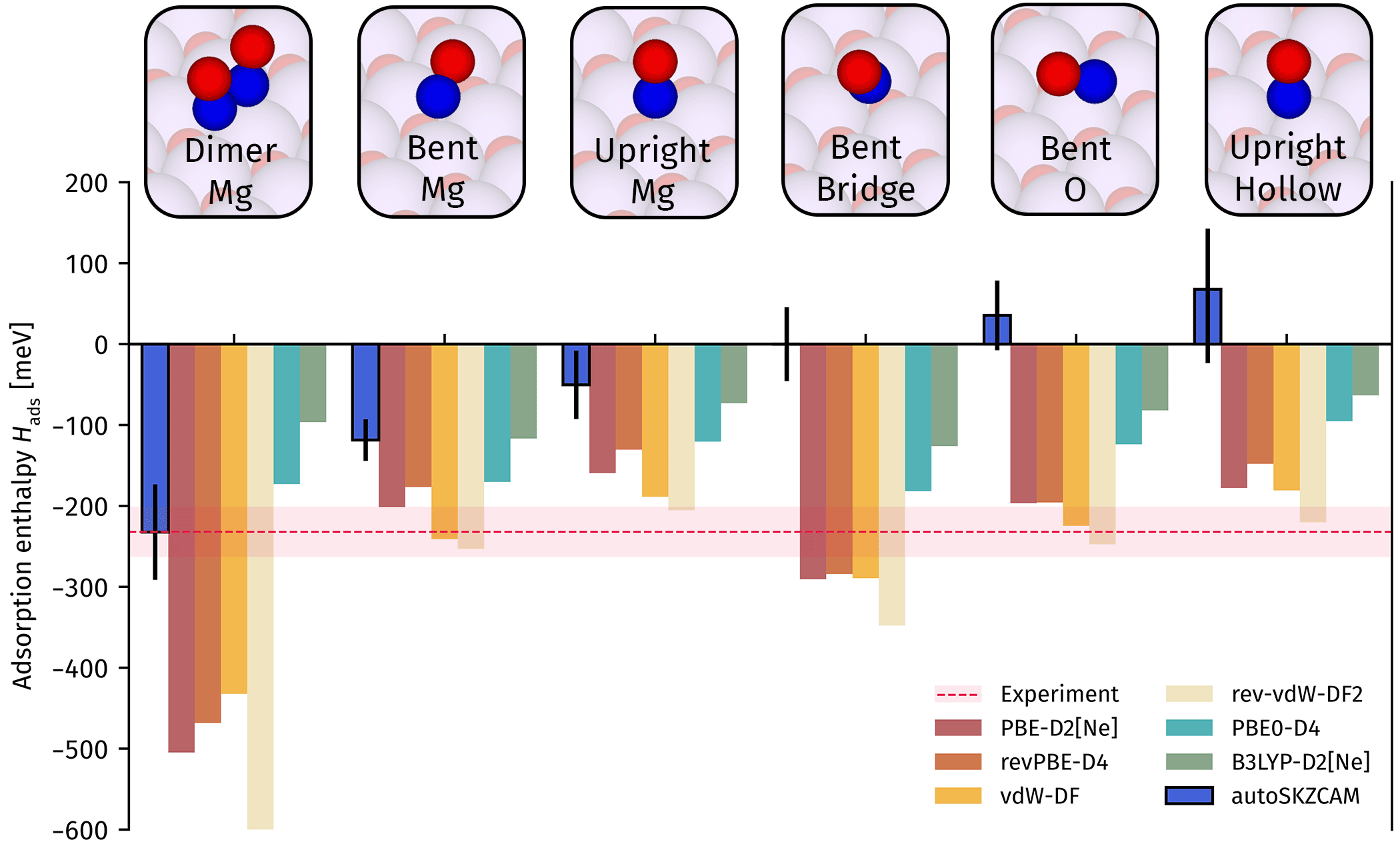}
    \caption{\label{fig:no_configurations}Comparison of $H_\textrm{ads}$ (per molecule) calculated with the autoSKZCAM framework and several DFAs (from the ensemble) for the various adsorption configurations of \ce{NO} on the MgO(001) surface. These are compared against experimental TPD measurements by Wichtendahl \etal{}~\cite{wichtendahlThermodesorptionCONO1999}. Conservative error estimates (corresponding to a 95\% confidence interval or more) are given for both experiment and autoSKZCAM, as discussed in Secs.~\ref{sec:final_hads} and~\ref{sec:exp_redhead_analysis}, respectively.}
\end{figure}

The adsorption of nitric oxide on the MgO(001) surface has been widely studied by both experiments and computational simulations.
Within the computational literature, a wide range of adsorption configurations have been proposed, which we show in Fig.~3 of the main text.
For each given geometry, most studies (see Table~\ref{tab:comp_lit}) employing DFT have found $E_\textrm{ads}$ in a range in agreement with the experimental $H_\textrm{ads}$ estimate~\cite{wichtendahlThermodesorptionCONO1999}.
In particular, most studies have looked at the adsorption of monomers on the surface.
On the other hand, experiments have largely pointed towards the absence of monomers on the surface.
For example, EPR~\cite{divalentinNOMonomersMgO2002a} indicates that only 0.5\% of sites contained NO monomers (bonded mostly to defects), with FTIR~\cite{plateroSpectroscopicStudyNO1985a} showing that most NO species exist as a closed-shell diamagnetic \textit{cis-}(NO)$_2$ species, hereafter dubbed the NO dimer.

In Fig.~\ref{fig:no_configurations}, we have studied all proposed (monomer and dimer) geometries of NO on the MgO(001) surface with the autoSKZCAM framework and the ensemble of DFAs (providing specific values in Table~\ref{tab:no_configurations_dft_hads}).
We find that the monomer species are not as stable as many DFAs predict, being $100\,$meV higher in $H_\textrm{ads}$ than the NO dimer. 
On top of being the configuration with the lowest $H_\textrm{ads}$, the autoSKZCAM framework also predicts an $H_\textrm{ads}$ which reproduces experiment, coming into agreement with previous experimental evidence that NO exists in its \textit{cis-}(NO)$_2$ geometry on top of the MgO surface.
In addition, our estimates show that monomers are unlikely to exist on the terrace sites at low temperatures, with only the NO dimer expected to form.

\begin{table}
\caption{\label{tab:no_configurations_dft_hads}Comparison between autoSKZCAM and 6 DFAs in their predicted $H_\text{ads}$ (in meV) for the different configurations of NO on MgO(001). The experimental $H_\text{ads}$ is $-232 {\pm} 31\,$meV.}
\begin{tabular}{lrrrrrrr}
\toprule
 & \rotatebox{90}{autoSKZCAM} & \rotatebox{90}{PBE-D2[Ne]} & \rotatebox{90}{revPBE-D4} & \rotatebox{90}{vdW-DF} & \rotatebox{90}{rev-vdW-DF2} & \rotatebox{90}{PBE0-D4} & \rotatebox{90}{B3LYP-D2[Ne]} \\ 
\midrule
Dimer & -232 ± 59 & -505 ± 5 & -468 ± 5 & -432 ± 5 & -607 ± 5 & -173 ± 5 & -96 ± 5 \\
Bent-Mg & -119 ± 26 & -201 ± 3 & -177 ± 3 & -241 ± 3 & -253 ± 3 & -170 ± 3 & -117 ± 3 \\
Vertical-Mg & -50 ± 42 & -159 ± 3 & -130 ± 3 & -189 ± 3 & -205 ± 3 & -120 ± 3 & -73 ± 3 \\
Bent-Bridge & 0 ± 46 & -291 ± 5 & -284 ± 5 & -289 ± 5 & -348 ± 5 & -182 ± 5 & -126 ± 5 \\
Bent-O & 35 ± 43 & -197 ± 3 & -196 ± 3 & -225 ± 3 & -247 ± 3 & -124 ± 3 & -82 ± 3 \\
Vertical-Hollow & 68 ± 91 & -178 ± 6 & -148 ± 6 & -181 ± 6 & -220 ± 6 & -95 ± 6 & -63 ± 6 \\
\bottomrule
\end{tabular}
\end{table}

\subsection{\label{sec:co2_rutile_config}\ce{CO2} on \ce{TiO2} rutile(110)}

The adsorption of \ce{CO2} on \ce{TiO2} rutile(110) has been the subject of many theoretical and experimental studies over the years.
Besides one study~\cite{kubasSurfaceAdsorptionEnergetics2016c}, the majority of computational simulations have demonstrated \ce{CO2} on \ce{TiO2} rutile(110) to take on a tilted geometry on top of the surface.
Similarly, barring one experiment, the majority of experiments have pointed towards a tilted geometry on the rutile(110) surface.
A study by Sorescu \etal{}~\cite{sorescuCO2AdsorptionTiO22011} found that a simulated STM image of the parallel configuration was more consistent with experimental STM images.
However, a later study with some of the original authors showed a preference for a tilted geometry~\cite{leeDiffusionCO2Rutile2011} which rotates about the $z-$axis to diffuse across the \ce{TiO2} rutile(110) surface.
This was later confirmed by further independent studies with STM~\cite{linStructureDynamicsCO22012} and FTIR~\cite{caoAdsorptionInteractionCO22015}.

In Extended Data Fig.~4, we have compared the $H_\textrm{ads}$ computed with the autoSKZCAM framework (and several DFAs) against experiment for the parallel and tilted geometry.
We find that the tilted geometry is more stable than the parallel geometry by $45\,$meV, albeit with some overlap in their error bars.
With the tilted geometry, we are able to reach agreement with experiment, providing strong evidence that it is the expected geometry on the \ce{TiO2} rutile(110) surface, although the small energy difference with the parallel state suggests that under normal temperatures, it can easily move into this metastable state, commensurate with its easy diffusion observed in experiments.
We find that all of the DFAs also predict a more stable tilted geometry, although revPBE-D4 predicts the two geometries to be nearly isoenergetic with a $0.3\,$meV difference.


Our work has also allowed us to re-examine some of the previous experimental and computational work that have predicted a more stable parallel adsorption configuration.
The experiment of Sorescu \etal{}~\cite{sorescuCO2AdsorptionTiO22011} with STM had indicated a more consistent agreement towards a parallel geometry.
However, this is in disagreement with later STM studies and recent work~\cite{e.hamlynImagingOrderingWeakly2018} have actually simulated the STM images of a tilted, parallel and vertical geometry of \ce{CO2} on rutile(110).
The differences between the simulated STM images were found to be minor between the adsorption configurations and in fact, the low resolution of STM (in general) meant that they found the vertical geometry to be most consistent, in disagreement with Sorescu \etal{}, and found to be the least stable configuration from their simulations~\cite{e.hamlynImagingOrderingWeakly2018}.
This overall suggests that STM does not have the resolution to discern the geometry of \ce{CO2} on rutile(110).
The computational simulation by Kubas \etal{}~\cite{kubasSurfaceAdsorptionEnergetics2016c} was the only work we found which predicted a parallel geometry to be lower in energy than the tilted geometry, using not only DFT but also DLPNO-CCSD(T)
In Table~\ref{tab:co2_rutile_dft_literature}, we show that the surface model which they used - an embedded cluster approach - differs from previous DFT simulations as well as our simulations for the PBE-D3 geometry.
Specifically, we have compared PBE-D3 predictions using their embedded cluster model with periodic slab models from two previous DFT simulations as well as our own periodic DFT calculations.
All the slab calculations come into agreement that the tilted geometry is between $18\,$meV to $37\,$meV more stable than the parallel geometry, while Kubas \etal{} predicts it to be less stable by $49\,$meV.
Similarly, all other studied functionals within their work have predicted an unstable parallel geometry, while the autoSKZCAM framework calculations as well as periodic DFT calculations from this work suggest a tilted geometry.

\begin{table}
\caption{\label{tab:co2_rutile_dft_literature}Comparison of the $E_\text{ads}$ of the CO$_2$ on TiO$_2$ rutile(110) adsorption configurations from previous DFT studies using PBE-D3 (with zero-damping).}
\begin{tabular}{lrrr}
\toprule
 & \rotatebox{90}{Parallel} & \rotatebox{90}{Tilted} & \rotatebox{90}{$\Delta$} \\ 
\midrule
Sorescu et al.~\cite{sorescuCoadsorptionPropertiesCO22012} & -371 & -389 & -18 \\
Kubas et al.~\cite{kubasSurfaceAdsorptionEnergetics2016c} & -446 & -397 & 49 \\
Dohnalek et al.~\cite{linStructureDynamicsCO22012} & -420 & -450 & -30 \\
This work & -345 & -382 & -37 \\
\bottomrule
\end{tabular}
\end{table}

\subsection{\label{sec:n2o_configurations_si}\ce{N2O} on MgO(001)}
\ce{N2O} on MgO(001) is an example of a system that has been very sparsely studied within previous computational simulations, with Scagnelli \etal{}~\cite{scagnelliCatalyticDissociationN2O2006} predicting no binding for a geometry with O pointing towards the Mg atom on the MgO surface, while Huesges \etal{}~\cite{huesgesDispersionCorrectedDFT2014} found a tilted geometry with either N or O pointing towards the Mg atom on the MgO surface that are roughly isoenergetic (to within $10\,$meV).
With our DFT ensemble, we find that the tilted geometry with the O pointing towards the Mg atom relaxes to a geometry close to parallel to the surface, such that both N and O close to an Mg atom.
This was confirmed through an additional random structure search, which did not find a tilted O-down geometry,
Thus, we compare the parallel configuration with a (N-down) tilted configuration in Extended Data Fig.~5.
Both the autoSKZCAM framework and the 6 studied DFAs predict the parallel configuration to be the lowest in energy, with autoSKZCAM coming into agreement with the experimental $H_\textrm{ads}$.
The variation between DFAs is relatively small (within $70\,$meV), with four of the six DFAs lying within the experimental $H_\textrm{ads}$ error bars.


\subsection{Previously debated systems}

Most of the other systems within this work have got geometries that are now well-resolved, however some of these systems were previously topics of debate and we briefly mention some of these systems.
For example, it was previously debated whether the CO molecule adsorbs with the C or O atom pointing towards the Mg atom within early simulations~\cite{causaMgO110SurfaceCO1987} but this was later understood to arise from errors with using Hartree-Fock theory with a small basis set~\cite{pacchioniMolecularOrbitalCluster1992a}.
Another system which has been previously under debate is the arrangement of the alkane monolayers on the MgO(001) surface.
In particular, for methane, it was not known whether the molecules take up a dipod or tripod~\cite{alaviMoleculardynamicsSimulationMethane1990,todnemMolecularAdsorptionMethane1999a} configuration, containing two and three H atoms pointing downwards respectively.
Furthermore, there were questions whether the methane molecules were rotated with respect to its neighbours~\cite{drummondDensityFunctionalInvestigation2006a}.
Both of these questions were resolved through a combination of experiments~\cite{lareseRotationalTunnelingMethane1991,lareseTrackingEvolutionInteratomic2001} and theory~\cite{drummondDensityFunctionalInvestigation2006a,tosoniAccurateQuantumChemical2010}.
Similarly, the ethane monolayer has been determined by LEED~\cite{sidoumouEthaneAdsorbedMgO1992,hoangStructureEthaneMonolayers1993,trabelsiStructuralStudyC2D62015} to take its current configuration in Fig.~\ref{fig:molsurf_systems}, which is in agreement with work by Alessio \etal{}~\cite{alessioChemicallyAccurateAdsorption2018}.
As seen in Fig.~2 of the main text, we are able achieve excellent agreement to experiment on the $H_\textrm{ads}$ for both alkane monolayers on MgO.

\clearpage

\section{\label{sec:final_eint_estimates}A benchmark dataset for non-covalent interactions of adsorbate--surface systems}

We have highlighted in Fig.~4 of the main text the possibility to use the numbers generated by the autoSKZCAM framework to benchmark the performance of density functional approximations for adsorbate--surface systems -- currently lacking~\cite{bligaardBenchmarkingCatalysisScience2016a}.
Specifically, the final $E_\textrm{int}$ estimates used in Fig.~4 of the main text are given in Table~\ref{tab:dft_xc_compare_eint}.
This forms a database covering a range of adsorbate--surface interactions on three prototypical ionic surfaces.
Such databases~\cite{rezacS66WellbalancedDatabase2011,goerigkLookDensityFunctional2017d} have been used to calibrate the performance of modern density functional approximations.
The poor performance of many sophisticated (hybrid) DFAs in Fig.~4 of the main text can be attributed to the lack of adsorbate--surface benchmarks to calibrate them.
The corresponding geometries used to calculate $E_\textrm{int}$ are available in the Github repository~\cite{shiResearchDataSupporting2024} at: \url{https://github.com/benshi97/Data_autoSKZCAM/tree/master/Data/Miscellaneous/DFT_Comparison/Geometries} for comparison to the benchmark values in Table~\ref{tab:dft_xc_compare_eint}.

\begin{table}[h]
\caption{\label{tab:dft_xc_compare_eint}Comparison of $E_\text{int}$ (in meV) from a set of DFAs against autoSKZCAM estimates.}
\begin{adjustbox}{max width=1\textwidth}
\begin{tabular}{lrrrrrrrrrrrrr}
\toprule
 & \rotatebox{90}{CH$_4$ on MgO(001)} & \rotatebox{90}{C$_2$H$_6$ on MgO(001)} & \rotatebox{90}{CO on MgO(001)} & \rotatebox{90}{Physisorbed CO$_2$ on MgO(001)} & \rotatebox{90}{Monomer H$_2$O on MgO(001)} & \rotatebox{90}{Parallel N$_2$O on MgO(001)} & \rotatebox{90}{NH$_3$ on MgO(001)} & \rotatebox{90}{CH$_4$ on TiO$_2$ rutile(110)} & \rotatebox{90}{Tilted CO$_2$ on TiO$_2$ rutile(110)} & \rotatebox{90}{H$_2$O on TiO$_2$ rutile(110)} & \rotatebox{90}{CH$_3$OH on TiO$_2$ rutile(110)} & \rotatebox{90}{H$_2$O on TiO$_2$ anatase(101)} & \rotatebox{90}{NH$_3$ on TiO$_2$ anatase(101)} \\ 
\midrule
autoSKZCAM & -122 & -175 & -207 & -308 & -703 & -256 & -657 & -269 & -493 & -1310 & -1634 & -1208 & -1377 \\
RPA+rSE & -141 & -200 & -294 & -328 & -689 & -269 & -698 & - & - & - & - & - & - \\
RPA & -96 & -137 & -98 & -236 & -614 & -204 & -630 & - & - & - & - & - & - \\
HSE06-D4 & -162 & -233 & -252 & -319 & -728 & -245 & -692 & -295 & -496 & -1397 & -1718 & -1252 & -1528 \\
PBE0-TS/HI & -169 & -262 & -245 & -289 & -719 & -231 & -686 & -333 & -489 & -1417 & -1784 & -1254 & -1528 \\
r$^2$SCAN-D4 & -173 & -244 & -296 & -380 & -784 & -294 & -734 & -306 & -528 & -1429 & -1743 & -1303 & -1527 \\
SCAN-rVV10 & -182 & -257 & -323 & -427 & -823 & -329 & -761 & -323 & -573 & -1492 & -1821 & -1363 & -1576 \\
rev-vdW-DF2 & -144 & -213 & -272 & -299 & -672 & -246 & -665 & -273 & -462 & -1286 & -1607 & -1163 & -1402 \\
PBE-MBD/FI & -121 & -192 & -325 & -319 & -689 & -258 & -687 & -291 & -472 & -1289 & -1608 & -1183 & -1414 \\
PBE-D3 & -257 & -375 & -320 & -334 & -770 & -314 & -764 & -357 & -439 & -1280 & -1611 & -1176 & -1441 \\
\bottomrule
\end{tabular}
\end{adjustbox}
\end{table}

\subsection{Computational details for the DFA benchmark} \label{sec:rpa_validation}

We have compared a set of 8 DFAs against the $E_\textrm{int}$ calculated by the SKZCAM protocol in Sec.~\ref{sec:skzcam_si_sec}.
This was performed for a set of adsorbate--surface systems which cover an $E_\textrm{int}$ of over $1.6\,$eV.
Discussion of the performance of these methods can be found in the main text and we describe the computational details here.

The same $k$-point grids were used as those in Table~\ref{tab:dft_parameters}. 
The GGA and meta-GGA calculations were performed with a $1000\,$eV energy cutoff together with small core (\texttt{Ti\_sv} and \texttt{Mg\_sv}) PAW potentials together with hard C, H, N and O PAW potentials.
The hybrid DFT calculations were performed with the \texttt{Ti\_pv} and \texttt{Mg\_pv} PAW potentials with standard C, H, N and O PAW potentials with an energy cutoff of 600 eV.
A correction for errors (typically less than $10\,$meV) in the PAW potential was calculated at the GGA (PBE-TS/HI for PBE0-TS/HI and PBE-D4 for HSE06-D4) level.

The random phase approximation (RPA) calculations were performed in VASP with an energy cutoff of $550\,$eV and a cutoff of $366\,$eV for the response function, using the same $k$-point grid as the DFT calculations.
Using PBE orbitals, we calculated RPA and RPA$+$rSE~\cite{klimesSinglesCorrelationEnergy2015} energies for MgO(001) in a (4$\times$4) supercell with a 2$\times$2$\times$1 $k$-point mesh.
In Table~\ref{tab:rpa_kpoint_conv}, we show that the RPA adsorption energy $E_\text{ads}$ for CO on MgO(001) changes by only $4\,$meV when compared to a 3$\times$3$\times$1 $k$-point mesh.
We used the \texttt{GW} variants of the \texttt{Mg\_sv}, \texttt{C}, \texttt{H}, \texttt{N} and \texttt{O} PAW potentials and employed VASP's low-scaling implementation~\cite{kaltakCubicScalingAlgorithm2014} with a plane-wave cutoff of $550\,$eV and a 12-point frequency grid.
The complete basis set limit was estimated via the built-in extrapolation technique.
We show in Table~\ref{tab:rpa_encut_conv} that the RPA correlation contribution to the CO on MgO(001) $E_\text{ads}$ changes by less than $10\,$meV for finite energy cutoffs (prior to extrapolation).
The above convergence tests were performed on $E_\text{ads}$ (using revPBE-D4 geometries) rather than $E_\text{int}$ to enable comparison to previous literature, as discussed below.

\begin{table}[h]
\caption{\label{tab:rpa_kpoint_conv}Convergence of the RPA energy contributions: exact exchange (EXX), RPA correlation (RPAc) and total RPA, to the adsorption energy $E_\text{ads}$ (in meV) of CO on MgO(001) as a function of $k$-point grid size. These tests were performed with a smaller energy cutoff of $414\,$eV.}
\begin{tabular}{@{}lrrr@{}}
\toprule
      & EXX & RPAc & RPA \\ \midrule
$2\times2\times1$ & 283 & -362 & -79 \\
$3\times3\times1$ & 282 & -365 & -83 \\ \bottomrule
\end{tabular}
\end{table}

\clearpage

\begin{table}[h]
\caption{\label{tab:rpa_encut_conv}Convergence of the RPA correlation (RPAc) contribution to the adsorption energy $E_\text{ads}$ of CO on MgO(001) as a function of the energy cutoff for the response function (to evaluate RPAc). In VASP, the RPAc energy cutoff is normally $\frac{2}{3}$ of the standard energy cutoff (of $550\,$eV) and extrapolated to the infinite ($\infty$) basis set limit from smaller energy cutoffs.}
\begin{tabular}{@{}lr@{}}
\toprule
RPAc energy cutoff (eV) & RPAc $E_\text{ads}$ (meV) \\ \midrule
$\infty$       & -373 \\
366     & -376 \\
349     & -376 \\
332     & -376 \\
316     & -376 \\
301     & -376 \\
287     & -377 \\
273     & -377 \\
260     & -378 \\ \bottomrule
\end{tabular}
\end{table}

In Table~\ref{tab:rpa_validation}, we compare our RPA estimates of the adsorption energy $E_\text{ads}$ for CO and H$_2$O on MgO(001) against estimates by Bajdich \etal{}~\cite{bajdichSurfaceEnergeticsAlkalineearth2015b}.
The agreement is to within $20\,$meV for both systems, confirming the validity and reproducibility of our chosen RPA settings.

\begin{table}[h]
\caption{\label{tab:rpa_validation}Comparison of the adsorption energy $E_\text{ads}$ in meV for CO and H$_2$O on MgO(001) computed in this work to the work of Bajdich \etal{}~\cite{bajdichSurfaceEnergeticsAlkalineearth2015b}.}
\begin{adjustbox}{center}
\begin{tabular}{lrr}
\toprule
 & This work & Bajdich \etal{}~\cite{bajdichSurfaceEnergeticsAlkalineearth2015b} \\ 
\midrule
CO & -92 & -72 \\
H$_2$O & -479 & -492 \\
\bottomrule
\end{tabular}
\end{adjustbox}
\end{table}

\cleardoublepage 
\thispagestyle{empty} 
\phantomsection 
\addcontentsline{toc}{section}{Part II: Supporting data} 

\begin{center}
    \vspace*{\fill} 
    {\LARGE Part II: Supporting data}
    \vspace*{\fill}
\end{center}

\clearpage 

\section{\label{sec:adsorption_enthalpy_def}The approach to calculating the adsorption enthalpy}

The adsorption enthalpy $H_\textrm{ads}$ is the central quantity in surface chemistry.
It represents the enthalpy released when a molecule binds to a surface, giving a physical description for the strength of this binding.
Adsorption and desorption represents primary processes within any chemical reaction happening on a surface and as such, $H_\textrm{ads}$ is a key quantity that can control the reaction rate, as empirically shown by the Sabatier principle~\cite{sabatierHydrogenationsDeshydrogenationsPar1911}.
%

The standard path towards computing $H_\textrm{ads}$ starts from $E_\textrm{ads}$ -- the adsorption energy under static conditions (i.e., zero temperature and pressure) -- and adds contributions for the zero-point vibrational (ZPV) energies $E_\textrm{ZPV}$ and vibrational temperature contributions $E_\textrm{T}$ alongside an $RT$ term for the work done due to pressure~\cite{sauerInitioCalculationsMolecule2019b}:
\begin{equation} \label{eq:hads_eq1}
    H_\textrm{ads} = E_\textrm{ads} + E_\textrm{ZPV} + E_\textrm{T} -  RT.
\end{equation}
$E_\textrm{ads}$ can be obtained from the total energy calculations from e.g., standard correlated wave-function theory (cWFT) and DFT calculations.
It is defined as the difference between the energy of the adsorbate--surface complex and the energies of the gas-phase molecule (M) and the pristine surface (S), all in their equilibrium geometries.
$E_\textrm{ZPV}$ and $E_\textrm{T}$ can be obtained from vibrational energies by considering a Taylor expansion of the potential energy surface (PES) around the energetically most stable equilibrium structure.
This approach assumes that the molecules are localised within a potential well to one specific adsorption site.
More challenging cases may require incorporating anharmonic effects, models that sample more adsorption sites or to perform global molecular dynamics to obtain ensemble averages~\cite{bergerMolecularDynamicsChemical2023a}: $ H_\textrm{ads} = \langle H_\textrm{MS} \rangle - \langle H_\textrm{M} \rangle - \langle H_\textrm{S} \rangle$.
However, for most surface phenomena, these effects are not expected to contribute significantly to the adsorption enthalpy, especially at lower temperatures.

\begin{figure}
    \includegraphics[width=\textwidth]{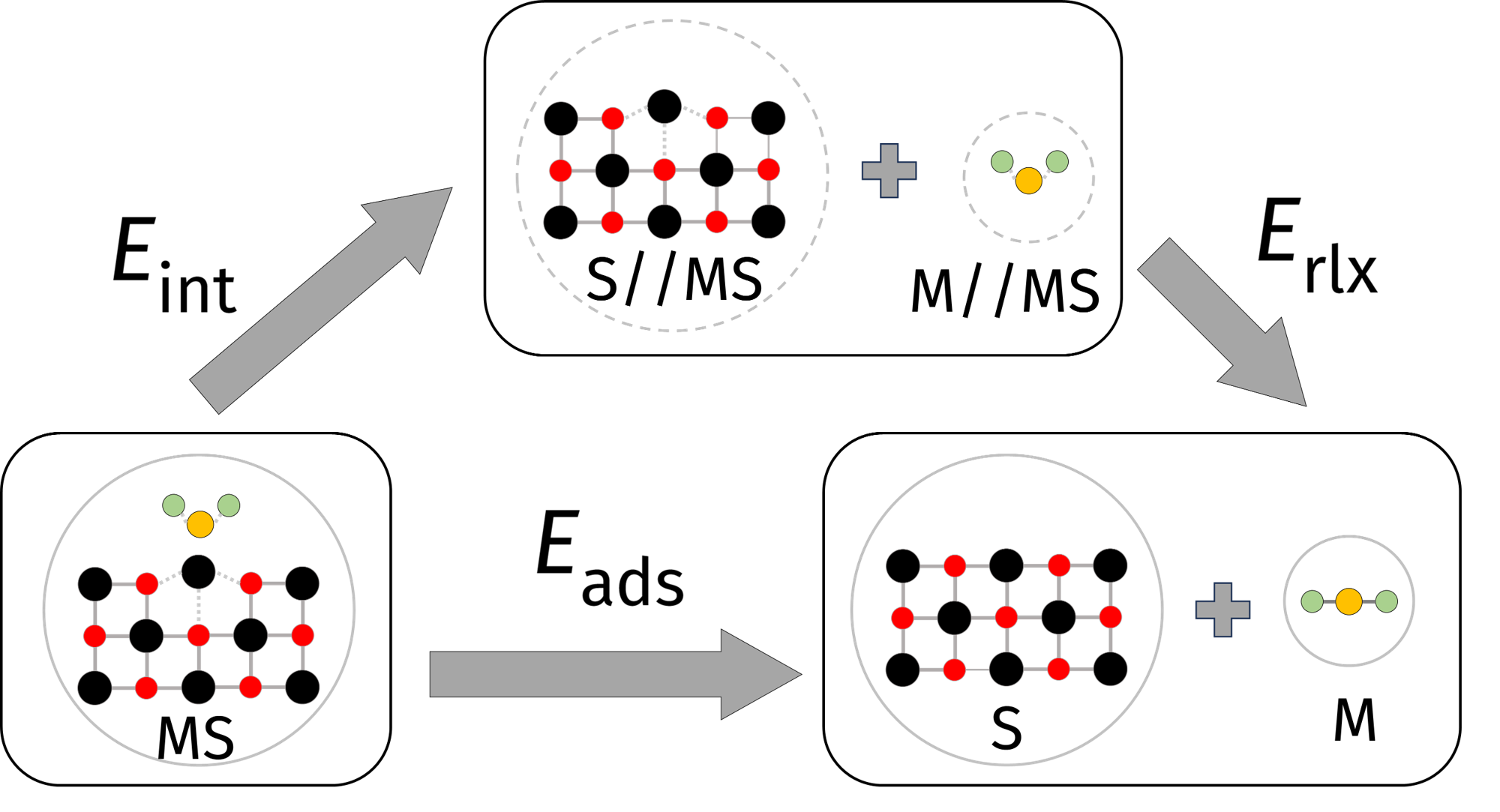}
    \caption{\label{fig:eads_definition} The physical description of the adsorption energy $E_\textrm{ads}$ in terms of the total energy of the adsorbate--surface complex (MS), pristine surface (S) and gas-phase molecule (M) in their equilibrium positions. This can be further broken down in a thermodynamic cycle into an interaction energy $E_\textrm{int}$ contribution, defined as the energetic difference between MS against M and S fixed to their geometries in MS (i.e., M//MS and S//MS respectively), followed by a relaxation term $E_\textrm{rlx}$ bring these two systems into their equilibrium geometries. The circles represent a single system/calculation, with a dashed-line circle indicating a geometry fixed to that found in the adsorbate--surface complex while a line circle indicates an equilibrium geometry.} 
\end{figure}

Within this work, we make a further breakdown of $E_\textrm{ads}$ into a contribution from the interaction energy $E_\textrm{int}$ and the geometric relaxation energy $E_\textrm{rlx}$:
\begin{equation} \label{eq:hads_eq2}
    E_\textrm{ads} = E_\textrm{int} + E_\textrm{rlx}.
\end{equation}
The former aims to obtain only the interaction between the molecule and surface and is defined as the difference between the energy of the adsorbate--surface complex against the energies of the molecule (M//MS) and surface (M//S) constrained to their geometry in the complex.
The latter accounts for the geometric relaxation of the molecule and surface in the complex from their respective equilibrium positions:
\begin{equation}
    E_\textrm{rlx} = E_\textrm{M//MS} - E_\textrm{M} + E_\textrm{S//MS} - E_\textrm{S}
\end{equation}

In the subsequent sections, we will discuss how these individual terms are obtained in the autoSKZCAM framework, leading to the results in Secs.~\ref{sec:system_discuss} and~\ref{sec:final_eint_estimates}.

\section{\label{sec:adsorbate-surface_systems}The adsorbate--surface complexes studied in this work}

\begin{figure}
    \includegraphics[width=\textwidth]{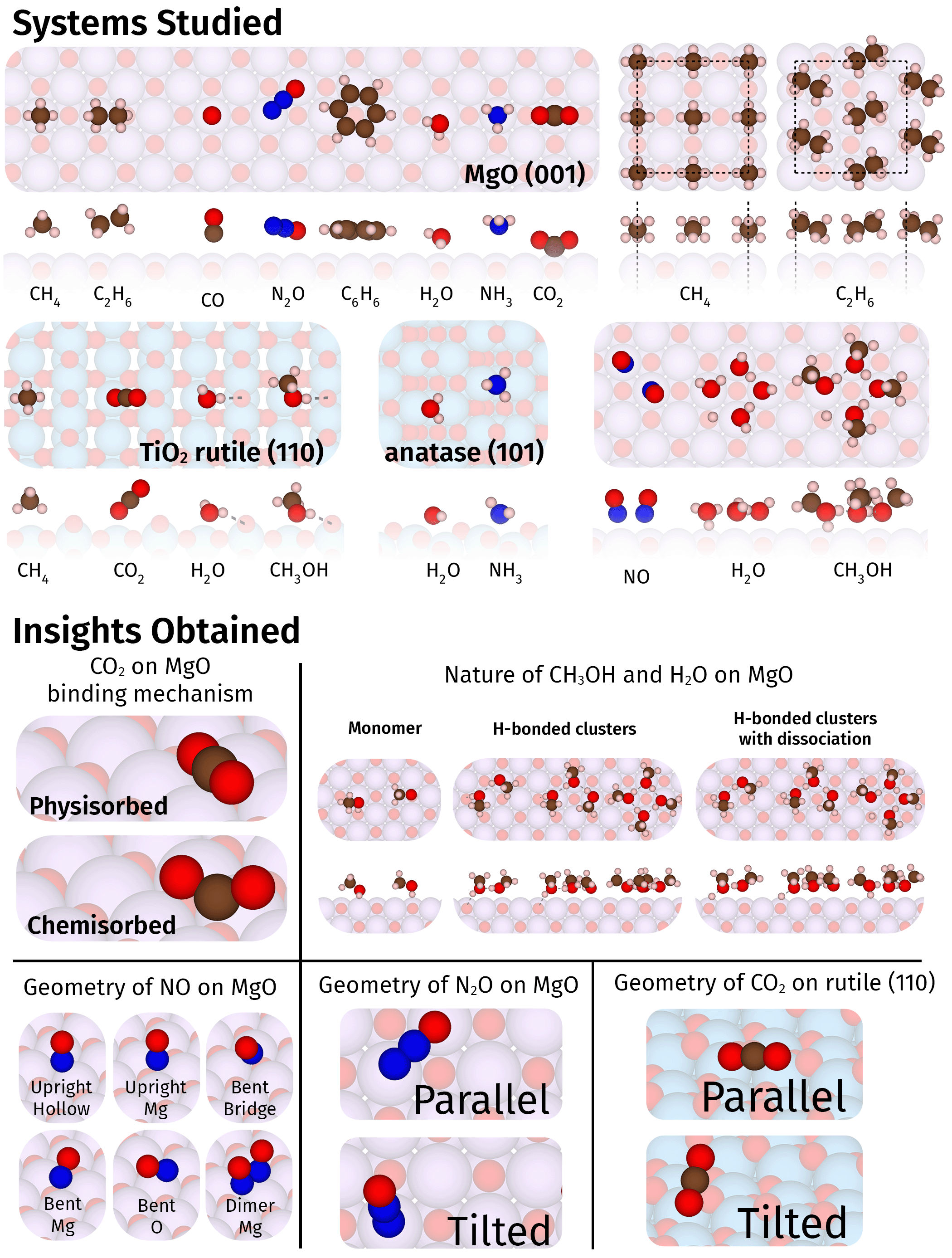}
    \caption{\label{fig:molsurf_systems} The adsorbate--surface systems that we study within this work. There are 19 adsorbate--surface systems in total. For several of these systems, we have examined multiple competing adsorption configurations to reveal new atomic-level insights on how the molecules bind onto the surface.} 
\end{figure}

Within this work, we have studied 19 adsorbate--surface systems in total, considering several adsorption configurations (29 in total) to obtain new insights into their binding mechanism as shown in Fig.~\ref{fig:molsurf_systems}.
The systems consists of several molecular adsorbates of technological relevance (CO, NO, \ce{N2O}, \ce{NH3}, \ce{H2O}, \ce{CO2}, \ce{CH3OH}, \ce{CH4}, \ce{C2H6} and \ce{C6H6}).
The adsorption of these molecules have been considered on the MgO(001) surface and on both the TiO\textsubscript{2} anatase(101) and rutile(110) surfaces, the prototypical metal-oxide surfaces, all with important technological applications.
For example \ce{TiO2} has been under heavy investigation for the photocatalytic conversion of \ce{H2O} to hydrogen~\cite{xiaEmergingCocatalystsTiO2based2022,wangParticulatePhotocatalystsLightDriven2020}, while MgO is being investigated to be used as an adsorbent of harmful \ce{CO2}~\cite{mcqueenAmbientWeatheringMagnesium2020b} and NO$_x$~\cite{wangHighlyEfficientOnepot2024} gases.
These systems have been chosen because high quality experimental estimates~\cite{campbellEnthalpiesEntropiesAdsorption2013} exist for their adsorption enthalpy, with open questions on their adsorption mechanism.
Some of these systems have also been well-studied with computational simulations, namely density functional theory (DFT), and only a handful have been studied with methods from correlated wave-function theory (cWFT).
There are often discrepancies in the predicted binding mechanism (see Sec.\ref{sec:system_discuss}, which we resolve with new predictions from the autoSKZCAM framework).

\clearpage

\section{\label{sec:cwft_details}Computational details for correlated wave-function theory}

The ability to reach an accurate $E_\textrm{int}$ and ultimately $H_\textrm{ads}$ value which agrees with experiments rests upon employing methods from correlated wave-function theory (cWFT).
Specifically, we make use of two levels of theory: second-order M{\o}ller-Plesset perturbation theory (MP2)~\cite{mollerNoteApproximationTreatment1934} and coupled cluster theory with single, double, and perturbative triple excitations [CCSD(T)]~\cite{raghavachariFifthorderPerturbationComparison1989}.
We leverage the developments within two two efficient quantum chemistry codes, ORCA 5.0.3~\cite{neeseORCAQuantumChemistry2020} and MRCC 2023~\cite{kallayMRCCProgramSystem2020}, to perform our cWFT calculations.
Here, the MP2 calculations were performed with the ORCA program using the resolution-of-identity approximation.
We leverage the developments in efficient local approximations to CCSD(T) - specifically the local natural orbital (LNO)~\cite{nagyOptimizationLinearScalingLocal2018,nagyApproachingBasisSet2019} approximation in MRCC and the domain-based local pair natural orbital (DLPNO)~\cite{riplingerEfficientLinearScaling2013,riplingerNaturalTripleExcitations2013,riplingerEfficientLinearScaling2013a,riplingerSparseMapsSystematic2016} approximation in ORCA.
The former was used for the majority of CCSD(T) calculations while the latter was used for treating the open-shell NO monomers adsorbed on MgO.
In addition, while all other systems started from a restricted Hartree-Fock (HF) reference determinant, the open-shell NO monomers on MgO(001) used an unrestricted Hartree-Fock determinant (of doublet multiplicity) for the adsorbate and adsorbate--slab complex.

We have used the Dunning family~\cite{petersonAccurateCorrelationConsistent2002} of correlation consistent basis sets, where aug-cc-pV$X$Z are used on the non-metal (C, H, N, O) atoms, with $X$ represent its size in terms of double (DZ), triple (TZ) or quadruple (QZ) zeta.
For the metal cations (Mg and Ti), we do not include augmentation and use either the cc-pV$X$Z or cc-pwCV$X$Z basis sets alongside an associated treatment of correlation from semicore electrons on the metal cations.
The former treats only valence (i.e., 3s$^2$ or 3d$^2$4s$^2$) electrons while the latter includes weighted core-valence basis functions~\cite{balabanovSystematicallyConvergentBasis2005} to incorporate  sub-valence $s$ and $p$ electron contributions to the electron correlation treatment~\cite{bistoniTreatingSubvalenceCorrelation2017}.
The combinations that we specifically compute for $E_\textrm{int}$ will be shortened to: aVDZ, aV(DZ/TZ), aV(TZ/QZ), awCVDZ, awCV(DZ/TZ), awCV(TZ/QZ).
Here, the `a' in front indicates that there is an augmentation treatment (only on the non-metal atoms), while the inclusion of wC indicates the use of core-valence basis sets on the metal atoms.
Those with (DZ/TZ) or (TZ/QZ) involve a two-point complete basis set (CBS) extrapolation, using parameters taken from Neese and Valeev~\cite{neeseRevisitingAtomicNatural2011}, for the enclosed pair of basis functions.

We use the def2-QZVPP-RI-JK auxiliary basis function for density-fitting/resolution-of-identity  Hartree--Fock (HF) computations, and the RI auxiliary basis sets from Weigend~\cite{weigendRIMP2OptimizedAuxiliary1998,hellwegOptimizedAccurateAuxiliary2007} corresponding to the AO basis sets for subsequent cWFT calculations.
For the metal atoms, where RI basis functions were not available, we generated automatic auxiliary basis functions using the approach of Stoychev \etal{}~\cite{stoychevAutomaticGenerationAuxiliary2017,lehtolaStraightforwardAccurateAutomatic2021}.
The interaction energy calculations all employed counterpoise corrections to overcome basis set superposition errors.
We used \texttt{TightPNO} settings for the DLPNO-CCSD(T) treatment in ORCA and \texttt{tight} LNO thresholds together with setting \texttt{bpedo=0.99999} for improved correlation energy capture in MRCC.
The corresponding DLPNO-MP2 and LMP2 estimates with the same settings were used when calculating the $\Delta$CC quantity.
Within LNO-CCSD(T), the LMP2 energy comes out naturally as the MP2 amplitudes are used to construct local natural orbitals and (local) truncation errors are corrected at the MP2 level~\cite{nagyApproachingBasisSet2019}.
On the other hand, DLPNO-MP2 is evaluated in a separate calculation from DLPNO-CCSD(T).

\section{\label{sec:skzcam_si_sec}Interaction energies computed with the SKZCAM protocol}

\begin{figure}
    \includegraphics[width=\textwidth]{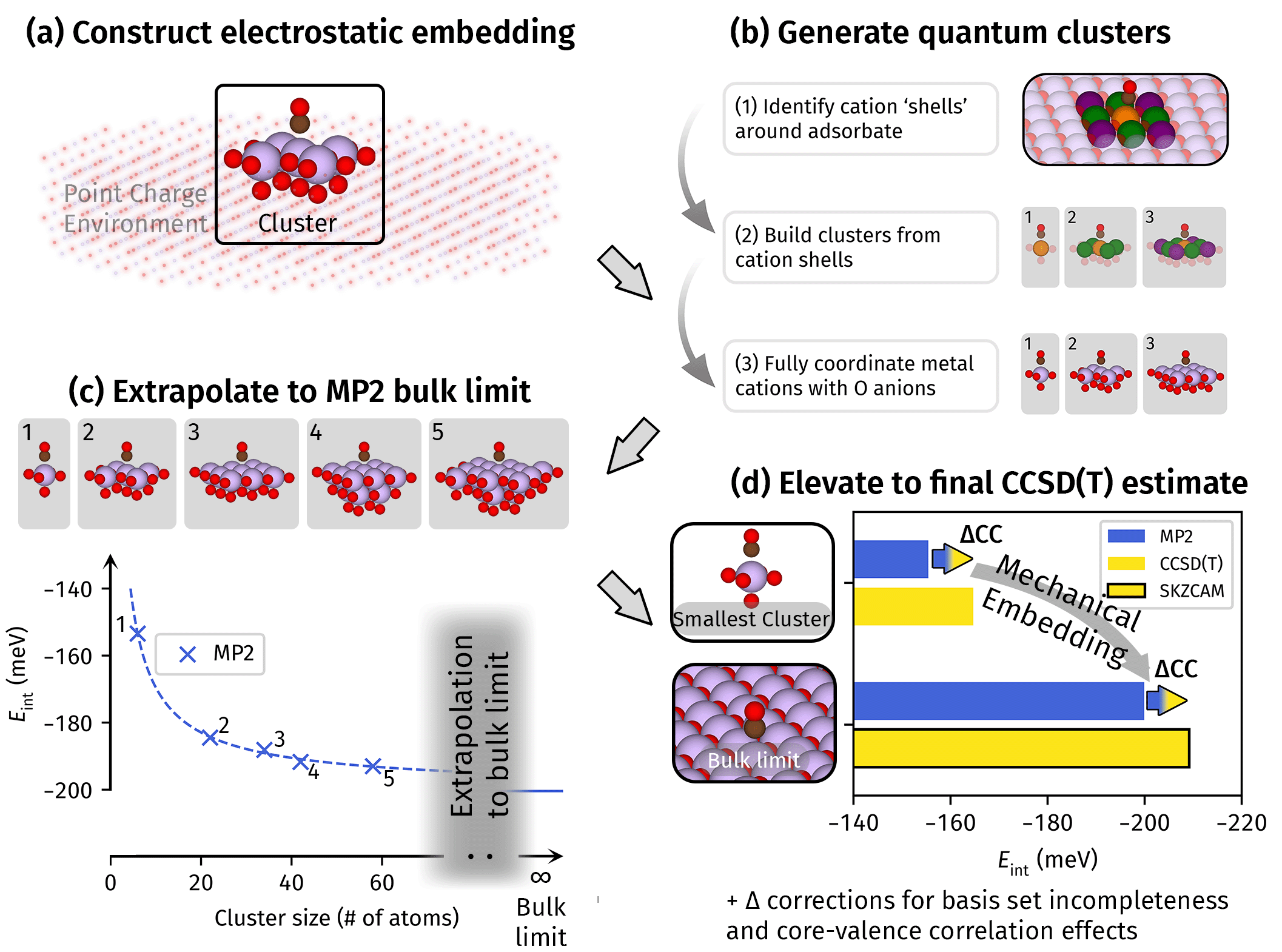}
    \caption{\label{fig:skzcam_details} Schematic of the SKZCAM protocol to calculate $E_\textrm{int}$. It starts by (a) generating the electrostatic embedding environment, followed by (b) creating the series of systematically converging clusters. From these clusters, the (c) MP2 bulk limit can be obtained through extrapolation. Finally, we reach the CCSD(T) estimate by adding a $\Delta$CC correction (from the smallest cluster) onto this MP2 bulk limit.} 
\end{figure}

The interaction energy $E_\text{int}$ forms the majority of the contribution towards $H_\text{ads}$ and it can be challenging for DFT to predict correctly (see Fig.~4 of the main text).
The SKZCAM protocol was developed in Refs.~\citenum{shiGeneralEmbeddedCluster2022b} and~\citenum{shiManyBodyMethodsSurface2023a} to reach a CCSD(T)-quality $E_\text{int}$ at low cost for adsorbate--surface systems involving ionic materials.
There are four major steps in the SKZCAM protocol (illustrated in Fig.~\ref{fig:skzcam_details}), starting with (a) constructing an electrostatic embedding environment, (b) generating the (embedded) clusters and (c) extrapolating $E_\text{int}$ to the bulk limit at the lower-level second-order M{\o}ller-Plesset perturbation theory before (d) elevating $E_\text{int}$ to the CCSD(T) level with a $\Delta$CC contribution through mechanical embedding.
We will describe each of these steps in the next few sub-sections, providing tables with the contributions towards the final $E_\text{int}$.

\subsection{\label{sec:skzcam_protocol_details}Generating a systematic series of clusters within electrostatic embedding}

The SKZCAM protocol models the surface through an electrostatic embedding scheme~\cite{pacchioniClusterModelsSurface2013}, where a (quantum) cluster that has been cleaved out of the surface is coupled to an environment that represents the interactions arising from the atoms outside the cluster.
This environment consists of point charges placed at the positions of the metal cations and oxygen anions, taking formal values of +4 and +2 for Ti and Mg respectively, while taking a value of -2 for O.
We have used the py-ChemShell 2020 package~\cite{luMultiscaleQMMM2023} to generate the embedding environment, placing point charges within a hemisphere (of radius \texttt{radius\_cluster=}$50\,$\AA{} for \ce{TiO2} [both rutile(110) and anatase(101)] and $60\,$\AA{} for the MgO(001) surface, respectively) around the adsorbate molecule, with a further set of (fitting) point charges placed (\texttt{bq\_layer=$6\,$\AA{}}) on the outer edge to reproduce the Madelung potential within (\texttt{radius\_active=})40\AA{} the adsorbate.
Subsequently, after cleaving the quantum cluster, capped effective core potentials (ECP) are placed on positive (metal cation) point charges within a radius of \texttt{cutoff\_boundary=}6\AA{} and 4\AA{} for the \ce{TiO2} and $60\,$\AA{} for MgO surfaces, respectively, to prevent spurious charge leakage.

With the electrostatic embedding environment created, the key remaining question is how to design the quantum cluster with particular attention required on controlling its size.
High-level methods from cWFT, such as CCSD(T), can quickly become intractable for larger clusters while small clusters are not typically converged towards the bulk (infinite size) limit.
The SKZCAM protocol provides a set of rubrics - hence its description as a `protocol' - to automatically generate a set of small clusters [as shown in part (b) of Fig.~\ref{fig:skzcam_details}].
It involves a simple two-step process, whereby cation metal shells (from e.g., a radial distribution function around the adsorbate molecule) are used to progressively build up the metal cations in clusters of growing size.
This is followed by fully coordinating the metal cations with O anions to complete each cluster.
The first step is intuitive since it is expected that atoms closer to the adsorbate contribute most significantly to $E_\textrm{int}$ while we have found that convergence of properties such as vacancy formation energies~\cite{shiGeneralEmbeddedCluster2022b} and adsorption energies~\cite{shiManyBodyMethodsSurface2023a} become significantly faster when the metal cations are fully coordinated by O anions in the second step to prevent any dangling bonds on the metal cations.

\begin{figure}
    \includegraphics[]{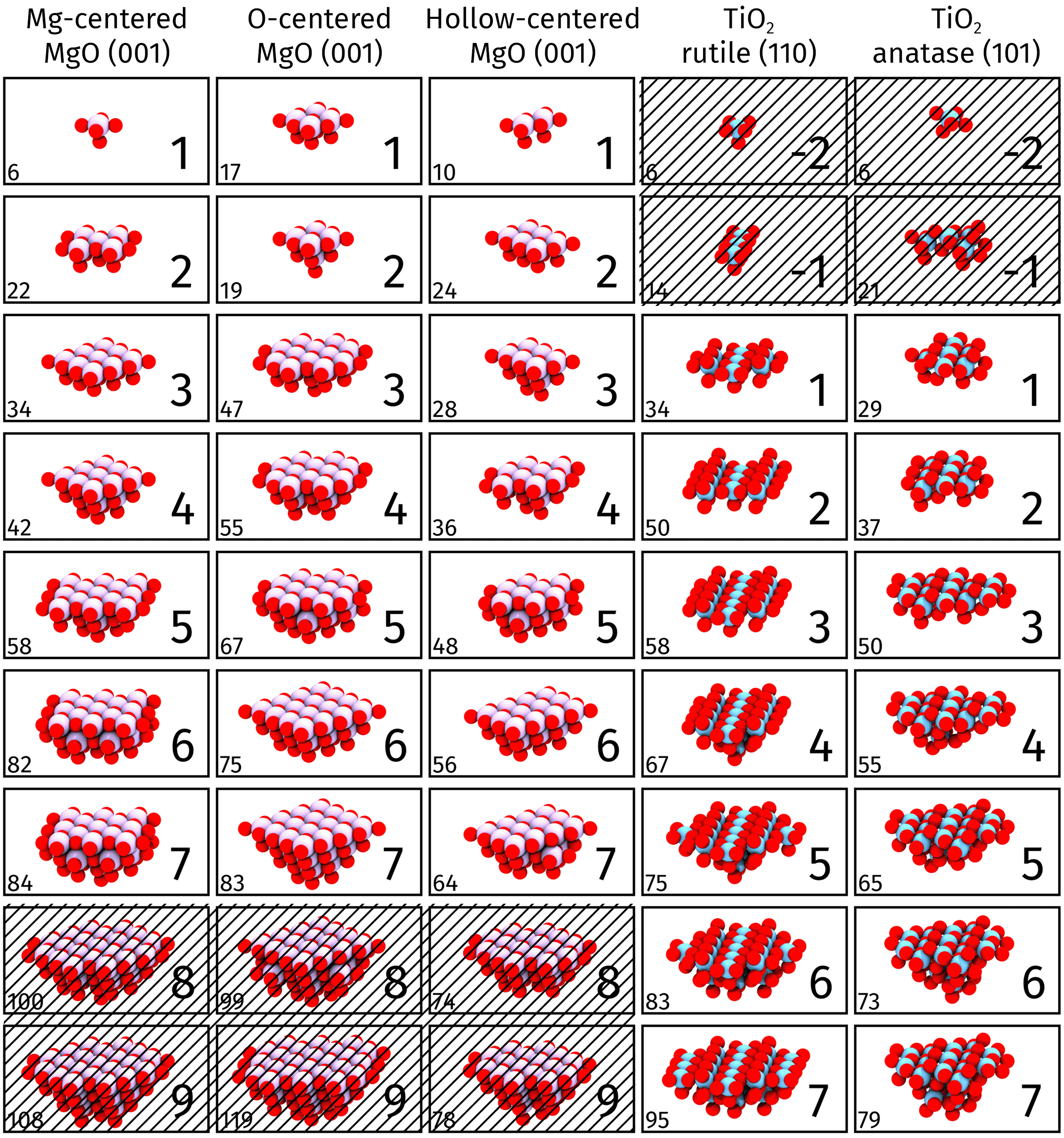}
    \caption{\label{fig:skzcam_clusters} The series of clusters generated by the SKZCAM protocol for the systems studied within this work. We have generated the first 9 clusters from the SKZCAM protocol and shade in hatched lines those which we do not study within this work, as explained within the text.}
\end{figure}

In Fig.~\ref{fig:skzcam_clusters}, we visualise the set of clusters generated by the SKZCAM protocol for the MgO(001), \ce{TiO2} rutile(110) and anatase(101) surfaces respectively.
Based on the rubrics explained above, the generated clusters will be adapted to be efficient and centered around each specific adsorbate--surface system.
For the MgO(001) surface, there are three different sets of clusters which will be generated based on the adsorption site - an `Mg-centered', `O-centered' or `Hollow-centered' cluster.
The adsorbate--surface systems for the rutile(110) and anatase(101) surfaces have their adsorbates all centred on the Ti-site, so only one set of unique clusters ended up being generated for each of the system.

We specifically show the first 9 clusters generated by the SKZCAM protocol (although clusters up to arbitrary size can be created).
We find that only the first 7 clusters are required to reach accurate estimates of $E_\textrm{int}$, so only the first 7 clusters were used for the MgO(001) surface as described in the subsequent sections.
For the clusters used in the \ce{TiO2} rutile(110) and anatase(101) surfaces, we find that the first two clusters (dubbed `-2' and `-1' clusters) were too small to provide sensible $E_\textrm{int}$ values, instead starting the SKZCAM protocol from the third cluster onwards (dubbed `1', `2', `3',\dots,`7')


\subsection{Extrapolating towards the bulk limit with the series of clusters}

The series of clusters generated by the SKZCAM protocol achieves several qualities that can be exploited to reach the CCSD(T) bulk limit accurately and efficient, which we discuss in this section and the next.
One of the key properties is that there is a smooth and fast convergence of $E_\textrm{int}$ with cluster size (as a function of the number of atoms $N$ in the cluster).
For the MgO surfaces, we use this smooth convergence to extrapolate towards the bulk limit $E_\textrm{int}^\textrm{bulk}$, using a formula of the form:
\begin{equation}
    E_\textrm{int}(N) = E_\textrm{int}^\textrm{bulk} + \dfrac{A}{N},
\end{equation}
where $A$ and $E_\textrm{int}^\textrm{bulk}$ can be obtained by fitting $E_\textrm{int}$ to the series of clusters.
This formula was inspired by the $N^{-1}$ scaling of finite-effects observed in cWFT methods~\cite{drummondFinitesizeErrorsContinuum2008} and the dependence of $E_\textrm{int}$ on a pairwise dispersive additive interaction is $N^{-2}$, which has been verified for \ce{H2O} adsorption on the 2D hBN surface~\cite{al-hamdaniTheoreticalApproachAccurate2016,al-hamdaniPropertiesWaterBoron2017a}.
Regardless of the formula, given the systematic convergence of $E_\textrm{int}$ to $E_\textrm{int}^\textrm{bulk}$, we expect to reach the correct extrapolated value once sufficient number of clusters are included within the extrapolation.
As a rule of thumb, we find that fits involving the first 5 clusters from the SKZCAM protocol are sufficient at reaching a converged estimate to within $5\,$meV, which we observe for most of the systems (see Table~\ref{tab:skzcam_system_eint}).
The largest cluster may involve sizes up to 70 atoms, and while this size may be too large for CCSD(T), it can be affordably tackled with MP2 at the complete basis set (CBS) limit using a two-point extrapolation with the double-zeta (DZ) and triple-zeta (TZ) basis sets.
Furthermore, we can estimate the error on this fit by using MP2 with the smaller DZ basis set to observe how $E_\textrm{int}^\textrm{bulk}$ changes as more clusters are included.
We have specifically estimated the error of our CBS(DZ/TZ) estimate of $E_\textrm{int}^\textrm{bulk}$ with the fifth cluster from the SKZCAM protocol by finding its maximum deviation at the DZ level when compared to incorporating 6 or 7 clusters in the fit.

For both the \ce{TiO2} surfaces, we find that the convergence is not as smooth as with the MgO surface, with $E_\textrm{int}$ oscillating as a function of cluster size.
Such behaviour is reminiscent of the odd-even oscillations common to \ce{TiO2} studies~\cite{bredowElectronicPropertiesRutile2004a} and what we observed for oxygen vacancy formation energies~\cite{shiGeneralEmbeddedCluster2022b}.
For these systems, we simply set $E_\textrm{int}^\textrm{bulk}$ to be the $E_\textrm{int}$ value from the fifth cluster, and set the error bar to the maximum deviation observed with respect to larger clusters with the DZ basis set.
We find errors that are typically below $20\,$meV for most of the studied \ce{TiO2} adsorbate--surface systems, with only the \ce{H2O} and \ce{CH3OH} molecules on rutile(110) leading to slightly larger errors that are still within $40\,$meV.

\LTcapwidth=\textwidth
\begin{longtable}{lrrrrrrr}
\caption{\label{tab:skzcam_system_eint} $E_\text{int}$ (in meV) of the clusters generated by the SKZCAM protocol for the 19 adsorbate--surface systems and their studied adsorption configurations. The type of clusters used is given within the brackets and the corresponding size for each cluster is provided in Figure~\ref{fig:skzcam_clusters}.} \\

\toprule
CH$_4$ on MgO(001) (Mg-centered) & 1 & 2 & 3 & 4 & 5 & 6 & 7 \\ 
\midrule
\endfirsthead

\caption[]{(continued)} \\
\endhead

\multicolumn{8}{r}{{Continued on next page}} \\
\endfoot

\bottomrule
\endlastfoot

$E_\text{int}$ MP2 awCVDZ & -25 & -55 & -62 & -65 & -69 & -72 & -73 \\
$E_\text{int}^\text{bulk}$ MP2 awCVDZ & - & - & - & -70 & -71 & -72 & -73 \\
$E_\text{int}$ MP2 CBS(awCVDZ/awCVTZ) & -71 & -96 & -101 & -103 & -105 & - & - \\
$E_\text{int}^\text{bulk}$ MP2 CBS(awCVDZ/awCVTZ) & - & - & - & -107 & -108 & - & - \\
$E_\text{int}$ MP2 CBS(awCVTZ/awCVQZ) & -68 & -95 & -99 & - & - & - & - \\
$E_\text{int}$ LMP2 CBS(awCVDZ/awCVTZ) & -70 & -91 & -93 & - & - & - & - \\
$E_\text{int}$ LNO-CCSD(T) CBS(awCVDZ/awCVTZ) & -86 & -107 & -109 & - & - & - & - \\
\toprule
Monolayer CH$_4$ on MgO(001) (Mg-centered) & 1 & 2 & 3 & 4 & 5 & 6 & 7 \\ 
\midrule
$E_\text{int}$ MP2 awCVDZ & -24 & -54 & -62 & -64 & -68 & -71 & -73 \\
$E_\text{int}^\text{bulk}$ MP2 awCVDZ & - & - & - & -69 & -70 & -72 & -73 \\
$E_\text{int}$ MP2 CBS(awCVDZ/awCVTZ) & -70 & -95 & -100 & -102 & -104 & - & - \\
$E_\text{int}^\text{bulk}$ MP2 CBS(awCVDZ/awCVTZ) & - & - & - & -106 & -107 & - & - \\
$E_\text{int}$ MP2 CBS(awCVTZ/awCVQZ) & -67 & -94 & -98 & - & - & - & - \\
$E_\text{int}$ LMP2 CBS(awCVDZ/awCVTZ) & -69 & -90 & -94 & - & - & - & - \\
$E_\text{int}$ LNO-CCSD(T) CBS(awCVDZ/awCVTZ) & -85 & -106 & -110 & - & - & - & - \\
\toprule
C$_2$H$_6$ on MgO(001) (Mg-centered) & 1 & 2 & 3 & 4 & 5 & 6 & 7 \\ 
\midrule
$E_\text{int}$ MP2 awCVDZ & -30 & -79 & -93 & -97 & -104 & -108 & -110 \\
$E_\text{int}^\text{bulk}$ MP2 awCVDZ & - & - & - & -105 & -107 & -109 & -110 \\
$E_\text{int}$ MP2 CBS(awCVDZ/awCVTZ) & -90 & -136 & -146 & -149 & -153 & - & - \\
$E_\text{int}^\text{bulk}$ MP2 CBS(awCVDZ/awCVTZ) & - & - & - & -157 & -158 & - & - \\
$E_\text{int}$ MP2 CBS(awCVTZ/awCVQZ) & -87 & -133 & -141 & - & - & - & - \\
$E_\text{int}$ LMP2 CBS(awCVDZ/awCVTZ) & -92 & -131 & -140 & - & - & - & - \\
$E_\text{int}$ LNO-CCSD(T) CBS(awCVDZ/awCVTZ) & -112 & -151 & -160 & - & - & - & - \\
\toprule
Monolayer C$_2$H$_6$ on MgO(001) (Mg-centered) & 1 & 2 & 3 & 4 & 5 & 6 & 7 \\ 
\midrule
$E_\text{int}$ MP2 awCVDZ & -21 & -71 & -86 & -89 & -96 & -101 & -102 \\
$E_\text{int}^\text{bulk}$ MP2 awCVDZ & - & - & - & -98 & -100 & -102 & -103 \\
$E_\text{int}$ MP2 CBS(awCVDZ/awCVTZ) & -73 & -121 & -131 & -134 & -139 & - & - \\
$E_\text{int}^\text{bulk}$ MP2 CBS(awCVDZ/awCVTZ) & - & - & - & -142 & -144 & - & - \\
$E_\text{int}$ MP2 CBS(awCVTZ/awCVQZ) & -71 & -118 & -127 & - & - & - & - \\
$E_\text{int}$ LMP2 CBS(awCVDZ/awCVTZ) & -72 & -115 & -124 & - & - & - & - \\
$E_\text{int}$ LNO-CCSD(T) CBS(awCVDZ/awCVTZ) & -91 & -135 & -144 & - & - & - & - \\
\toprule
CO on MgO(001) (Mg-centered) & 1 & 2 & 3 & 4 & 5 & 6 & 7 \\ 
\midrule
$E_\text{int}$ MP2 awCVDZ & -47 & -100 & -111 & -117 & -122 & -127 & -128 \\
$E_\text{int}^\text{bulk}$ MP2 awCVDZ & - & - & - & -125 & -127 & -128 & -129 \\
$E_\text{int}$ MP2 CBS(awCVDZ/awCVTZ) & -155 & -184 & -192 & -195 & -197 & - & - \\
$E_\text{int}^\text{bulk}$ MP2 CBS(awCVDZ/awCVTZ) & - & - & - & -199 & -200 & - & - \\
$E_\text{int}$ MP2 CBS(awCVTZ/awCVQZ) & -155 & -186 & -189 & - & - & - & - \\
$E_\text{int}$ LMP2 CBS(awCVDZ/awCVTZ) & -153 & -183 & -185 & - & - & - & - \\
$E_\text{int}$ LNO-CCSD(T) CBS(awCVDZ/awCVTZ) & -160 & -190 & -194 & - & - & - & - \\
\toprule
C$_6$H$_6$ on MgO(001) (O-centered) & 1 & 2 & 3 & 4 & 5 & 6 & 7 \\ 
\midrule
$E_\text{int}$ MP2 awCVDZ & -197 & -203 & -305 & -313 & -322 & -330 & -335 \\
$E_\text{int}^\text{bulk}$ MP2 awCVDZ & - & - & - & -367 & -367 & -368 & -369 \\
$E_\text{int}$ MP2 CBS(awCVDZ/awCVTZ) & -338 & -344 & -414 & -422 & -428 & - & - \\
$E_\text{int}^\text{bulk}$ MP2 CBS(awCVDZ/awCVTZ) & - & - & - & -460 & -460 & - & - \\
$E_\text{int}$ MP2 CBS(awCVTZ/awCVQZ) & -354 & -359 & -416 & - & - & - & - \\
$E_\text{int}$ LMP2 CBS(awCVDZ/awCVTZ) & -335 & -340 & -404 & - & - & - & - \\
$E_\text{int}$ LNO-CCSD(T) CBS(awCVDZ/awCVTZ) & -311 & -316 & -379 & - & - & - & - \\
\toprule
Parallel N$_2$O on MgO(001) (Hollow-centered) & 1 & 2 & 3 & 4 & 5 & 6 & 7 \\ 
\midrule
$E_\text{int}$ MP2 awCVDZ & -143 & -175 & -179 & -185 & -188 & -193 & -195 \\
$E_\text{int}^\text{bulk}$ MP2 awCVDZ & - & - & - & -200 & -200 & -201 & -202 \\
$E_\text{int}$ MP2 CBS(awCVDZ/awCVTZ) & -206 & -229 & -232 & -238 & -242 & - & - \\
$E_\text{int}^\text{bulk}$ MP2 CBS(awCVDZ/awCVTZ) & - & - & - & -248 & -249 & - & - \\
$E_\text{int}$ MP2 CBS(awCVTZ/awCVQZ) & -207 & -231 & -230 & - & - & - & - \\
$E_\text{int}$ LMP2 CBS(awCVDZ/awCVTZ) & -203 & -225 & -226 & - & - & - & - \\
$E_\text{int}$ LNO-CCSD(T) CBS(awCVDZ/awCVTZ) & -209 & -231 & -233 & - & - & - & - \\
\toprule
Tilted N$_2$O on MgO(001) (Mg-centered) & 1 & 2 & 3 & 4 & 5 & 6 & 7 \\ 
\midrule
$E_\text{int}$ MP2 awCVDZ & -44 & -98 & -111 & -116 & -122 & -126 & -129 \\
$E_\text{int}^\text{bulk}$ MP2 awCVDZ & - & - & - & -125 & -127 & -128 & -130 \\
$E_\text{int}$ MP2 CBS(awCVDZ/awCVTZ) & -122 & -159 & -168 & -172 & -175 & - & - \\
$E_\text{int}^\text{bulk}$ MP2 CBS(awCVDZ/awCVTZ) & - & - & - & -178 & -179 & - & - \\
$E_\text{int}$ MP2 CBS(awCVTZ/awCVQZ) & -129 & -165 & -171 & - & - & - & - \\
$E_\text{int}$ LMP2 CBS(awCVDZ/awCVTZ) & -120 & -155 & -159 & - & - & - & - \\
$E_\text{int}$ LNO-CCSD(T) CBS(awCVDZ/awCVTZ) & -104 & -139 & -144 & - & - & - & - \\
\toprule
Vertical-Hollow NO on MgO(001) (Hollow-centered) & 1 & 2 & 3 & 4 & 5 & 6 & 7 \\ 
\midrule
$E_\text{int}$ MP2 aVDZ & -159 & -177 & -183 & -185 & -186 & -185 & -186 \\
$E_\text{int}^\text{bulk}$ MP2 aVDZ & - & - & - & -194 & -194 & -193 & -192 \\
$E_\text{int}$ MP2 CBS(aVDZ/aVTZ) & -224 & -236 & -233 & -240 & -241 & - & - \\
$E_\text{int}^\text{bulk}$ MP2 CBS(aVDZ/aVTZ) & - & - & - & -243 & -244 & - & - \\
$E_\text{int}$ MP2 CBS(awCVDZ/awCVTZ) & -242 & -259 & -257 & - & - & - & - \\
$E_\text{int}$ MP2 CBS(aVTZ/aVQZ) & -213 & -227 & -225 & - & - & - & - \\
$E_\text{int}$ DLPNO-MP2 CBS(aVDZ/aVTZ) & -219 & - & - & - & - & - & - \\
$E_\text{int}$ DLPNO-CCSD(T) CBS(aVDZ/aVTZ) & 71 & - & - & - & - & - & - \\
\toprule
Vertical-Mg NO on MgO(001) (Mg-centered) & 1 & 2 & 3 & 4 & 5 & 6 & 7 \\ 
\midrule
$E_\text{int}$ MP2 aVDZ & 135 & 94 & 91 & 86 & 82 & 80 & 77 \\
$E_\text{int}^\text{bulk}$ MP2 aVDZ & - & - & - & 79 & 78 & 78 & 77 \\
$E_\text{int}$ MP2 CBS(aVDZ/aVTZ) & 65 & 35 & 33 & 33 & 30 & - & - \\
$E_\text{int}^\text{bulk}$ MP2 CBS(aVDZ/aVTZ) & - & - & - & 26 & 26 & - & - \\
$E_\text{int}$ MP2 CBS(awCVDZ/awCVTZ) & 43 & 9 & 5 & - & - & - & - \\
$E_\text{int}$ MP2 CBS(aVTZ/aVQZ) & 67 & 42 & 41 & - & - & - & - \\
$E_\text{int}$ DLPNO-MP2 CBS(aVDZ/aVTZ) & 64 & - & - & - & - & - & - \\
$E_\text{int}$ DLPNO-CCSD(T) CBS(aVDZ/aVTZ) & -4 & - & - & - & - & - & - \\
\toprule
Bent-Bridge NO on MgO(001) (Mg-centered) & 1 & 2 & 3 & 4 & 5 & 6 & 7 \\ 
\midrule
$E_\text{int}$ MP2 aVDZ & -460 & -486 & -494 & -503 & -505 & -507 & -510 \\
$E_\text{int}^\text{bulk}$ MP2 aVDZ & - & - & - & -503 & -505 & -507 & -508 \\
$E_\text{int}$ MP2 CBS(aVDZ/aVTZ) & -620 & -649 & -654 & -656 & -658 & - & - \\
$E_\text{int}^\text{bulk}$ MP2 CBS(aVDZ/aVTZ) & - & - & - & -661 & -661 & - & - \\
$E_\text{int}$ MP2 CBS(awCVDZ/awCVTZ) & -639 & -675 & -682 & - & - & - & - \\
$E_\text{int}$ MP2 CBS(aVTZ/aVQZ) & -618 & -654 & -658 & - & - & - & - \\
$E_\text{int}$ DLPNO-MP2 CBS(aVDZ/aVTZ) & -620 & - & - & - & - & - & - \\
$E_\text{int}$ DLPNO-CCSD(T) CBS(aVDZ/aVTZ) & 6 & - & - & - & - & - & - \\
\toprule
Bent-Mg NO on MgO(001) (Mg-centered) & 1 & 2 & 3 & 4 & 5 & 6 & 7 \\ 
\midrule
$E_\text{int}$ MP2 aVDZ & 52 & 9 & 5 & 2 & -2 & -4 & -7 \\
$E_\text{int}^\text{bulk}$ MP2 aVDZ & - & - & - & -6 & -7 & -7 & -8 \\
$E_\text{int}$ MP2 CBS(aVDZ/aVTZ) & -15 & -43 & -46 & -47 & -50 & - & - \\
$E_\text{int}^\text{bulk}$ MP2 CBS(aVDZ/aVTZ) & - & - & - & -53 & -53 & - & - \\
$E_\text{int}$ MP2 CBS(awCVDZ/awCVTZ) & -37 & -70 & -75 & - & - & - & - \\
$E_\text{int}$ MP2 CBS(aVTZ/aVQZ) & -17 & -42 & -44 & - & - & - & - \\
$E_\text{int}$ DLPNO-MP2 CBS(aVDZ/aVTZ) & -14 & - & - & - & - & - & - \\
$E_\text{int}$ DLPNO-CCSD(T) CBS(aVDZ/aVTZ) & -62 & - & - & - & - & - & - \\
\toprule
Bent-O NO on MgO(001) (O-centered) & 1 & 2 & 3 & 4 & 5 & 6 & 7 \\ 
\midrule
$E_\text{int}$ MP2 aVDZ & -496 & -500 & -502 & -502 & -504 & -503 & -503 \\
$E_\text{int}^\text{bulk}$ MP2 aVDZ & - & - & - & -505 & -506 & -505 & -505 \\
$E_\text{int}$ MP2 CBS(aVDZ/aVTZ) & -640 & -641 & -643 & -645 & -646 & - & - \\
$E_\text{int}^\text{bulk}$ MP2 CBS(aVDZ/aVTZ) & - & - & - & -646 & -647 & - & - \\
$E_\text{int}$ MP2 CBS(awCVDZ/awCVTZ) & -649 & -656 & -661 & - & - & - & - \\
$E_\text{int}$ MP2 CBS(aVTZ/aVQZ) & -639 & -638 & -648 & - & - & - & - \\
$E_\text{int}$ DLPNO-MP2 CBS(aVDZ/aVTZ) & -630 & - & - & - & - & - & - \\
$E_\text{int}$ DLPNO-CCSD(T) CBS(aVDZ/aVTZ) & 29 & - & - & - & - & - & - \\
\toprule
Dimer NO on MgO(001) (Hollow-centered) & 1 & 2 & 3 & 4 & 5 & 6 & 7 \\ 
\midrule
$E_\text{int}$ MP2 awCVDZ & -138 & -172 & -177 & -184 & -188 & -191 & -193 \\
$E_\text{int}^\text{bulk}$ MP2 awCVDZ & - & - & - & -200 & -200 & -201 & -202 \\
$E_\text{int}$ MP2 CBS(awCVDZ/awCVTZ) & -222 & -244 & -246 & -251 & -253 & - & - \\
$E_\text{int}^\text{bulk}$ MP2 CBS(awCVDZ/awCVTZ) & - & - & - & -261 & -261 & - & - \\
$E_\text{int}$ MP2 CBS(awCVTZ/awCVQZ) & -226 & -246 & -246 & - & - & - & - \\
$E_\text{int}$ LMP2 CBS(awCVDZ/awCVTZ) & -217 & -239 & - & - & - & - & - \\
$E_\text{int}$ LNO-CCSD(T) CBS(awCVDZ/awCVTZ) & -199 & -222 & - & - & - & - & - \\
\toprule
Monomer H$_2$O on MgO(001) (Mg-centered) & 1 & 2 & 3 & 4 & 5 & 6 & 7 \\ 
\midrule
$E_\text{int}$ MP2 awCVDZ & -514 & -542 & -546 & -552 & -556 & -559 & -559 \\
$E_\text{int}^\text{bulk}$ MP2 awCVDZ & - & - & - & -555 & -557 & -558 & -559 \\
$E_\text{int}$ MP2 CBS(awCVDZ/awCVTZ) & -641 & -667 & -671 & -671 & -674 & - & - \\
$E_\text{int}^\text{bulk}$ MP2 CBS(awCVDZ/awCVTZ) & - & - & - & -677 & -677 & - & - \\
$E_\text{int}$ MP2 CBS(awCVTZ/awCVQZ) & -641 & -670 & -668 & - & - & - & - \\
$E_\text{int}$ LMP2 CBS(awCVDZ/awCVTZ) & -639 & -663 & -664 & - & - & - & - \\
$E_\text{int}$ LNO-CCSD(T) CBS(awCVDZ/awCVTZ) & -665 & -688 & -689 & - & - & - & - \\
\toprule
Tetramer H$_2$O on MgO(001) (O-centered) & 1 & 2 & 3 & 4 & 5 & 6 & 7 \\ 
\midrule
$E_\text{int}$ MP2 awCVDZ & -296 & -299 & -333 & -335 & -339 & -344 & -347 \\
$E_\text{int}^\text{bulk}$ MP2 awCVDZ & - & - & - & -354 & -354 & -355 & -357 \\
$E_\text{int}$ MP2 CBS(awCVDZ/awCVTZ) & -392 & -394 & -409 & -412 & -414 & - & - \\
$E_\text{int}^\text{bulk}$ MP2 CBS(awCVDZ/awCVTZ) & - & - & - & -420 & -421 & - & - \\
$E_\text{int}$ MP2 CBS(awCVTZ/awCVQZ) & -411 & -412 & -418 & - & - & - & - \\
$E_\text{int}$ LMP2 CBS(awCVDZ/awCVTZ) & -390 & -392 & -407 & - & - & - & - \\
$E_\text{int}$ LNO-CCSD(T) CBS(awCVDZ/awCVTZ) & -417 & -419 & -434 & - & - & - & - \\
\toprule
Tilted CH$_3$OH on MgO(001) (Mg-centered) & 1 & 2 & 3 & 4 & 5 & 6 & 7 \\ 
\midrule
$E_\text{int}$ MP2 awCVDZ & -569 & -603 & -613 & -619 & -625 & -630 & -632 \\
$E_\text{int}^\text{bulk}$ MP2 awCVDZ & - & - & - & -623 & -625 & -628 & -629 \\
$E_\text{int}$ MP2 CBS(awCVDZ/awCVTZ) & -703 & -745 & -754 & -756 & -760 & - & - \\
$E_\text{int}^\text{bulk}$ MP2 CBS(awCVDZ/awCVTZ) & - & - & - & -764 & -765 & - & - \\
$E_\text{int}$ MP2 CBS(awCVTZ/awCVQZ) & -704 & -749 & -752 & - & - & - & - \\
$E_\text{int}$ LMP2 CBS(awCVDZ/awCVTZ) & -701 & -739 & -746 & - & - & - & - \\
$E_\text{int}$ LNO-CCSD(T) CBS(awCVDZ/awCVTZ) & -723 & -759 & -768 & - & - & - & - \\
\toprule
Parallel CH$_3$OH on MgO(001) (Mg-centered) & 1 & 2 & 3 & 4 & 5 & 6 & 7 \\ 
\midrule
$E_\text{int}$ MP2 awCVDZ & -395 & -424 & -423 & -429 & -427 & -429 & -428 \\
$E_\text{int}^\text{bulk}$ MP2 awCVDZ & - & - & - & -433 & -432 & -432 & -432 \\
$E_\text{int}$ MP2 CBS(awCVDZ/awCVTZ) & -443 & -493 & -499 & -496 & -499 & - & - \\
$E_\text{int}^\text{bulk}$ MP2 CBS(awCVDZ/awCVTZ) & - & - & - & -509 & -508 & - & - \\
$E_\text{int}$ MP2 CBS(awCVTZ/awCVQZ) & -426 & -474 & -481 & - & - & - & - \\
$E_\text{int}$ LMP2 CBS(awCVDZ/awCVTZ) & -442 & -488 & -491 & - & - & - & - \\
$E_\text{int}$ LNO-CCSD(T) CBS(awCVDZ/awCVTZ) & -458 & -505 & -507 & - & - & - & - \\
\toprule
Tetramer CH$_3$OH on MgO(001) (O-centered) & 1 & 2 & 3 & 4 & 5 & 6 & 7 \\ 
\midrule
$E_\text{int}$ MP2 awCVDZ & -320 & -323 & -370 & -373 & -378 & -386 & -389 \\
$E_\text{int}^\text{bulk}$ MP2 awCVDZ & - & - & - & -399 & -399 & -401 & -403 \\
$E_\text{int}$ MP2 CBS(awCVDZ/awCVTZ) & -423 & -425 & -451 & -455 & -457 & - & - \\
$E_\text{int}^\text{bulk}$ MP2 CBS(awCVDZ/awCVTZ) & - & - & - & -469 & -469 & - & - \\
$E_\text{int}$ MP2 CBS(awCVTZ/awCVQZ) & -440 & -443 & -462 & - & - & - & - \\
$E_\text{int}$ LMP2 CBS(awCVDZ/awCVTZ) & -420 & -422 & -449 & - & - & - & - \\
$E_\text{int}$ LNO-CCSD(T) CBS(awCVDZ/awCVTZ) & -447 & -449 & -476 & - & - & - & - \\
\toprule
NH$_3$ on MgO(001) (Mg-centered) & 1 & 2 & 3 & 4 & 5 & 6 & 7 \\ 
\midrule
$E_\text{int}$ MP2 awCVDZ & -377 & -449 & -475 & -482 & -493 & -500 & -501 \\
$E_\text{int}^\text{bulk}$ MP2 awCVDZ & - & - & - & -492 & -497 & -500 & -502 \\
$E_\text{int}$ MP2 CBS(awCVDZ/awCVTZ) & -551 & -582 & -602 & -606 & -608 & - & - \\
$E_\text{int}^\text{bulk}$ MP2 CBS(awCVDZ/awCVTZ) & - & - & - & -608 & -610 & - & - \\
$E_\text{int}$ MP2 CBS(awCVTZ/awCVQZ) & -570 & -602 & -611 & - & - & - & - \\
$E_\text{int}$ LMP2 CBS(awCVDZ/awCVTZ) & -547 & -580 & -599 & - & - & - & - \\
$E_\text{int}$ LNO-CCSD(T) CBS(awCVDZ/awCVTZ) & -578 & -610 & -630 & - & - & - & - \\
\toprule
Physisorbed CO$_2$ on MgO(001) (Hollow-centered) & 1 & 2 & 3 & 4 & 5 & 6 & 7 \\ 
\midrule
$E_\text{int}$ MP2 awCVDZ & -147 & -180 & -184 & -186 & -188 & -189 & -192 \\
$E_\text{int}^\text{bulk}$ MP2 awCVDZ & - & - & - & -202 & -201 & -200 & -200 \\
$E_\text{int}$ MP2 CBS(awCVDZ/awCVTZ) & -240 & -264 & -265 & -269 & -272 & - & - \\
$E_\text{int}^\text{bulk}$ MP2 CBS(awCVDZ/awCVTZ) & - & - & - & -280 & -280 & - & - \\
$E_\text{int}$ MP2 CBS(awCVTZ/awCVQZ) & -239 & -262 & -261 & - & - & - & - \\
$E_\text{int}$ LMP2 CBS(awCVDZ/awCVTZ) & -237 & -258 & -261 & - & - & - & - \\
$E_\text{int}$ LNO-CCSD(T) CBS(awCVDZ/awCVTZ) & -267 & -288 & -293 & - & - & - & - \\
\toprule
Chemisorbed CO$_2$ on MgO(001) (O-centered) & 1 & 2 & 3 & 4 & 5 & 6 & 7 \\ 
\midrule
$E_\text{int}$ MP2 awCVDZ & -2843 & -2857 & -2760 & -2767 & -2774 & -2754 & -2743 \\
$E_\text{int}^\text{bulk}$ MP2 awCVDZ & - & - & - & -2767 & -2774 & -2754 & -2743 \\
$E_\text{int}$ MP2 CBS(awCVDZ/awCVTZ) & -3334 & -3332 & -3292 & -3295 & -3309 & - & - \\
$E_\text{int}^\text{bulk}$ MP2 CBS(awCVDZ/awCVTZ) & - & - & - & -3295 & -3309 & - & - \\
$E_\text{int}$ MP2 CBS(awCVTZ/awCVQZ) & -3311 & -3301 & -3272 & - & - & - & - \\
$E_\text{int}$ LMP2 CBS(awCVDZ/awCVTZ) & -3330 & -3328 & -3283 & - & - & - & - \\
$E_\text{int}$ LNO-CCSD(T) CBS(awCVDZ/awCVTZ) & -3548 & -3547 & -3506 & - & - & - & - \\
\toprule
CH$_4$ on TiO$_2$ rutile(110) & 1 & 2 & 3 & 4 & 5 & 6 & 7 \\ 
\midrule
$E_\text{int}$ MP2 aVDZ & -219 & -202 & -208 & -208 & -211 & -210 & -213 \\
$E_\text{int}^\text{bulk}$ MP2 aVDZ & - & - & - & -208 & -211 & -210 & -213 \\
$E_\text{int}$ MP2 CBS(aVDZ/aVTZ) & -282 & -262 & -267 & -265 & -267 & - & - \\
$E_\text{int}^\text{bulk}$ MP2 CBS(aVDZ/aVTZ) & - & - & - & -265 & -267 & - & - \\
$E_\text{int}$ MP2 CBS(aVTZ/aVQZ) & -281 & -261 & - & - & - & - & - \\
$E_\text{int}$ MP2 CBS(awCVTZ/awCVQZ) & -283 & -263 & - & - & - & - & - \\
$E_\text{int}$ LMP2 CBS(aVDZ/aVTZ) & -282 & -263 & - & - & - & - & - \\
$E_\text{int}$ LNO-CCSD(T) CBS(aVDZ/aVTZ) & -283 & -263 & - & - & - & - & - \\
\toprule
Parallel CO$_2$ on TiO$_2$ rutile(110) & 1 & 2 & 3 & 4 & 5 & 6 & 7 \\ 
\midrule
$E_\text{int}$ MP2 aVDZ & -337 & -315 & -320 & -323 & -327 & -326 & -331 \\
$E_\text{int}^\text{bulk}$ MP2 aVDZ & - & - & - & -323 & -327 & -326 & -331 \\
$E_\text{int}$ MP2 CBS(aVDZ/aVTZ) & -425 & -401 & -405 & -405 & -408 & - & - \\
$E_\text{int}^\text{bulk}$ MP2 CBS(aVDZ/aVTZ) & - & - & - & -405 & -408 & - & - \\
$E_\text{int}$ MP2 CBS(aVTZ/aVQZ) & -425 & -404 & - & - & - & - & - \\
$E_\text{int}$ MP2 CBS(awCVTZ/awCVQZ) & -433 & -411 & - & - & - & - & - \\
$E_\text{int}$ LMP2 CBS(aVDZ/aVTZ) & -421 & -398 & - & - & - & - & - \\
$E_\text{int}$ LNO-CCSD(T) CBS(aVDZ/aVTZ) & -413 & -393 & - & - & - & - & - \\
\toprule
Tilted CO$_2$ on TiO$_2$ rutile(110) & 1 & 2 & 3 & 4 & 5 & 6 & 7 \\ 
\midrule
$E_\text{int}$ MP2 aVDZ & -441 & -390 & -401 & -395 & -399 & -392 & -396 \\
$E_\text{int}^\text{bulk}$ MP2 aVDZ & - & - & - & -395 & -399 & -392 & -396 \\
$E_\text{int}$ MP2 CBS(aVDZ/aVTZ) & -521 & -467 & -476 & -468 & -470 & - & - \\
$E_\text{int}^\text{bulk}$ MP2 CBS(aVDZ/aVTZ) & - & - & - & -468 & -470 & - & - \\
$E_\text{int}$ MP2 CBS(aVTZ/aVQZ) & -527 & -475 & - & - & - & - & - \\
$E_\text{int}$ MP2 CBS(awCVTZ/awCVQZ) & -519 & -466 & - & - & - & - & - \\
$E_\text{int}$ LMP2 CBS(aVDZ/aVTZ) & -516 & -461 & - & - & - & - & - \\
$E_\text{int}$ LNO-CCSD(T) CBS(aVDZ/aVTZ) & -539 & -486 & - & - & - & - & - \\
\toprule
H$_2$O on TiO$_2$ rutile(110) & 1 & 2 & 3 & 4 & 5 & 6 & 7 \\ 
\midrule
$E_\text{int}$ MP2 aVDZ & -1256 & -1141 & -1161 & -1139 & -1146 & -1113 & -1119 \\
$E_\text{int}^\text{bulk}$ MP2 aVDZ & - & - & - & -1139 & -1146 & -1113 & -1119 \\
$E_\text{int}$ MP2 CBS(aVDZ/aVTZ) & -1420 & -1304 & -1319 & -1292 & -1299 & - & - \\
$E_\text{int}^\text{bulk}$ MP2 CBS(aVDZ/aVTZ) & - & - & - & -1292 & -1299 & - & - \\
$E_\text{int}$ MP2 CBS(aVTZ/aVQZ) & -1436 & -1324 & - & - & - & - & - \\
$E_\text{int}$ MP2 CBS(awCVTZ/awCVQZ) & -1399 & -1280 & - & - & - & - & - \\
$E_\text{int}$ LMP2 CBS(aVDZ/aVTZ) & -1418 & -1301 & - & - & - & - & - \\
$E_\text{int}$ LNO-CCSD(T) CBS(aVDZ/aVTZ) & -1447 & -1340 & - & - & - & - & - \\
\toprule
CH$_3$OH on TiO$_2$ rutile(110) & 1 & 2 & 3 & 4 & 5 & 6 & 7 \\ 
\midrule
$E_\text{int}$ MP2 aVDZ & -1555 & -1427 & -1451 & -1433 & -1441 & -1405 & -1413 \\
$E_\text{int}^\text{bulk}$ MP2 aVDZ & - & - & - & -1433 & -1441 & -1405 & -1413 \\
$E_\text{int}$ MP2 CBS(aVDZ/aVTZ) & -1752 & -1622 & -1641 & -1616 & -1624 & - & - \\
$E_\text{int}^\text{bulk}$ MP2 CBS(aVDZ/aVTZ) & - & - & - & -1616 & -1624 & - & - \\
$E_\text{int}$ MP2 CBS(aVTZ/aVQZ) & -1768 & -1643 & - & - & - & - & - \\
$E_\text{int}$ MP2 CBS(awCVTZ/awCVQZ) & -1741 & -1608 & - & - & - & - & - \\
$E_\text{int}$ LMP2 CBS(aVDZ/aVTZ) & -1748 & -1617 & - & - & - & - & - \\
$E_\text{int}$ LNO-CCSD(T) CBS(aVDZ/aVTZ) & -1767 & -1644 & - & - & - & - & - \\
\toprule
H$_2$O on TiO$_2$ anatase(101) & 1 & 2 & 3 & 4 & 5 & 6 & 7 \\ 
\midrule
$E_\text{int}$ MP2 aVDZ & -1273 & -1266 & -1118 & -1107 & -1074 & -1061 & -1059 \\
$E_\text{int}^\text{bulk}$ MP2 aVDZ & - & - & - & -1107 & -1074 & -1061 & -1059 \\
$E_\text{int}$ MP2 CBS(aVDZ/aVTZ) & -1406 & -1396 & -1249 & -1236 & -1202 & - & - \\
$E_\text{int}^\text{bulk}$ MP2 CBS(aVDZ/aVTZ) & - & - & - & -1236 & -1202 & - & - \\
$E_\text{int}$ MP2 CBS(aVTZ/aVQZ) & -1407 & -1399 & - & - & - & - & - \\
$E_\text{int}$ MP2 CBS(awCVTZ/awCVQZ) & -1394 & -1384 & - & - & - & - & - \\
$E_\text{int}$ LMP2 CBS(aVDZ/aVTZ) & -1401 & -1391 & - & - & - & - & - \\
$E_\text{int}$ LNO-CCSD(T) CBS(aVDZ/aVTZ) & -1419 & -1409 & - & - & - & - & - \\
\toprule
NH$_3$ on TiO$_2$ anatase(101) & 1 & 2 & 3 & 4 & 5 & 6 & 7 \\ 
\midrule
$E_\text{int}$ MP2 aVDZ & -1348 & -1333 & -1272 & -1262 & -1223 & -1209 & -1206 \\
$E_\text{int}^\text{bulk}$ MP2 aVDZ & - & - & - & -1262 & -1223 & -1209 & -1206 \\
$E_\text{int}$ MP2 CBS(aVDZ/aVTZ) & -1511 & -1491 & -1422 & -1410 & -1368 & - & - \\
$E_\text{int}^\text{bulk}$ MP2 CBS(aVDZ/aVTZ) & - & - & - & -1410 & -1368 & - & - \\
$E_\text{int}$ MP2 CBS(aVTZ/aVQZ) & -1523 & -1506 & - & - & - & - & - \\
$E_\text{int}$ MP2 CBS(awCVTZ/awCVQZ) & -1511 & -1492 & - & - & - & - & - \\
$E_\text{int}$ LMP2 CBS(aVDZ/aVTZ) & -1506 & -1485 & - & - & - & - & - \\
$E_\text{int}$ LNO-CCSD(T) CBS(aVDZ/aVTZ) & -1514 & -1493 & - & - & - & - & - \\
\end{longtable}

\clearpage

\subsection{Multilevel $\Delta$CC contribution through mechanical embedding with small clusters}

Another quality of the quantum clusters generated by the SKZCAM protocol is that the difference between methods from cWFT, namely CCSD(T) and MP2, is relatively independent of the chosen cluster size.
Thus, from the small quantum clusters generated by the SKZCAM protocol where local CCSD(T) is affordable, one can obtain a $\Delta$CC contribution which can be used to elevate the MP2 bulk limit effectively up to the CCSD(T) bulk limit.
This mechanical embedding procedure is based upon Morokuma’s ONIOM method~\cite{svenssonONIOMMultilayeredIntegrated1996} as well as the hybrid high-level:low-level approach of Sauer~\cite{sauerInitioCalculationsMolecule2019b}.
For each of the studied adsorbate--surface systems and adsorption configurations, we perform local CCSD(T) calculations up to the third SKZCAM cluster (where possible) as shown in Table~\ref{tab:deltacc}.
We set $\Delta$CC to be the average of the three cluster, with its error taken to be the maximum deviation from the mean.
We find that the deviations are very small; being $1\,$meV or less for most of the systems involving the MgO surface and \ce{TiO2} anatase surface, while this error is typically below $5\,$meV for the \ce{TiO2} rutile surface systems.

\begin{table}[h]
\caption{\label{tab:deltacc}Comparison of the $\Delta$CC values for the clusters generated from SKZCAM protocol for all of the studied systems. The mean is calculated from the set of clusters used, with the error being the maximum deviation from the mean.}
\begin{adjustbox}{center,max width=1\textwidth}
\begin{tabular}{lrrrrr}
\toprule
Cluster & 1 & 2 & 3 & Mean & Error \\ 
System &  &  &  &  &  \\
\midrule
CH$_4$ on MgO(001) & -15 & -16 & -16 & -16 & 0 \\
Monolayer CH$_4$ on MgO(001) & -15 & -16 & -16 & -16 & 0 \\
C$_2$H$_6$ on MgO(001) & -20 & -20 & -20 & -20 & 0 \\
Monolayer C$_2$H$_6$ on MgO(001) & -19 & -19 & -20 & -20 & 1 \\
CO on MgO(001) & -7 & -7 & -9 & -8 & 1 \\
C$_6$H$_6$ on MgO(001) & 24 & 24 & 25 & 24 & 1 \\
Parallel N$_2$O on MgO(001) & -6 & -7 & -7 & -6 & 1 \\
Tilted N$_2$O on MgO(001) & 17 & 16 & 16 & 16 & 1 \\
Vertical-Hollow NO on MgO(001) & 289 & - & - & 289 & 0 \\
Vertical-Mg NO on MgO(001) & -68 & - & - & -68 & 0 \\
Bent-Bridge NO on MgO(001) & 626 & - & - & 626 & 0 \\
Bent-Mg NO on MgO(001) & -48 & - & - & -48 & 0 \\
Bent-O NO on MgO(001) & 659 & - & - & 659 & 0 \\
Dimer NO on MgO(001) & 18 & 17 & - & 18 & 1 \\
Monomer H$_2$O on MgO(001) & -26 & -25 & -26 & -25 & 0 \\
Tetramer H$_2$O on MgO(001) & -27 & -27 & -27 & -27 & 0 \\
Tilted CH$_3$OH on MgO(001) & -22 & -21 & -22 & -21 & 1 \\
Parallel CH$_3$OH on MgO(001) & -16 & -17 & -17 & -17 & 0 \\
Tetramer CH$_3$OH on MgO(001) & -27 & -27 & -27 & -27 & 0 \\
NH$_3$ on MgO(001) & -31 & -30 & -31 & -31 & 1 \\
Physisorbed CO$_2$ on MgO(001) & -30 & -31 & -32 & -31 & 1 \\
Chemisorbed CO$_2$ on MgO(001) & -218 & -219 & -224 & -220 & 4 \\
CH$_4$ on TiO$_2$ rutile(110) & -1 & 0 & - & -1 & 1 \\
Parallel CO$_2$ on TiO$_2$ rutile(110) & 8 & 5 & - & 7 & 2 \\
Tilted CO$_2$ on TiO$_2$ rutile(110) & -23 & -24 & - & -24 & 1 \\
H$_2$O on TiO$_2$ rutile(110) & -29 & -39 & - & -34 & 5 \\
CH$_3$OH on TiO$_2$ rutile(110) & -19 & -27 & - & -23 & 4 \\
H$_2$O on TiO$_2$ anatase(101) & -18 & -18 & - & -18 & 0 \\
NH$_3$ on TiO$_2$ anatase(101) & -8 & -8 & - & -8 & 0 \\
\bottomrule
\end{tabular}
\end{adjustbox}
\end{table}

It should be noted that within this $\Delta$CC contribution, there are also errors arising from using a local approximation for the CCSD(T) treatment.
In Table~\ref{tab:deltacc_lno_errors}, we have performed canonical CBS(DZ/TZ) CCSD(T) on the smallest cluster (corresponding to cluster `-2' for the \ce{TiO2} clusters) for a selection of adsorbate--surface systems.
Together with the canonical MP2 estimate we have already computed for these clusters, we compare these against the LNO-CCSD(T) and local-MP2 (LMP2) estimates of $E_\textrm{int}$.
It can be seen that the errors in LMP2 and LNO-CCSD(T) are all below $5\,$meV, attesting to the accuracy of our final $\Delta$CC contribution (using LNO-CCSD(T) and LMP2), where $\Delta$CC  itself has an MAD of $3\,$meV from the canonical $\Delta$CC for this selection of systems.
The final error which we compute for our $\Delta$CC contribution is the root squared sum of the error from the mechanical embedding procedure and $5\,$meV for the maximum observed error from using the local approximation.

\begin{table}
\caption{\label{tab:deltacc_lno_errors}Comparing canonical (C-)MP2, canonical CCSD(T) and canonical $\Delta$CC (in meV) against their local variants (i.e., LMP2, LNO-CCSD(T), L-$\Delta$CC). This was compared for the first cluster generated by the SKZCAM protocol for the specific MgO adsorption site in Fig.~\ref{fig:skzcam_clusters}, while it corresponds to cluster `${-}2$' for the \ce{TiO2} surfaces.}
\begin{adjustbox}{center,max width=1\textwidth}
\begin{tabular}{lrrrrrrrrr}
\toprule
 & \rotatebox{90}{C-MP2} & \rotatebox{90}{L-MP2} & \rotatebox{90}{L-MP2 Error} & \rotatebox{90}{C-CCSD(T)} & \rotatebox{90}{LNO-CCSD(T)} & \rotatebox{90}{LNO-CCSD(T) Error} & \rotatebox{90}{C-DeltaCC} & \rotatebox{90}{(L-)DeltaCC} & \rotatebox{90}{(L-)DeltaCC Error} \\ 
\midrule
CH$_4$ on MgO(001) & -71 & -70 & 0 & -91 & -86 & 5 & -20 & -15 & 4 \\
C$_2$H$_6$ on MgO(001) & -90 & -92 & 2 & -116 & -112 & 4 & -26 & -20 & 6 \\
CO on MgO(001) & -155 & -153 & 2 & -165 & -160 & 5 & -9 & -7 & 3 \\
Parallel N$_2$O on MgO(001) & -206 & -203 & 2 & -215 & -209 & 6 & -9 & -6 & 3 \\
Tilted N$_2$O on MgO(001) & -122 & -120 & 2 & -107 & -104 & 4 & 15 & 17 & 2 \\
Dimer NO on MgO(001) & -222 & -217 & 5 & -205 & -199 & 6 & 17 & 18 & 2 \\
Monomer H$_2$O on MgO(001) & -641 & -639 & 2 & -670 & -665 & 5 & -29 & -26 & 3 \\
Tilted CH$_3$OH on MgO(001) & -703 & -701 & 2 & -729 & -723 & 7 & -27 & -22 & 5 \\
Parallel CH$_3$OH on MgO(001) & -443 & -442 & 1 & -466 & -458 & 8 & -23 & -16 & 6 \\
NH$_3$ on MgO(001) & -551 & -547 & 4 & -583 & -578 & 5 & -32 & -31 & 1 \\
Physisorbed CO$_2$ on MgO(001) & -240 & -237 & 3 & -265 & -267 & 2 & -25 & -30 & 5 \\
CH$_4$ on TiO$_2$ rutile(110) & -395 & -395 & 0 & -398 & -394 & 5 & -3 & 2 & 4 \\
Parallel CO$_2$ on TiO$_2$ rutile(110) & -454 & -451 & 3 & -467 & -463 & 4 & -13 & -12 & 1 \\
Tilted CO$_2$ on TiO$_2$ rutile(110) & -830 & -828 & 3 & -851 & -850 & 0 & -20 & -23 & 2 \\
H$_2$O on TiO$_2$ rutile(110) & -2469 & -2466 & 3 & -2477 & -2477 & 0 & -9 & -12 & 3 \\
CH$_3$OH on TiO$_2$ rutile(110) & -2921 & -2916 & 5 & -2922 & -2921 & 1 & -1 & -5 & 4 \\
H$_2$O on TiO$_2$ anatase(101) & -1874 & -1871 & 3 & -1896 & -1895 & 1 & -22 & -23 & 1 \\
NH$_3$ on TiO$_2$ anatase(101) & -1801 & -1799 & 3 & -1822 & -1824 & 3 & -20 & -26 & 5 \\
MAD & 0 & 0 & 3 & 0 & 0 & 4 & 0 & 0 & 3 \\
\bottomrule
\end{tabular}
\end{adjustbox}
\end{table}

\clearpage

\subsection{Further multilevel contributions for basis set and semi-core electron correlation}

Besides the $\Delta$CC contribution, it is possible to make further multilevel contributions (especially now that it has been automated in Section~\ref{sec:quacc}) in the vein of Pople's `model chemistry'~\cite{popleQuantumChemicalModels1999}.
This lowers the cost of reaching an accurate estimate as these contributions (e.g., basis set size or frozen core treatment) need to only be evaluated on the small affordable clusters generated by the SKZCAM protocol.
Specifically we consider further contributions to: (1) fix our basis set extrapolation treatment to more accurately reach the basis set limit and (2) incorporate electron correlation contributions beyond the valence electrons from semi-core electrons.
Adding these contributions can lower the cost as it means that we can use cheaper settings (i.e., smaller basis sets while correlating only valence electrons) in the $E_\textrm{int}^\textrm{bulk}$ and $\Delta$CC calculations.
For the adsorbate--surface systems involving the \ce{TiO2} surfaces and the open-shell NO monomers on the MgO surface, we have computed the MP2 $E_\textrm{int}^\textrm{bulk}$ and $\Delta$CC with semi-core electrons (i.e., 2s2p on Mg or 3s3p on Ti) frozen in the correlation treatment.
This has been particularly important towards making it possible to tackle these systems, involving more electrons or requiring unrestricted MP2/CCSD(T), more affordable.
We use a $\Delta_\textrm{core}$ contribution to account for these missing effects at the MP2 level from the smaller and more affordable clusters of the SKZCAM protocol (up to cluster 3).
The $\Delta_\textrm{core}$ contributions are shown in Table~\ref{tab:deltacore}.
We take the mean of the clusters as our final $\Delta_\textrm{core}$ estimate and use the maximum deviation from the mean as the error.
The maximum deviation are all below $5\,$meV.

\begin{table}[h]
\caption{\label{tab:deltacore}Comparison of the $\Delta_\text{core}$ values for the clusters generated from SKZCAM protocol for all of the studied systems. The mean is calculated from the set of clusters used, with the error being the maximum deviation from the mean.}
\begin{adjustbox}{center,max width=1\textwidth}
\begin{tabular}{lrrrrr}
\toprule
Cluster & 1 & 2 & 3 & Mean & Error \\ 
System &  &  &  &  &  \\
\midrule
Vertical-Hollow NO on MgO(001) & -18 & -24 & -24 & -22 & 4 \\
Vertical-Mg NO on MgO(001) & -22 & -26 & -28 & -25 & 3 \\
Bent-Bridge NO on MgO(001) & -19 & -26 & -28 & -24 & 5 \\
Bent-Mg NO on MgO(001) & -22 & -27 & -29 & -26 & 3 \\
Bent-O NO on MgO(001) & -9 & -15 & -18 & -14 & 5 \\
CH$_4$ on TiO$_2$ rutile(110) & -2 & -3 & - & -2 & 0 \\
Parallel CO$_2$ on TiO$_2$ rutile(110) & -8 & -7 & - & -7 & 0 \\
Tilted CO$_2$ on TiO$_2$ rutile(110) & 8 & 10 & - & 9 & 1 \\
H$_2$O on TiO$_2$ rutile(110) & 37 & 44 & - & 41 & 4 \\
CH$_3$OH on TiO$_2$ rutile(110) & 28 & 35 & - & 31 & 4 \\
H$_2$O on TiO$_2$ anatase(101) & 13 & 15 & - & 14 & 1 \\
NH$_3$ on TiO$_2$ anatase(101) & 13 & 14 & - & 13 & 1 \\
\bottomrule
\end{tabular}
\end{adjustbox}
\end{table}

For all of the adsorbate--surface systems, we add a further $\Delta_\textrm{basis}$ contribution which corrects for potentials errors in our extrapolation treatment to reach the complete basis set limit; we used a two-point CBS(aVDZ/aVTZ) or CBS(awCDZ/awCVTZ) extrapolation.
This contribution is calculated by considering the difference between the more accurate CBS(aVTZ/aVQZ) or CBS(awCVTZ/awCVQZ) extrapolation treatment with the CBS(aVDZ/aVTZ) or CBS(awCVDZ/awCVTZ) extrapolation treatment for the smallest clusters.
The calculated $\Delta_\textrm{basis}$ contributions for the SKZCAM protocol clusters are shown in Table~\ref{tab:deltabasis}.
We take the mean of the clusters as our final $\Delta_\textrm{basis}$ estimate and use the maximum deviation from the mean as the error, with most systems having errors below $5\,$meV, besides the larger systems (e.g., \ce{C6H6} and the tetramers on MgO(001)), where deviations may go up to $10\,$meV.

Both $\Delta_\textrm{core}$ and $\Delta_\textrm{basis}$ are important contributions which enable our final $H_\textrm{ads}$ estimate to better match experiments while maintaining a low cost, with contributions up to $40\,$meV for the former and $13\,$meV in the latter.

\begin{table}[h]
\caption{\label{tab:deltabasis}Comparison of the $\Delta_\text{basis}$ values for the clusters generated from SKZCAM protocol for all of the studied systems. The mean is calculated from the set of clusters used, with the error being the maximum deviation from the mean.}
\begin{adjustbox}{center,max width=1\textwidth}
\begin{tabular}{lrrrrr}
\toprule
Cluster & 1 & 2 & 3 & Mean & Error \\ 
System &  &  &  &  &  \\
\midrule
CH$_4$ on MgO(001) & 3 & 0 & 2 & 2 & 1 \\
Monolayer CH$_4$ on MgO(001) & 3 & 1 & 2 & 2 & 1 \\
C$_2$H$_6$ on MgO(001) & 3 & 3 & 4 & 3 & 1 \\
Monolayer C$_2$H$_6$ on MgO(001) & 2 & 2 & 4 & 3 & 1 \\
CO on MgO(001) & 0 & -1 & 3 & 1 & 2 \\
C$_6$H$_6$ on MgO(001) & -15 & -15 & -2 & -11 & 9 \\
Parallel N$_2$O on MgO(001) & -2 & -2 & 2 & -1 & 2 \\
Tilted N$_2$O on MgO(001) & -6 & -6 & -3 & -5 & 2 \\
Vertical-Hollow NO on MgO(001) & 10 & 8 & 8 & 9 & 1 \\
Vertical-Mg NO on MgO(001) & 2 & 7 & 8 & 5 & 4 \\
Bent-Bridge NO on MgO(001) & 2 & -5 & -4 & -2 & 4 \\
Bent-Mg NO on MgO(001) & -2 & 2 & 2 & 1 & 3 \\
Bent-O NO on MgO(001) & 1 & 2 & -6 & -1 & 5 \\
Dimer NO on MgO(001) & -4 & -3 & -1 & -2 & 2 \\
Monomer H$_2$O on MgO(001) & 0 & -3 & 3 & 0 & 3 \\
Tetramer H$_2$O on MgO(001) & -19 & -18 & -9 & -15 & 6 \\
Tilted CH$_3$OH on MgO(001) & -1 & -3 & 2 & -1 & 3 \\
Parallel CH$_3$OH on MgO(001) & 17 & 19 & 18 & 18 & 1 \\
Tetramer CH$_3$OH on MgO(001) & -18 & -18 & -11 & -16 & 4 \\
NH$_3$ on MgO(001) & -19 & -20 & -9 & -16 & 7 \\
Physisorbed CO$_2$ on MgO(001) & 1 & 2 & 4 & 3 & 2 \\
Chemisorbed CO$_2$ on MgO(001) & 23 & 32 & 20 & 25 & 7 \\
CH$_4$ on TiO$_2$ rutile(110) & 1 & 1 & - & 1 & 0 \\
Parallel CO$_2$ on TiO$_2$ rutile(110) & -1 & -3 & - & -2 & 1 \\
Tilted CO$_2$ on TiO$_2$ rutile(110) & -6 & -8 & - & -7 & 1 \\
H$_2$O on TiO$_2$ rutile(110) & -16 & -20 & - & -18 & 2 \\
CH$_3$OH on TiO$_2$ rutile(110) & -16 & -21 & - & -19 & 2 \\
H$_2$O on TiO$_2$ anatase(101) & 0 & -3 & - & -2 & 1 \\
NH$_3$ on TiO$_2$ anatase(101) & -13 & -15 & - & -14 & 1 \\
\bottomrule
\end{tabular}
\end{adjustbox}
\end{table}

\clearpage

\subsection{The final $E_\textrm{int}$ estimates and their error bars} \label{sec:skzcam_error_bar_estimates}
To summarise the above sections, the final CCSD(T) interaction energy bulk-limit estimate, $E_\textrm{int}^\textrm{CCSD(T)}$, is given as the sum of the following contributions:
\begin{equation}
    E_\textrm{int}^\textrm{bulk CCSD(T)} = E_\textrm{int}^\textrm{bulk MP2} + \Delta_\textrm{basis} + \Delta_\textrm{core} + \Delta\textrm{CC}.
\end{equation}
The contributions are shown in Table~\ref{tab:final_eint} for all of the adsorbate--surface systems studied within this work.
The series of clusters generated by the SKZCAM protocol provides a means to estimate the error from each of these contributions.
For $E_\textrm{int}^\textrm{bulk MP2}$, we calculate the error in this term by finding the maximum deviation when a larger set of clusters are used to extrapolate $E_\textrm{int}^\textrm{bulk MP2}$ (with only the awCVDZ/aVDZ basis set) for the systems with the MgO surface (barring the chemisorbed \ce{CO2} system).
For the \ce{TiO2} rutile(110) and anatase(101) surfaces, as well as the chemisorbed \ce{CO2} on MgO(001), the larger clusters are used to estimate the maximum deviation.
For $\Delta$CC, $\Delta_\textrm{basis}$ and $\Delta_\textrm{core}$, we estimate the error by taking the maximum deviation from the mean of the clusters used to calculate these terms.
The final error on $E_\textrm{int}^\textrm{bulk CCSD(T)}$ is the root squared sum of the errors from each of these terms.
It can be seen in Table~\ref{tab:final_eint} that the errors are typically below $10\,$meV for all of the MgO adsorbate--surface systems, with some of the \ce{TiO2} adsorbate--surface systems having a larger error, but all are within chemical accuracy of $43\,$meV.

\begin{table}
\caption{\label{tab:final_eint}Final E$_\text{int}$ values (in meV) for the systems studied in this work. We show the individual terms which make up these final $E_\text{int}$ values and also give final MP2, CCSD and CCSD(T) estimates, where the latter two are obtained by adding the $\Delta$CC values to the final MP2 $E_\text{int}$ value.}
\begin{adjustbox}{center,max width=0.9\textwidth}
\begin{tabular}{lrrrrrrrr}
\toprule
 & \rotatebox{90}{$E_\text{int}^\text{bulk MP2}$} & \rotatebox{90}{$\Delta$CC [CCSD(T)]} & \rotatebox{90}{$\Delta$CC [CCSD]} & \rotatebox{90}{$\Delta_\text{basis}$} & \rotatebox{90}{$\Delta_\text{core}$} & \rotatebox{90}{$E_\text{int}^\text{autoSKZCAM}$ [MP2]} & \rotatebox{90}{$E_\text{int}^\text{autoSKZCAM}$ [CCSD]} & \rotatebox{90}{$E_\text{int}^\text{autoSKZCAM}$ [CCSD(T)]} \\ 
System &  &  &  &  &  &  &  &  \\
\midrule
CH$_4$ on MgO(001) & -108 $\pm$ 2 & -16 $\pm$ 0 & 18 $\pm$ 7 & 2 $\pm$ 1 & 0 $\pm$ 0 & -106 $\pm$ 2 & -88 $\pm$ 7 & -122 $\pm$ 2 \\
Monolayer CH$_4$ on MgO(001) & -107 $\pm$ 2 & -16 $\pm$ 0 & 18 $\pm$ 7 & 2 $\pm$ 1 & 0 $\pm$ 0 & -105 $\pm$ 3 & -87 $\pm$ 7 & -121 $\pm$ 3 \\
C$_2$H$_6$ on MgO(001) & -158 $\pm$ 3 & -20 $\pm$ 0 & 32 $\pm$ 13 & 3 $\pm$ 1 & 0 $\pm$ 0 & -155 $\pm$ 4 & -122 $\pm$ 14 & -175 $\pm$ 4 \\
Monolayer C$_2$H$_6$ on MgO(001) & -144 $\pm$ 3 & -20 $\pm$ 1 & 29 $\pm$ 13 & 3 $\pm$ 1 & 0 $\pm$ 0 & -141 $\pm$ 3 & -112 $\pm$ 13 & -161 $\pm$ 3 \\
CO on MgO(001) & -200 $\pm$ 3 & -8 $\pm$ 1 & 33 $\pm$ 8 & 1 $\pm$ 2 & 0 $\pm$ 0 & -199 $\pm$ 3 & -166 $\pm$ 9 & -207 $\pm$ 4 \\
C$_6$H$_6$ on MgO(001) & -460 $\pm$ 2 & 24 $\pm$ 1 & 164 $\pm$ 24 & -11 $\pm$ 9 & 0 $\pm$ 0 & -470 $\pm$ 9 & -307 $\pm$ 26 & -446 $\pm$ 9 \\
Parallel N$_2$O on MgO(001) & -249 $\pm$ 2 & -6 $\pm$ 1 & 40 $\pm$ 6 & -1 $\pm$ 2 & 0 $\pm$ 0 & -250 $\pm$ 3 & -210 $\pm$ 7 & -256 $\pm$ 3 \\
Tilted N$_2$O on MgO(001) & -179 $\pm$ 3 & 16 $\pm$ 1 & 64 $\pm$ 11 & -5 $\pm$ 2 & 0 $\pm$ 0 & -184 $\pm$ 4 & -120 $\pm$ 11 & -168 $\pm$ 4 \\
Vertical-Hollow NO on MgO(001) & -244 $\pm$ 1 & 289 $\pm$ 0 & 334 $\pm$ 0 & 9 $\pm$ 1 & -22 $\pm$ 4 & -257 $\pm$ 4 & 76 $\pm$ 4 & 32 $\pm$ 4 \\
Vertical-Mg NO on MgO(001) & 26 $\pm$ 2 & -68 $\pm$ 0 & -49 $\pm$ 0 & 5 $\pm$ 4 & -25 $\pm$ 3 & 6 $\pm$ 5 & -43 $\pm$ 5 & -62 $\pm$ 5 \\
Bent-Bridge NO on MgO(001) & -661 $\pm$ 3 & 626 $\pm$ 0 & 693 $\pm$ 0 & -2 $\pm$ 4 & -24 $\pm$ 5 & -688 $\pm$ 8 & 6 $\pm$ 8 & -62 $\pm$ 8 \\
Bent-Mg NO on MgO(001) & -53 $\pm$ 1 & -48 $\pm$ 0 & -23 $\pm$ 0 & 1 $\pm$ 3 & -26 $\pm$ 3 & -78 $\pm$ 5 & -102 $\pm$ 5 & -126 $\pm$ 5 \\
Bent-O NO on MgO(001) & -647 $\pm$ 0 & 659 $\pm$ 0 & 720 $\pm$ 0 & -1 $\pm$ 5 & -14 $\pm$ 5 & -661 $\pm$ 7 & 59 $\pm$ 7 & -3 $\pm$ 7 \\
Dimer NO on MgO(001) & -261 $\pm$ 2 & 18 $\pm$ 1 & 67 $\pm$ 5 & -2 $\pm$ 2 & 0 $\pm$ 0 & -263 $\pm$ 2 & -197 $\pm$ 5 & -246 $\pm$ 3 \\
Monomer H$_2$O on MgO(001) & -677 $\pm$ 2 & -25 $\pm$ 0 & 22 $\pm$ 9 & 0 $\pm$ 3 & 0 $\pm$ 0 & -677 $\pm$ 4 & -655 $\pm$ 10 & -703 $\pm$ 4 \\
Tetramer H$_2$O on MgO(001) & -421 $\pm$ 3 & -27 $\pm$ 0 & 3 $\pm$ 7 & -15 $\pm$ 6 & 0 $\pm$ 0 & -436 $\pm$ 7 & -433 $\pm$ 10 & -463 $\pm$ 7 \\
Tilted CH$_3$OH on MgO(001) & -765 $\pm$ 4 & -21 $\pm$ 1 & 45 $\pm$ 14 & -1 $\pm$ 3 & 0 $\pm$ 0 & -765 $\pm$ 5 & -720 $\pm$ 14 & -787 $\pm$ 5 \\
Parallel CH$_3$OH on MgO(001) & -508 $\pm$ 1 & -17 $\pm$ 0 & 50 $\pm$ 10 & 18 $\pm$ 1 & 0 $\pm$ 0 & -490 $\pm$ 1 & -440 $\pm$ 10 & -506 $\pm$ 1 \\
Tetramer CH$_3$OH on MgO(001) & -469 $\pm$ 4 & -27 $\pm$ 0 & 18 $\pm$ 10 & -16 $\pm$ 4 & 0 $\pm$ 0 & -485 $\pm$ 6 & -466 $\pm$ 12 & -511 $\pm$ 6 \\
NH$_3$ on MgO(001) & -610 $\pm$ 5 & -31 $\pm$ 1 & 9 $\pm$ 16 & -16 $\pm$ 7 & 0 $\pm$ 0 & -626 $\pm$ 9 & -618 $\pm$ 18 & -657 $\pm$ 9 \\
Physisorbed CO$_2$ on MgO(001) & -280 $\pm$ 1 & -31 $\pm$ 1 & 21 $\pm$ 7 & 3 $\pm$ 2 & 0 $\pm$ 0 & -277 $\pm$ 2 & -256 $\pm$ 7 & -308 $\pm$ 2 \\
Chemisorbed CO$_2$ on MgO(001) & -3309 $\pm$ 32 & -220 $\pm$ 4 & -447 $\pm$ 7 & 25 $\pm$ 7 & 0 $\pm$ 0 & -3284 $\pm$ 32 & -3731 $\pm$ 33 & -3504 $\pm$ 32 \\
CH$_4$ on TiO$_2$ rutile(110) & -267 $\pm$ 2 & -1 $\pm$ 1 & 53 $\pm$ 0 & 1 $\pm$ 0 & -2 $\pm$ 0 & -268 $\pm$ 2 & -216 $\pm$ 2 & -269 $\pm$ 2 \\
Parallel CO$_2$ on TiO$_2$ rutile(110) & -408 $\pm$ 4 & 7 $\pm$ 2 & 71 $\pm$ 1 & -2 $\pm$ 1 & -7 $\pm$ 0 & -417 $\pm$ 4 & -346 $\pm$ 4 & -410 $\pm$ 5 \\
Tilted CO$_2$ on TiO$_2$ rutile(110) & -470 $\pm$ 6 & -24 $\pm$ 1 & 8 $\pm$ 1 & -7 $\pm$ 1 & 9 $\pm$ 1 & -469 $\pm$ 7 & -460 $\pm$ 7 & -493 $\pm$ 7 \\
H$_2$O on TiO$_2$ rutile(110) & -1299 $\pm$ 33 & -34 $\pm$ 5 & -68 $\pm$ 6 & -18 $\pm$ 2 & 41 $\pm$ 4 & -1276 $\pm$ 33 & -1344 $\pm$ 34 & -1310 $\pm$ 33 \\
CH$_3$OH on TiO$_2$ rutile(110) & -1624 $\pm$ 36 & -23 $\pm$ 4 & -31 $\pm$ 5 & -19 $\pm$ 2 & 31 $\pm$ 4 & -1612 $\pm$ 36 & -1642 $\pm$ 37 & -1634 $\pm$ 37 \\
H$_2$O on TiO$_2$ anatase(101) & -1202 $\pm$ 15 & -18 $\pm$ 0 & -32 $\pm$ 0 & -2 $\pm$ 1 & 14 $\pm$ 1 & -1190 $\pm$ 16 & -1222 $\pm$ 16 & -1208 $\pm$ 16 \\
NH$_3$ on TiO$_2$ anatase(101) & -1368 $\pm$ 18 & -8 $\pm$ 0 & -35 $\pm$ 1 & -14 $\pm$ 1 & 13 $\pm$ 1 & -1369 $\pm$ 18 & -1404 $\pm$ 18 & -1377 $\pm$ 18 \\
\bottomrule
\end{tabular}
\end{adjustbox}
\end{table}
\clearpage

\subsection{\label{sec:lih_nacl_test}Validating the SKZCAM protocol beyond metal-oxides}

To check the validity of the SKZCAM protocol and thus, autoSKZCAM framework, beyond metal-oxides towards surfaces of traditional ionic materials, we also performed additional calculations to obtain the $E_\text{int}$ for H$_2$O on LiH(001) and acetylene on NaCl(001).
These two systems follow the same set of clusters and calculations as MgO(001) described in the previous subsections.
An $E_\text{int}$ of $-265 \pm 9\,$meV for H$_2$O on LiH(001) computed with the SKZCAM protocol is shown in Table~\ref{tab:lih_eint} and compared to previous calculations by Tsatsoulis \etal{}~\cite{tsatsoulisComparisonQuantumChemistry2017}, being within the statistical error bars of DMC ($-250 \pm 7\,$meV) and an alternative (gas-phase) cluster CCSD(T) approach ($-256\,$meV).
\begin{table}[h]
\caption{\label{tab:lih_eint}Final $E_\text{int}$ values (in meV) with the autoSKCAM framework for the water adsorbed on LiH(001). The DMC and CCSD(T) values are taken from Ref.~\citenum{tsatsoulisComparisonQuantumChemistry2017}.}
\begin{adjustbox}{center,max width=1\textwidth}
\begin{tabular}{lr}
\toprule
 H$_2$O on LiH(001) & Contribution [meV] \\ 
\midrule
$E_\text{int}^\text{bulk MP2}$ & -247$\pm$1 \\
$\Delta$CC & -17$\pm$5 \\
$\Delta_\text{basis}$ & -2$\pm$8 \\
Final autoSKZCAM $E_\text{int}$ & -265$\pm$9 \\
Cluster CCSD(T)~\cite{tsatsoulisComparisonQuantumChemistry2017} & -256 \\
DMC~\cite{tsatsoulisComparisonQuantumChemistry2017} & -250$\pm$7 \\
\bottomrule
\end{tabular}
\end{adjustbox}
\end{table}

For acetylene on NaCl(001), we have inferred the experimental estimate of $E_\text{int}$ from previous experimental estimates by Cabello-Cartagena \etal{}~\cite{cabello-cartagenaStructureInfraredAbsorption2010} and Dunn and Ewing~\cite{dunnInfraredSpectraStructure1992}.
This is shown in Table~\ref{tab:nacl_expt_hads_eint}, where the lateral interactions~\cite{alloucheQuantumInitioStudy1997,picaudPhononlibronDynamicsAcetylene1998} are first removed from the $H_\text{ads}$ estimates by the experiments, followed by a correction to convert the dilute limit $H_\text{ads}$ to $E_\text{int}$ using revPBE-D4.
The final estimates of $E_\text{int}$ for the dilute limit go from $-130\,$meV to $-237\,$meV by Cabello-Cartagena \etal{} and Dunn and Ewing, respectively, covering a large range of $107\,$meV.
The $E_\text{int}$ estimate by the SKZCAM protocol of $-181\pm 3$ lies in the middle of this range, confirming its applicability to this system.

\begin{table}[h]
\caption{\label{tab:nacl_expt_hads_eint}Converting to the experimental values for the adsorption enthalpy $H_\text{ads}$ to an interaction energy $E_\text{int}$ by removing lateral interaction and the contributions to convert from an $H_\textrm{ads}$ to $E_\textrm{int}$ at the revPBE-D4 level (encapsulated in $\Delta^{H_\text{ads}}_{E_\text{int}}$[revPBE-D4]).}
\begin{adjustbox}{center,max width=1\textwidth}
\begin{tabular}{lrr}
\toprule
 & Cabello-Cartagena et al.~\cite{cabello-cartagenaStructureInfraredAbsorption2010} & Dunn and Ewing~\cite{dunnInfraredSpectraStructure1992} \\ 
\midrule
Coverage ($\theta$) & 1.0 & 0.5 \\
$H_\text{ads}$ & -249 & -311 \\
Lateral Interaction~\cite{alloucheQuantumInitioStudy1997,picaudPhononlibronDynamicsAcetylene1998} & -140 & -95 \\
$\Delta^{H_\text{ads}}_{E_\text{int}}$[revPBE-D4] & 18 & 18 \\
$E_\text{int}$ & -127 & -234 \\
\bottomrule
\end{tabular}
\end{adjustbox}
\end{table}
\begin{table}[h]
\caption{\label{tab:nacl_eint}Final $E_\text{int}$ values (in meV) with the autoSKCAM framework for acetylene adsorbed on NaCl(001). The (inferred) experimental $E_\text{int}$ is explained in the text and shown in Table~\ref{tab:nacl_expt_hads_eint}.}
\begin{adjustbox}{center,max width=1\textwidth}
\begin{tabular}{lr}
\toprule
 Acetylene on NaCl(001) & Contribution [meV] \\ 
\midrule
$E_\text{int}^\text{bulk MP2}$ & -203$\pm$1 \\
$\Delta$CC & 22$\pm$1 \\
$\Delta_\text{basis}$ & 0$\pm$2 \\
Final autoSKZCAM $E_\text{int}$ & -181$\pm$3 \\
Experiment [from Table~\ref{tab:nacl_expt_hads_eint}] & -127 to -234 \\
\bottomrule
\end{tabular}
\end{adjustbox}
\end{table}

\clearpage

\subsection{\label{sec:improvements_skzcam}Improvements to SKZCAM protocol in present work}

We highlight here that because the SKZCAM protocol has been automatised in the present work (described in Sec.~\ref{sec:quacc}), it has allowed for a significant lowering of computational cost compared to previous applications of the SKZCAM protocol~\cite{shiManyBodyMethodsSurface2023a,shiGoingGoldStandardAttaining2024}.
The improvements are as follows:
\begin{enumerate}
    \item More intermediate layers in the ONIOM mechanical embedding treatment. For example, we have added the $\Delta_\textrm{basis}$ and $\Delta_\textrm{core}$ terms to the final $E_\textrm{int}$, enabling cheaper calculations of $\Delta$CC and $E_\textrm{int}^\textrm{bulk MP2}$ -- the two dominant contributions to the overall computational cost -- including only valence, no longer requiring semicore, electrons in the correlation treatment (due to $\Delta_\textrm{core}$) and a smaller basis set treatment (due to $\Delta_\textrm{basis}$).
    \item The use of multiple different codes that are efficient for different types of calculations. We can utilise the ORCA code to perform efficient MP2 calculations~\cite{neeseORCAQuantumChemistry2020} while still leveraging the LNO-CCSD(T) implementation~\cite{nagyOptimizationLinearScalingLocal2018,nagyApproachingBasisSet2019} in MRCC. Previously, we were limited towards only using one code due to the additional manual overhead.
\end{enumerate}

Overall, these improvements enable significant lowering of costs.
To put these improvements into context, our previous simulation~\cite{shiManyBodyMethodsSurface2023a} of the CO on MgO(001) required arond 20,000 CPUh to perform while these improvements enable a cost to be lowered by two orders of magnitude to around $600\,$CPUh.
Without these improvements, the application to the TiO$_2$ surface systems would have been largely unfeasible as well.

\clearpage

\section{\label{sec:wft_ecoh_econf} Contributions for the cohesive and conformational energy from cWFT in selected systems}

For the adsorption of monomers on the MgO surface, the definition of $E_\textrm{int}$ has been defined in Fig.~\ref{fig:eads_definition}, where its difference with the adsorption energy $E_\textrm{ads}$ is the energy to relax the monomer molecule and surface from their geometries in the adsorbate--surface complex to their equilibrium geometry, given by the quantity $E_\textrm{rlx}$.
When computing the adsorption energy of clusters and the monolayers on the MgO(001) surface, there are additional contributions towards $E_\textrm{ads}$ arising from the cohesive energy $E_\textrm{coh}$ (or lateral interaction) that binds the molecules together.
In general, $E_\textrm{rlx}$ does not form a large contribution towards $E_\textrm{ads}$, with $E_\textrm{int}$ being the dominant contribution, oftentimes more than 80\% of $H_\textrm{ads}$.
This is one of the reasons why we have computed $E_\textrm{rlx}$ with DFT.
However, it is not the case for systems where the molecule undergoes significant conformational changes (due to e.g., charge transfer).
This is specifically the case for the chemisorbed \ce{CO2} on MgO(001), where there is a large energy change of almost ${\sim}2\,$eV to bring the \ce{CO2} from its bent geometry back into its linear geometry.
This large conformational energy change $E_\textrm{conf}$ is a quantity that is highly sensitive to the DFA.
As such, for this system, we have computed $E_\textrm{conf}$ with CCSD(T) - a negligible overall computational cost for the \ce{CO2} molecule.
We elaborate further on the value of these terms in the next sections.

\subsection{\label{sec:co2_econf_contributions} Conformational energy of the chemisorbed \ce{CO2} on MgO(001)}

\begin{figure}
    \includegraphics[width=\textwidth]{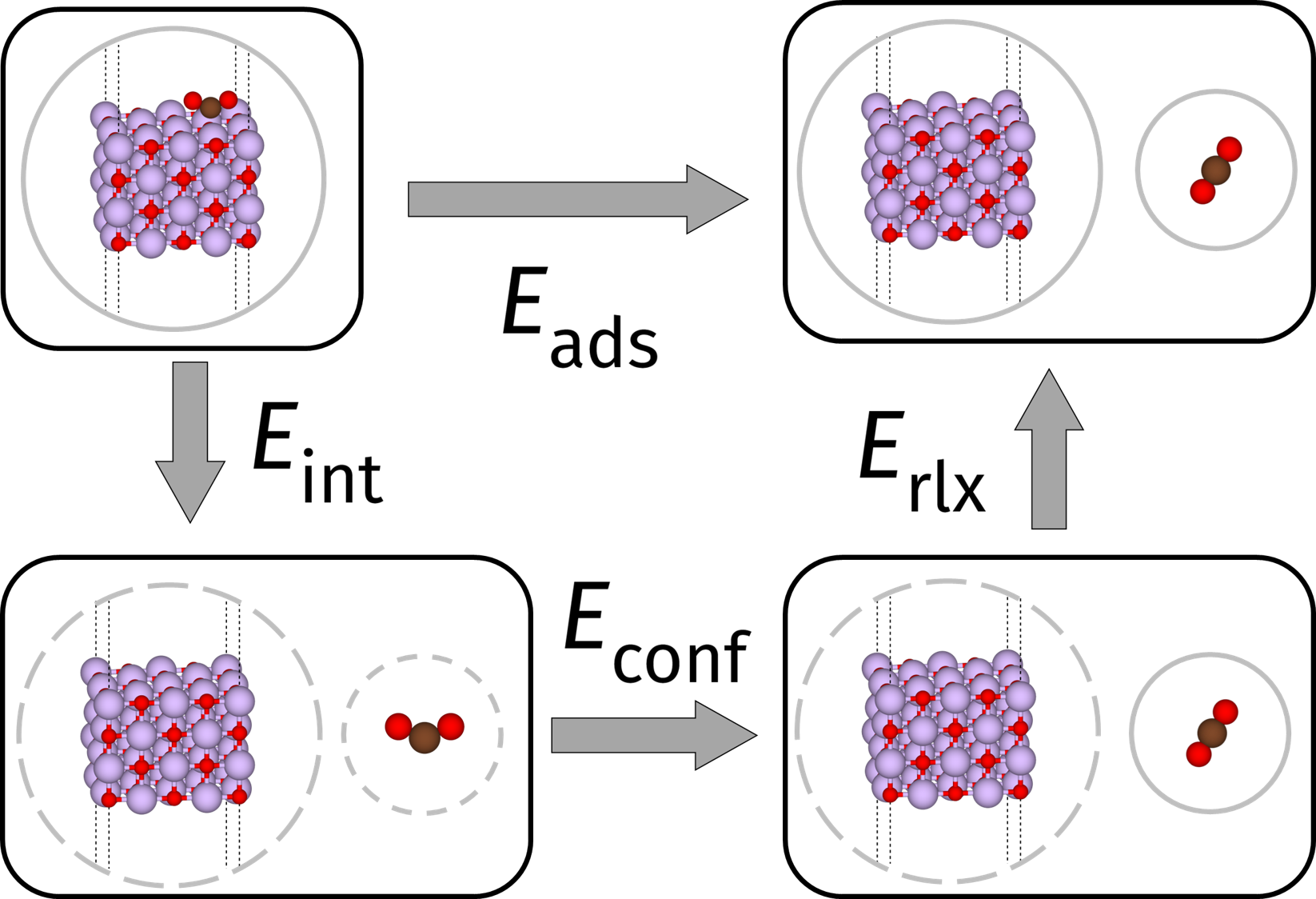}
    \caption{\label{fig:co2_conform} The contributions to the adsorption energy $E_\textrm{ads}$ for the chemisorbed \ce{CO2} on MgO(001). Beyond the interaction energy $E_\textrm{int}$, there is $E_\textrm{conf}$ - the relaxation energy to return the molecule in its geometry within the adsorbate--surface complex to its equilibrium geometry, leaving $E_\textrm{rlx}$ to be the relaxation energy to bring the surface (at the geometry of the adsorbate--surface complex) into its equilibrium geometry. The circles represent a single system/calculation, with a dashed circle indicating a geometry fixed to that found in the adsorbate--surface complex while a line circle indicates an equilibrium geometry.} 
\end{figure}

As illustrated in Fig.~\ref{fig:co2_conform}, for the chemisorbed \ce{CO2} on MgO(001), we have effectively broken up the original $E_\textrm{rlx}$, containing the relaxation energy for both the molecule and surface, such that it only pertains to the surface, with $E_\textrm{conf}$ corresponding to the relaxation energy for the molecule.
The importance of this contribution is shown in Table~\ref{tab:econf_co2} where the conformational energy change is predicted by several DFAs (as well as MP2, CCSD) and find that they can differ by more than $300\,$meV from CCSD(T).
These are well-known effects arising from the delocalisation error inherent in most DFAs~\cite{bryentonDelocalizationErrorGreatest2023}.
The canonical MP2, CCSD and CCSD(T) calculations were performed in \texttt{MRCC} with the CBS(aVTZ/aVQZ) basis set treatment.

\begin{table}
\caption{\label{tab:econf_co2}Comparison of the $E_\text{conf}$ values (in meV) for chemisorbed CO$_2$ chemisorbed on MgO for several DFAs as well as MP2, CCSD and CCSD(T).}
\begin{tabular}{lr}
\toprule
 & $E_\text{conf}$ [meV] \\ 
\midrule
PBE-D2[Ne] & 1841 \\
revPBE-D4 & 1797 \\
vdW-DF & 1763 \\
rev-vdW-DF2 & 1836 \\
PBE0-D4 & 2153 \\
B3LYP-D2[Ne] & 2075 \\
MP2 & 2045 \\
CCSD & 2304 \\
CCSD(T) & 2094 \\
\bottomrule
\end{tabular}
\end{table}

\clearpage
\subsection{\label{sec:ml_ecoh_contributions} Cohesive energy in \ce{CH4} and \ce{C2H6} monolayers on MgO(001)}

\begin{figure}[h]
    \includegraphics[width=\textwidth]{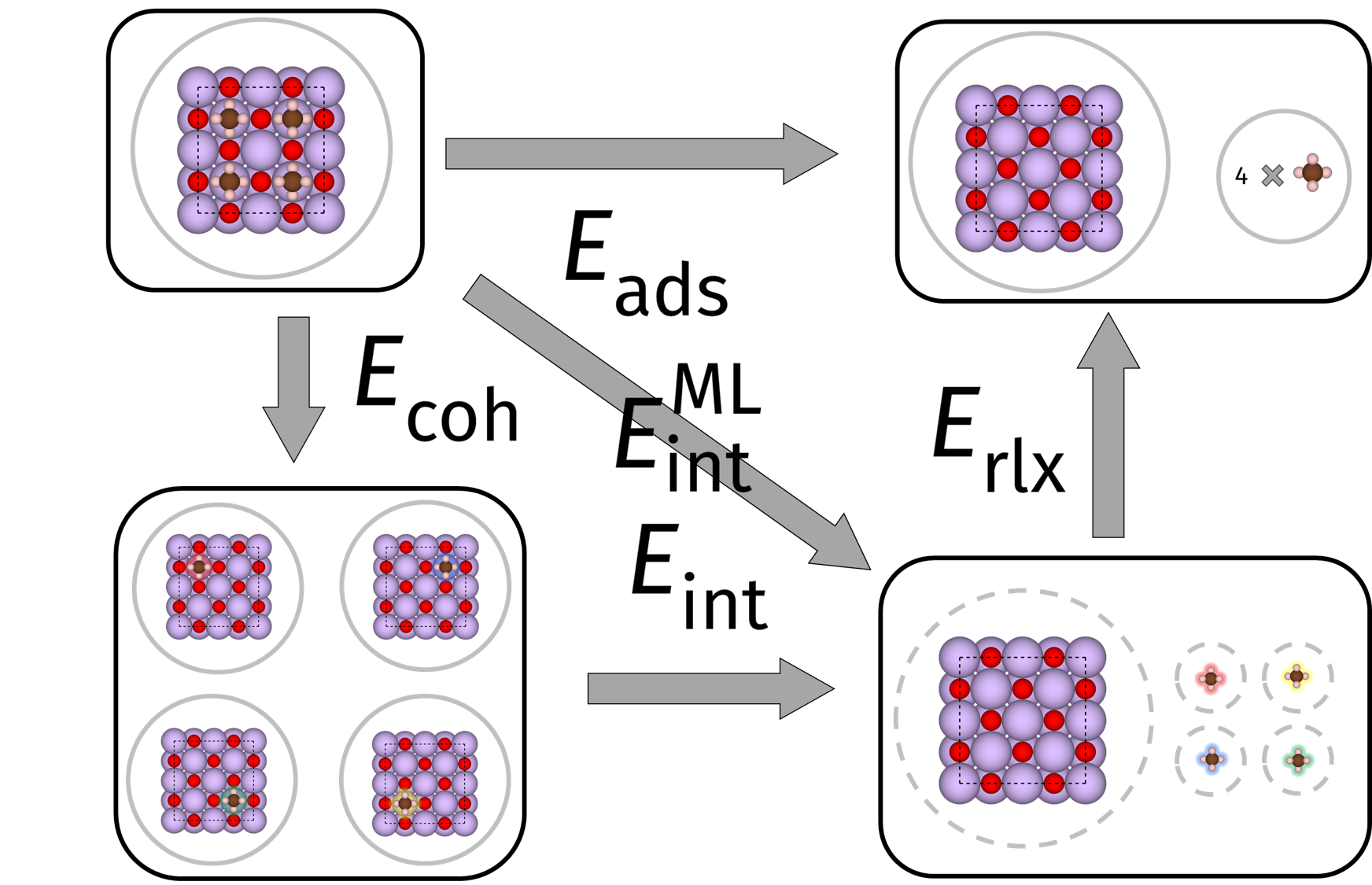}
    \caption{\label{fig:ml_ecoh_definition} The contributions to the adsorption energy $E_\textrm{ads}$ for the monolayer \ce{CH4} on MgO(001). Here, the $E_\textrm{rlx}$ contribution remains the same as in Fig.~\ref{fig:eads_definition}. There is now an additional cohesive energy $E_\textrm{coh}$ term which represents the lateral interactions between the molecules within the monolayer (in the presence of the MgO surface). Under this definition, the interaction energy $E_\textrm{int}$ is calculated by treating the four molecules of the monolayer as individual monomers on the MgO(001) surface. The circles represent a single system/calculation, with a dashed circle indicating a geometry fixed to that found in the adsorbate--surface complex while a line circle indicates an equilibrium geometry.} 
\end{figure}

As illustrated in Fig.~\ref{fig:ml_ecoh_definition} for monolayer \ce{CH4} on MgO(001), the $E_\textrm{coh}$ cohesive energy term corresponds to the energetic contribution to $E_\textrm{ads}$ arising from the lateral interactions of the molecules on the surface, with $E_\textrm{int}$ corresponding to the interaction energy of the individual molecules (as monomers) on the MgO surface.
The choice to compute $E_\textrm{coh}$ and $E_\textrm{int}$ separately, as opposed to computing their sum: $E_\textrm{int}^\textrm{ML}$ directly is because the SKZCAM protocol, as an embedding approach, can only calculate $E_\textrm{int}$ for localised phenomena such as the adsorption of monomers.
Practically, $E_\textrm{coh}$ is computed by first computing $E_\textrm{int}$ and subtracting it from $E_\textrm{int}^\textrm{ML}$.
The computed $E_\textrm{coh}$ is shown in Table~\ref{tab:monolayer_ecoh} and it can be seen that across the different functionals, it can vary by over $40\,$meV, clearly requiring the need to go beyond DFT to treat this interaction.

\begin{table}
\caption{\label{tab:monolayer_ecoh}Table showcasing how the final CCSD(T)-quality $E_\text{coh}$ (in meV) is computed. CCSD(T) (and other WFT methods) is used to compute the two-body (2B) contribution to the many-body expansion of the $E_\text{coh}^\text{gas}$ - the cohesive energy of the alkane monolayer in the absence of the MgO surface. This contribution is used to correct the 2B contribution to $E_\text{coh}^\text{gas}$ for revPBE-D4 (i.e., it is used to account for 3B and beyond contributions). Finally, we reach the final $E_\text{coh}$ by incorporating the effect of the MgO surface at the revPBE-D4 level (i.e., the difference between $E_\text{coh}$ and $E_\text{coh}^\text{gas}$).}
\begin{adjustbox}{center}
\begin{tabular}{lrrrrrr}
\toprule
 & \multicolumn{3}{c}{\ce{CH4} Monolayer} & \multicolumn{3}{c}{\ce{C2H6} Monolayer} \\ 
 & $E_\text{coh}^\text{2B gas}$ & $E_\text{coh}^\text{gas}$ & $E_\text{coh}$ & $E_\text{coh}^\text{2B gas}$ & $E_\text{coh}^\text{gas}$ & $E_\text{coh}$ \\
\midrule
PBE-D2[Ne] & - & -53 & -40 & - & -125 & -106 \\
revPBE-D4 & -46 & -42 & -31 & -79 & -62 & -48 \\
vdW-DF & - & -72 & -58 & - & -88 & -67 \\
rev-vdW-DF2 & - & -40 & -43 & - & -69 & -83 \\
PBE0-D4 & - & -43 & -34 & - & -77 & -64 \\
B3LYP-D2[Ne] & - & -39 & -37 & - & -124 & -124 \\
MP2 & -15 & -11 & - & -58 & -41 & -27 \\
CCSD & -33 & -29 & -18 & -59 & -43 & -29 \\
CCSD(T) & -40 & -37 & -25 & -92 & -75 & -61 \\
\bottomrule
\end{tabular}
\end{adjustbox}
\end{table}

\begin{figure}
    \includegraphics[width=\textwidth]{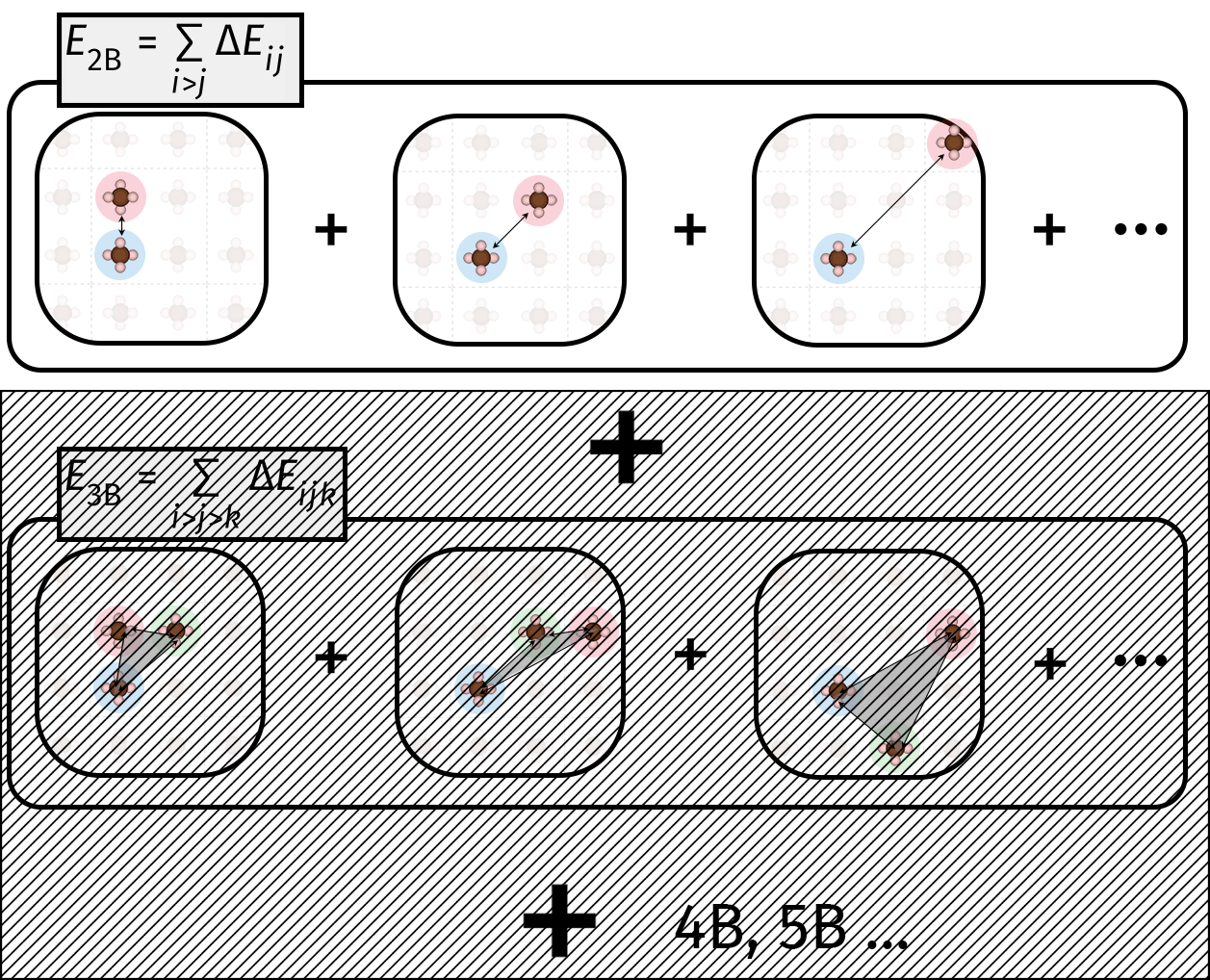}
    \caption{\label{fig:ecoh_mbe} Showcasing the many-body expansion to obtain $E_\textrm{coh}^\textrm{gas}$ of alkane monolayers (using \ce{CH4} as an example). We compute the 2B contribution at the CCSD(T) level and treat the higher-body contributions with revPBE-D4.}
\end{figure}

It is not directly possible to elevate $E_\textrm{coh}$ up to CCSD(T) quality as it involves (minor effects) arising from the presence of the MgO surface.
Instead, we have to consider the cohesive energy $E^\textrm{gas}_\textrm{coh}$ of monolayer in the absence of the surface (i.e., in the gas phase).
Specifically, it is possible to make a many-body expansion of $E^\textrm{gas}_\textrm{coh}$, incorporating 2-body (2B) contributions (see Fig.~\ref{fig:ecoh_mbe} and beyond.
As we show for revPBE-D4, the 2B contribution ($E^\textrm{2B gas}_\textrm{coh}$) makes up the most dominant contribution of $E^\textrm{gas}_\textrm{coh}$, differing by $4\,$meV and $17\,$meV for the \ce{CH4} and \ce{C2H6} monolayers respectively.
We have computed the CCSD(T) $E^\textrm{2B gas}_\textrm{coh}$ contribution and from this, we can correct this (major) part of $E^\textrm{gas}_\textrm{coh}$ and subsequently $E_\textrm{coh}$.
We also make use of the frozen-natural orbital (FNO) approximation to speed up these calculations.
This treatment has been performed before by Alessio \etal{}~\cite{alessioChemicallyAccurateAdsorption2018} and Tosoni \etal{}~\cite{tosoniAccurateQuantumChemical2010} and we come in excellent agreement; differences could arise from differing geometries or CCSD(T) being more accurate than MP2.

\clearpage

\subsection{\label{sec:tetramer_ecoh_contributions} Cohesive and dissociation energy in \ce{H2O} and \ce{CH3OH} clusters on MgO(001)}

\begin{figure}[h]
    \includegraphics[width=\textwidth]{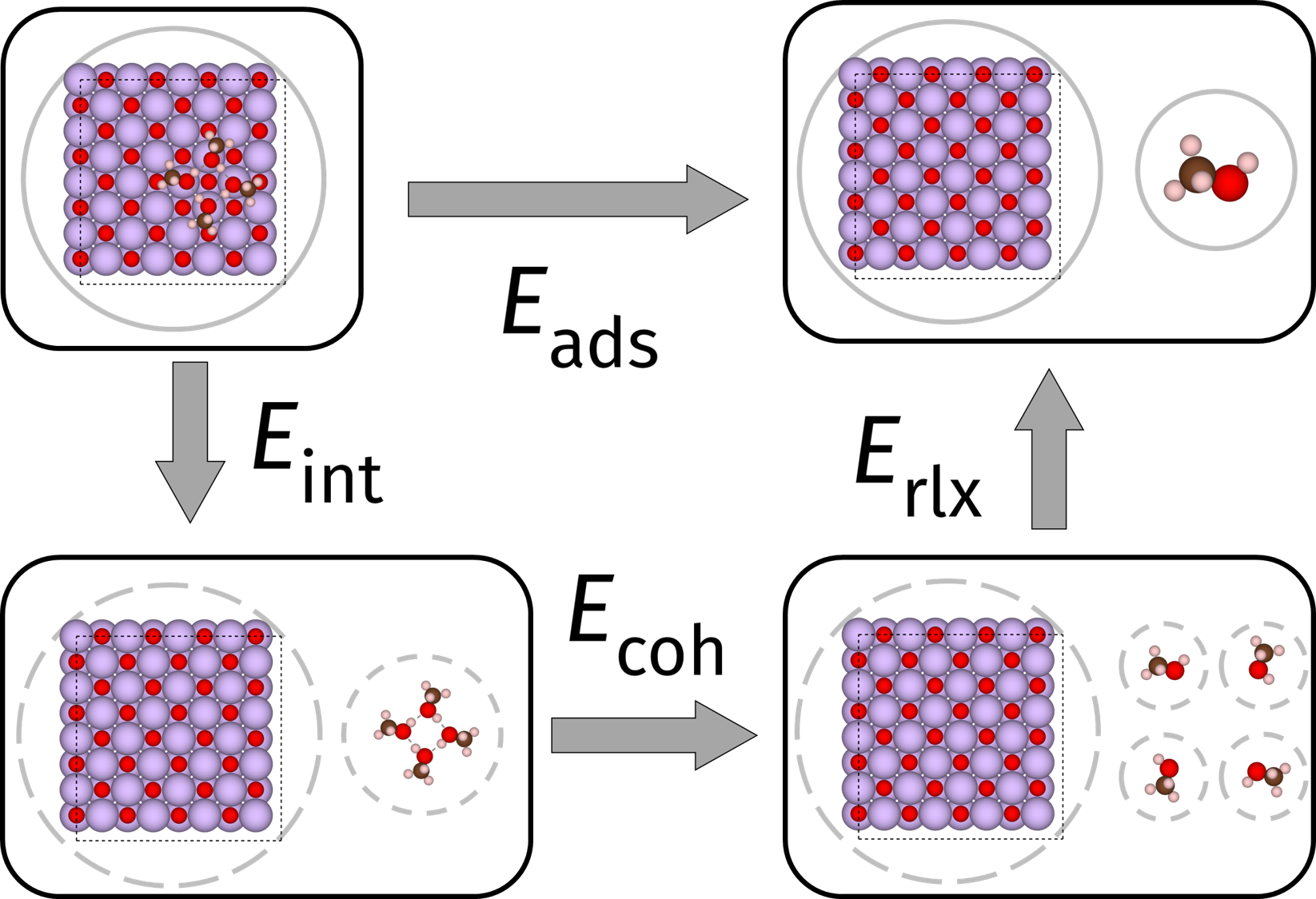}
    \caption{\label{fig:tetramer_ecoh_definition} The contributions to the adsorption energy $E_\textrm{ads}$ for the tetramer \ce{CH3OH} cluster on MgO(001). Here, the $E_\textrm{rlx}$ contribution remains the same as in Fig.~\ref{fig:eads_definition}. There is now an additional cohesive energy $E_\textrm{coh}$ term which represents the lateral interactions between the molecules within the cluster (in the gas phase). Under this definition, the interaction energy $E_\textrm{int}$ is calculated by treating the cluster as a single `molecule' that first desorbs from the surface. The circles represent a single system/calculation, with a dashed circle indicating a geometry fixed to that found in the adsorbate--surface complex while a line circle indicates an equilibrium geometry.} 
\end{figure}

For the (non-dissociated) molecular \ce{H2O} and \ce{CH3OH} clusters on MgO(001), we partition the $E_\textrm{ads}$ into contributions from $E_\textrm{int}$, $E_\textrm{rlx}$ and $E_\textrm{coh}$, as seen in Fig.~\ref{fig:tetramer_ecoh_definition}
Here, $E_\textrm{int}$ is defined to be the interaction energy required to remove the entire cluster from the surface, taking the cluster to be the `molecule'.
The cohesive energy is then the binding energy to break the cluster into its separate molecules, all within the gas phase and having their geometries fixed to that within the cluster-surface complex.
Finally $E_\textrm{rlx}$ is the energy to relax the molecules and surface into their equilibrium geometries.

The cohesive energy is computed for the revPBE-D4 geometry for the ensemble of DFAs as well as MP2, CCSD and CCSD(T) in Table~\ref{tab:tetramer_ecoh}.
There is a range of $90\,$meV and $70\,$meV between the DFAs, highlighting the need to move towards CCSD(T) for high accuracy.

\begin{table}
\caption{\label{tab:tetramer_ecoh}Comparison between several DFAs, MP2 and CCSD against CCSD(T) for the cohesive energy $E_\text{coh}$ per monomer (in meV) of the tetramer \ce{CH3OH} and \ce{H2O} cluster.}
\begin{adjustbox}{center}
\begin{tabular}{lrr}
\toprule
 & \ce{CH3OH} & \ce{H2O} \\ 
\midrule
PBE-D2[Ne] & -380 & -309 \\
revPBE-D4 & -328 & -261 \\
vdW-DF & -291 & -240 \\
rev-vdW-DF2 & -357 & -289 \\
PBE0-D4 & -367 & -302 \\
B3LYP-D2[Ne] & -371 & -303 \\
MP2 & -341 & -276 \\
CCSD & -311 & -269 \\
CCSD(T) & -336 & -281 \\
\bottomrule
\end{tabular}
\end{adjustbox}
\end{table}

\begin{figure}[h]
    \centering
    \includegraphics[width=0.9\linewidth]{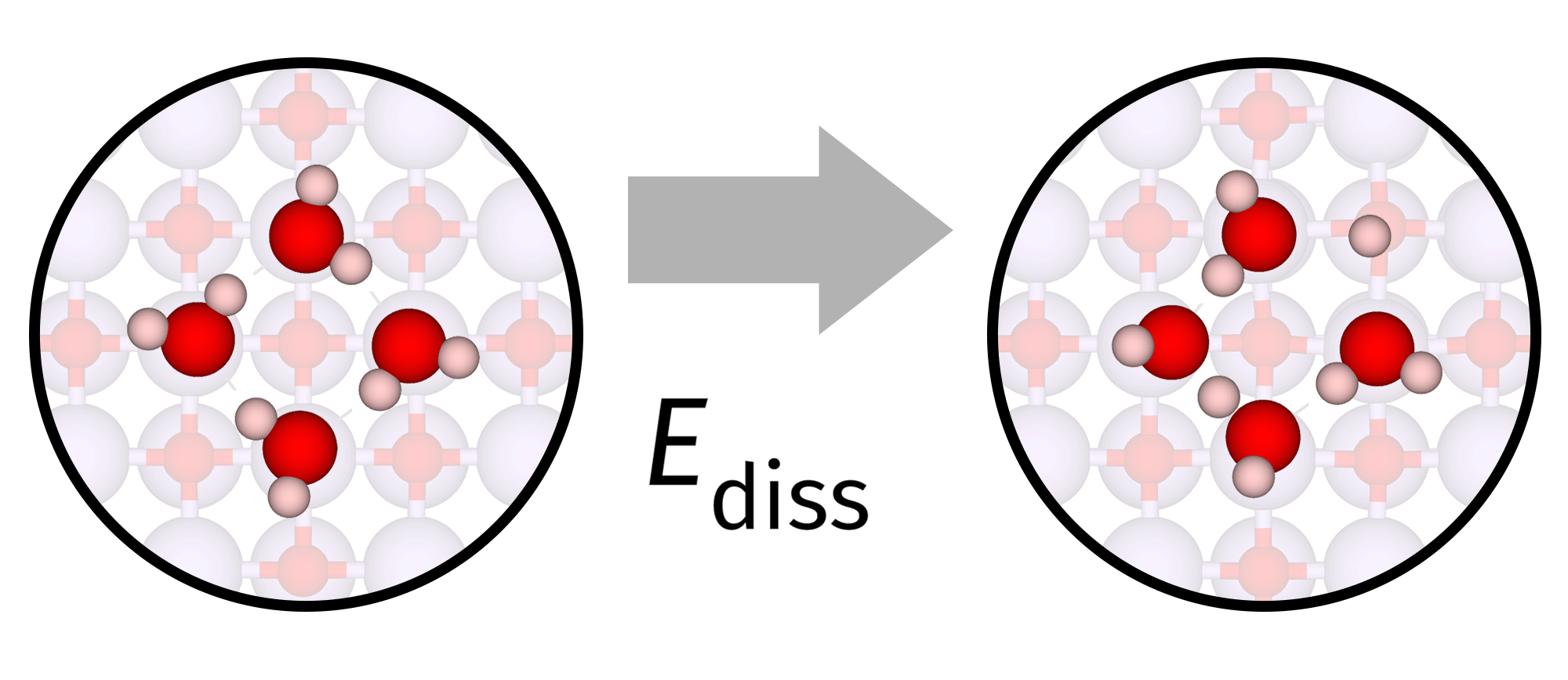}
    \caption{The dissociation energy $E_\textrm{diss}$ is defined as the energetic stabilisation to form a dissociated cluster from the original molecularly adsorbed cluster. We use the \ce{H2O} tetramer cluster as the example.}
    \label{fig:ediss_energy}
\end{figure}

To calculate the adsorption energy $E_\textrm{ads}$ of the partially dissociated clusters of \ce{H2O} and \ce{CH3OH} on MgO(001), we compute an additional term $E_\textrm{diss}$ which accounts for the energy difference/stabilisation between the molecularly adsorbed and partially-dissociated forms of the \ce{H2O} and \ce{CH3OH} tetramers (see Fig.~\ref{fig:ediss_energy}).
We use the revPBE-D4 geometry of the molecular and partially-dissociated forms and have computed this energy using the ensemble of DFAs, as shown in Sec.~\ref{sec:ediss_h2o_ch3oh}.

\subsection{\label{sec:no_ecoh_contributions} Cohesive energy of NO dimer on MgO(001)}

\begin{figure}[h] 
    \includegraphics[width=\textwidth]{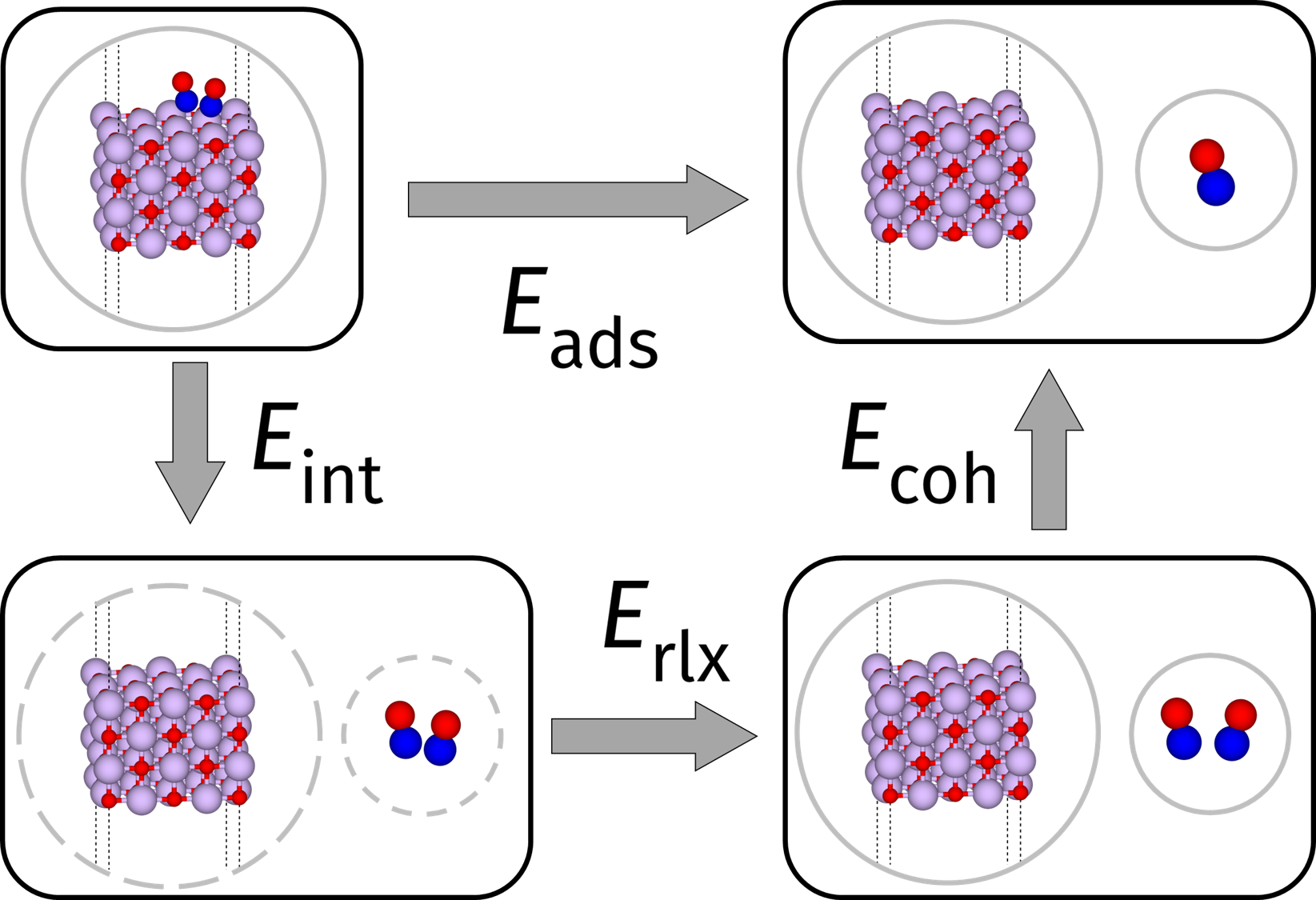}
    \caption{\label{fig:no_ecoh_definition} The contributions to the adsorption energy $E_\textrm{ads}$ for the NO dimer on MgO(001). Here, the interaction energy $E_\textrm{int}$ is calculated by treating the dimer as a single `molecule' that first desorbs from the surface. Then, $E_\textrm{rlx}$ represents the relaxation energy for the surface into its equilibrium geometry and the dimer into its equilibrium geometry (as a dimer). The cohesive energy $E_\textrm{coh}$ term which represents the binding energy to bring the dimer to two separate monomers, all in their equilibrium geometry. The circles represent a single system/calculation, with a dashed circle indicating a geometry fixed to that found in the adsorbate--surface complex while a line circle indicates an equilibrium geometry.} 
\end{figure}

As illustrated in Fig.~\ref{fig:no_ecoh_definition}, the contributions to the $E_\textrm{ads}$ of the NO dimer on MgO(001) differ slightly from those of the \ce{H2O} and \ce{CH3OH} tetramers.
The definition of $E_\textrm{int}$ remains the same, treating the dimer as the `molecule'.
However, the relaxation energy $E_\textrm{rlx}$ now pertains to the the energy to bring the dimer from its geometry in the dimer-surface complex into its equilibrium geometry (together with the surface) and $E_\textrm{coh}$ now represents the binding energy of the dimer against the separate monomers.
This new definition of $E_\textrm{coh}$ was chosen because the binding energy of the NO dimer is a well-studied topic both with experiments~\cite{dkhissiNODimer1997,wadeDirectMeasurementBinding2002a} and computational simulations~\cite{gonzalez-luqueTheoreticalDeterminationDissociation1994,sayosTheoreticalInvestigationEight2000}, with a complete summary of previous work found in Ref.~\citenum{ivanicHighlevelTheoreticalStudy2012a}.

\begin{table}
\caption{\label{tab:dimer_ecoh}Comparison of experiment and MRMP2 against several DFAs and CCSD(T) for the cohesive energy $E_\text{coh}$ per monomer (in meV) of the NO dimer.}
\begin{adjustbox}{center}
\begin{tabular}{lr}
\toprule
Method & $E_\text{coh}$ [meV] \\ 
\midrule
PBE-D2[Ne] & -239 \\
revPBE-D4 & -192 \\
vdW-DF & -160 \\
rev-vdW-DF2 & -281 \\
PBE0-D4 & 53 \\
B3LYP-D2[Ne] & 63 \\
CCSD(T)~\cite{tobitaCriticalComparisonSinglereference2003} & -22 \\
Experiment~\cite{ivanicHighlevelTheoreticalStudy2012a} & -61 to -82 \\
MRMP2(18,14)~\cite{ivanicHighlevelTheoreticalStudy2012a} & -75 \\
\bottomrule
\end{tabular}
\end{adjustbox}
\end{table}

As a multireference problem, DFT has trouble with getting $E_\textrm{coh}$ right, clearly seen in Table~\ref{tab:dimer_ecoh} where there is an $344\,$meV range in $E_\textrm{coh}$ across the DFAs that we have studied.
Relative to the experimental $E_\textrm{coh}$ window of ${-}61$ to ${-}82\,$meV per monomer (or $121{-}165\,$meV in terms of binding energy), CCSD(T) has been shown to underbind this quantity with a value of ${-}22\,$meV.
Previous work with multireference second-order perturbation theory (MRMP2) using a large (18,14) active space have demonstrated much better agreement~\cite{ivanicHighlevelTheoreticalStudy2012a} with an $E_\textrm{coh}$ of ${-}75\,$meV.
One of the advantages of the partitioning of the $E_\textrm{ads}$ in Fig.~\ref{fig:no_ecoh_definition} is that we have broken down $E_\textrm{ads}$ into various contributions which can be tackled with different methods.
Specifically, $E_\textrm{int}$ can be reliably treated with CCSD(T), while $E_\textrm{rlx}$ can be treated with DFT and $E_\textrm{coh}$ can be treated with multireference methods.
This allows us to use the previously computed MRMP2 value~\cite{ivanicHighlevelTheoreticalStudy2012a} of $E_\textrm{coh}$ in our final $E_\textrm{ads}$ and $H_\textrm{ads}$.

\clearpage

\section{\label{sec:dft_geom_error} Geometry relaxation and enthalpic contributions from a DFT ensemble}

The autoSKZCAM framework extends the accurate $E_\textrm{int}$ computed with the SKZCAM protocol towards computing an accurate $H_\textrm{ads}$ estimate that can be compared directly with experiments.
The aim is to make the calculation of $H_\textrm{ads}$ affordable and to decrease the number of CCSD(T)-level calculations required, especially on the contributions where it isn't needed.
Specifically, the autoSKZCAM framework uses DFT to compute the remaining geometrical relaxation $E_\textrm{rlx}$, zero-point vibrational $E_\textrm{ZPV}$ and temperature $E_\textrm{T}$ contributions to enable the calculation of $H_\textrm{ads}$ within Eqs.~\ref{eq:hads_eq1} and~\ref{eq:hads_eq2}.
These terms make overall small contributions to the final $H_\textrm{ads}$, thus errors due to the approximate nature of DFT are not expected to alter the numbers significantly.
Furthermore, these are quantities that do not depend on the absolute value of the potential energy surface (as in the case of $E_\textrm{int}$) but on `relative' changes around the minima to e.g., compute vibrational frequencies or relaxation energies, which DFT can perform accurately for.

To ensure reliable estimates with error bars, we use an ensemble/set of density functional approximations (DFAs).
For MgO, we used the 6 high-accuracy DFAs within the ensemble: PBE-D2[Ne]~\cite{perdewGeneralizedGradientApproximation1996d,tosoniAccurateQuantumChemical2010}, revPBE-D4~\cite{zhangCommentGeneralizedGradient1998,caldeweyherGenerallyApplicableAtomiccharge2019}, vdW-DF~\cite{leeHigheraccuracyVanWaals2010}, rev-vdW-DF2~\cite{hamadaVanWaalsDensity2014}, PBE0-D4~\cite{adamoReliableDensityFunctional1999d,caldeweyherGenerallyApplicableAtomiccharge2019} and B3LYP-D2[Ne]~\cite{beckeDensityFunctionalThermochemistry1993,tosoniAccurateQuantumChemical2010}, where[Ne] denotes that the Neon D2 parameters~\cite{grimmeSemiempiricalGGAtypeDensity2006} have been used on the Mg atom~\cite{tosoniAccurateQuantumChemical2010}, or subsets of them as we discuss later.
For the \ce{TiO2} systems, we used the: PBE-TS/HI~\cite{perdewGeneralizedGradientApproximation1996d,buckoExtendingApplicabilityTkatchenkoScheffler2014}, revPBE-D4, vdW-DF, rev-vdw-DF2, r$^2$SCAN-rVV10~\cite{ningWorkhorseMinimallyEmpirical2022} and HSE06-D4~\cite{heydHybridFunctionalsBased2003,caldeweyherGenerallyApplicableAtomiccharge2019} functionals.
These DFAs were chosen because they are generally expected to perform well for their respective surfaces for both $E_\textrm{int}$ and unit cell lattice parameters (see Tables~\ref{tab:lattice_parametersmgo},~\ref{tab:lattice_parametersr-tio2} and~\ref{tab:lattice_parametersa-tio2} for the MgO, \ce{TiO2} rutile(110) and anatase(101) surfaces respectively), allowing us to probe a sensible range of values as close as possible to the true answer.
The use of an ensemble of DFAs allows us to take averages for improved estimates and to give error estimates on $E_\textrm{rlx}$ and $E_\textrm{ZPV}$ and $E_\textrm{T}$ which we will discuss in Section~\ref{sec:egeom_errors} and~\ref{sec:ezpv_etherm} respectively.
Their specific computational details are provided in Section~\ref{sec:dft_details}.

\begin{table}
\caption{\label{tab:lattice_parametersmgo}Lattice parameter (in \AA{}) and \ce{H2O} $E_\text{int}$ for MgO(001) obtained from the DFT ensemble. These are compared to experiment~\cite{singhSynthesisCharacterizationAlkalineEarthOxide2018} for the lattice parameter and the autoSKZCAM $E_\text{int}$.}
\begin{tabular}{lrr}
\toprule
 & Lattice Parameter a & \ce{H2O} $E_\text{int}$ \\ 
\midrule
PBE-D2[Ne] & 4.234 & -693 \\
revPBE-D4 & 4.220 & -637 \\
vdW-DF & 4.273 & -567 \\
rev-vdW-DF2 & 4.220 & -711 \\
PBE0-D4 & 4.175 & -695 \\
B3LYP-D2[Ne] & 4.202 & -618 \\
Reference & 4.214 & -702 \\
\bottomrule
\end{tabular}
\end{table}
\begin{table}
\caption{\label{tab:lattice_parametersr-tio2}Lattice parameter (in \AA{}) and \ce{H2O} $E_\text{int}$ for TiO$_2$ rutile(110) obtained from the DFT ensemble. These are compared to experiment~\cite{burdettStructuralelectronicRelationshipsInorganic1987} for the lattice parameter and the autoSKZCAM $E_\text{int}$.}
\begin{tabular}{lrrr}
\toprule
 & Lattice Parameter a & Lattice Parameter c & \ce{H2O} $E_\text{int}$ \\ 
\midrule
PBE-TS/HI & 4.611 & 2.970 & -1282 \\
revPBE-D4 & 4.598 & 2.958 & -1214 \\
vdW-DF & 4.685 & 2.995 & -1026 \\
rev-vdW-DF2 & 4.618 & 2.961 & -1344 \\
r$^2$SCAN-rVV10 & 4.590 & 2.957 & -1552 \\
HSE06-D4 & 4.559 & 2.940 & -1410 \\
Reference & 4.587 & 2.954 & -1310 \\
\bottomrule
\end{tabular}
\end{table}

\begin{table}
\caption{\label{tab:lattice_parametersa-tio2}Lattice parameter (in \AA{}) and \ce{H2O} $E_\text{int}$ for TiO$_2$ anatase(101) obtained from the DFT ensemble. These are compared to experiment~\cite{burdettStructuralelectronicRelationshipsInorganic1987} for the lattice parameter and the autoSKZCAM $E_\text{int}$.}
\begin{tabular}{lrrr}
\toprule
 & Lattice Parameter a & Lattice Parameter c & \ce{H2O} $E_\text{int}$ \\ 
\midrule
PBE-TS/HI & 3.789 & 9.659 & -1095 \\
revPBE-D4 & 3.790 & 9.548 & -1096 \\
vdW-DF & 3.839 & 9.767 & -917 \\
rev-vdW-DF2 & 3.798 & 9.590 & -1179 \\
r$^2$SCAN-rVV10 & 3.785 & 9.531 & -1392 \\
HSE06-D4 & 3.751 & 9.540 & -1224 \\
Reference & 3.782 & 9.502 & -1207 \\
\bottomrule
\end{tabular}
\end{table}

\subsection{\label{sec:dft_details} Computational details for periodic density functional theory}

Periodic DFT were performed with the Vienna \textit{Ab-Initio} Simulation Package 6.3.0~\cite{kresseEfficiencyAbinitioTotal1996a,kresseEfficientIterativeSchemes1996a} (VASP).
For each of the three surfaces, we used an ensemble of 6 DFAs to calculate the terms that make up $H_\textrm{ads}$.
The electronic structure parameters [$k$-point grid, energy cutoff and projected augmented wave (PAW) potentials] are provided in Table~\ref{tab:dft_parameters}.
An energy cutoff of 600 eV was used for most of the DFAs, although this was reduced to $520\,$eV for the hybrid HSE06-D4 calculations on the TiO\textsubscript{2} surface systems.
For the HSE06-D4 calculation relaxations, we lower its cost by using a reduced $\Gamma$-point grid (via \texttt{NKRED}) for the exact exchange contribution to the total energy and use a 18 electron core PAW potential on the Ti atoms.
A $\Gamma$-centered $k$-point mesh was used in all the systems, with the $k$-point grids chosen to converge $E_\textrm{int}$ to $1\,$meV.

\begin{figure}[h]
    \includegraphics[width=\textwidth]{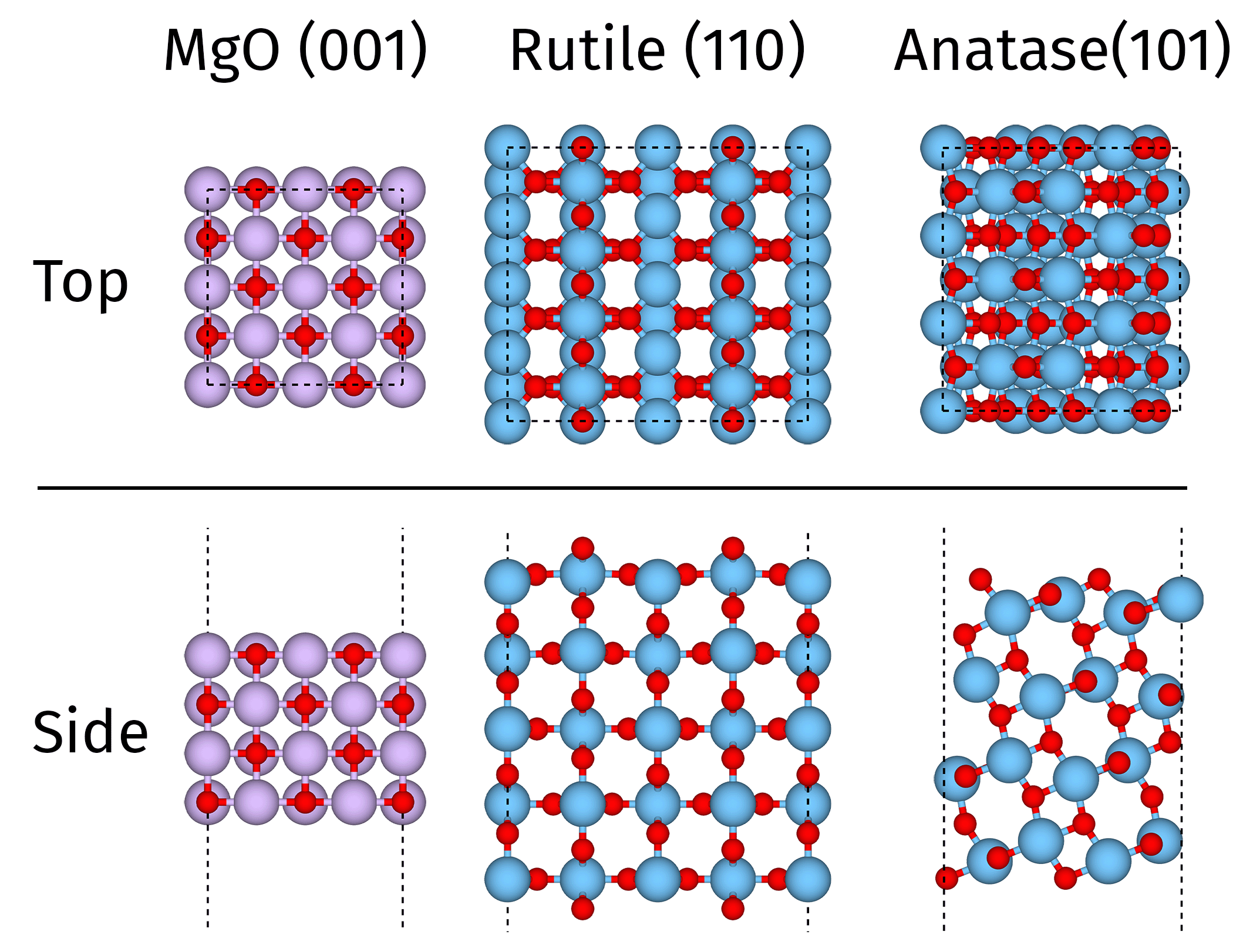}
    \caption{\label{fig:surface_top_side}Top and side views for the MgO(001), \ce{TiO2} rutile(110) and \ce{TiO2} anatase(101) surfaces.}
\end{figure}

The three surfaces used to model the adsorbate--surface systems are shown in Fig.~\ref{fig:surface_top_side}.
For the MgO system, we perform calculations on a 4 layer slab, where the bottom two layers are fixed.
The majority of systems used a $4{\times}4$, with an $8{\times}8$ supercell used for \ce{C6H6} and the \ce{CH3OH} and \ce{H2O} clusters.
The $2{\times}2{\times}1$ $k$-point grid used with the $4{\times}4$ supercell was reduced to a $\Gamma$-point grid for the $8{\times}8$ supercell.
The TiO\textsubscript{2} rutile(110) surface slab consisted of a p($4 {\times} 2$) supercell with 5 tri-layers (and the bottom three fixed), while the anatase(101) surface slab consisted of a ($3 {\times} 1$) supercell with 4 tri-layers and the bottom layer fixed.
All surfaces incorporated 15\AA{} of vacuum and were generated through a geometrical relaxation performed with a force convergence cutoff of $0.01\,$eV/\AA{}.
Subsequently, a molecule was added onto the surface and relaxed with the same force convergence cutoff.
Vibrational frequency calculations were performed for a subset of the DFT ensemble (see Section~\ref{sec:ezpv_etherm}) using a finite-differences approach with \texttt{POTIM=0.01\AA{}} displacements, one in the positive and negative direction along the three Cartesian directions (i.e., \texttt{NFREE=2}).
The self-consistent field cycles were set to an energy convergence of $10^{-8}\,$eV for the vibrational frequency calculations, with $10^{-7}\,$eV as standard for geometry optimisations.

\begin{turnpage}
\begin{table}

\caption{\label{tab:dft_parameters}DFT parameters used for the three different surfaces. The parameters for hybrids are also different from those for the metaGGA, GGA and vdW-inclusive functionals, grouped as (meta)GGA in the table. The number of layers in parentheses indicates the number of layers fixed at the bottom of the slab. The $k$-points in parenthesis indicates the $k$-point mesh used for the exact exchange potential. We used the PBE version 54 set of PAW potentials for all the calculations.}
\begin{tabular}{lrrrrrr}
\toprule
 & \multicolumn{2}{c}{MgO(001)} & \multicolumn{2}{c}{TiO$_2$ rutile(110)} & \multicolumn{2}{c}{TiO$_2$ anatase(101)} \\ \cmidrule(lr){2-3}  \cmidrule(lr){4-5} \cmidrule(lr){6-7}
 & (meta)GGA & hybrid & (meta)GGA & hybrid & (meta)GGA & hybrid \\
\midrule
Supercell Size & 4${\times}$4 & 4${\times}$4 & p(4${\times}$2) & p(4${\times}$2) & 3${\times}$1 & 3${\times}$1 \\
Number of Layers & 4(2) & 4(2) & 5(3) O-Ti-O & 5(2) O-Ti-O & 8(2) O-Ti-O layers & 8(2) O-Ti-O \\
$k$-point Mesh & 2$\times$2$\times$1 & 2$\times$2$\times$1(2$\times$2$\times$1) & 2$\times$2$\times$1 & 2$\times$2$\times$1(1$\times$1$\times$1) & 3$\times$3$\times$1 & 3$\times$3$\times$1(1$\times$1$\times$1) \\
Energy cutoff & 600 & 600 & 600 & 520 & 600 & 520 \\
Vacuum & 15 & 15 & 15 & 13 & 15 & 13 \\
Metal PAW potential & Mg\_pv (4e core) & Mg\_pv (4e core) & Ti\_pv (12e core) & Ti (18e core) & Ti\_pv (12e core) & Ti (18e core) \\
\bottomrule
\end{tabular}

\end{table}
\end{turnpage}

\subsection{\label{sec:erlx_def} The relaxation energy}

For the adsorption of monomers on the surfaces, the relaxation energy $E_\textrm{rlx}$ is defined as the energy change when the surface and molecule are relaxed from their geometry in the adsorbate--surface complex to their equilibrium geometries.
This definition persists for the tetramer \ce{CH3OH} and \ce{H2O} clusters and the monolayer alkanes on MgO(001), where the individual monomers in the cluster or monolayer are compared to their equilibrium geometires in the gas-phase.
For the NO dimer, this definition slightly changes to be the energy change from the dimer on the MgO(001) surface to the dimer's equilibrium geometry for reasons discussed in Section~\ref{sec:no_ecoh_contributions}.
For the chemisorbed \ce{CO2} on the MgO(001) surface, $E_\textrm{rlx}$ now only corresponds to the relaxation of the surface, with the relaxation of the molecule encapsulated in the $E_\textrm{conf}$ term that is treated at the CCSD(T) level, discussed in Section~\ref{sec:co2_econf_contributions}.
The result is that $E_\textrm{rlx}$ is a small quantity relative to the overall $E_\textrm{ads}$, as we show for the DFT ensemble in Tables~\ref{tab:eads_dft_ensemble_errors_mgo} and~\ref{tab:eads_dft_ensemble_errors_tio2} for the MgO and TiO$_2$ surfaces respectively.
As discussed in the next section, we use revPBE-D4 to generate the adsorbate--surface geometries, and hence use the $E_\textrm{rlx}$ term generated by this DFA.
For the NO dimer system, we have opted to use the geometry and $E_\textrm{rlx}$ from B3LYP-D2[Ne] as revPBE-D4 strongly overestimates the binding of the NO dimer.

\subsection{\label{sec:egeom_errors} Estimating geometrical errors}

Obtaining energy gradients (for e.g., forces) is challenging with methods from cWFT such as CCSD(T) and even for codes where this is possible, it would be highly expensive.
As such, it is common to use geometries generated by a lower level of theory such as DFT.
There is thus an error in the resulting $E_\textrm{ads}$ that arises from the use of a DFT geometry.
Besides this geometrical error in $E_\textrm{ads}$, there is also an additional error because we use DFT to calculate $E_\textrm{rlx}$ in Eq.~\ref{eq:hads_eq2}; specifically with the DFA used to generate the geometry.
There is an additional error associated with using this DFT value of $E_\textrm{rlx}$ and we aim to capture the combination of these two errors within an error term dubbed $\epsilon_\textrm{geom}$.

This $\epsilon_\textrm{geom}$ error can be estimated from the ensemble of DFAs.
The application of a method, whether another DFA or one from cWFT, on the revPBE-D4 geometry can be denoted as $\textrm{Method}//\textrm{revPBE-D4}$ and the resulting approximate adsorption energy $E_\textrm{ads}^\textrm{approx}$ can be defined as:
\begin{align}
\label{eq:eads_approx}
\begin{split}
    E_\textrm{ads}^\textrm{approx} [\textrm{Method}//\textrm{revPBE-D4}] = & E_\textrm{int} [\textrm{Method}//\textrm{revPBE-D4}] + \\
    & E_\textrm{rlx}[\textrm{revPBE-D4}//\textrm{revPBE-D4}] + \\
    & \left ( E_\textrm{coh}[\textrm{Method}//\textrm{revPBE-D4}] + E_\textrm{conf}[\textrm{Method}//\textrm{revPBE-D4}] \right ),
\end{split}
\end{align}
where $E_\textrm{coh}$ and $E_\textrm{conf}$ are included for the systems that have those terms, discussed in Section~\ref{sec:wft_ecoh_econf}.
The true adsorption energy $E_\textrm{ads}^\textrm{true}$ corresponds to the $E_\textrm{ads}$ evaluated with the method with its corresponding geometry (i.e., \textrm{Method}//\textrm{Method}) and it is the quantity that we ultimately aim to approximate with $E_\textrm{ads}^\textrm{approx}$.

\begin{figure}
    \includegraphics[width=\textwidth]{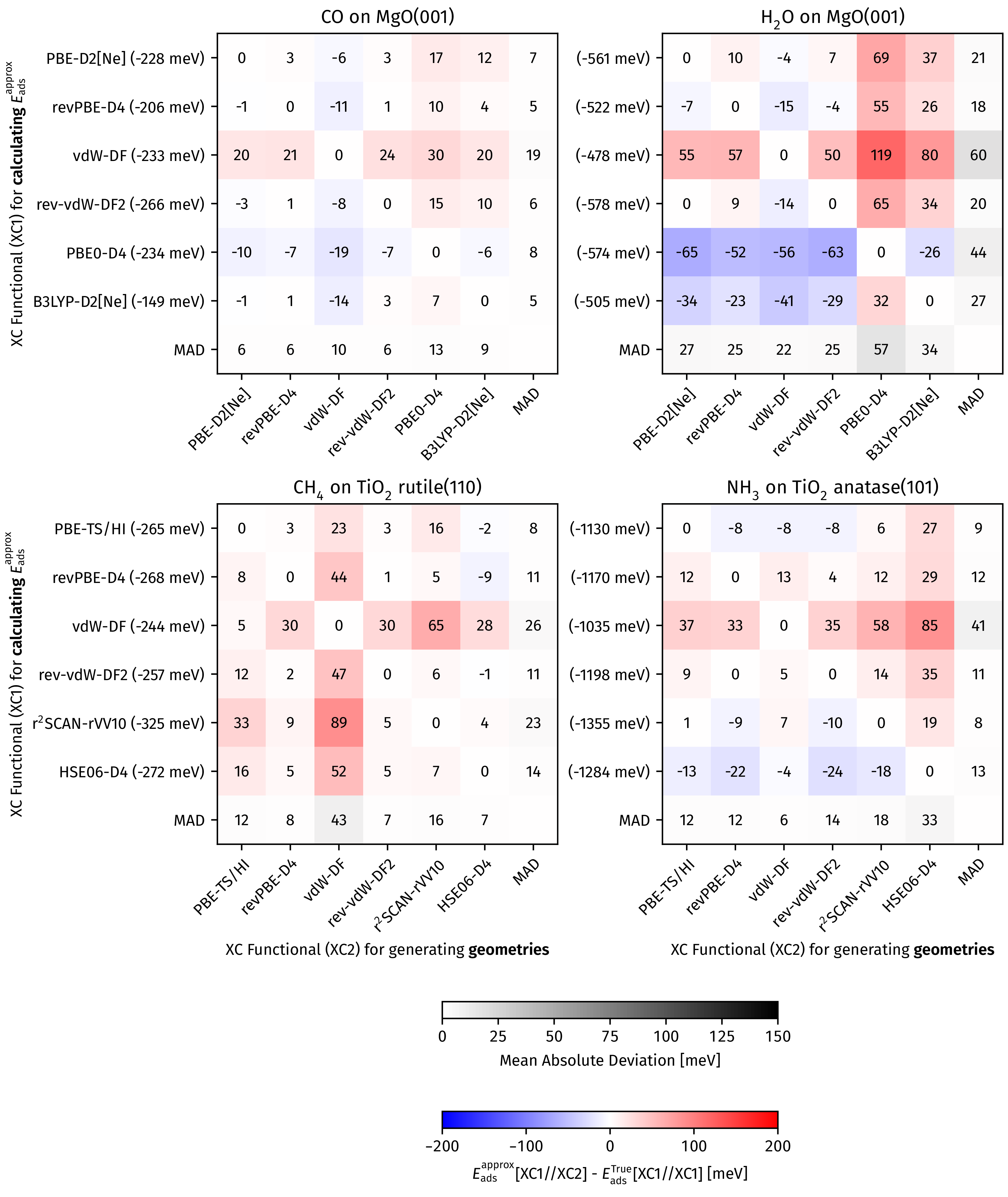}
    \caption{\label{fig:relax_dft_error} Estimating the errors for using geometries generated by a DFA. The CO on MgO(001), \ce{H2O} on MgO(001), \ce{CH4} on \ce{TiO2} rutile(110) and \ce{NH3} on \ce{TiO2} anatase(101) adsorbate--surface systems were used as illustration. For the geometry generated by each DFA (on the x-axis), an approximate $E_\textrm{ads}^\textrm{approx}$ is calculated for the DFT ensemble along each column as defined in Eq.~\ref{eq:eads_approx}. This is compared to the true $E_\textrm{ads}^\textrm{true}$ from using the corresponding geometry of each DFA. The difference between $E_\textrm{ads}^\textrm{approx}$ and $E_\textrm{ads}^\textrm{true}$ is plotted, with a corresponding mean absolute deviation for each DFA's geometry is given in the bottom row.}
\end{figure}

We chose revPBE-D4 as the functional to generate the geometries for subsequent $E_\textrm{int}$ with the SKZCAM protocol because we find that it provides a low error on $E_\textrm{ads}^\textrm{approx}$ relative to $E_\textrm{ads}^\textrm{true}$ when evaluated across the other functionals in the ensemble.
In Fig.~\ref{fig:relax_dft_error}, we have compared the 6 DFAs within the ensemble for their performance in reproducing $E_\textrm{ads}^\textrm{true}$.
The chosen systems highlight a range of binding and surfaces and revPBE-D4 performs well across all the adsorbate--surface systems, with an MAD that are all less than 5\% of $E_\textrm{ads}^\textrm{true}$.
We note that the only system where we have chosen not to use the revPBE-D4 geometry is NO dimer as a hybrid such as B3LYP-D2[Ne] predicts the correct ground state and does not strongly overbind its cohesive energy $E_\textrm{coh}$ like the GGAs~\cite{ivanicHighlevelTheoreticalStudy2012a}.

In Tables~\ref{tab:eads_dft_ensemble_errors_mgo} and~\ref{tab:eads_dft_ensemble_errors_tio2}, we have computed $E_\textrm{ads}^\textrm{approx}$ (including the corresponding $E_\textrm{int}$) and $E_\textrm{ads}^\textrm{true}$ across the entire DFT ensemble for the monomers adsorbed on MgO(001) and \ce{TiO2} surfaces respectively.
We take $\epsilon_\textrm{geom}$ to be 2 times the root mean squared error (2RMSE) of $E_\textrm{ads}^\textrm{approx}$ against $E_\textrm{ads}^\textrm{true}$ for the DFAs in the ensemble, excluding revPBE-D4 (which should have no error by definition).
Assuming an even/normal distribution around $E_\textrm{ads}^\textrm{true}$ from using the revPBE-D4 geometry for $E_\textrm{ads}^\textrm{approx}$ , this error choice gives a 95\% confidence interval on the final estimate.

We also compute this error for the monolayer and clusters on MgO in Table~\ref{tab:eads_dft_ensemble_errors_ml_clus_mgo}, where an additional cohesive energy term $E_\textrm{coh}$ has evaluated with the corresponding DFA on the revPBE-D4 geometry.
Table~\ref{tab:eads_dft_ensemble_errors_ml_clus_mgo} also includes chemisorbed \ce{CO2} on MgO(001), where there is an additional $E_\textrm{conf}$ term to estimate $E_\textrm{ads}^\textrm{approx}$ and obtain $\epsilon_\textrm{geom}$.

\LTcapwidth=\textwidth
    
\begin{longtable}{llrrrrrrrrrrrrrrrr}
\caption{\label{tab:eads_dft_ensemble_errors_mgo}For the monomers on the MgO(001) surface, we estimate the errors for using the revPBE-D4 geometry and $E_\text{rlx}$ in the final $E_\text{ads}$ of the autoSKZCAM protocol using an ensemble of 6 different DFAs. The errors are calculated as the difference between the true $E_\text{ads}^\text{true}$ (using the appropriate DFA) and the approximated $E_\text{ads}^\text{approx}$ using the revPBE-D4 geometry and $E_\text{rlx}$.} \\

\toprule
 &  & \rotatebox{90}{\ce{CH4}} & \rotatebox{90}{\ce{C2H6}} & \rotatebox{90}{\ce{CO}} & \rotatebox{90}{\ce{C6H6}} & \rotatebox{90}{Parallel \ce{N2O}} & \rotatebox{90}{Tilted \ce{N2O}} & \rotatebox{90}{Vertical-Hollow \ce{NO}} & \rotatebox{90}{Vertical-Mg \ce{NO}} & \rotatebox{90}{Bent-Bridge \ce{NO}} & \rotatebox{90}{Bent-Mg \ce{NO}} & \rotatebox{90}{Bent-O \ce{NO}} & \rotatebox{90}{Monomer \ce{H2O}} & \rotatebox{90}{Tilted \ce{CH3OH}} & \rotatebox{90}{Parallel \ce{CH3OH}} & \rotatebox{90}{\ce{NH3}} & \rotatebox{90}{Physisorbed \ce{CO2}} \\ 
\midrule
\endfirsthead

\caption[]{(continued)} \\
\endhead

\multicolumn{18}{r}{{Continued on next page}} \\
\endfoot

\bottomrule
\endlastfoot

\multirow[]{5}{*}{\rotatebox{90}{PBE-D2[Ne]}} & $E_\text{int}$ & -115 & -157 & -233 & -278 & -180 & -137 & -263 & -170 & -353 & -207 & -242 & -666 & -735 & -460 & -617 & -238 \\
 & $E_\text{rlx}$ & 2 & 1 & 8 & 26 & 3 & 10 & 29 & 9 & 51 & 6 & 39 & 114 & 145 & 46 & 89 & 14 \\
 & $E_\text{ads}^\text{approx}$ & -114 & -156 & -224 & -252 & -177 & -127 & -234 & -161 & -302 & -201 & -203 & -552 & -590 & -414 & -528 & -225 \\
 & $E_\text{ads}^\text{true}$ & -115 & -158 & -228 & -261 & -179 & -132 & -184 & -162 & -301 & -203 & -196 & -561 & -599 & -417 & -540 & -227 \\
 & Error & 2 & 2 & 3 & 9 & 2 & 5 & -50 & 1 & -1 & 2 & -7 & 10 & 9 & 3 & 12 & 2 \\
\cline{1-18}
\multirow[]{5}{*}{\rotatebox{90}{revPBE-D4}} & $E_\text{int}$ & -143 & -219 & -215 & -565 & -192 & -133 & -184 & -142 & -346 & -184 & -234 & -637 & -770 & -502 & -636 & -268 \\
 & $E_\text{rlx}$ & 2 & 1 & 8 & 26 & 3 & 10 & 29 & 9 & 51 & 6 & 39 & 114 & 145 & 46 & 89 & 14 \\
 & $E_\text{ads}^\text{approx}$ & -142 & -218 & -206 & -538 & -189 & -123 & -154 & -133 & -294 & -178 & -195 & -522 & -625 & -457 & -547 & -254 \\
 & $E_\text{ads}^\text{true}$ & -142 & -218 & -206 & -538 & -189 & -123 & -154 & -133 & -294 & -178 & -195 & -522 & -625 & -457 & -547 & -254 \\
 & Error & 0 & 0 & 0 & 0 & 0 & 0 & 0 & 0 & 0 & 0 & 0 & 0 & 0 & 0 & 0 & 0 \\
\cline{1-18}
\multirow[]{5}{*}{\rotatebox{90}{vdW-DF}} & $E_\text{int}$ & -137 & -197 & -221 & -391 & -246 & -181 & -174 & -179 & -306 & -227 & -221 & -535 & -643 & -405 & -590 & -258 \\
 & $E_\text{rlx}$ & 2 & 1 & 8 & 26 & 3 & 10 & 29 & 9 & 51 & 6 & 39 & 114 & 145 & 46 & 89 & 14 \\
 & $E_\text{ads}^\text{approx}$ & -135 & -196 & -212 & -365 & -242 & -171 & -145 & -170 & -255 & -221 & -182 & -421 & -498 & -360 & -502 & -245 \\
 & $E_\text{ads}^\text{true}$ & -156 & -233 & -233 & -466 & -261 & -220 & -187 & -191 & -300 & -243 & -224 & -478 & -576 & -408 & -543 & -277 \\
 & Error & 21 & 37 & 21 & 101 & 19 & 50 & 42 & 21 & 45 & 22 & 42 & 57 & 78 & 48 & 41 & 33 \\
\cline{1-18}
\multirow[]{5}{*}{\rotatebox{90}{rev-vdW-DF2}} & $E_\text{int}$ & -141 & -207 & -273 & -460 & -247 & -192 & -293 & -254 & -411 & -260 & -290 & -683 & -794 & -490 & -671 & -303 \\
 & $E_\text{rlx}$ & 2 & 1 & 8 & 26 & 3 & 10 & 29 & 9 & 51 & 6 & 39 & 114 & 145 & 46 & 89 & 14 \\
 & $E_\text{ads}^\text{approx}$ & -140 & -206 & -265 & -433 & -244 & -182 & -264 & -245 & -360 & -254 & -251 & -568 & -649 & -444 & -583 & -289 \\
 & $E_\text{ads}^\text{true}$ & -140 & -207 & -266 & -449 & -246 & -183 & -227 & -208 & -358 & -255 & -247 & -578 & -663 & -451 & -584 & -288 \\
 & Error & 0 & 1 & 1 & 16 & 2 & 1 & -37 & -37 & -2 & 0 & -5 & 9 & 14 & 6 & 1 & -2 \\
\cline{1-18}
\multirow[]{5}{*}{\rotatebox{90}{PBE0-D4}} & $E_\text{int}$ & -159 & -227 & -250 & -523 & -246 & -166 & -103 & -126 & -259 & -171 & -167 & -741 & -853 & -536 & -701 & -325 \\
 & $E_\text{rlx}$ & 2 & 1 & 8 & 26 & 3 & 10 & 29 & 9 & 51 & 6 & 39 & 114 & 145 & 46 & 89 & 14 \\
 & $E_\text{ads}^\text{approx}$ & -158 & -226 & -241 & -497 & -243 & -156 & -74 & -117 & -208 & -165 & -128 & -626 & -708 & -491 & -613 & -312 \\
 & $E_\text{ads}^\text{true}$ & -158 & -229 & -234 & -521 & -251 & -160 & -102 & -123 & -192 & -172 & -123 & -574 & -659 & -469 & -576 & -304 \\
 & Error & 0 & 4 & -7 & 24 & 8 & 4 & 28 & 6 & -16 & 6 & -5 & -52 & -49 & -22 & -37 & -7 \\
\cline{1-18}
\multirow[]{5}{*}{\rotatebox{90}{B3LYP-D2[Ne]}} & $E_\text{int}$ & -89 & -127 & -156 & -210 & -168 & -98 & -37 & -66 & -171 & -108 & -100 & -643 & -714 & -427 & -599 & -230 \\
 & $E_\text{rlx}$ & 2 & 1 & 8 & 26 & 3 & 10 & 29 & 9 & 51 & 6 & 39 & 114 & 145 & 46 & 89 & 14 \\
 & $E_\text{ads}^\text{approx}$ & -87 & -126 & -148 & -184 & -164 & -88 & -7 & -57 & -120 & -102 & -61 & -528 & -569 & -382 & -510 & -216 \\
 & $E_\text{ads}^\text{true}$ & -88 & -131 & -149 & -219 & -173 & -101 & -70 & -76 & -136 & -118 & -81 & -505 & -552 & -371 & -492 & -215 \\
 & Error & 1 & 5 & 1 & 35 & 8 & 13 & 62 & 18 & 16 & 16 & 20 & -23 & -17 & -11 & -18 & -1 \\
\cline{1-18}
\rotatebox{90}{} & 2RMSE & 19 & 34 & 20 & 100 & 20 & 46 & 91 & 42 & 45 & 25 & 42 & 73 & 85 & 49 & 53 & 30 \\
\cline{1-18}
\end{longtable}
\LTcapwidth=\textwidth
    
\begin{longtable}{llrrrrrrr}
\caption{\label{tab:eads_dft_ensemble_errors_tio2} For the monomers on the \ce{TiO2} rutile(110) and anatase(101) surfaces, we estimate the errors for using the revPBE-D4 geometry and $E_\text{rlx}$ in the final $E_\text{ads}$ of the autoSKZCAM protocol using an ensemble of 6 different DFAs. The errors are calculated as the difference between the true $E_\text{ads}^\text{true}$ (using the appropriate DFA) and the approximated $E_\text{ads}^\text{approx}$ using the revPBE-D4 geometry and $E_\text{rlx}$.} \\

\toprule
 &  & \rotatebox{90}{\ce{CH4} on TiO$_2$ rutile(110)} & \rotatebox{90}{Parallel \ce{CO2} on TiO$_2$ rutile(110)} & \rotatebox{90}{Tilted \ce{CO2} on TiO$_2$ rutile(110)} & \rotatebox{90}{\ce{H2O} on TiO$_2$ rutile(110)} & \rotatebox{90}{\ce{CH3OH} on TiO$_2$ rutile(110)} & \rotatebox{90}{\ce{H2O} on TiO$_2$ anatase(101)} & \rotatebox{90}{\ce{NH3} on TiO$_2$ anatase(101)} \\ 
\midrule
\endfirsthead

\caption[]{(continued)} \\
\endhead

\multicolumn{9}{r}{{Continued on next page}} \\
\endfoot

\bottomrule
\endlastfoot

\multirow[]{5}{*}{\rotatebox{90}{PBE-TS/HI}} & $E_\text{int}$ & -283 & -333 & -400 & -1236 & -1560 & -1108 & -1351 \\
 & $E_\text{rlx}$ & 22 & 14 & 50 & 238 & 302 & 225 & 212 \\
 & $E_\text{ads}^\text{approx}$ & -262 & -318 & -350 & -998 & -1258 & -883 & -1138 \\
 & $E_\text{ads}^\text{true}$ & -265 & -329 & -366 & -988 & -1231 & -860 & -1130 \\
 & Error & -3 & -11 & -16 & 10 & 27 & 23 & 8 \\
\cline{1-9}
\multirow[]{5}{*}{\rotatebox{90}{revPBE-D4}} & $E_\text{int}$ & -289 & -402 & -441 & -1214 & -1549 & -1096 & -1382 \\
 & $E_\text{rlx}$ & 22 & 14 & 50 & 238 & 302 & 225 & 212 \\
 & $E_\text{ads}^\text{approx}$ & -268 & -388 & -390 & -976 & -1247 & -871 & -1170 \\
 & $E_\text{ads}^\text{true}$ & -268 & -388 & -390 & -976 & -1247 & -871 & -1170 \\
 & Error & 0 & 0 & 0 & 0 & 0 & 0 & 0 \\
\cline{1-9}
\multirow[]{5}{*}{\rotatebox{90}{vdW-DF}} & $E_\text{int}$ & -235 & -366 & -437 & -1080 & -1370 & -969 & -1214 \\
 & $E_\text{rlx}$ & 22 & 14 & 50 & 238 & 302 & 225 & 212 \\
 & $E_\text{ads}^\text{approx}$ & -214 & -352 & -387 & -841 & -1068 & -744 & -1002 \\
 & $E_\text{ads}^\text{true}$ & -244 & -381 & -395 & -849 & -1090 & -759 & -1035 \\
 & Error & -30 & -28 & -8 & -8 & -23 & -15 & -33 \\
\cline{1-9}
\multirow[]{5}{*}{\rotatebox{90}{rev-vdW-DF2}} & $E_\text{int}$ & -276 & -409 & -472 & -1300 & -1619 & -1171 & -1410 \\
 & $E_\text{rlx}$ & 22 & 14 & 50 & 238 & 302 & 225 & 212 \\
 & $E_\text{ads}^\text{approx}$ & -255 & -395 & -422 & -1062 & -1317 & -946 & -1198 \\
 & $E_\text{ads}^\text{true}$ & -257 & -402 & -430 & -1050 & -1312 & -933 & -1198 \\
 & Error & -2 & -7 & -7 & 12 & 5 & 13 & 0 \\
\cline{1-9}
\multirow[]{5}{*}{\rotatebox{90}{r$^2$SCAN-rVV10}} & $E_\text{int}$ & -338 & -533 & -584 & -1501 & -1832 & -1367 & -1577 \\
 & $E_\text{rlx}$ & 22 & 14 & 50 & 238 & 302 & 225 & 212 \\
 & $E_\text{ads}^\text{approx}$ & -316 & -519 & -533 & -1263 & -1529 & -1142 & -1364 \\
 & $E_\text{ads}^\text{true}$ & -325 & -537 & -557 & -1232 & -1491 & -1124 & -1355 \\
 & Error & -9 & -18 & -23 & 31 & 39 & 18 & 9 \\
\cline{1-9}
\multirow[]{5}{*}{\rotatebox{90}{HSE06-D4}} & $E_\text{int}$ & -288 & -406 & -481 & -1368 & -1688 & -1229 & -1518 \\
 & $E_\text{rlx}$ & 22 & 14 & 50 & 238 & 302 & 225 & 212 \\
 & $E_\text{ads}^\text{approx}$ & -267 & -392 & -431 & -1130 & -1386 & -1004 & -1306 \\
 & $E_\text{ads}^\text{true}$ & -272 & -400 & -447 & -1092 & -1333 & -967 & -1284 \\
 & Error & -5 & -8 & -16 & 38 & 52 & 36 & 22 \\
\cline{1-9}
\rotatebox{90}{} & 2RMSE & 29 & 33 & 31 & 46 & 66 & 45 & 37 \\
\cline{1-9}
\end{longtable}
\LTcapwidth=\textwidth
    
\begin{longtable}{llrrrrrr}
\caption{\label{tab:eads_dft_ensemble_errors_ml_clus_mgo} For the clusters and monolayer systems as well as chemisorbed \ce{CO2} on MgO(001) surface, we estimate the errors for using the DFT geometry and $E_\text{rlx}$ in the final $E_\text{ads}$ of the autoSKZCAM protocol using an ensemble of 6 different DFAs. The errors are calculated as the difference between the true $E_\text{ads}^\text{true}$ (using the appropriate DFA) and the approximated $E_\text{ads}^\text{approx}$ using the revPBE-D4 geometry and $E_\text{rlx}$. For the NO dimer, this is done with respect to the B3LYP-D2[N2] geometry and $E_\text{rlx}$. There is an additional cohesive energy $E_\text{coh}$ term for the monolayer and cluster systems. For the chemisorbed \ce{CO2}, there is an additional conformational energy $E_\text{conf}$ term.} \\

\toprule
 &  & \rotatebox{90}{Monolayer \ce{CH4}} & \rotatebox{90}{Monolayer \ce{C2H6}} & \rotatebox{90}{Dimer \ce{NO}} & \rotatebox{90}{Tetramer \ce{H2O}} & \rotatebox{90}{Tetramer \ce{CH3OH}} & \rotatebox{90}{Chemisorbed \ce{CO2}} \\ 
\midrule
\endfirsthead

\caption[]{(continued)} \\
\endhead

\multicolumn{8}{r}{{Continued on next page}} \\
\endfoot

\bottomrule
\endlastfoot

\multirow[]{6}{*}{\rotatebox{90}{PBE-D2[Ne]}} & $E_\text{int}$ & -114 & -142 & -340 & -397 & -416 & -2993 \\
 & $E_\text{rlx}$ & 1 & 4 & 26 & 52 & 68 & 663 \\
 & $E_\text{coh}$ or $E_\text{conf}$ & -40 & -106 & -234 & -309 & -380 & 1841 \\
 & $E_\text{ads}^\text{approx}$ & -153 & -245 & -548 & -654 & -729 & -489 \\
 & $E_\text{ads}^\text{true}$ & -153 & -248 & -551 & -660 & -733 & -518 \\
 & Error & 0 & -3 & -3 & -6 & -5 & -29 \\
\cline{1-8}
\multirow[]{6}{*}{\rotatebox{90}{revPBE-D4}} & $E_\text{int}$ & -142 & -207 & -351 & -404 & -483 & -2981 \\
 & $E_\text{rlx}$ & 1 & 4 & 26 & 52 & 68 & 663 \\
 & $E_\text{coh}$ or $E_\text{conf}$ & -31 & -48 & -186 & -261 & -328 & 1797 \\
 & $E_\text{ads}^\text{approx}$ & -171 & -251 & -511 & -613 & -743 & -521 \\
 & $E_\text{ads}^\text{true}$ & -171 & -251 & -514 & -613 & -743 & -521 \\
 & Error & 0 & 0 & -3 & 0 & 0 & 0 \\
\cline{1-8}
\multirow[]{6}{*}{\rotatebox{90}{vdW-DF}} & $E_\text{int}$ & -136 & -200 & -337 & -376 & -447 & -2638 \\
 & $E_\text{rlx}$ & 1 & 4 & 26 & 52 & 68 & 663 \\
 & $E_\text{coh}$ or $E_\text{conf}$ & -58 & -67 & -145 & -240 & -291 & 1763 \\
 & $E_\text{ads}^\text{approx}$ & -192 & -263 & -457 & -564 & -670 & -212 \\
 & $E_\text{ads}^\text{true}$ & -219 & -319 & -478 & -597 & -719 & -361 \\
 & Error & -27 & -55 & -21 & -33 & -48 & -149 \\
\cline{1-8}
\multirow[]{6}{*}{\rotatebox{90}{rev-vdW-DF2}} & $E_\text{int}$ & -140 & -197 & -398 & -452 & -519 & -3082 \\
 & $E_\text{rlx}$ & 1 & 4 & 26 & 52 & 68 & 663 \\
 & $E_\text{coh}$ or $E_\text{conf}$ & -43 & -83 & -276 & -289 & -357 & 1836 \\
 & $E_\text{ads}^\text{approx}$ & -181 & -277 & -648 & -689 & -809 & -583 \\
 & $E_\text{ads}^\text{true}$ & -182 & -282 & -654 & -693 & -811 & -605 \\
 & Error & -1 & -5 & -5 & -4 & -2 & -21 \\
\cline{1-8}
\multirow[]{6}{*}{\rotatebox{90}{PBE0-D4}} & $E_\text{int}$ & -158 & -211 & -290 & -465 & -526 & -3713 \\
 & $E_\text{rlx}$ & 1 & 4 & 26 & 52 & 68 & 663 \\
 & $E_\text{coh}$ or $E_\text{conf}$ & -34 & -64 & 62 & -302 & -367 & 2153 \\
 & $E_\text{ads}^\text{approx}$ & -191 & -271 & -202 & -715 & -825 & -898 \\
 & $E_\text{ads}^\text{true}$ & -193 & -274 & -219 & -692 & -795 & -764 \\
 & Error & -3 & -2 & -17 & 23 & 30 & 134 \\
\cline{1-8}
\multirow[]{6}{*}{\rotatebox{90}{B3LYP-D2[Ne]}} & $E_\text{int}$ & -87 & -111 & -231 & -400 & -429 & -3250 \\
 & $E_\text{rlx}$ & 1 & 4 & 26 & 52 & 68 & 663 \\
 & $E_\text{coh}$ or $E_\text{conf}$ & -37 & -124 & 63 & -303 & -371 & 2075 \\
 & $E_\text{ads}^\text{approx}$ & -124 & -231 & -143 & -651 & -732 & -512 \\
 & $E_\text{ads}^\text{true}$ & -127 & -238 & -143 & -643 & -717 & -460 \\
 & Error & -4 & -7 & 0 & 8 & 16 & 52 \\
\cline{1-8}
\rotatebox{90}{} & 2RMSE & 25 & 50 & 25 & 38 & 53 & 188 \\
\cline{1-8}
\end{longtable}

In general, we find that $\epsilon_\textrm{geom}$ is relatively small, being less than $40\,$meV for the majority of systems.
For the monomers on MgO, the $\epsilon_\textrm{geom}$ range is between $20{-}100\,$meV, while this error range lowers to $29{-}66\,$meV for the \ce{TiO2} surfaces.
We find that this geometrical error is particularly dependent on the strength of the binding, going from $20\,$ meV for CO on MgO - a system with weak physisorption to $66\,$meV for \ce{CH3OH} on \ce{TiO2} rutile(110), a system with much stronger adsorption behaviour.
In particular, the largest error is found for the chemisorbed \ce{CO2} system with a value of $188\,$meV, arising from the large change to the surface and molecule electronic structure and geometry due to charge transfer.

\subsection{\label{sec:ezpv_etherm} Zero-point vibrational and enthalpic contributions}

The DFT ensemble was also used to calculate the zero-point vibrational $E_\textrm{ZPV}$ and thermal $E_\textrm{T}$ contributions to the adsorption enthalpy $H_\textrm{ads}$.
As both $E_\textrm{ZPV}$ and $E_\textrm{T}$ can be computed from vibrational modes, we will particular discuss overall enthalpic contributions: $\Delta H = E_\textrm{ZPV} + E_\textrm{T} - RT$,  to add onto $E_\textrm{ads}$ to make $H_\textrm{ads}$
We utilised the rigid-rotor quasi-harmonic approximation (quasi-RRHO) to compute these terms based on the vibrational frequencies computed by DFT.
These were first proposed by Grimme \etal{}~\cite{grimmeSupramolecularBindingThermodynamics2012} for entropies and then adapted for enthalpies by Li \etal{}~\cite{liImprovedForceFieldParameters2015}. 
Within the simple rigid-rotor harmonic oscillator (RRHO) models, the contribution of each vibrational mode $i$ with frequency $\nu_i$ contributes $V_i^{RRHO} = \frac{1}{2}h\nu_i + \frac{h\nu_i}{1+e^{h\nu_i/k_B T}}$ to ${\Delta}H$ (including the contribution to $E_\textrm{ZPV}$ in the first term and $E_\textrm{T}$ in the second term).
One deficiency of this model is that it is wrong for the zero-frequency (translation + rotation) modes as it predicts a $k_B T$ contribution to ${\Delta}H$ rather than $\frac{1}{2}k_B T$.
The quasi-RRHO fixes this issue by creating an interpolation between the RRHO model and the free rotor model (whereby each mode contributes $\frac{1}{2}k_B T$) at low frequencies:
\begin{equation}
    V_i^\textrm{quasi-RRHO} = \omega(\nu_i) \times V_i^\textrm{RRHO}  + (1 - \omega(\nu_i)) \times \dfrac{1}{2} k_B T,
\end{equation}
\begin{equation}
 \omega(\nu_i) = \dfrac{1}{1+(\nu_0/\nu_i)^4} 
\end{equation}
where we have set the interpolation to start at around $\nu_0$ = 100 cm\textsuperscript{-1}.

\begin{turnpage}
\begin{table}
\caption{\label{tab:ethermal}The zero-point vibrational energy ($E_\text{ZPV}$), thermal energy ($E_\text{therm}$), and overall enthalpy ($\Delta H$) contributions (in meV) to the adsorption enthalpy for all studied systems using an ensemble of 4 different DFAs. The errors are calculated as the 2$\sigma$ standard deviation of the $\Delta H$ values for the 4 DFAs.}
\begin{adjustbox}{center,max width=1.4\textwidth}
\begin{tabular}{lrrrrrrrrrrrrrrrrrrrrrrrrrrrrr}
\toprule
 & \rotatebox{90}{CH$_4$ on MgO(001)} & \rotatebox{90}{Monolayer CH$_4$ on MgO(001)} & \rotatebox{90}{C$_2$H$_6$ on MgO(001)} & \rotatebox{90}{Monolayer C$_2$H$_6$ on MgO(001)} & \rotatebox{90}{CO on MgO(001)} & \rotatebox{90}{C$_6$H$_6$ on MgO(001)} & \rotatebox{90}{Parallel N$_2$O on MgO(001)} & \rotatebox{90}{Tilted N$_2$O on MgO(001)} & \rotatebox{90}{Vertical-Hollow NO on MgO(001)} & \rotatebox{90}{Vertical-Mg NO on MgO(001)} & \rotatebox{90}{Bent-Bridge NO on MgO(001)} & \rotatebox{90}{Bent-Mg NO on MgO(001)} & \rotatebox{90}{Bent-O NO on MgO(001)} & \rotatebox{90}{Dimer NO on MgO(001)} & \rotatebox{90}{Monomer H$_2$O on MgO(001)} & \rotatebox{90}{Tetramer H$_2$O on MgO(001)} & \rotatebox{90}{Tilted CH$_3$OH on MgO(001)} & \rotatebox{90}{Parallel CH$_3$OH on MgO(001)} & \rotatebox{90}{Tetramer CH$_3$OH on MgO(001)} & \rotatebox{90}{NH$_3$ on MgO(001)} & \rotatebox{90}{Physisorbed CO$_2$ on MgO(001)} & \rotatebox{90}{Chemisorbed CO$_2$ on MgO(001)} & \rotatebox{90}{CH$_4$ on TiO$_2$ rutile(110)} & \rotatebox{90}{Parallel CO$_2$ on TiO$_2$ rutile(110)} & \rotatebox{90}{Tilted CO$_2$ on TiO$_2$ rutile(110)} & \rotatebox{90}{H$_2$O on TiO$_2$ rutile(110)} & \rotatebox{90}{CH$_3$OH on TiO$_2$ rutile(110)} & \rotatebox{90}{H$_2$O on TiO$_2$ anatase(101)} & \rotatebox{90}{NH$_3$ on TiO$_2$ anatase(101)} \\ 
\midrule
Temperature [K] & 47 & 47 & 75 & 75 & 61 & 162 & 77 & 77 & 80 & 80 & 80 & 80 & 80 & 80 & 203 & 235 & 286 & 286 & 286 & 160 & 120 & 230 & 85 & 177 & 177 & 303 & 370 & 257 & 410 \\
$RT$ [meV] & 4 & 4 & 6 & 6 & 5 & 14 & 7 & 7 & 7 & 7 & 7 & 7 & 7 & 7 & 17 & 20 & 25 & 25 & 25 & 14 & 10 & 20 & 7 & 15 & 15 & 26 & 32 & 22 & 35 \\
$E_\text{ZPV}$ [XC 1] & 29 & 29 & 12 & 17 & 31 & 1 & 1 & 3 & 18 & 13 & 23 & 11 & 10 & 63 & 84 & 98 & 36 & 22 & 61 & 72 & 2 & 47 & 17 & 7 & 9 & 110 & 68 & 113 & 111 \\
$E_\text{T}$ [XC 1] & -5 & -5 & -2 & -4 & -6 & 2 & -1 & 0 & -5 & -3 & -6 & -3 & -4 & -10 & -22 & -19 & 3 & 1 & 1 & -13 & 0 & -8 & -3 & 3 & 3 & -18 & 10 & -18 & -2 \\
$\Delta H$ [XC 1] & 20 & 20 & 3 & 7 & 20 & -12 & -6 & -4 & 7 & 3 & 10 & 1 & 0 & 47 & 45 & 58 & 14 & -1 & 38 & 46 & -8 & 19 & 6 & -5 & -3 & 66 & 46 & 73 & 74 \\
$E_\text{ZPV}$ [XC 2] & 23 & 28 & 13 & 37 & 28 & 9 & 2 & 6 & 22 & 14 & 26 & 13 & 11 & 64 & 85 & 100 & 38 & 24 & 63 & 72 & 1 & 47 & 25 & 9 & 12 & 111 & 74 & 112 & 115 \\
$E_\text{T}$ [XC 2] & -4 & -5 & -2 & -5 & -5 & 2 & -1 & -1 & -5 & -4 & -6 & -3 & -4 & -10 & -21 & -19 & 3 & 1 & 1 & -12 & 0 & -8 & -4 & 3 & 3 & -15 & 9 & -15 & 1 \\
$\Delta H$ [XC 2] & 15 & 20 & 4 & 26 & 18 & -3 & -5 & -2 & 10 & 4 & 12 & 3 & 0 & 47 & 47 & 61 & 17 & 0 & 39 & 46 & -9 & 19 & 13 & -3 & 0 & 69 & 51 & 75 & 80 \\
$E_\text{ZPV}$ [XC 3] & 10 & 10 & 0 & 28 & 28 & -1 & 0 & 0 & 12 & 10 & 18 & 9 & 7 & 58 & 84 & 97 & 41 & 24 & 62 & 69 & -2 & 44 & 6 & 5 & 7 & 96 & 71 & 90 & 105 \\
$E_\text{T}$ [XC 3] & -2 & -2 & 0 & -5 & -5 & 2 & 0 & 0 & -3 & -3 & -5 & -2 & -3 & -9 & -18 & -16 & 5 & 1 & 5 & -12 & 1 & -8 & -1 & 3 & 3 & -7 & 13 & -9 & 3 \\
$\Delta H$ [XC 3] & 3 & 4 & -7 & 17 & 17 & -14 & -7 & -7 & 2 & 0 & 6 & 0 & -3 & 42 & 48 & 60 & 21 & 1 & 43 & 43 & -11 & 16 & -3 & -7 & -5 & 63 & 52 & 59 & 72 \\
$E_\text{ZPV}$ [XC 4] & 15 & 17 & 4 & 22 & 32 & -1 & 1 & 4 & 19 & 14 & 25 & 12 & 10 & 65 & 85 & 100 & 37 & 21 & 61 & 68 & 0 & 48 & 19 & 11 & 13 & 108 & 67 & 107 & 107 \\
$E_\text{T}$ [XC 4] & -4 & -4 & -1 & -5 & -6 & 2 & -1 & 0 & -5 & -4 & -6 & -3 & -4 & -10 & -21 & -20 & 3 & 1 & 0 & -13 & 0 & -9 & -5 & 3 & 3 & -17 & 9 & -15 & 0 \\
$\Delta H$ [XC 4] & 8 & 8 & -4 & 10 & 21 & -13 & -6 & -3 & 7 & 4 & 12 & 2 & 0 & 48 & 46 & 60 & 15 & -3 & 37 & 42 & -10 & 19 & 7 & -2 & 0 & 64 & 44 & 70 & 71 \\
Final $\Delta H$ & 11 & 13 & -1 & 15 & 19 & -10 & -6 & -4 & 7 & 2 & 10 & 1 & -1 & 46 & 47 & 60 & 17 & -1 & 39 & 44 & -10 & 18 & 6 & -4 & -2 & 65 & 48 & 69 & 75 \\
Error & 13 & 14 & 10 & 14 & 3 & 9 & 1 & 4 & 6 & 3 & 5 & 3 & 3 & 5 & 3 & 2 & 6 & 3 & 5 & 3 & 2 & 2 & 11 & 4 & 4 & 5 & 7 & 13 & 7 \\
\bottomrule
\end{tabular}
\end{adjustbox}
\end{table}
\end{turnpage}

We have calculated the vibrational frequencies of the molecule (both on the surface and in the gas-phase) to calculate the ${\Delta}H$ contribution to $H_\textrm{ads}$.
This is shown in Table~\ref{tab:ethermal}, where we have computed $E_\textrm{ZPV}$, $E_\textrm{T}$ and ${\Delta}H$ using four DFAs from the ensemble.
For the MgO surface, the four DFAs were: XC 1 = PBE-D2[Ne], XC 2 = revPBE-D4, XC 3 = vdW-DF, XC 4 = rev-vdW-DF2.
For the \ce{TiO2} surfaces, the DFAs selected were: XC 1 = PBE-TS/HI, XC 2 = revPBE-D4, XC 3 = vdW-DF, XC 4 = rev-vdW-DF2.
The final ${\Delta}H$ contribution was taken to be the average of the four functionals with the error estimated to be the 2$\sigma$ standard deviation of the four estimates to ensure 95\% confidence interval in the estimate.
It can be seen that the ${\Delta}H$ contribution, relative to $E_\textrm{ads}$ is overall small to $H_\textrm{ads}$, ranging from ${-}10$ to $75\,$meV.
Moreover the error estimates are also small, ranging from $1$ to $15\,$meV.

We expect nuclear quantum~\cite{alfeInitioStatisticalMechanics2010} contributions to ${\Delta}H$ to be small while anharmonic effects can have a potential effect for weakly binding molecules such as \ce{CH4} on MgO(001)~\cite{picciniEffectAnharmonicityAdsorption2014}.
This effect is well-captured by the error bar, where \ce{CH4} on MgO(001) gives a larger error bar on ${\Delta}H$ than the stronger binding systems.
Another source of error is the freezing of surface vibrational modes when calculating ${\Delta}H$.
In Table~\ref{tab:ethermal_slab_effect}, we have calculated ${\Delta}H$ for a select number of systems which include the surface degrees of freedom in the top two layers of MgO(001).
Overall, there is negligible change of maximum $2\,$meV for the majority of systems.
The only system where there is a significant effect is for chemisorbed \ce{CO2} where it changes from $18{\pm}5\,$meV to $31{\pm}5\,$meV, due to changes in the surface electronic structure from charge transfer.
In our final $H_\textrm{ads}$ analysis, we will include the surface degrees of freedom for \ce{CO2} on MgO(001) only, while neglecting it for all other systems.

\begin{table}
\caption{\label{tab:ethermal_slab_effect}Comparing the effect of only the molecule vibrational degrees of freedom and the inclusion of surface degrees of freedom on the enthalpy ($\Delta H$) contribution (in meV) to the adsorption enthalpy for a select few molecules adsorbed on MgO(001). $\Delta H$ is calculated as the mean from an ensemble of 4 DFAs (neglecting the hybrid DFAs) with 2$\sigma$ error included.}
\begin{adjustbox}{center,max width=1.4\textwidth}
\begin{tabular}{lrrrr}
\toprule
 & \rotatebox{90}{CO} & \rotatebox{90}{H$_2$O Monomer} & \rotatebox{90}{CO$_2$ Physisorbed} & \rotatebox{90}{CO$_2$ Chemisorbed} \\ 
\midrule
Molecule & 19 $\pm$ 6 & 47 $\pm$ 5 & -10 $\pm$ 5 & 18 $\pm$ 5 \\
Molecule+Surface & 19 $\pm$ 1 & 48 $\pm$ 5 & -8 $\pm$ 2 & 31 $\pm$ 3 \\
\bottomrule
\end{tabular}
\end{adjustbox}
\end{table}

\subsection{\label{sec:ediss_h2o_ch3oh} Dissociation energy for the \ce{H2O} and \ce{CH3OH} clusters}

As discussed in Sec.~\ref{sec:ch3oh_discuss}, we use the autoSKZCAM framework to obtain the adsorption enthalpy for a molecularly adsorbed (tetramer) cluster of \ce{CH3OH} and \ce{H2O}.
The lowest energy geometry involves partial dissocation of the \ce{CH3OH} and \ce{H2O} clusters.
In our final $H_\textrm{ads}$ for these two systems, we compute an additional term $E_\textrm{diss}$ which accounts for the (electronic) energetic stabilisation to form the partially dissociated cluster.
This is performed using the DFT ensemble -- using the revPBE-D4 geometry -- shown in Table~\ref{tab:ediss} for the 6 DFAs, where we obtain an average and calculate the error as the $2{\sigma}$ standard deviation.
We have also computed the enthalpic difference (${-}31\,$meV and ${-}18\,$meV for \ce{H2O} and \ce{CH3OH}) between the dissociated and molecular configurations (with geometries taken from revPBE-D4) and added this contribution to the original $\Delta H$ calculated for the molecular configuration in Table~\ref{tab:ethermal}; this leads to a final $\Delta H$ of $42{\pm}2\,$meV and $8{\pm}5\,$meV for the dissociated configuration of \ce{H2O} and \ce{CH3OH} respectively, as shown in Table~\ref{tab:autoskzcam_hads_terms}.

\begin{table}
\caption{\label{tab:ediss}The dissociation energy $E_\text{diss}$ (in meV) for the \ce{CH3OH} and \ce{H2O} tetramer calculated for the DFT ensemble. This is defined as the energy difference between the dissociated and molecular configurations of the tetramer.}
\begin{adjustbox}{center}
\begin{tabular}{lrr}
\toprule
 & \ce{CH3OH} Tetramer & \ce{H2O} Tetramer \\ 
\midrule
PBE-D2[Ne] & -76 & -89 \\
revPBE-D4 & -78 & -81 \\
vdW-DF & -48 & -52 \\
rev-vdW-DF2 & -65 & -75 \\
PBE0-D4 & -80 & -83 \\
B3LYP-D2[Ne] & -81 & -98 \\
Average & -71 & -80 \\
Error & 24 & 29 \\
Final & -71 ± 24 & -79 ± 29 \\
\bottomrule
\end{tabular}
\end{adjustbox}
\end{table}

\section{\label{sec:final_hads} Final autoSKZCAM estimates}

In Table~\ref{tab:autoskzcam_hads_terms}, we show the terms which make up the final $E_\textrm{ads}$ and $H_\textrm{ads}$ estimates in our autoSKZCAM framework.
Robust error bars have been estimated which aim to encapsulate the major sources of errors (to at least a 95\% confidence interval) within this estimate with respect to a fully converged CCSD(T) estimate.
The $E_\textrm{int}^\textrm{SKZCAM}$ error estimate covers the potential finite-size errors from utilising the SKZCAM protocol in Sec.~\ref{sec:skzcam_si_sec}, while $\epsilon_\textrm{geom}$ (discussed in Sec.~\ref{sec:dft_geom_error}) covers the error from computing $E_\textrm{rlx}^\textrm{DFT}$ with DFT and the error from using a DFT geometry in all the individual terms of $E_\textrm{ads}$.
We reach a final error estimate as the root squared sum of the errors within the individual terms.
As seen in Table~\ref{tab:autoskzcam_hads_terms}, $\epsilon_\textrm{geom}$ is in general the largest source of error out of all of the terms and serves as the major term to target for future improvements.

\begin{table}
\caption{\label{tab:autoskzcam_hads_terms}A summary of the terms which make up the final autoSKZCAM $E_\text{ads}$ and $H_\text{ads}$ estimate (in meV). Here, the final $E_\text{ads}$ value is the sum of $E_\text{int}^\text{SKZCAM}$, $E_\text{rlx}^\text{DFT}$, $E_\text{coh}^\text{CCSD(T)}$ and $E_\text{conf}^\text{CCSD(T)}$ values, where the last two terms are only included for systems where they are calculated in Sec.~\ref{sec:wft_ecoh_econf}. The errors due to $E_\text{rlx}^\text{DFT}$ and from using the revPBE-D4 geometry for all the $E_\text{ads}$ terms is encapsulated in $\epsilon_\text{geom}$ from an ensemble of DFAs. An additional $\Delta H$ term is also calculated, further calculated from a DFT ensemble.}
\begin{adjustbox}{center,max width=0.95\textwidth}
\begin{tabular}{lrrrrrrrrr}
\toprule
 & \rotatebox{90}{$E_\text{int}^\text{SKZCAM}$} & \rotatebox{90}{$E_\text{rlx}^\text{DFT}$} & \rotatebox{90}{$E_\text{coh}^\text{CCSD(T)}$} & \rotatebox{90}{$E_\text{conf}^\text{CCSD(T)}$} & \rotatebox{90}{$\epsilon_\text{geom}$} & \rotatebox{90}{$E_\text{diss.}^\text{DFT}$} & \rotatebox{90}{$E_\text{ads}^\text{autoSKZCAM}$} & \rotatebox{90}{$\Delta H^\text{DFT}$} & \rotatebox{90}{$H_\text{ads}^\text{autoSKZCAM}$} \\ 
\midrule
CH$_4$ on MgO(001) & -122 ± 2 & 2 & - & - & 19 & - & -120 ± 19 & 11 ± 13 & -109 ± 23 \\
Monolayer CH$_4$ on MgO(001) & -121 ± 3 & 1 & -25 & - & 25 & - & -145 ± 25 & 13 ± 14 & -132 ± 28 \\
C$_2$H$_6$ on MgO(001) & -175 ± 4 & 1 & - & - & 34 & - & -174 ± 34 & -1 ± 10 & -175 ± 35 \\
Monolayer C$_2$H$_6$ on MgO(001) & -161 ± 3 & 4 & -61 & - & 50 & - & -218 ± 50 & 15 ± 14 & -203 ± 52 \\
CO on MgO(001) & -207 ± 4 & 8 & - & - & 20 & - & -198 ± 20 & 19 ± 3 & -180 ± 20 \\
C$_6$H$_6$ on MgO(001) & -446 ± 9 & 26 & - & - & 100 & - & -420 ± 100 & -10 ± 9 & -430 ± 100 \\
Parallel N$_2$O on MgO(001) & -256 ± 3 & 3 & - & - & 20 & - & -253 ± 20 & -6 ± 1 & -259 ± 20 \\
Tilted N$_2$O on MgO(001) & -168 ± 4 & 10 & - & - & 46 & - & -157 ± 46 & -4 ± 4 & -161 ± 47 \\
Vertical-Hollow NO on MgO(001) & 32 ± 4 & 29 & - & - & 91 & - & 61 ± 91 & 7 ± 6 & 68 ± 91 \\
Vertical-Mg NO on MgO(001) & -62 ± 5 & 9 & - & - & 42 & - & -53 ± 42 & 2 ± 3 & -50 ± 42 \\
Bent-Bridge NO on MgO(001) & -62 ± 8 & 51 & - & - & 45 & - & -10 ± 45 & 10 ± 5 & 0 ± 46 \\
Bent-Mg NO on MgO(001) & -126 ± 5 & 6 & - & - & 25 & - & -120 ± 26 & 1 ± 3 & -119 ± 26 \\
Bent-O NO on MgO(001) & -3 ± 7 & 39 & - & - & 42 & - & 36 ± 43 & -1 ± 3 & 35 ± 43 \\
Dimer NO on MgO(001) & -246 ± 3 & 42 & -75 & - & 59 & - & -278 ± 59 & 46 ± 5 & -232 ± 59 \\
Monomer H$_2$O on MgO(001) & -703 ± 4 & 114 & - & - & 73 & - & -588 ± 73 & 47 ± 3 & -542 ± 73 \\
Dissociated Tetramer H$_2$O on MgO(001) & -463 ± 7 & 52 & -281 & - & 38 & -80 ± 29 & -772 ± 48 & 42 ± 2 & -730 ± 48 \\
Tilted CH$_3$OH on MgO(001) & -787 ± 5 & 145 & - & - & 85 & - & -642 ± 86 & 17 ± 6 & -625 ± 86 \\
Parallel CH$_3$OH on MgO(001) & -506 ± 1 & 46 & - & - & 49 & - & -461 ± 49 & -1 ± 3 & -461 ± 49 \\
Dissociated Tetramer CH$_3$OH on MgO(001) & -511 ± 6 & 68 & -336 & - & 53 & -71 ± 24 & -851 ± 59 & 8 ± 5 & -843 ± 59 \\
NH$_3$ on MgO(001) & -657 ± 9 & 89 & - & - & 53 & - & -568 ± 53 & 44 ± 3 & -524 ± 54 \\
Physisorbed CO$_2$ on MgO(001) & -308 ± 2 & 14 & - & - & 30 & - & -294 ± 30 & -10 ± 2 & -304 ± 30 \\
Chemisorbed CO$_2$ on MgO(001) & -3504 ± 32 & 663 & - & 2094 & 188 & - & -747 ± 191 & 18 ± 2 & -729 ± 191 \\
CH$_4$ on TiO$_2$ rutile(110) & -269 ± 2 & 22 & - & - & 29 & - & -247 ± 29 & 6 ± 11 & -241 ± 31 \\
Parallel CO$_2$ on TiO$_2$ rutile(110) & -410 ± 5 & 14 & - & - & 33 & - & -396 ± 33 & -4 ± 4 & -400 ± 34 \\
Tilted CO$_2$ on TiO$_2$ rutile(110) & -493 ± 7 & 50 & - & - & 31 & - & -442 ± 31 & -2 ± 4 & -445 ± 32 \\
H$_2$O on TiO$_2$ rutile(110) & -1310 ± 33 & 238 & - & - & 46 & - & -1072 ± 57 & 65 ± 5 & -1007 ± 57 \\
CH$_3$OH on TiO$_2$ rutile(110) & -1634 ± 37 & 302 & - & - & 66 & - & -1332 ± 76 & 48 ± 7 & -1284 ± 76 \\
H$_2$O on TiO$_2$ anatase(101) & -1208 ± 16 & 225 & - & - & 45 & - & -983 ± 48 & 69 ± 13 & -913 ± 50 \\
NH$_3$ on TiO$_2$ anatase(101) & -1377 ± 18 & 212 & - & - & 37 & - & -1165 ± 41 & 75 ± 7 & -1090 ± 42 \\
\bottomrule
\end{tabular}
\end{adjustbox}
\end{table}

\clearpage

\subsection{\label{sec:error_validation}Validating autoSKZCAM error estimates}

In this section, we validate our chosen procedure for estimating errors on $H_\text{ads}$ in the autoSKZCAM framework, using the chemisorbed \ce{CO2} on MgO(001) as an example due to the relatively large estimated error bars for this system.
As seen in Table~\ref{tab:autoskzcam_hads_terms}, the major source of error comes from $\epsilon_\textrm{geom}$ -- the error for utilising DFT to generate the geometry to calculate the adsorption energy $E_\textrm{ads}$, encapsulating the interaction energy $E_\textrm{int}$, relaxation energy $E_\textrm{rlx}$ and additional terms (e.g., $E_\textrm{coh}$ and $E_\textrm{conf}$).
To understand how well our error estimates cover the errors from utilising DFT to generate the geometry, we have calculated $H_\textrm{ads}$ with the autoSKZCAM framework using the geometries from the 6 different density functional approximations in the DFT ensemble in Table~\ref{tab:co2_error_validation}.
Our final $H_\textrm{ads}$ estimate (in Table~\ref{tab:autoskzcam_hads_terms}) of $-729{\pm}191\,$meV covers a range from $-538,$meV to $-920\,$meV.
The actual range of values for using the six different DFAs goes from $-599\,$meV for PBE0-D4 to $-840,$meV for vdW-DF.
There is overall a faithful (and conservative) representation of the range of possible values of $H_\textrm{ads}$ in Table~\ref{tab:co2_error_validation}, with no values lying outside this error bar.

\begin{table}[h]
\caption{\label{tab:co2_error_validation}Final $H_\text{ads}$ (in meV) for CO$_2$ chemisorbed on MgO using various DFAs as the geometry within the subsequent autoSKZCAM framework treatment. The final $H_\text{ads}$ is the sum of $E_\text{int}^\text{MP2,bulk}$, $\Delta_\text{CC}$, $\Delta_\text{basis}$, $E_\text{rlx}^\text{DFT}$, $E_\text{conf}^\text{CCSD(T)}$ and $\Delta H^\text{DFT}$ terms as described in Sec.~\ref{sec:dft_geom_error}.}
\begin{adjustbox}{center}
\begin{tabular}{lrrrrrrr}
\toprule
 & $E_\text{int}^\text{MP2,bulk}$ & $\Delta_\text{CC}$ & $\Delta_\text{basis}$ & $E_\text{rlx}^\text{DFT}$ & $E_\text{conf}^\text{CCSD(T)}$ & $\Delta H^\text{DFT}$ & $H_\text{ads}^\text{final}$ \\ 
\midrule
PBE-D2[Ne] & -3358 & -225 & 24 & 693 & 2085 & 18 & -763 \\
revPBE-D4 & -3309 & -220 & 25 & 663 & 2094 & 18 & -729 \\
vdW-DF & -3457 & -228 & 23 & 704 & 2100 & 18 & -840 \\
rev-vdW-DF2 & -3296 & -223 & 24 & 676 & 2055 & 18 & -746 \\
PBE0-D4 & -3223 & -224 & 24 & 773 & 2033 & 18 & -599 \\
B3LYP-D2[Ne] & -3338 & -227 & 24 & 810 & 2029 & 18 & -685 \\
\bottomrule
\end{tabular}
\end{adjustbox}
\end{table}

\clearpage

\section{\label{sec:quacc}Automation of the autoSKZCAM framework}
The autoSKZCAM framework is freely available and open-source on Github~\cite{shiBenshi97AutoSKZCAM2025a}.
We have developed a set of functions that automates the entire SKZCAM protocol to generate the clusters and inputs to obtain an accurate interaction energy $E_\textrm{int}$.
The remaining contributions to the adsorption enthalpy is calculated with the DFT ensemble and this can involve a significant number of calculations/book-keeping.
We make extensive use of the QuAcc computational materials science workflow library~\cite{rosenQuaccQuantumAccelerator2024} to manage these calculations.
All of this can be achieved within a single Jupyter Notebook, requiring minimal user intervention.

\subsection{QuAcc computational workflow details}

The QuAcc workflow library contains (user-defined) `recipes' that represent specific tasks (i.e., calculating the static energy or performing a geometry optimisation) and `flows' which represent workflows that combine these recipes to e.g., generate slabs from the bulk or make an adsorbed surface geometry.
This allows for the study of surfaces to be automated and efficiently dispatched to any computing environment.
We make use of many of the pre-existing recipes and flows within QuAcc and as part of this work, we have also developed new recipes and flows to enable calculations using a DFT ensemble (or arbitrary number of DFAs).

\begin{figure}[ht]
    \includegraphics[width=\textwidth,trim=0 70 0 70, clip]{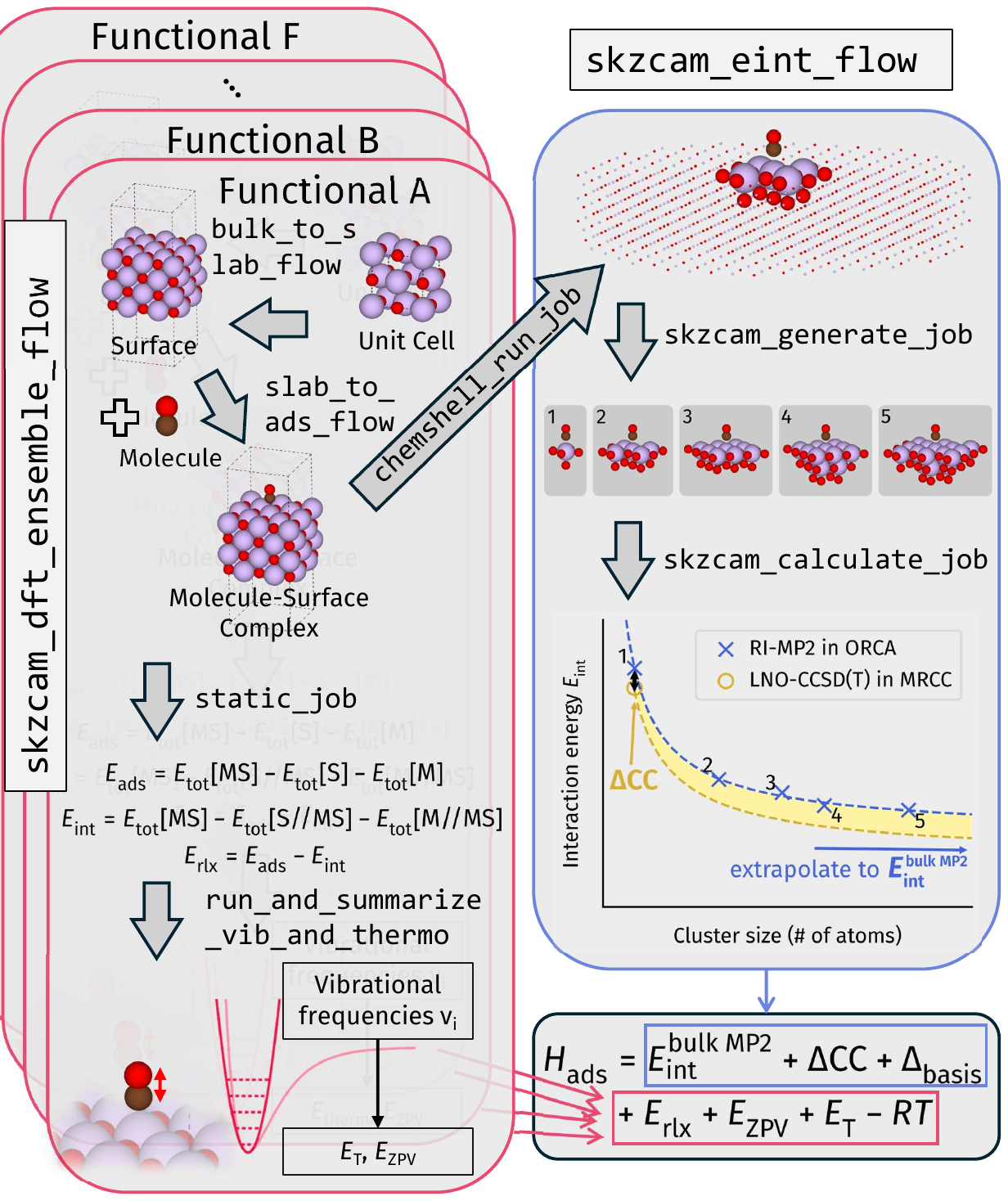}
    \caption{\label{fig:quacc_procedure} The computational workflow for calculating an $H_\textrm{ads}$ using the autoSKZCAM framework. It makes heavy use of the QuAcc computational materials science workflow library~\cite{rosenQuaccQuantumAccelerator2024}. We make use of pre-existing `flows' to generate the adsorbate--surface complex for the density functional approximations within the DFT ensemble and to subsequently calculate $E_\textrm{ads}$, $E_\textrm{int}$ and $E_\textrm{rlx}$ as well as $E_\textrm{ZPV}$ and $E_\textrm{T}$. We have developed new Python modules to generate the series of clusters and inputs necessary to calculate a CCSD(T)-quality $E_\textrm{int}$ with the SKZCAM protocol. The inputs can be either submitted directly on a computing cluster or managed through QuAcc. These terms are all combined to reach the final $H_\textrm{ads}$ estimate.}
\end{figure}

In the left panel of Fig.~\ref{fig:quacc_procedure}, we highlight the processes executed within QuAcc to calculate the DFT ensemble contributions to $H_\textrm{ads}$.
Starting from a unit cell, the \texttt{bulk\_to\_slab\_flow} will generate the necessary surface termination and relax the surface to its equilibrium geometry.
Subsequently, the adsorbate is added onto the surface within the \texttt{slab\_to\_ads\_flow} to generate the relaxed adsorbate--surface complex.
Currently the flow accepts placing the adsorbates on the ``ontop'', ``bridge'', ``hollow'' and ``subsurface'' sites if the position of the molecule is known or expected to follow chemical intuition.
We also provide a short function to perform random structure search to produce starting geometries that can be relaxed to identify unbiased low-energy adsorbate--surface adsorption configurations.

From the candidate adsorbate--surface structure(s), the QuAcc workflow will then submit a series of \texttt{static\_job} calculations which aims to calculate $E_\textrm{ads}$, $E_\textrm{int}$ and $E_\textrm{rlx}$ across the ensemble of DFAs.
In particular, it will calculate both $E_\textrm{ads}^\textrm{true}$ and $E_\textrm{ads}^\textrm{approx}$ (see Section~\ref{sec:dft_geom_error}) to estimate the error with using a DFT geometry for $E_\textrm{ads}$.
Subsequently the vibrational frequencies of the molecule (on the surface and in the gas-phase) are calculated in order to determine $E_\textrm{ZPV}$ and $E_\textrm{T}$ across the DFT ensemble.

\subsection{Automated SKZCAM protocol}

One of the key developments within the present work is the automatisation of the SKZCAM protocol, which significantly lowers its cost and requires minimal user intervention to operate.
From an adsorbate-surface structure, either generated by the one of the DFAs in the DFT ensemble or taken from the literature, it will generate all of the inputs necessary to come to a final estimate of the $E_\textrm{int}$, with the choice of several (ONIOM) embedding layers within an intuitive interface.

We have developed a set of Python modules which runs through the key steps of the SKZCAM protocol.
Firstly, it interfaces with the py-ChemShell~\cite{luMultiscaleQMMM2023} program to generate the necessary point charge environment using the \texttt{chemshell\_run\_job}.
The generated output (a .\texttt{pun} file) is then read and used to generate the set of embedded clusters with \texttt{skzcam\_calculate\_job}. 
This will generate the approriate set of inputs (for the different levels of theory, basis set sizes and frozen core treatments, described in Sec.~\ref{sec:skzcam_si_sec}) that define all the individual calculations that need to be performed.
At present, these inputs can be generated for either ORCA~\cite{neeseORCAQuantumChemistry2020} or MRCC~\cite{kallayMRCCProgramSystem2020}.
The inputs can be copied to the computing cluster of choice for the calculations to be performed.
Alternatively, one can make use of the QuAcc computational workflow library to directly manage and submit the jobs.
Once the MP2 and CCSD(T) calculations are complete, we provide analysis scripts to calculate the MP2 bulk limit $E_\textrm{int}^\textrm{bulk MP2}$, $\Delta$CC contribution and further contributions for basis set $\Delta_\textrm{basis}$ and core contributions $\Delta_\textrm{core}$.


\clearpage

\section{\label{sec:exp_redhead_analysis}Analysing experimental estimates and techniques}

Experimental techniques can obtain the adsorption enthalpy $H_\textrm{ads}$ through methods such as single crystal adsorption calorimetry (SCAC), equilibrium adsorption isotherms (EAI) and temperature programmed desorption  (TPD) experiments~\cite{campbellEnthalpiesEntropiesAdsorption2013,campbellEnergiesAdsorbedCatalytic2019a}.
SCAC are considered to provide the most reliable measurements of $H_\textrm{ads}$ while EAI provides a straightforward means for measuring this quantity for reversible adsorption-desorption experiments using the Clausius-Clapeyron expression.
Despite requiring more analysis than the other two methods, TPD experiments are by far the most common technique for measuring $H_\textrm{ads}$ for metal-oxide surfaces~\cite{campbellEnthalpiesEntropiesAdsorption2013} due to the simplicity and ready availability of TPD equipment.

\begin{figure}[h]
    \includegraphics[width=0.5\textwidth]{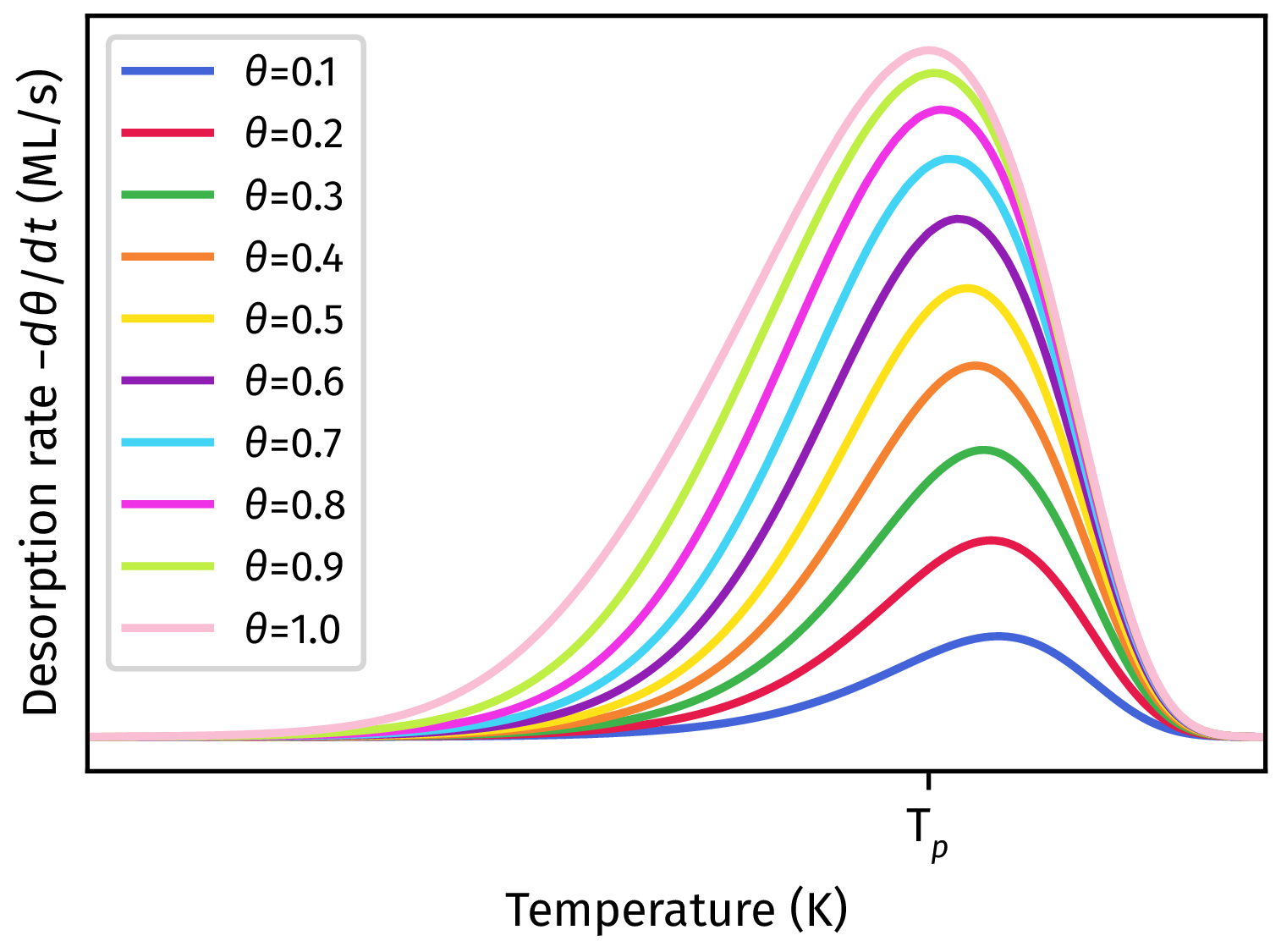}
    \caption{\label{fig:tpd_schematic} A schematic of a typical temperature programmed desorption spectra as a function of concentration $\theta$. Created with tools provided within Ref.~\citenum{schmidAnalysisTemperatureProgrammedDesorption2023a}.}
\end{figure}

In TPD experiments, the surface with preadsorbed molecules is heated at a constant heating rate.
The rate of appearance of the gas desorbing from the surface is then monitored using a mass spectrometer.
The desorption rate is measured as a function of the temperature to create plots such as shown in Fig.~\ref{fig:tpd_schematic}.
The central equation for first order desorption TPD spectra is an Arrhenius type relationship which relates the desorption rate $-\frac{\dd{\theta}}{\dd{t}}$ to the activation energy $E_\textrm{d}$ for desorption:
\begin{equation}
    -\frac{\dd{\theta}}{\dd{t}} = \nu \theta \exp(-\frac{E_\textrm{d}}{RT_p}),
\end{equation}
where $\nu$ is the pre-exponential factor, $\theta$ is the concentration and $T_p$ is the temperature where the desorption rate is a maximum.
One can then obtain $E_\textrm{d}$ by inverting this first-order Wigner-Polanyi equation:
\begin{equation}
    E_\textrm{d}(\theta) = - RT_p \ln(-\frac{\dd{\theta}}{\dd{t}}/(\nu \theta)).
\end{equation}
From TPD curves plotted at various surface concentrations, the value of $E_\textrm{d}$ and $\nu$ can also be obtained by finding the best fit to simulated TPDs.
The errors on $\log(\nu)$ with this approach can be of the order of ${\pm}2$~\cite{dohnalekPhysisorptionCOMgO2001}.

The simplest method to determine $E_\textrm{d}$ is through the Redhead equation~\cite{redheadThermalDesorptionGases1962}, which relates the heating rate $\beta$ and $T_p$ to $E_\textrm{d}$ by the following equation:
\begin{equation}
    \frac{E_\textrm{d}}{RT_p^2} = \left ( \frac{\nu}{\beta} \right )  \exp(-\frac{E_\textrm{d}}{RT_p}),
\end{equation}
Here, $T_p$ and $\beta$ are known but $\nu$ is normally estimated to be $10^{13}$.
Rearranging the equation and taking an empirical linear relationship between $E_\textrm{d}$ and $T_p$ gives the following relationship:
\begin{equation}
    E_\textrm{d} = RT_p  \ln(\frac{\nu \textrm{$T_p$}}{\beta} - 3.64  ).
\end{equation}
Thus, a ten order of magnitude change in $\nu$ is expected to change $E_\textrm{d}$ by ${\sim}\ln(10)RT_p = 2.3RT_p$.

There can be three sources of discrepancy in reaching an experimental measurement of $H_\text{ads}$.
The major source of error arises in the prescribed value of $\nu$ in the determination of $E_\textrm{d}$ from the TPD spectra.
This is not known but it can span between $\log(\nu)=12$ and $\log(\nu)=19$~\cite{campbellEntropiesAdsorbedMolecules2012} and most experiments will simply set $\log(\nu)=13$, which can introduce large errors.
Recently, Sellers and Campbell~\cite{campbellEntropiesAdsorbedMolecules2012} have demonstrated a relationship between the entropy of the gas when adsorbed on a surface and in the gas phase, which has allowed for predictions of $\log(\nu)$ to a $2\sigma$ standard deviation of ${\sim}1.72$. 
The majority of experiments are taken from Campbell and Sellers~\cite{campbellEnthalpiesEntropiesAdsorption2013}, with $H_\textrm{ads}$ using the predicted values of $\nu$.
We perform re-analysis of $E_\textrm{d}$ with the predicted $\nu$ for systems not included within their work, indicated by systems where $\log(\nu)=13.0 \to ...$ in Table~\ref{tab:expt_hads}.
Second, there is also an error in $E_\textrm{d}$ as it is not directly equal to $H_\textrm{ads}$.
$E_\textrm{d}$ is an activation energy, but its relationship to $H_\textrm{ads}$ has differed between different studies.
For example, the collection of experimental estimates~\cite{tosoniAccurateQuantumChemical2010,boeseAccurateAdsorptionEnergies2013a,sauerInitioCalculationsMolecule2019b,alessioChemicallyAccurateAdsorption2019a} re-analysed by Sauer and co-workers has used the relationship:
\begin{equation}
    H_\textrm{ads} = -E_\textrm{d} + RT_p,
\end{equation}
to compare to experiments, following on the approximate relation between the enthalpy and activation energy of reaction barriers~\cite{espensonChemicalKineticsReaction1995}.
On the other hand, the collection of work by Campbell~\cite{campbellEnthalpiesEntropiesAdsorption2013,wellendorffBenchmarkDatabaseAdsorption2015} and co-workers as well as others~\cite{rangarajanComparativeAnalysisDifferent2023a} have used the relationship:
\begin{equation}
    H_\textrm{ads} = -E_\textrm{d} - \frac{1}{2}RT_p,
\end{equation}
which was derived from relating the isosteric heat of adsorption (the negative of $H_\textrm{ads}$) from SCAC experiments to $E_\textrm{d}$ in Ref.~\citenum{brownFemtomoleAdsorptionCalorimetry1998}.
In this work we opt to set $H_\textrm{ads}$ to ${-}E_\textrm{d}$, and add an $RT_p$ error contribution to account for the two potential directions which $H_\textrm{ads}$ can point to with respect to $E_\textrm{d}$.
The final source of error arises in differences in the concentration between simulation and experiment.
For most work, for example from the re-analysis by Campbell and Sellers, a concentration dependence of $E_\textrm{d}$ is provided and we opt to take the smallest concentration for the adsorption of monomers, while we take the highest concentration for the monolayers.
For the clusters, we aim to take the average of the low and high concentration.
We add the error bars as half the difference between the low and high concentration for the clusters.
At the end, the final estimate of $H_\textrm{ads}$ will have its error be the root squared sum of these three sources of errors.

\begin{turnpage}
\begin{table}

\caption{\label{tab:expt_hads}Experimental adsorption enthalpies (in meV) for the systems studied within this work.}
\begin{adjustbox}{max width=1.4\textwidth}
\begin{tabular}{lrrrrrp{12cm}}
\toprule
Surface & Adsorbate & Temperature & $\log(\nu)$ & $H_\textrm{ads}$ (meV) & Error & Details \\ 
\midrule
\ce{MgO}(001) & \ce{CH4} & 47 & 13.1 $\pm$ 2.0 & -115 & 19 & Dilute limit $E_\textrm{d}$ estimate by Tait \etal{}~\cite{taitNalkanesMgO100Coveragedependent2005,taitNalkanesMgO1002005b}. \\
\ce{MgO}(001) & \ce{C2H6} & 75 & 14.9 $\pm$ 2.0 & -221 & 30 & Dilute limit $E_\textrm{d}$ estimate by Tait \etal{}~\cite{taitNalkanesMgO100Coveragedependent2005,taitNalkanesMgO1002005b}. \\
\ce{MgO}(001) & \ce{CO} & 61 & 13.8 $\pm$ 1.6 & -176 & 21 & Average of the $H_\textrm{ads}$ re-analysis by C\&S for Refs.~\citenum{dohnalekPhysisorptionCOMgO2001} and~\citenum{wichtendahlThermodesorptionCONO1999} at low coverage, with 0.5$R$T removed. \\
\ce{MgO}(001) & \ce{N2O} & 77 & 13.0 $\to$ 14.0 $\pm$ 2.0 & -239 & 31 & $E_\textrm{d}$ measured by Lian \etal{}~\cite{lianN2OAdsorptionSurface2010b} with subsequent conversion to $H_\textrm{ads}$. \\
\ce{MgO}(001) & \ce{C6H6} & 161.5 & 15.1 $\pm$ 1.6 & -481 & 72 & Average taken of $H_\textrm{ads}$ re-analyzed by C\&S between low and high coverage in Ref.~\citenum{streetAdsorptionElectronicStates1996a}, with 0.5$R$T removed. \\
\ce{MgO}(001) & \ce{H2O} & 203 & - & -520 & 121 & $H_\textrm{ads}$ from Ferry \etal{}~\cite{alessioChemicallyAccurateAdsorption2019a,ferryPropertiesTwodimensionalWater1997a,ferryWaterMonolayersMgO1001998} estimated by subtracting lateral molecule-molecule interactions (-35.1 $\pm$ 9.6 kJ/mol) from the $H_\textrm{ads}$ of \ce{H2O} monolayer on MgO, both obtained from LEED adsorption isotherms. \\
\ce{MgO}(001) & \ce{NH3} & 160 & 13.0 $\to$ 14.0 $\pm$ 2.0 & -613 & 65 & $E_\textrm{d}$ measurement from Arthur \etal{}~\cite{arthurAdsorptionDesorptionSurface1991a} with subsequent conversion to $H_\textrm{ads}$. \\
\ce{MgO}(001) & Physisorbed \ce{CO2} & 120 & 13.0 $\to$ 14.0 $\pm$ 2.0 & -431 & 49 & $E_\textrm{d}$ measurement from Meixner \etal{}~\cite{meixnerKineticsDesorptionAdsorption1992b} with subsequent conversion to $H_\textrm{ads}$. \\
\ce{MgO}(001) & Chemisorbed \ce{CO2} & 230 & 13.0 $\to$ 14.0 $\pm$ 2.0 & -664 & 125 & $E_\textrm{d}$ measurement from Chakradhar and Burghaus~\cite{chakradharCarbonDioxideAdsorption2013a} as the average of the $\alpha$ and $\beta$ peaks with subsequent conversion to $H_\textrm{ads}$. \\
\ce{MgO}(001) & Monolayer \ce{CH4} & 47 & 13.1 $\pm$ 2.0 & -131 & 19 & Monolayer $E_\textrm{d}$ estimate by Tait \etal{}~\cite{taitNalkanesMgO100Coveragedependent2005,taitNalkanesMgO1002005b}. \\
\ce{MgO}(001) & Monolayer \ce{C2H6} & 75 & 14.9 $\pm$ 2.0 & -236 & 30 & Monolayer $E_\textrm{d}$ estimate by Tait \etal{}~\cite{taitNalkanesMgO100Coveragedependent2005,taitNalkanesMgO1002005b}. \\
\ce{MgO}(001) & Cluster \ce{NO} & 79.5 & 14.0 $\pm$ 1.6 & -232 & 31 & Average taken of $H_\textrm{ads}$ re-analyzed by C\&S between low and high coverage in Ref.~\citenum{wichtendahlThermodesorptionCONO1999}, with 0.5$R$T removed. \\
\ce{MgO}(001) & Cluster \ce{H2O} & 235 & 14.5 $\pm$ 1.6 & -694 & 83 & $H_\textrm{ads}$ re-analysis by C\&S for Ref.~\citenum{stirnimanAdsorptionDesorptionWater1996}, with 0.5$R$T removed. \\
\ce{MgO}(001) & Cluster \ce{CH3OH} & 285.5 & 15.3 $\pm$ 1.6 & -890 & 106 & $H_\textrm{ads}$ re-analysis by C\&S for Ref.~\citenum{gunsterInteractionMethanolWater1999}, with 0.5$R$T removed. \\
\ce{TiO2} rutile(110) & \ce{CH4} & 85 & 14.9 $\pm$ 2.0 & -249 & 34 & Low-coverage estimate of $E_\textrm{d}$ by Chen \etal{}~\cite{chenAdsorptionSmallHydrocarbons2016a} with subsequent conversion to $H_\textrm{ads}$. \\
\ce{TiO2} rutile(110) & \ce{CO2} & 177 & 13.6 $\pm$ 2.0 & -493 & 62 & $H_\textrm{ads}$ re-analysis by C\&S for Ref.~\citenum{thompsonCO2ProbeMonitoring2003}, with 0.5$R$T removed. \\
\ce{TiO2} rutile(110) & \ce{H2O} & 303 & 14.7 $\pm$ 1.6 & -917 & 111 & Average taken of $H_\textrm{ads}$ re-analysis by C\&S for Refs.~\citenum{hugenschmidtInteractionH2OTiO21101994} and~\citenum{dohnalekPhysisorptionN2O22006} in low coverage limit \\
\ce{TiO2} rutile(110) & \ce{CH3OH} & 370 & 15.5 $\pm$ 1.6 & -1197 & 130 & $H_\textrm{ads}$ re-analysis by C\&S for Ref.~\citenum{liDeterminationAbsoluteCoverages2011}, with 0.5$R$T removed. \\
\ce{TiO2} anatase(101) & \ce{H2O} & 257 & 14.6 $\pm$ 1.6 & -786 & 90 & $H_\textrm{ads}$ re-analysis by C\&S for Ref.~\citenum{hermanExperimentalInvestigationInteraction2003}, with 0.5$R$T removed. \\
\ce{TiO2} anatase(101) & \ce{NH3} & 410 & 13.3 $\pm$ 2.0 & -1180 & 182 & $E_\textrm{d}$ estimate taken from Koust \etal{}~\cite{koustNH3AdsorptionAnataseTiO21012018} at the lowest studied coverage with subsequent conversion to $H_\textrm{ads}$, with 0.5$R$T removed. \\
\bottomrule
\end{tabular}
\end{adjustbox}

\end{table}
\end{turnpage}

\clearpage
\section{\label{sec:autoskzcam_exp_compare}Comparison of $H_\textrm{ads}$ between autoSKZCAM and experiments}

With the autoSKZCAM approach, we have computed a final $H_\textrm{ads}$ that aims to reach a converged CCSD(T)-level of accuracy.
The error bars have been designed to capture the major sources of error to at least a 95\% confidence interval: (1) finite size errors for calculating $E_\textrm{int}$ with the SKZCAM protocol, (2) errors for using a DFT geometry and DFT $E_\textrm{rlx}$ and (3) errors from using DFT to calculate zero-point vibrational and temperature contributions.
Similarly, we have also analysed experimental TPD experiments to obtain $H_\textrm{ads}$ estimates with reliable error bars that account for: (1) errors in the $\nu$ estimate, (2) errors between the measured Arrhenius activation energy and $H_\textrm{ads}$ and (3) errors arising from concentration dependence.

\begin{figure}[p]
    \centering
    \includegraphics[width=\textwidth]{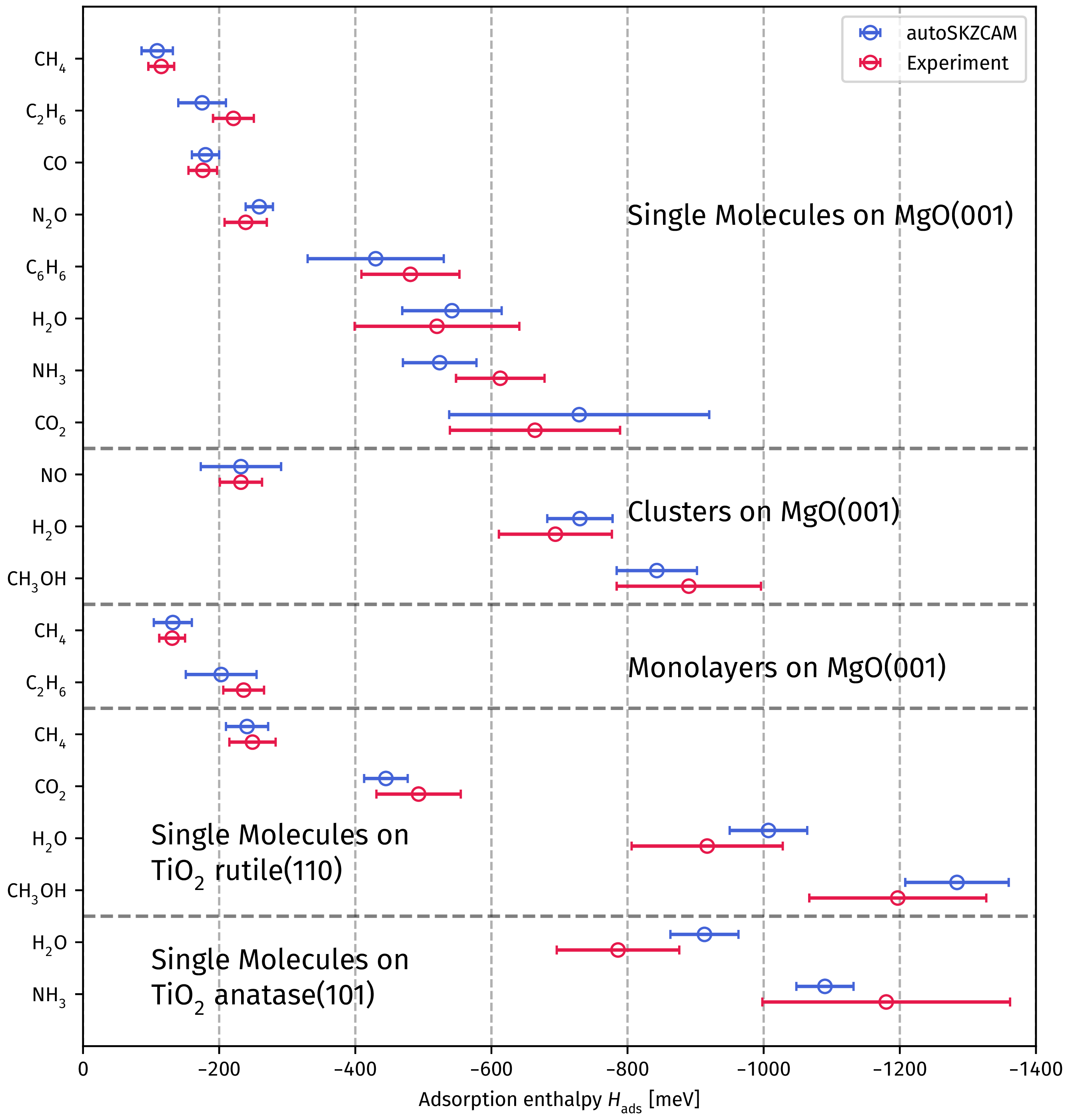}
    \caption{\label{fig:hads_all}A comparison of adsorption enthalpies computed with the autoSKZCAM framework against high-quality temperature programmed desorption experiments for a set of 19 adsorbate--surface combinations. These include single molecules adsorbed on the MgO(001) surface, monolayers adsorbed on MgO(001), single molecules adsorbed on \ce{TiO2} rutile(110) and anatase(101) as well as clusters adsorbed on MgO(001). We discuss how we calculate the error bars (corresponding to 95\% confidence intervals or more) on $H_\textrm{ads}$ for the simulations and experiments (including references to the experimental data) in Secs.~\ref{sec:final_hads} and~\ref{sec:exp_redhead_analysis}, respectively.}
    \label{fig:landscape}
\end{figure}

We compare the autoSKZCAM predicted $H_\textrm{ads}$ with experiment in Table~\ref{tab:hads_comparison} (and in Fig.~\ref{fig:hads_all}).
In particular, we calculate $\Delta_\textrm{min}$, the smallest (absolute) deviation between autoSKZCAM and experiment within the limits of their error bars.
If $\Delta_\textrm{min}=0$, we expect the two estimates to be indistinguishable and to agree within their respective error bars.
For all of the systems, we find the autoSKZCAM protocol to be indistinguishable from experiments.

In general, we find that the error bars from the autoSKZCAM framework are lower than experiment.
For both the autoSKZCAM framework and experiment, the error bars increase when $H_\textrm{ads}$ increases.
For experiments, this arises because the errors are typically a function of $RT_p$ and stronger binding correlates with a higher $T_p$.
On the other hand, for the autoSKZCAM protocol, the errors rest mostly on $\epsilon_\textrm{geom}$ - the error arising from using a DFT geometry.
A stronger binding typically means that the molecule has a stronger effect on the surface (electronic structure), hence leading to larger changes (and in turn $E_\textrm{rlx}$) to the geometry of the surface.

For several of the systems: \ce{CO2}, \ce{N2O}, \ce{NO}, \ce{H2O}, \ce{CH3OH} on MgO(001) and \ce{CO2} on \ce{TiO2} rutile(110), we have studied multiple geometries using the autoSKZCAM framework in Section~\ref{sec:system_discuss} and show in Table~\ref{tab:hads_comparison} the structure with the most negative $H_\textrm{ads}$ value - the stable geometry we expect will be observed within experiments.
Similarly, some systems have had several experimental $H_\textrm{ads}$ estimates.
For most of these systems, the experimental $H_\textrm{ads}$ from different studies agree with each other, barring \ce{CO2} on MgO(001), where two experiments have given $H_\textrm{ads}$ that differ by more than $300\,$meV with one~\cite{meixnerKineticsDesorptionAdsorption1992b} suggesting a physisorbed structure while the other~\cite{chakradharCarbonDioxideAdsorption2013a} predicting a chemisorbed structure.
Our autoSKZCAM estimates have helped to shed further light on the discrepancies between these two experiments and provide evidence for the accuracy of one over the other, as discussed in Section~\ref{sec:co2_configurations}.

\begin{table}
\caption{\label{tab:hads_comparison}Comparison of the experimental and autoSKZCAM $H_\text{ads}$ values (in meV) for the systems studied in this work. The $\Delta_\text{min}$ column shows the minimum difference between the experimental $H_\text{ads}$ value and the autoSKZCAM $H_\text{ads}$ value accounting for their error bars.}
\begin{adjustbox}{center,max width=0.9\textwidth}
\begin{tabular}{lrrrrrr}
\toprule
System & Expt. $H_\text{ads}$ & Expt. $\epsilon$ & autoSKZCAM $H_\text{ads}$ & autoSKZCAM $\epsilon$ & $\Delta_\text{min}$ \\ 
\midrule
CH$_4$ on MgO(001) & -115 & 19 & -109 & 23 & 0 \\
C$_2$H$_6$ on MgO(001) & -221 & 30 & -175 & 35 & 0 \\
CO on MgO(001) & -176 & 21 & -180 & 20 & 0 \\
Parallel N$_2$O on MgO(001) & -239 & 31 & -259 & 20 & 0 \\
C$_6$H$_6$ on MgO(001) & -481 & 72 & -430 & 100 & 0 \\
Monomer H$_2$O on MgO(001) & -520 & 121 & -542 & 73 & 0 \\
NH$_3$ on MgO(001) & -613 & 65 & -524 & 54 & 0 \\
Chemisorbed CO$_2$ on MgO(001) & -664 & 125 & -729 & 191 & 0 \\
Monolayer CH$_4$ on MgO(001) & -131 & 19 & -132 & 28 & 0 \\
Monolayer C$_2$H$_6$ on MgO(001) & -236 & 30 & -203 & 52 & 0 \\
Dimer NO on MgO(001) & -232 & 31 & -232 & 59 & 0 \\
Dissociated Tetramer H$_2$O on MgO(001) & -694 & 83 & -730 & 48 & 0 \\
Dissociated Tetramer CH$_3$OH on MgO(001) & -890 & 106 & -843 & 59 & 0 \\
CH$_4$ on TiO$_2$ rutile(110) & -249 & 34 & -241 & 31 & 0 \\
Tilted CO$_2$ on TiO$_2$ rutile(110) & -493 & 62 & -445 & 32 & 0 \\
H$_2$O on TiO$_2$ rutile(110) & -917 & 111 & -1007 & 57 & 0 \\
CH$_3$OH on TiO$_2$ rutile(110) & -1197 & 130 & -1284 & 76 & 0 \\
H$_2$O on TiO$_2$ anatase(101) & -786 & 90 & -913 & 50 & 0 \\
NH$_3$ on TiO$_2$ anatase(101) & -1180 & 182 & -1090 & 42 & 0 \\
\bottomrule
\end{tabular}
\end{adjustbox}
\end{table}

We show in Table~\ref{tab:hads_nu13_comparison} the importance in using the correct analysis of experimental results, where the RMSD is increased from $58\,$meV to $102\,$meV when going from the system-specific $\nu$ approach suggested by Campbell and Sellers~\cite{campbellEnthalpiesEntropiesAdsorption2013} to the standard approach of using $\nu=10^{13}$.

\begin{table}
\caption{\label{tab:hads_nu13_comparison}Comparison of the experimental $H_\text{ads}$ values in meV with the autoSKZCAM $H_\text{ads}$ values for the systems studied in this work. The $H_\text{ads}$ values are compared with the re-analysed values in Table~\ref{tab:expt_hads} and the values obtained by using the conventional $\log(\nu)=13$ in temperature programmed desorption (TPD) experiments. The root mean squared deviation (RMSD) is also calculated against the autoSKZCAM $H_\text{ads}$ values.}
\begin{adjustbox}{center,max width=1\textwidth}
\begin{tabular}{lrrrr}
\toprule
System & autoSKZCAM & Experiment (Table S1) & Experiment ($\nu{=}10^{13}$) \\ 
\midrule
CH$_4$ on MgO(001) & -109 $\pm$ 23 & -115 & -114 \\
C$_2$H$_6$ on MgO(001) & -175 $\pm$ 35 & -221 & -192 \\
CO on MgO(001) & -180 $\pm$ 20 & -176 & -166 \\
Parallel N$_2$O on MgO(001) & -259 $\pm$ 20 & -239 & -223 \\
C$_6$H$_6$ on MgO(001) & -430 $\pm$ 100 & -481 & -413 \\
Monomer H$_2$O on MgO(001) & -542 $\pm$ 73 & -520 & -520 \\
NH$_3$ on MgO(001) & -524 $\pm$ 54 & -613 & -581 \\
Chemisorbed CO$_2$ on MgO(001) & -729 $\pm$ 191 & -664 & -618 \\
Monolayer CH$_4$ on MgO(001) & -132 $\pm$ 28 & -131 & -130 \\
Monolayer C$_2$H$_6$ on MgO(001) & -203 $\pm$ 52 & -236 & -207 \\
Dimer NO on MgO(001) & -232 $\pm$ 59 & -232 & -216 \\
Dissociated Tetramer H$_2$O on MgO(001) & -730 $\pm$ 48 & -694 & -624 \\
Dissociated Tetramer CH$_3$OH on MgO(001) & -843 $\pm$ 59 & -890 & -759 \\
CH$_4$ on TiO$_2$ rutile(110) & -241 $\pm$ 31 & -249 & -216 \\
Tilted CO$_2$ on TiO$_2$ rutile(110) & -445 $\pm$ 32 & -493 & -471 \\
H$_2$O on TiO$_2$ rutile(110) & -1007 $\pm$ 57 & -917 & -814 \\
CH$_3$OH on TiO$_2$ rutile(110) & -1284 $\pm$ 76 & -1197 & -1013 \\
H$_2$O on TiO$_2$ anatase(101) & -913 $\pm$ 50 & -786 & -704 \\
NH$_3$ on TiO$_2$ anatase(101) & -1090 $\pm$ 42 & -1180 & -1155 \\
RMSD &  & 58 & 102 \\
\bottomrule
\end{tabular}
\end{adjustbox}
\end{table}

\clearpage

\section{\label{sec:comp_literature}Previous computational literature}

Many of the adsorbate--surface systems within this work have been studied before by both DFT and cWFT.
In particular, owing to its affordable nature and popularity, DFT has been widely used, providing several predictions for each system and we have collated some of this previous literature in Table~\ref{tab:comp_lit}.
Much of the early work in the late 1990's and early 2000's have utilised embedded cluster calculations (often with hybrid functionals such as B3LYP), with the current DFT workhorse being planewave periodic codes.
For each system, we observe large deviations across the DFT literature, varying by over $300\,$meV for most systems.
These variations mostly arise from differences in the exchange-correlation functional but even for the same functional, predictions can significantly vary due to differences in the basis set and pseudopotential treatment, which are discussed within each cited literature.

Importantly, for some of these systems, even the predicted adsorption configuration of the adsorbate is not known and different studies may suggest different geometries.
For example, whether \ce{CO2} adopts a chemisorbed (bent) or physisorbed (linear) geometry on the MgO(001) has been under debate, with studies which point towards either.
Similarly, a myriad of geometries have been predicted for NO on MgO(001) by different studies, many of which cite agreement to experiment.
With DFT, this agreement can be fortuitous many times due to errors in the surface model, DFA, electronic structures or neglect of thermal contributions.
These discrepancies have highlighted the need for accurate approaches such as the autoSKZCAM framework that can predict the correct geometries (with $H_\textrm{ads}$ that match experiments) that get the right answers for the right reasons; we highlight its success in predicting the ground-state configuration for several systems in Section~\ref{sec:system_discuss}.

While less common, calculations with cWFT have also been applied before to a selection of the studied systems and we highlight these previous work (and the method used) in Table~\ref{tab:comp_lit}.
As we showed in Fig.~2 of the main text, the deviations between methods from cWFT can be significant (to as large as $500\,$meV for some systems).
These deviations can arise from errors in the method used and also from electronic structure parameters; methods from cWFT are severely affected by basis set and surface model, as has been discussed extensively for CO on MgO(001) in Ref.~\citenum{shiManyBodyMethodsSurface2023a}.
Overall, these deviations highlight the need for going to a high level of theory, namely CCSD(T) and accurate surface models --- the targets of the autoSKZCAM framework.

\begin{table}

\caption{\label{tab:comp_lit}Compilation of $E_\text{ads}$ values (in meV) of previous density functional theory (DFT) and correlated wave-function theory (cWFT) literature for the systems studied within this work.}
\begin{adjustbox}{center,max width=0.85\textwidth}
\begin{tabular}{lp{7cm}p{7cm}}
\toprule
System & DFT & WFT \\ 
\midrule
CH$_4$ on MgO(001) & -50~\cite{trevethanBuildingBlocksMolecular2007}, 2~\cite{todnemMolecularAdsorptionMethane1999a}, 17~\cite{ferrariFTIRSpectroscopicDensity1998}, -300~\cite{manaeActivationCO2CH42022}, -120~\cite{mazheikaNiSubstitutionalDefects2016a}, -152~\cite{picciniEffectAnharmonicityAdsorption2014} & -134 [MP2+$\Delta$CC]~\cite{boeseAccurateAdsorptionEnergies2016} \\
C$_2$H$_6$ on MgO(001) & -127~\cite{boeseAccurateAdsorptionEnergies2016}, -154~\cite{boeseAccurateAdsorptionEnergies2016} & -196 [MP2+$\Delta$CC]~\cite{boeseAccurateAdsorptionEnergies2016} \\
CO on MgO(001) & -9 to -282~\cite{valeroGoodPerformanceM062008e}, -175 to -408~\cite{r.rehakIncludingDispersionDensity2020} & -207 [LNO-CCSD(T)]~\cite{yeAdsorptionVibrationalSpectroscopy2024a}, -199 [LNO-CCSD(T)]~\cite{shiManyBodyMethodsSurface2023a}, -230 [MP2+$\Delta$CC]~\cite{alessioChemicallyAccurateAdsorption2018}, 70 [CCSD(T)]~\cite{mazheikaNiSubstitutionalDefects2016a}, -398 [CCSD]~\cite{mitraPeriodicDensityMatrix2022a}, -72 [RPA@PBE]~\cite{bajdichSurfaceEnergeticsAlkalineearth2015b}, -310 [RPA@PBE+rSE]~\cite{bajdichSurfaceEnergeticsAlkalineearth2015b} \\
N$_2$O on MgO(001) & -258~\cite{huesgesDispersionCorrectedDFT2014}, $>$0~\cite{scagnelliCatalyticDissociationN2O2006}, -137~\cite{huesgesDispersionCorrectedDFT2014} &  \\
C$_6$H$_6$ on MgO(001) & -20~\cite{trevethanBuildingBlocksMolecular2007} &  \\
Monomer H$_2$O on MgO(001) & -342~\cite{giordanoPartialDissociationWater1998a}, -500~\cite{carrascoDynamicIonPairs2008}, -340~\cite{huTrendsWaterMonomer2011}, -422 to -667~\cite{kebedeComparingVanWaals2017} & -574 [$\Delta$CC]~\cite{alessioChemicallyAccurateAdsorption2018}, -480 [DMC]~\cite{karaltiAdsorptionWaterMolecule2012}, -608 [RPA@PBE+rSE]~\cite{bajdichSurfaceEnergeticsAlkalineearth2015b}, -492 [RPA@PBE]~\cite{bajdichSurfaceEnergeticsAlkalineearth2015b} \\
NH$_3$ on MgO(001) & -668~\cite{pughEnergeticsNH3Adsorption1994}, -867~\cite{nakajimaAmmoniaAdsorptionMgO1001996} & -608 [PCT]~\cite{alloucheVibrationalInfraredSpectrum1995} \\
CO$_2$ on MgO(001) & Chemisorbed: -1000~\cite{manaeActivationCO2CH42022}, 135~\cite{jensenCO2SorptionMgO2005b}, -680~\cite{downingReactivityCO2MgO2013}, -640~\cite{mazheikaNiSubstitutionalDefects2016a}, -380~\cite{baltrusaitisPeriodicDFTStudy2012a}, Physisorbed: -320~\cite{lvCO2AdsorptionKpromoted2024}, -90~\cite{cornuLewisAcidoBasicInteractions2012b}, -126~\cite{hammamiCO2Adsorption0012008a}, -120~\cite{manaeActivationCO2CH42022} & Chemisorbed: 870 [MP2]~\cite{pacchioniInitioClusterModel1994b}, -492 [RPA@PBE] [This work], -494 [RPA@PBE+rSE] [This work], Physisorbed: -340 [MP2]~\cite{pacchioniPhysisorbedChemisorbedCO21993a} \\
NO on MgO(001) & Hollow: -312~\cite{songRemarkablyStrongChemisorption2017a}, Bent-Mg: -520~\cite{yanagisawaThreeTypesAdsorptions1999b}, Upright-Mg: -260~\cite{rodriguezStudiesBehaviorMixedmetal2001d}, -297~\cite{anezNONO2Adsorption2017a}, Bent-O: -464~\cite{mileticFirstPrinciplesCharacterizationNOx2003a}, -312~\cite{schneiderDramaticCooperativeEffects2002a,schneiderQualitativeDifferencesAdsorption2004a}, Dimer: -30~\cite{luAdsorptionDecompositionNO1999b}, -40~\cite{divalentinNOMonomersMgO2002a}, Bent-Bridge: -150~\cite{limDensityFunctionalTheory2019} & Bent-Mg: 29 [RPA@PBE]~\cite{bajdichSurfaceEnergeticsAlkalineearth2015b}, -360 [RPA@PBE+rSE]~\cite{bajdichSurfaceEnergeticsAlkalineearth2015b} \\
Cluster CH$_3$OH on MgO(001) & -718~\cite{rodriguezAdsorptionMethanolMolecules2007b}, -466~\cite{gayDensityFunctionalStudy2005a}, -640~\cite{manTheoreticalAspectsMethyl2017}, -964~\cite{a.sainnaCombinedPeriodicDFT2021a}, -508~\cite{petitjeanQuantitativeInvestigationMgO2010a} & -598 [MP2]~\cite{brandaTheoreticalStudyCharge2002b} \\
Cluster H$_2$O on MgO(001) & -610~\cite{huProtonTransferAdsorbed2010a}, -592~\cite{alvimDensityFunctionalTheorySimulation2012}, -720~\cite{dingHydrationStructureFlat2021} &  \\
Monolayer CH$_4$ on MgO(001) & -18~\cite{drummondDensityFunctionalInvestigation2006a,stimacSimulatingCH4Physisorption2008}, -124 to -353~\cite{r.rehakIncludingDispersionDensity2020} & -79 [PCI-80]~\cite{todnemMolecularAdsorptionMethane1999a}, -114 [LMP2]~\cite{pisaniPeriodicLocalMP22008d}, -138 [MP2+$\Delta$CC]~\cite{tosoniAccurateQuantumChemical2010}, -145 [MP2+$\Delta$CC]~\cite{alessioChemicallyAccurateAdsorption2018}, -87 [RPA@PBE]~\cite{bajdichSurfaceEnergeticsAlkalineearth2015b}, -140 [RPA@PBE+rSE]~\cite{bajdichSurfaceEnergeticsAlkalineearth2015b} \\
Monolayer C$_2$H$_6$ on MgO(001) & -234 to -568~\cite{r.rehakIncludingDispersionDensity2020} & -242 [MP2+$\Delta$CC]~\cite{alessioChemicallyAccurateAdsorption2018} \\
CH$_4$ on TiO$_2$ rutile(110) & -48 to -412~\cite{kubasSurfaceAdsorptionEnergetics2016c}, -360~\cite{tillotsonAdsorptionOrganicMolecules2015} & -408 [CCSD(T)]~\cite{kubasSurfaceAdsorptionEnergetics2016c} \\
CO$_2$ on TiO$_2$ rutile(110) & -95 to -559~\cite{kubasSurfaceAdsorptionEnergetics2016c}, -640~\cite{kovacicElectronicPropertiesRutile2022} & -542 [CCSD(T)]~\cite{kubasSurfaceAdsorptionEnergetics2016c} \\
H$_2$O on TiO$_2$ rutile(110) & -442 to -950~\cite{kubasSurfaceAdsorptionEnergetics2016c}, -840~\cite{liuStructureDynamicsLiquid2010b}, -1638~\cite{banduraAdsorptionWaterTiO22004a}, -1032~\cite{sorescuCoadsorptionPropertiesCO22012} & -1492 [MP2]~\cite{stefanovichInitioStudyWater1999}, -984 [DLPNO-CCSD(T)]~\cite{kubasSurfaceAdsorptionEnergetics2016c}, -964 [LNO-CCSD(T)]~\cite{yeInitioSurfaceChemistry2024}, -1390 [CCSD]~\cite{schaferLocalEmbeddingCoupled2021b} \\
CH$_3$OH on TiO$_2$ rutile(110) & -468 to -1145~\cite{kubasSurfaceAdsorptionEnergetics2016c}, -1490~\cite{kieuTrendsAdsorptionEnergy2002}, -1234~\cite{batesAdsorptionDissociationROH1998}, -760~\cite{langFirstPrinciplesStudyMethanol2014} & -1106 [DLPNO-CCSD(T)]~\cite{kubasSurfaceAdsorptionEnergetics2016c} \\
H$_2$O on TiO$_2$ anatase(101) & -740~\cite{vittadiniStructureEnergeticsWater1998}, -650~\cite{millerEffectsWaterFormic2011}, -1088~\cite{onalAdsorptionWaterAmmonia2006}, -977~\cite{zhaoStructurePropertiesWater2012} & -1170 [DLPNO-CCSD(T)]~\cite{petersenWaterAdsorptionIdeal2020} \\
NH$_3$ on TiO$_2$ anatase(101) & -1200~\cite{koustNH3AdsorptionAnataseTiO21012018}, -1113~\cite{onalAdsorptionWaterAmmonia2006}, -1153~\cite{wanbayorAdsorptionCOH22010}, -1193~\cite{changAdsorptionConfigurationDissociative2009} &  \\
\bottomrule
\end{tabular}
\end{adjustbox}

\end{table}

\clearpage
\section{\label{sec:cost_benchmark}Benchmarking the cost of the autoSKZCAM framework}

We perform an analysis of the cost for computing the interaction energy $E_\textrm{int}$ from the autoSKZCAM framework against a GGA-based (PBE-D3) and hybrid-based (PBE0-TS/HI) DFA.
In both cases, we have attempted to make the calculations efficient to the best of our abilities.
For example, for periodic hybrid calculations, we make use of the Adaptively Compressed Exchange Operator~\cite{linAdaptivelyCompressedExchange2016} and use an initial wave-function coming from the GGA calculation to enable efficient self-consistent field (SCF) energy convergence to the cutoff of $10^{-5}\,$eV (looser than our standard settings in Sec.~\ref{sec:dft_details}).
We also use a smaller energy cutoff of $520\,$eV for both the GGA and hybrid DFT calculations.
The cost for the autoSKZCAM framework is the sum of all the individual contributions to $E_\textrm{int}$ outlined in Table~\ref{tab:skzcam_system_eint}, with the same corresponding computational details described in Sec.~\ref{sec:cwft_details}.
The efficiency of the autoSKZCAM framework calculations for \ce{CO2} on MgO(001) is further enhanced by performing both MP2 and CCSD(T) calculations using the aV$X$Z basis sets rather than awCV$X$Z and only treating the valence electrons in the correlation treatment.
A further $\Delta_\textrm{core}$ contribution is calculated by performing additional CBS(awCVTZ/awCVQZ) calculations for the first three clusters generated by the SKZCAM protocol, commensurate with the procedure used for H$_2$O on TiO$_2$ rutile(110).

The costs in CPU-hours (CPUh) are compared in Table~\ref{tab:computational_cost} for \ce{CO2} on MgO(001) and \ce{H2O} on \ce{TiO2} rutile(110).
Both periodic DFT and autoSKZCAM costs were assessed on 2.1 GHz, 18-core Intel Xeon E5-2695 (Broadwell) series processors on the Cirrus high-performance computing (HPC) cluster [\href{https://www.cirrus.ac.uk/}{https://www.cirrus.ac.uk/}], barring hybrid DFT calculations for H$_2$O on TiO$_2$ rutile(110), which were evaluated on improved hardware, involving 3.1 (turbo-boosted to 3.9) GHz Intel Xeon Platinum 8174 (Skylake) processors on the Vienna Scientific Cluster (VSC-4) [\href{https://vsc.ac.at//systems/vsc-4/}{https://vsc.ac.at//systems/vsc-4/}].
In addition, for the hybrid DFT calculations involving H$_2$O on TiO$_2$ rutile(110), we have provided costs to perform to perform $1 \times 1 \times 1 $ and $2 \times 2 \times 1$ $k$-point grids are provided, as the former can achieve significant speed-ups while trading some accuracy (${\sim}20\,$meV).

\begin{table}[h]
\caption{\label{tab:computational_cost}Computational cost in CPU hours for periodic DFT, both a GGA (PBE-D3) and hybrid (PBE0-TS/HI), compared to autoSKZCAM for the CO$_2$ on MgO(001) and H$_2$O on TiO$_2$ rutile(110) adsorbate--surface systems. Details of these calculations are described in the text.}
\begin{tabular}{lrrr}
\toprule
 & GGA & hybrid & autoSKZCAM \\ 
\midrule
CO$_2$ on MgO(001) & 200 & 1500 & 2900 \\
H$_2$O on TiO$_2$ rutile(110) & 300 & 700 to 10300 & 6700 \\
\bottomrule
\end{tabular}
\end{table}

We find that the cost with the autoSKZCAM framework for \ce{CO2} on MgO(001) is slightly higher (less than 2 times) than hybrid DFT, which is in turn an order of magnitude more expensive than the GGA calculation.
However, when moving towards the more complex \ce{H2O} on \ce{TiO2} rutile(110) -- involving heavier atoms and a larger number of atoms in the periodic model of its surface -- the cost of the autoSKZCAM framework becomes comparable to periodic hybrid DFT calculation (with a $2\times 2 \times 1$ $k$-point grid).
The GGA calculation remains an order of magnitude cheaper than the autoSKZCAM framework for this system.
It should be highlighted that the cost moving from \ce{CO2} on MgO(001) to \ce{H2O} on \ce{TiO2} rutile(110) does not change significantly (less than 3 times increase) for the autoSKZCAM framework despite its increased complexity.
This feature of the autoSKZCAM framework arises because the size of the clusters selected by the SKZCAM protocol (described in Sec.~\ref{sec:skzcam_protocol_details}) does not depend on the complexity of the surface, and should remain similar in size between different types of surfaces since it is generated based on radial cutoffs.

It is also useful to highlight the cost of the autoSKZCAM framework in relation to previous high-level calculations.
In particular, CO on MgO(001) has been the prototypical surface system for cWFT methods, as highlighted in Ref.~\citenum{shiManyBodyMethodsSurface2023a}.
We gather previous estimates in Table~\ref{tab:cwft_computational_cost}; these should be taken as rough estimates as they were all performed on different computing systems.
Out of all of these previous works, the autoSKZCAM is by far the cheapest, with a cost of ${\sim}600\,$CPUh.
This is improved over the previous SKZCAM protocol calculations due to the described improvements in Sec.~\ref{sec:improvements_skzcam}.
Ye and Berkelbach previously performed periodic LNO-CCSD(T) calculations for this system and arrived at a cost of ${\sim}18,000\,$CPUh.
Compared to this, periodic CCSD(T) (without the LNO approximation) was shown to take ${\sim}200,000\,$CPUh to perform in Ref.~\citenum{shiManyBodyMethodsSurface2023a}, with periodic DMC being even more costly at ${\sim}1,000,000\,$CPUh.
In the present study, the RPA calculations took ${\sim}4,000\,$CPUh, relatively comparable to hybrid DFT at ${\sim}1,000\,$CPUh.

\begin{table}[h]
\caption{\label{tab:cwft_computational_cost}Rough computational cost in CPU hours for several methods from correlated wave-function theory applied to CO on MgO(001). Details are provided in their respective references. As a guide, hybrid DFT costs ${\sim}1,000\,$CPUh~\cite{shiManyBodyMethodsSurface2023a} for this system.}
\begin{tabular}{@{}lr@{}}
\toprule
Method                           & Cost {[}CPUh{]} \\ \midrule
autoSKZCAM framework (This work) & $\sim$600       \\
RPA (This work)                              & $\sim$4000      \\
Periodic LNO-CCSD(T)~\cite{yeAdsorptionVibrationalSpectroscopy2024a}             & $\sim$18000     \\
Cluster CCSD(T)~\cite{shiManyBodyMethodsSurface2023a} [SKZCAM protocol]                  & $\sim$20000     \\
Periodic CCSD(T)~\cite{shiManyBodyMethodsSurface2023a}                & $\sim$200000    \\
Periodic DMC~\cite{shiManyBodyMethodsSurface2023a}                     & $\sim$1000000   \\ \bottomrule
\end{tabular}
\end{table}

\clearpage